\documentclass[fleqn,usenatbib,useAMS]{mnras}

\usepackage{graphicx}	
\usepackage{amsmath}	
\usepackage{amssymb}	
\usepackage{multicol}        
\usepackage{bm}		
\usepackage{pdflscape}	

\usepackage{longtable}

\newcommand{\kms}{\,km\,s$^{-1}$} 
\newcommand{\cms}{\,(cm\,s$^{-2}$)} 
\newcommand{\cd}{d\,$^{-1}$} 
\newcommand{\ms}{m\,s$^{-1}$} 
\newcommand{\Teff}{$T_{\rm eff}$}

\usepackage[dvipsnames,usenames]{color}

\usepackage[T1]{fontenc}
\usepackage{ae,aecompl}

\usepackage{newtxtext,newtxmath}

\title[TESS Cycle\,1 roAp stars]{TESS Cycle\,1 observations of roAp stars with 2-min cadence data}

\author[D.~L.~Holdsworth et al.]{D.~L.~Holdsworth,$^{1}$\thanks{e-mail:dlholdsworth@uclan.ac.uk}
M.~S.~Cunha,$^{2}$
D.~W.~Kurtz,$^{3,1}$
V.~Antoci,$^{4,5}$
D.~R.~Hey,$^{6,5}$
D.~M.~Bowman,$^{7}$\and
O.~Kobzar,$^{8}$
D.~L.~Buzasi,$^{9}$ 
O.~Kochukhov,$^{10}$
E.~Niemczura,$^{11}$
D.~Ozuyar,$^{12}$
F.~Shi,$^{13,14}$
R.~Szab\'o,$^{15,16,17}$\and
A.~Samadi-Ghadim,$^{18,19}$
Zs.~Bogn\'ar,$^{15,16,17}$
L.~Fox-Machado,$^{20}$
V.~Khalack,$^{8}$
M.~Lares-Martiz,$^{21}$\and
C.~C.~Lovekin,$^{22}$
P.~Miko\l{}ajczyk,$^{11}$
D.~Mkrtichian,$^{23}$
J.~Pascual-Granado,$^{21}$
E.~Paunzen,$^{24}$\and
T.~Richey-Yowell,$^{25}$
\'A.~S\'odor,$^{15,16}$
J.~Sikora,$^{26}$
T.~Z.~Yang,$^{27}$
E.~Brunsden,$^{28,29}$
A.~David-Uraz,$^{30,31}$\and
A.~Derekas,$^{32,33}$
A.~Garc\'{\i}a Hern\'andez,$^{34}$
J.~A.~Guzik,$^{35}$
N.~Hatamkhani,$^{36}$
R.~Handberg,$^{5}$
T.~S.~Lambert,$^{37}$\and
P.~Lampens,$^{38}$
S.~J.~Murphy,$^{6,5}$
R.~Monier,$^{39}$
K.~R.~Pollard,$^{29}$
P.~Quitral-Manosalva,$^{40}$\and
A.~Ram\'on-Ballesta,$^{21}$
B.~Smalley,$^{41}$
I.~Stateva$^{42}$
and R.~Vanderspek$^{43}$\\
Author affiliations shown in Appendix\,\ref{app:affiliations}.
}
\date{\today}

\pubyear{2021}

\begin{document}
\label{firstpage}
\pagerange{\pageref{firstpage}--\pageref{lastpage}}
\maketitle

\begin{abstract}
We present the results of a systematic search for new rapidly oscillating Ap (roAp) stars using the 2-min cadence data collected by the \textit{Transiting Exoplanet Survey Satellite} (TESS) during its Cycle\,1 observations. We identify 12 new roAp stars. Amongst these stars we discover the roAp star with the longest pulsation period, another with the shortest rotation period, and six with multiperiodic variability. In addition to these new roAp stars, we present an analysis of 44 known roAp stars observed by TESS during Cycle\,1, providing the first high-precision and homogeneous sample of a significant fraction of the known roAp stars. The TESS observations have shown that almost 60\,per\,cent (33) of our sample of stars are multiperiodic, providing excellent cases to test models of roAp pulsations, and from which the most rewarding asteroseismic results can be gleaned. We report four cases of the occurrence of rotationally split frequency multiplets that imply different mode geometries for the same degree modes in the same star. This provides a conundrum in applying the oblique pulsator model to the roAp stars. Finally, we report the discovery of non-linear mode interactions in $\alpha$\,Cir (TIC\,402546736, HD\,128898) around the harmonic of the principal mode -- this is only the second case of such a phenomenon.
 \end{abstract}




\section{Introduction}

The rapidly oscillating Ap (roAp) stars are a rare subset of the chemically peculiar, magnetic, Ap stars. They are found at the base of the classical instability strip where it intersects the main sequence, and range in age from the zero-age main sequence to beyond the terminal-age main sequence. Since their discovery \citep{1978IBVS.1436....1K,1982MNRAS.200..807K} there have been several attempts to enlarge, and study, the sample of roAp stars. These have included targeted and survey ground-based photometry \citep[e.g.,][]{1991MNRAS.250..666M,1993AJ....105.1903N,1994MNRAS.271..129M,2012A&A...542A..89P,2014MNRAS.439.2078H,2016A&A...590A.116J,2018RAA....18..135P}, time-resolved high resolution spectroscopic observations \citep[e.g.,][]{2008CoSka..38..317E,2008MNRAS.385.1402F,2013MNRAS.431.2808K} and most recently space photometry \citep[e.g.,][]{2009MNRAS.396.1189B,2011A&A...530A.135G,2011MNRAS.414.2550K,2011MNRAS.413.2651B,2016A&A...588A..54W,2019MNRAS.488...18H,2019MNRAS.487.2117B,2019MNRAS.487.3523C,10.3389/fspas.2021.626398}. To date, including the results presented here, there are 88 roAp stars known in the literature. 

The overarching class of Ap stars hosts strong, stable, global magnetic fields that can have field strengths of up to 34\,kG \citep{1960ApJ...132..521B,2017A&A...601A..14M}, with the magnetic axis misaligned with respect to the rotation axis. The magnetic field is thought to be the main factor in braking the rotation velocity of the star \citep{2000A&A...353..227S}, such that the Ap stars show rotation periods of between a few days and perhaps centuries \citep{1984A&A...141..328N,2019MNRAS.487.4695S,2020A&A...639A..31M}. The presence of the magnetic field also serves to suppress near-surface convection, allowing for the stratification of elements in the stellar atmosphere \citep{1981A&A...103..244M,2009A&A...495..937L,2010A&A...516A..53A} through the effects of radiative levitation and gravitational settling. In particular, the radiative levitation leads to surface inhomogeneities of elements such as Ce, Pr, Nd, Sm, Eu and Tb, which can be overabundant by over a million times the solar value \citep[e.g.,][]{2010A&A...509A..71L}.

These chemical spots, which are either bright or dark depending on the wavelength of observation, cause non-sinusoidal modulation of the light curve over the stellar rotation period. This allows for precise rotation periods to be measured, and is well represented by the oblique rotator model \citep{1950MNRAS.110..395S}. Such a modulation is also seen in the magnetic field strength \citep[e.g.,][]{2017A&A...601A..14M,2018MNRAS.477.3791H}, and spectral line strength. These rotationally modulated stars are commonly known as $\alpha^2$\,CVn stars \citep{2017ARep...61...80S}. \citet{2011IAUS..273..249K} concluded that the distribution of the chemical spots of the Ap stars is diverse, rarely showing a connection with the magnetic field structure.

The pulsations in the roAp stars are high overtone ($n\gtrsim15$), low degree ($\ell\lesssim3$) pressure (p) modes with frequencies in the range $0.7-3.6$\,mHz ($60-310$\,\cd; $P=4.7-23.6$\,min). The pulsation axis in these stars is misaligned to both the rotation and magnetic axes, leading to the development of the oblique pulsator model \citep{1982MNRAS.200..807K,1985PASJ...37..601S,2002A&A...391..235B,2011A&A...536A..73B}. Such a configuration means that the pulsation mode is viewed from varying aspects over the rotation cycle of the star, leading to an apparent amplitude modulation of the mode. In a Fourier spectrum of a light curve, a multiplet is seen with $2\ell+1$ components for a non-distorted mode, with the components split from the central mode frequency by exactly the stellar rotation frequency. 

The driving mechanism for the pulsations in roAp stars is still an area of active research. In many cases, it is currently thought that the pulsations are driven by the opacity ($\kappa$) mechanism in the hydrogen ionisation layers in regions of the star where the magnetic field suppresses convection \citep{2001MNRAS.323..362B}. However, this theory cannot reproduce observations of the highest frequency modes in some stars. In these cases \citet{2013MNRAS.436.1639C} proposed that a mechanism linked to turbulent pressure plays a role in the excitation of these highest frequency modes.

A further conundrum linked to the excitation mechanism of the pulsations in roAp stars comes with the inspection of the theoretical instability strip of these stars. \citet{2002MNRAS.333...47C}, based on the models of \citet{2001MNRAS.323..362B}, calculated the extent of the theoretical instability strip for roAp stars and found a much hotter blue edge than the observations mapped, and that some known roAp stars were cooler than the red edge. It is still unclear why the theoretical models do not match the observational evidence.

Of particular scientific interest are the roAp stars which show multiple pulsation modes as these stars provide the ability to constrain structure models of Ap stars. In the absence of a magnetic field, the pulsation modes of an roAp star would form a series of alternating degree modes that are equally spaced in frequency, as is expected in the asymptotic regime for high radial-order acoustic oscillations \citep{1979PASJ...31...87S,1980ApJS...43..469T}. The acoustic waves in the outer layers of the roAp stars, in the presence of the strong magnetic field, become magnetoacoustic in nature which consequently changes the mode frequencies \citep{2000MNRAS.319.1020C,2006MNRAS.365..153C,2004MNRAS.350..485S}.  In most cases, the frequency separation between consecutive modes is not significantly affected, still enabling the use of asteroseismic techniques to help constrain the stellar properties as would be done for non-magnetic stars \citep{2003MNRAS.343..831C,2021arXiv210408097D}. Nevertheless, at some particular frequencies the coupling between the magnetic field and pulsations is optimal and the pulsation spectrum can be significantly modified. It is the multiperiodic stars, which in some cases show significant shifts from the expected non-magnetic frequency pattern, that provide the most scientific return \citep[e.g.][]{2001MNRAS.325..373C,2008A&A...480..223G}.

The \textit{Transiting Exoplanet Survey Satellite} \citep[TESS;][]{2015JATIS...1a4003R} is surveying almost the entire sky, in 27-d long strips known as \textit{sectors}, in the search for transiting exoplanets. As such, it is collecting high precision photometric data on millions of stars at 2-min and 30-min cadences (with a 20-s cadence added in the extended mission and reducing the full frame image cadence to 10\,min). The Cycle\,1 observations, conducted in the first year of the primary mission, covered the southern ecliptic hemisphere, with Cycle\,2 observing the northern ecliptic hemisphere. Such an observational data set provides the opportunity to perform an homogeneous earch for high-frequency pulsations in a significant selection of stars.

This work utilises the 2-min cadence TESS Cycle\,1 observations (sectors 1 through 13) to search for high-frequency pulsations in main sequence stars hotter than $6000$\,K. We present the observational results here, and will, in a future work, use these results to perform an ensemble study of the properties of the roAp stars. The paper is laid out as follows: in Section\,\ref{sec:datasample} we introduce our data sample; in Section\,\ref{sec:spec} we provide details of spectroscopic follow up observations; in Section\,\ref{sec:results} we present the results of our search. In Section\,\ref{sec:cand} we present candidate roAp stars where we have insufficient data to confirm their true nature, and in Section\,\ref{sec:conc} we draw our conclusions.

\section{Data sample and search strategy}
\label{sec:datasample}

To construct our catalogue of target stars, we took the 2-min target list (constructed from proposals from the exoplanet and asteroseismic communities) for each sector of observations from TESS\footnote{https://tess.mit.edu/observations/target-lists/} which consisted of 20\,000 stars per sector. These lists were then crossmatched with version 8 of the TESS Input Catalog \citep[TIC;][]{2019AJ....158..138S} with a 3\,arcsec radius. This gave us access to the secondary information in the TIC to refine our targets. Using only the \Teff\ parameter, we selected stars with temperatures of 6000\,K or hotter. This temperature cut is to ensure we do not exclude around the red edge of the roAp star instability strip, and to account for any inaccuracies in the TIC temperature values. This amounted to about 7500 stars per sector, with 101\,847 light curves in total, corresponding to 50\,703 unique stars (with 15\,363 stars observed in more than one sector). The light curves for each star in all available sectors were downloaded from the Mikulski Archive for Space Telescopes (MAST) server. These data have been processed with the Science Processing Operations Center (SPOC) pipeline \citep{2016SPIE.9913E..3EJ}. In the following we used the Pre-search Data Conditioning Simple Aperture Photometry (PDC\_SAP) data unless otherwise stated.

Initially, three independent analysis techniques were implemented for the search of pulsational variability. For one team, each individual light curve was automatically prewhitened to a frequency of 0.23\,mHz (20\,\cd) to an amplitude limit of the approximate noise level between $2.3-3.3$\,mHz ($200-300$\,\cd) to remove instrumental artefacts and any low-frequency signal whose window function affects the noise at high frequency. If multiple sectors for a target were available, the prewhitened sectors were then combined. An amplitude spectrum of the light curve was calculated to the Nyquist frequency of 4.2\,mHz (360\,\cd), which also included a calculation of the false alarm probability (FAP). 

From these amplitude spectra, we selected stars which showed peaks with frequencies $>0.52$\,mHz ($>45$\,\cd) and a corresponding FAP of $<0.1$. The amplitude spectra of these stars were then plotted for visual inspection. This totalled 6713 stars, of which 2125 stars have multi-sector observations. 

A second team used a complementary method which calculated the skewness of the amplitude spectrum \citep{2019MNRAS.485.2380M} of the MAST PDC\_SAP data above 0.46\,mHz (40\,\cd). Where multiple sectors of data were available, they were combined to a single light curve. If the skewness was greater than 5, then all peaks in the amplitude spectrum with FAP values below 0.05 were extracted and the star flagged as variable for later human inspection. This method produced a total of 189 variable star candidates.

The third team used the TESS-AP automated analysis procedure created by \citet{2019MNRAS.490.2102K}. The light curves of stars with TIC effective temperatures in the range $7000-10\,000$\,K were analysed using the Period04 package \citep{2005CoAst.146...53L} in an automated way. A star was flagged as a candidate if a signal was found in the $1.16-4.17$\,mHz ($100-360$\,\cd) range with an amplitude 3 times that of the local noise. These candidates were then checked by eye before being included in a final list. The final count of variable star candidates was 78.

These three lists were then combined to produce a master list where each star and light curve was subjected to human inspection. An initial pass was made to remove false positive detections. A detection was determined to be false positive if the S/N of the high-frequency signal was below 4.0; if there was obvious contamination from a low-frequency harmonic series; or if the amplitude spectrum displayed obvious characteristics of either $\delta$\,Sct stars (with many modes in the $\approx0.2-1.0$\,mHz range), sdBV stars (stars with pulsations in low and high-frequency ranges) or pulsating white dwarfs (modes with amplitudes significantly greater than those known in roAp stars). The final sample consisted of 163 stars.

This final sample was distributed amongst the members of the TESS Asteroseismic Science Consortium (TASC) Working Group 4 (WG4) for detailed analysis to confirm the presence of a positive roAp detection, and to extract rotation periods from the light curves. This process identified 12 new roAp stars previously unreported in the literature and 10 roAp stars discovered through TESS observations and reported by either \citet{2019MNRAS.487.3523C} or \citet{2019MNRAS.487.2117B}. There are also positive detections of pulsations in 31 roAp stars known prior to the launch of TESS, with 3 roAp stars where TESS did not detect pulsational variability as their low amplitude modes are below the TESS detection limit. Finally, we present 5 roAp candidate stars where there is inconclusive evidence as to whether the star is truly an roAp pulsator.

\section{Spectroscopic observations}
\label{sec:spec}

For a subset of our sample, we have made new spectroscopic observations to confirm or detect the chemically peculiar nature of the star. Most of these observations were obtained with the SpUpNIC long-slit spectrograph \citep{2016SPIE.9908E..27C} mounted on the 1.9-m telescope of the South African Astronomical Observatory (SAAO). The observations were made in the blue part of the spectrum ($\sim3900-5100$\,\AA) with grating 4 which achieves a resolution of about 0.6\,\AA\ per pixel, with some stars also having a red spectrum ($\sim6100-7150$\,\AA) obtained with grating 6 with a resolution of 1.4\,\AA\ per pixel. The data were reduced in the standard way\footnote{\url{http://www.starlink.ac.uk/docs/sc7.htx/sc7.html}}, with wavelength calibrations provided by arc spectra obtained directly after the science observation.

Where possible, we also observed some stars with the Southern African Large Telescope \citep[SALT;][]{SALT}, utilising the High Resolution Spectrograph \citep[HRS;][]{HRS1,HRS2}. HRS is a dual-beam spectrograph with wavelength coverage of $3700-5500$\,\AA\ and $5500-8900$\,\AA. The observations were obtained in either the medium ($R\sim43000$) or high ($R\sim63000$) resolution modes, and automatically reduced using the SALT custom pipeline, which is based on the European Southern Observatory's (ESO) {\sc{midas}} pipeline \citep{pyhrs3,pyhrs2}.

Finally, we also used the High Efficiency and Resolution Canterbury University Large \'Echelle Spectrograph (HERCULES) at the 1.0-m McLellan Telescope of the University of Canterbury Mt John Observatory \citep{2003ASPC..289...11H}. HERCULES has a wavelength coverage of $3800-8800$\,\AA\ with observations made at $R = 41000$. The observations were automatically reduced using the HERCULES custom pipeline, MEGARA. We provide a log of all spectroscopic observations in Table\,\ref{tab:spec_obs}.

Spectral classification with these observations was made by comparison with MK standard stars, following the procedure of \citet{2009ssc..book.....G}. The observations were normalised using an automated spline fitting procedure in the i{\sc{spec}} spectral analysis software \citep{2014A&A...569A.111B,2019MNRAS.486.2075B}, with the MK standards plotted for comparison. With the classification as a starting point, we computed synthetic spectra using i{\sc{spec}} to estimate effective temperatures. The synthesis used the {\sc{spectrum}} code \citep{1994AJ....107..742G} with a MARCS Gaia-ESO survey model atmosphere \citep{2008A&A...486..951G,2012Msngr.147...25G} and a VALD line list \citep{2015PhyS...90e4005R}. Spectra were synthesised with a fixed $\log g =4.0$\,\cms, solar metallicity and zero microturbulence and macroturbulence. Although these parameters are not refined for each star, they are adequate given the low resolution of most of the data. Spectra were synthesised in temperature steps of 100\,K, with the result taken as the best fit to the wings of the observed Balmer lines. 

\section{Results}
\label{sec:results}

Here we present the results of our search and analysis of new and known roAp stars in TESS Cycle\,1 data. We provide an overview of the targets to be discussed in Table\,\ref{tab:stars} and divide this section, and the table, into subsections addressing new discoveries reported in this work, previous discoveries reliant on the TESS data, and TESS observations of known roAp stars. Finally, we provide a list of candidate roAp stars where the currently available data do not allow us to confirm the nature of these stars.

\begin{table*}
    \centering
    \caption{Details of the stars analysed in this paper. The columns provide the TIC identifier and star name, the TIC\,v8.1 TESS magnitude, the spectral type as referenced in the stars discussion section, where `*' denotes a classification derived in this work, the effective temperature, as provided in the TIC, the sectors in which TESS observed the target in Cycle\,1. The final three columns provide the stellar rotation period derived in this work, with a $^\dagger$ denoting the first detection of this period, the pulsation frequency(ies) found in this work (note that sidelobes that arise from oblique pulsation are not listed), and the pulsation amplitudes seen in the TESS data. A`**' denotes rotation periods derived from rotationally split sidelobes. A `***' denotes a tentative detection.}
    \label{tab:stars}
    \begin{tabular}{rcccccrrr}
        \hline
         \multicolumn{1}{c}{TIC} & {HD/TYC}    	& TESS & {Spectral} 	 & $T_{\rm eff}^{\rm TIC}$   & {Sectors} &  \multicolumn{1}{c}{{$P_{\rm rot}$}}  	& \multicolumn{1}{c}{Pulsation frequency}	&  \multicolumn{1}{c}{Pulsation amplitude}     \\
                       		             &   {name}          	&  mag & {type}            &   {(K)}                               & {} &  \multicolumn{1}{c}{(d)}         		&   \multicolumn{1}{c}{(mHz)}			& \multicolumn{1}{c}{(mmag)}  \\
        \hline

         \multicolumn{6}{l}{\textit{New TESS roAp stars found in this work}}\\
	96315731	&	51203	& 10.18	& Ap\,SrEuCr	& 7100	& 6,7		&$6.6713\pm0.0007$	& $1.91271\pm0.00001$	& $0.125\pm0.011$	\\
                                                                                
	119327278 & 45698		& 8.08	& A2p\,SrEu	& 8540	& 6,7		    &$1.08457\pm0.00003$  & $0.20583\pm0.00001$	& $0.036\pm0.004$	\\
			  &			&		&			&		&		&					                & $2.43755\pm0.00001$	& $0.041\pm0.004$	\\
                                                                                
	170586794 & 107619	& 8.47	& F5\,p\,EuCr*	& 6590	&10		  & $10.31\pm0.04^\dagger$	& $1.76151\pm0.00003$	& $0.045\pm0.006$	\\
                                                                                
	176516923 & 38823		& 7.12	& A5p\,SrEuCr	& 7660 	& 6  		&$8.782\pm0.003$	    &$1.76019\pm0.00005$	& $0.023\pm0.004$	\\
			  &			&		&			&		&		&				                	&$1.85820\pm0.00004$	& $0.006\pm0.004$	\\
			&			&		&			&		&		&				                	&$1.91302\pm0.00005$	& $0.023\pm0.004$	\\
                                                                                
	178575480 &   55852	& 8.83	& Ap\,SrEuCr	& 7720 	& 7 		&$4.7788\pm0.0005$	    &$2.23870\pm0.00004$	& $0.010\pm0.007$	\\
			  &			&		&			&		&		&					                &$2.39524\pm0.00004$	& $0.052\pm0.008$	\\
                                                                                
	294266638 & 6021-415-1	& 9.74	& A7p\,SrEu*	& 7230	& 8 		& No signature			    &$1.56421\pm0.00002$	& $0.119\pm0.012$	\\
			&			&		&			&		&		&					                &$1.62882\pm0.00002$	& $0.149\pm0.012$	\\
                                                                                
	294769049 & 161423	& 9.17	& Ap\,SrEu(Cr)	& 7760	& 12,13 & $10.4641\pm0.0007^\dagger$	&$2.26044\pm0.00001$	& $0.054\pm0.006$	\\
                                                                                
	310817678 & 88507		& 9.60	& Ap\,SrEu(Cr)	& 8230	& 9,10& $2.75003\pm0.00008^\dagger$ &$1.170979\pm0.000019$	& $0.049\pm0.032$	\\
			&			&		&			&		&		&					                &$1.205272\pm0.000003$	& $0.337\pm0.032$	\\
			&			&		&			&		&		&					                &$1.229890\pm0.000030$	& $0.133\pm0.032$	\\
                                                                                
	356088697 & 76460		& 9.53 	& A8p\,SrEuCr*	& 7110	& 9-11 	& No signature			    &$0.646584\pm0.000008$	& $0.053\pm0.006$	\\
                                                                                
	380651050 & 176384	& 8.19	& F0/F2		& 6490 	& 13		& $4.19\pm0.01^\dagger$		&$1.85817\pm0.00002$	& $0.059\pm0.004$	\\
			&			&		&			&		&		&					                &$1.87771\pm0.00002$	& $0.045\pm0.004$	\\	
                                                                                
	387115314 & 9462-347-1	& 9.56	& A5			& 7660	& 13 & $5.264\pm0.002^\dagger$	&$1.319020\pm0.000039$	& $0.051\pm0.009$	\\
			&			&		&			&		&		&					                &$1.328382\pm0.000020$	& $0.101\pm0.009$	\\	
			&			&		&			&		&		&					                &$1.356225\pm0.000006$	& $0.375\pm0.009$	\\	
			&			&		&			&		&		&					                &$1.384083\pm0.000007$	& $0.274\pm0.009$	\\	
                                                                                
	466260580 & 9087-1516-1& 11.64	& Ap\,EuCr*	& 6830 	& 13 		& No signature			    &$1.340060\pm0.00002$	& $0.453\pm0.035$	\\

        \hline

        \multicolumn{6}{l}{\textit{roAp stars previously discovered by TESS}}\\
        12968953 		& 217704 	& 10.09	& A7p\,SrEuCr*	& 7880	& 2 & No signature			&$1.27047\pm0.00005$	& $0.071\pm0.013$	\\
				&		&		&			&		&		&				                	&$1.25875\pm0.00004$	& $0.083\pm0.013$	\\
				&		&		&			&	&			&					                &$1.26961\pm0.00003$	& $0.064\pm0.013$	\\
				&		&		&			&	&			&					                &$1.33782\pm0.00003$	& $0.099\pm0.013$	\\       
                                                  	                                                           
        17676722 		& 63773 	& 8.61	& A3p\,SrEuCr*	& 8320 	&7	& $1.5995\pm0.0003$	&$1.941545\pm0.000002$	& $0.671\pm0.006$	\\
                                                 	                                                           
        41259805 	& 43226 &8.84	& A6p\,SrEu(Cr)*	& 8360 	&1-8,10-13 & $1.714489\pm0.000002$ 	&$2.311039\pm0.000001$	& $0.025\pm0.002$	\\
				&		&		&			&	&			&						            &$2.326368\pm0.000003$	& $0.013\pm0.002$	\\     	
                                                                                                                
        49818005 		& 19687 	& 9.36	& F2pSrEu(Cr)*	& 7300 	& 4		& No signature	    &$1.641910\pm0.000008$ 	& $0.267\pm0.009$		\\
                                                                                                                
        152808505 	& 216641 	& 7.88     	& F3p\,EuCr*	   & 6430	&1	& $1.877\pm0.006$**	&$1.36260\pm0.00005$	& $0.019\pm0.004$	\\
				&		&		&			&	&			&						            &$1.38593\pm0.00002$	& $0.054\pm0.004$\\      
				&		&		&			&	&			&					            	&$1.39119\pm0.00002$	& $0.041\pm0.004$\\

        156886111 	& 47284 	& 9.11	& A8p\,SrEuCr*	& 7540 	& 6,7	&$6.8580\pm0.0003$	&$0.063251\pm0.000002$	& $0.366\pm0.006$	\\
				&		&		&			&	&			&						            &$1.274574\pm0.000021$	& $0.039\pm0.006$	\\
				&		&		&			&	&			&					            	&$1.302738\pm0.000014$	& $<0.020$	\\
				&		&		&			&	&			&					            	&$1.317816\pm0.000008$	& $0.107\pm0.006$	\\
                                                                                                                
        259587315 	& 30849 	& 8.69	& Ap\,SrCrEu	    & 7720	&4,5 	&$15.776\pm0.005$	&$0.904064\pm0.000009$	& $0.071\pm0.005$	\\
				&		&		&			&	&			&						            &$0.905469\pm0.000013$	& $0.050\pm0.005$	\\
				&		&		&			&	&			&					            	&$0.931050\pm0.000003$	& $0.154\pm0.005$	\\
				&		&		&			&	&			&					            	&$0.950735\pm0.000022$	& $0.029\pm0.005$	\\
				                                                                                
        349945078 	& 57040 	& 8.92	& A2p\,EuCr	  & 7200 	& 6-9	&$13.4256\pm0.0008$	&$2.126381\pm0.000002$ &$0.140\pm0.004$	\\
				&		&		&			&	&			&						            &$2.186645\pm0.000005$ & $0.063\pm0.004$	\\	
                                                                                                                
        350146296 	& 63087 	&9.17	& F0p\,EuCr*& 7680 	&1-13  &$2.66387\pm0.00002$	    &$3.4031468\pm0.0000078$ & $0.005\pm0.002$	\\
				&		&		&			&	&			&						            &$3.4424674\pm0.0000021$	& $0.020\pm0.002$\\
				&		&		&			&	&			&						            &$3.4839854\pm0.0000015$	& $0.026\pm0.002$\\
				&		&		&			&	&			&						            &$3.5232955\pm0.0000005$	& $0.081\pm0.002$\\
				&		&		&			&	&			&						            &$3.5626041\pm0.0000031$	& $0.013\pm0.002$\\	
                                                                                                                
	431380369 	& 20880 	&7.82	& Ap\,Sr(EuCr)	  & 8230	& 2,6,13 & $5.19716\pm0.00005$	&$0.805252\pm0.000002$	& $0.022\pm0.003$	\\
				&		&		&			&	&			&					            	&$0.815837\pm0.000002$	& $0.016\pm0.003$	\\	
				&		&		&			&	&			&					            	&$0.860412\pm0.000001$	& $0.032\pm0.003$	\\		
				&		&		&			&	&			&					            	&$0.872662\pm0.000003$	& $0.013\pm0.003$	\\
        \hline
        
            \end{tabular}
\end{table*}

\begin{table*}
    \centering
    \contcaption{}
    \label{tab:stars_cont_2}
    \begin{tabular}{rcccccrrr}
        \hline
         \multicolumn{1}{c}{TIC} & {HD/TYC}    	& TESS & {Spectral} 	 & $T_{\rm eff}^{\rm TIC}$   & {Sectors} &  \multicolumn{1}{c}{{$P_{\rm rot}$}}  	& \multicolumn{1}{c}{Pulsation frequency}	&  \multicolumn{1}{c}{Pulsation amplitude}     \\
                       		             &   {name}          	&  mag & {type}            &   {(K)}                               & {} &  \multicolumn{1}{c}{(d)}         		&   \multicolumn{1}{c}{(mHz)}			& \multicolumn{1}{c}{(mmag)}  \\
        \hline
        \multicolumn{5}{l}{\textit{Known roAp stars prior to TESS launch}}\\
        6118924		& 116114   	& 6.77	& F0Vp\,SrCrEu  	& 7670 	&10		& No signature 			    & $0.76923\pm0.00004$	&$0.020\pm0.003$		\\

        33601621		& 42659  &6.60	& Ap\,SrCrEu        	& 7880 	& 6  & $2.6627\pm0.0003$	    & $1.74757\pm0.00003$	& $0.039\pm0.005$	\\

        35905913 		& 132205 		& 8.50	& Ap\,EuSrCr		& 7510	& 11,12	&  $7.513\pm0.001$	& No signal in TESS data	& 	$<0.025$	\\

        44827786 		& 150562 		&9.48	& F5Vp\,SrCrEu	& 7350	& 12 		& No signature		& $1.54700\pm0.00002$	& $0.181\pm0.013$	\\
                                	&			&		&	   			&		&		&				& $1.55040\pm0.00005$	& $0.068\pm0.013$	\\

        49332521 		& 119027 		& 9.66	& Ap\,SrEu(Cr)		& 6940	&11 	& No signature		& $1.57070\pm0.00005$	& $0.053\pm0.011$	\\
        				&			&		&	    			&		&		&					    & $1.83521\pm0.00004$	& $0.071\pm0.011$	\\     
        				&			&		&	    			&		&		&					    & $1.87490\pm0.00005$	& $0.059\pm0.011$	\\
        				&			&		&	    			&		&		&					    & $1.88779\pm0.00002$	& $0.174\pm0.011$	\\
        				&			&		&	    			&		&		&					    & $1.91338\pm0.00002$	& $0.160\pm0.011$	\\
        				&			&		&	    			&		&		&					    & $1.94029\pm0.00002$	& $0.184\pm0.012$	\\
				&			&		&	    			&		&		&					            & $1.94207\pm0.00002$	& $0.142\pm0.012$	\\	

        69855370 		& 213637 		& 9.19	& A\,(pEuSrCr)		& 6610	& 2 	    & No signature	& $1.42245\pm0.00001$	& $0.120\pm0.007$	\\
				&			&		&	    			&		&		&					            & $1.45237\pm0.00001$ 	& $0.172\pm0.007$	\\
        
        93522454 		& 143487 		& 9.11	& A3p\,SrEuCr		& 7110	&12 		& No signature	& No signal in TESS data	& $<0.030$	\\
        
        125297016	& 69013 	        & 9.21	& Ap\,EuSr		& 7010	& 7		& No signature			& $1.47519\pm0.00007${***} & $0.042\pm0.009$ \\
        
        136842396       	& 9289 		&9.20        & Ap\,SrEuCr		& 7750	& 3	& $8.660\pm0.006$ & $1.55361\pm0.00005$ 	& $<0.026$	\\
				&			&		&				&		&		&					                & $1.57327\pm0.00005$	& $0.059\pm0.009$	\\	        
				&			&		&				&		&		&					                & $1.58496\pm0.00001$	& $0.262\pm0.009$	\\	        
				&			&		&				&		&		&					                & $1.61549\pm0.00003$	& $<0.038$	\\	        

        139191168	 & 217522 	& 7.18	& A5p\,SrEuCr		& 6920	&1		& No signature			    & $1.20088\pm0.00002$	& $0.030\pm0.003$	\\
				&			&		&				&		&		&					                & $1.20169\pm0.00003$	& $0.028\pm0.003$	\\
				&			&		&				&		&		&					                & $1.20579\pm0.00002$	& $0.032\pm0.003$	\\
				&			&		&				&		&		&					                & $1.21513\pm0.00002$	& $0.031\pm0.003$	\\
 
        146715928       	& 92499 		& 8.72	& A2p\,SrEuCr		& 7730	& 9,10 	& No signature  & $1.54260\pm0.00002$	& $0.034\pm0.005$	\\
				&			&		&				&		&		&					                & $1.58221\pm0.00001$	& $0.043\pm0.005$	\\	        

        167695608       	& 8912-1407-1  &11.69 	& F0p\,SrEu(Cr)	& 7180	& 1-4,6-13	& No signature	& $1.512238\pm0.000004$ & $0.044\pm0.009$\\
				&			&		&				&		&		&					                & $1.532161\pm0.000001$	& $0.171\pm0.009$\\
				&			&		&				&		&		&				                	& $1.554816\pm0.000001$	& $0.135\pm0.009$\\
				&			&		&				&		&		&				                	& $1.557383\pm0.000002$	& $0.074\pm0.009$\\
				&			&		&				&		&		&				                	& $1.564767\pm0.000002$	& $0.079\pm0.009$\\
				&			&		&				&		&		&				                	& $1.586081\pm0.000003$	& $0.052\pm0.009$\\
        
        168383678 	& 96237 		& 9.25    	& A4p\,SrEuCr		&7410	& 9 & $\sim21$ or $\sim42$	& $1.19610\pm0.00003$	& $0.077\pm0.010$\\
        
        170419024 	& 151860 		&8.53       & Ap\,SrEu(Cr)		& 6630	& 12& No signature			& $0.85014\pm0.00002$	& $0.085\pm0.007$	\\
				&			&		&				&		&		&					                & $1.36330\pm0.00004$	& $0.045\pm0.007$	\\

        173372645 	& 154708		&8.43	& Ap\,SrEuCr		& 6920	&12 		& $5.363\pm0.001$	& No signal in TESS data & $<0.030$\\

        189996908 	& 75445 		& 6.89       & Ap\,SrEu(Cr)		& 7610	& 8,9 	& No signature		& $1.84163\pm0.00002${***} & $0.013\pm0.003$	\\
       
        211404370 	& 203932 		& 8.56	& Ap\,SrEu		&7540	& 1 		& $6.44\pm0.01$		& $2.69844\pm0.00005$ 		& $0.030\pm0.006$\\
				&			&		&				&		&		&					                & $2.80476\pm0.00002$ 		& $0.082\pm0.006$\\        
       
        237336864 	& 218495 		& 9.24	& Ap\,EuSr		&8120	& 1 		& $4.2006\pm0.0001$	    & $2.09827\pm0.00003$ &$0.056\pm0.008$ \\
				&			&		&				&		&		&					                & $2.22073\pm0.00002$ &$<0.035$	\\
				&			&		&				&		&		&					                & $2.24862\pm0.00002$ &$0.103\pm0.008$	\\
				&			&		&				&		&		&					                & $2.26123\pm0.00001$ &$0.159\pm0.008$	\\
       
        268751602 	& 12932 		& 9.98	& Ap\,SrEuCr 		& 7320	& 3		& No signature			& $1.436302\pm0.000006$	&$0.682\pm0.012$	\\
        
        279485093 	& 24712 		& 5.72	& A9p\,SrEuCr	& 7280	& 5 		& $12.578\pm0.008$	& $2.557764\pm0.000043$	&$0.013\pm0.002$	\\
        				&			&		&				&		&		&					        & $2.604135\pm0.000034$	&$0.018\pm0.002$	\\
				&			&		&				&		&		&					                & $2.619775\pm0.000013$	&$0.046\pm0.002$	\\
				&			&		&				&		&		&					                & $2.652959\pm0.000005$	&$0.113\pm0.002$	\\
				&			&		&				&		&		&					                & $2.687440\pm0.000003$	&$0.170\pm0.002$	\\
				&			&		&				&		&		&				                   	& $2.720639\pm0.000003$	&$0.199\pm0.002$	\\
				&			&		&				&		&		&				                   	& $2.755166\pm0.000025$	&$0.025\pm0.002$	\\
				&			&		&				&		&		&				                   	& $2.791338\pm0.000014$	&$0.040\pm0.002$	\\
				&			&		&				&		&		&				                	& $2.805580\pm0.000016$	&$0.036\pm0.002$	\\
				
        280198016	& 83368		& 6.06	& A8p\,SrCrEu		& 7710	& 9 		& $2.8517\pm0.0002$	& $ 1.42797\pm0.00001$	&$0.132\pm0.005$	\\

        315098995 	& 84041 		& 9.20	& Ap\,SrEuCr		& 7780	& 8,9 	& $3.6884\pm0.0004$	& $1.114117\pm0.000005$	&$<0.055$	\\

        322732889 	& 99563 		& 8.47	& F0p\,Sr			& 7940	& 9 	& $2.9114\pm0.0002$	& $1.557657\pm0.000002$	&$0.179\pm0.007$	\\

        326185137 	& 6532 		& 8.35	& Ap\,SrEuCr		& 8500	& 3 		& $1.9447\pm0.0001$	& $2.402163\pm0.000007$	&$0.063\pm0.006$	\\
    \hline
        
            \end{tabular}
\end{table*}

\begin{table*}
    \centering
    \contcaption{}
    \label{tab:stars_cont_1}
    \begin{tabular}{rcccccrrr}
        \hline
         \multicolumn{1}{c}{TIC} & {HD/TYC}    	& TESS & {Spectral} 	 & $T_{\rm eff}^{\rm TIC}$   & {Sectors} &  \multicolumn{1}{c}{{$P_{\rm rot}$}}  	& \multicolumn{1}{c}{Pulsation frequency}	&  \multicolumn{1}{c}{Pulsation amplitude}     \\
                       		             &   {name}          	&  mag & {type}            &   {(K)}                               & {} &  \multicolumn{1}{c}{(d)}         		&   \multicolumn{1}{c}{(mHz)}			& \multicolumn{1}{c}{(mmag)}  \\
        \hline
        \multicolumn{5}{l}{\textit{Known roAp stars prior to TESS launch -- continued}}\\
        
        340006157 	& 60435 		& 8.72	&A3p\,SrEu			& 8160	& 3,6-10,13     & $7.6797\pm0.0001$	&$1.29573\pm0.00063$	&$0.012\pm0.003$\\
				&			&		&				&		&		                    &					& $1.31293\pm0.00019$	&$0.042\pm0.003$	\\
				&			&		&				&		&		                    &					& $1.32446\pm0.00030$	&$0.025\pm0.003$	\\
				&			&		&				&		&		                    &					& $1.35251\pm0.00006$	&$0.126\pm0.003$	\\
				&			&		&				&		&		                    &					& $1.38056\pm0.00017$	&$0.043\pm0.003$	\\
				&			&		&				&		&		                    &					& $1.40721\pm0.00012$	&$0.064\pm0.003$	\\
				&			&		&				&		&		                    &					& $1.43233\pm0.00012$	&$0.064\pm0.003$	\\
				&			&		&				&		&	                    	&					& $1.45727\pm0.00036$	&$0.021\pm0.003$	\\

        348717688 	& 19918 		& 9.12	& A5p\,SrEuCr			& 7790	& 1,12,13 & No signature	& $1.37159815\pm0.00000146$	& $0.039\pm0.005$\\
				&			&		&					&		&					&					& $1.37673847\pm0.00000098$	& $0.058\pm0.005$\\
				&			&		&					&		&					&					& $1.39872517\pm0.00000194$	& $0.029\pm0.005$\\
				&			&		&					&		&					&					& $1.44003439\pm0.00000199$	& $0.029\pm0.005$\\
				&			&		&					&		&					&					& $1.48119405\pm0.00000130$	& $0.044\pm0.005$\\
				&			&		&					&		&					&					& $1.51008560\pm0.00000006$	& $0.997\pm0.005$\\
				&			&		&					&		&					&					& $1.53991119\pm0.00000191$	& $0.030\pm0.005$\\

        363716787 	& 161459 & 10.01 & Ap\,EuSrCr		& 7330	& 13 			    & $5.966\pm0.001$	& $1.39089\pm0.00003$	& $0.081\pm0.017$\\
				&			&		&					&		&					&					& $1.42253\pm0.00004$	& $0.092\pm0.017$\\

        368866492 	& 166473 		& 7.73	& A5p\,SrCrEu			& 7510	& 13	& No signature		& $1.884520\pm0.000017$		& $0.070\pm0.005$	\\
				&			&		&					&		&					&					& $1.888201\pm0.000030$		& $0.041\pm0.005$	\\
				&			&		&					&		&					&					& $1.891891\pm0.000005$		& $0.274\pm0.005$	\\

        369845536 	& 			& 12.90	& A7p\,Eu(Cr)			& 7930 	& 13		& $9.50\pm0.02$	& $2.041601\pm0.000003$ &$5.843\pm0.094$		\\
        
        394124612 	& 218994 		& 8.31	& A3p\,Sr				& 		& 1	    & $5.855 \pm 0.008$	& $1.14433\pm0.00001$ & $0.124\pm0.008$		 \\
       
        394272819 	& 115226 		& 8.27	& Ap\,Sr(Eu)		& 7630	& 11,12 & $2.98827\pm0.00008$   & $1.479334\pm0.000069$	& $<0.010$	\\
				&			&		&					&		&					&					& $1.508904\pm0.000002$	& $0.213\pm0.005$	\\
				&			&		&					&		&					&					& $1.538487\pm0.000016$	& $0.034\pm0.005$	\\
				&			&		&					&		&					&					& $1.568072\pm0.000007$	& $0.075\pm0.005$	\\

        402546736 	& 128898 		& 1.84	& A7p\,SrCrEu		&		& 11,12 & $4.4812\pm0.0005$	& $2.2213028\pm0.0000133$ & $0.009\pm0.001$		\\
				&			&		&					&		&					&					& $2.2680569\pm0.0000105$ & $0.012\pm0.001$	\\
				&			&		&					&		&					&					& $2.3692228\pm0.0000117$ & $0.011\pm0.001$		\\
				&			&		&					&		&					&					& $2.3817256\pm0.0000086$ & $0.015\pm0.001$	\\
				&			&		&					&		&					&					& $2.4118973\pm0.0000013$ & $0.100\pm0.001$		\\
				&			&		&					&		&					&					& $2.4420726\pm0.0000003$ & $0.428\pm0.001$	\\
				&			&		&					&		&					&					& $2.4722512\pm0.0000010$ & $0.118\pm0.001$		\\
				&			&		&					&		&					&					& $2.5024429\pm0.0000117$ & $0.011\pm0.001$		\\
				&			&		&					&		&					&					& $2.5326183\pm0.0000205$ & $0.006\pm0.001$	\\
				&			&		&					&		&					&					& $2.5666565\pm0.0000091$ & $0.014\pm0.001$	\\
   
        434449811 	& 80316 		& 7.69	& Ap\,Sr(Eu?)		& 8180	& 8		& $2.08862\pm0.00004$	& $0.112311\pm0.000022$	&$0.055\pm0.005$	\\
				&			&		&					&		&					&					& $0.131686\pm0.000014$	&$0.089\pm0.005$	\\
				&			&		&					&		&					&					& $2.251749\pm0.000007$	&$0.182\pm0.005$	\\
				
        469246567 	& 86181 		& 9.15	& Ap\,Sr			& 7210	& 9,10 	& $2.05115\pm0.00006$	& $2.663229\pm0.000015$ 	&$<0.020$	\\
				&			&		&					&		&					&					& $2.694099\pm0.000003$ 	&$0.276\pm0.007$	\\
				&			&		&					&		&					&					& $2.728458\pm0.000012$ 	&$0.071\pm0.007$	\\
				
    \hline
    \multicolumn{5}{l}{\textit{Candidate roAp}}\\
        1727745		& 113414		& 7.22	& F7/F8			& 6150		& 10& $3.172\pm0.001^\dagger$	& $2.29116\pm0.00004$		& $0.035\pm0.006$	\\
 
        3814749		& 3748		& 9.73	& A0/1\,IV/V		& 8420		& 3	& $1.689\pm0.002$	        & $1.63100\pm0.00006$ 		& $0.163\pm0.031$		\\
				&			&		&				&			&					&					& $2.75700\pm0.00002$ 		& $0.389\pm0.031$		\\

        158637987	& 10330		&8.91	& A9				& 7180		& 2,3		& No signature		& $2.72598\pm0.00002$		& $0.022\pm0.005$\\
				&			&		&				&			&					&					& $2.74240\pm0.00001$		& $0.031\pm0.005$\\

        324048193	& 85892		& 7.74	& Ap\,Si		& 11950		& 12,13		& $4.2953\pm0.0001$	& $3.15347\pm0.00002$		& $0.020\pm0.004$\\
 
        410163387	& 76276		& 9.50	& Ap\,SrEuCr		& 7480		& 8		& No signature 			& $1.43155\pm0.00003$		& $0.075\pm0.011$\\ 

        \hline 
    \end{tabular}
\end{table*}

Throughout this section, to keep homogeneity of the results, we present the numerical results of a single team, hereafter named the reference team. These results have been corroborated by the WG4 contribution to the analysis of the target stars. Where discrepancies were identified between the reference team and the WG4 members, secondary checks were conducted to ensure accurate results are presented.

To avoid repetition, we outline here the process by which each star was analysed in detail, unless a different method is detailed in the appropriate star's subsection. Where available, we combined multi-sector observations to produce a single light curve for analysis. Using the original PDC\_SAP data, we calculated a Discrete Fourier Transform, following \citet{1985MNRAS.213..773K} in the low-frequency range (to 0.11\,mHz; 10\,\cd) in the search for rotational variability. Where present, we fitted a harmonic series by non-linear least squares to determine the rotation frequency. Where we found a harmonic to have a stronger signal, we derived the rotation frequency from that signal as it can be more precisely determined. Where no rotational variability was found, we also checked the SAP light curve for signs of rotation as the SPOC pipeline can mistakenly remove low-frequency astrophysical signal. 

Subsequently, with the astrophysical information extracted, we removed the rotation signal and any instrumental artefacts through iterative pre-whitening of the light curve to a noise level determined in the range $2.3-3.3$\,mHz ($200-300$\,\cd). This serves to make the noise in the amplitude spectrum white when considering the errors on pulsation signals. Any pulsation signals found were extracted through non-linear least squares fitting. Both the derived rotation period and the pulsation frequency(ies) for each star are listed in Table\,\ref{tab:stars}. We provide only the presumed {\it{pulsation mode}} frequencies and not rotationally split sidelobes; any sidelobes detected are mentioned in the text, and indicated on the appropriate figure.

Since the launch of TESS, there have been two papers addressing the discovery of roAp stars in the new data set. The TESS Ap first light paper \citep{2019MNRAS.487.3523C} focused on the analysis of the first and second TESS sectors, while a paper by \citet{2019MNRAS.487.2117B} analysed data from sectors 1-7. We do not present an in depth analysis of the stars addressed in those papers again, unless there are more data available, or significant differences in the results were found. We do, however, include these stars in Table\,\ref{tab:stars} to provide a complete inventory of TESS observations of roAp stars in Cycle\,1.


\subsection{New TESS roAp stars}

We classify the following stars as new roAp stars based on the presence of high frequency pulsation in the target and either an Ap spectral classification, the presence of multiplets in the amplitude spectrum split by the stellar rotation frequency, or Str\"omgren-Crawford indices that indicate a chemically peculiar star.

\subsubsection{TIC\,96315731}
TIC\,96315731 (HD\,51203) is identified in the literature as an $\alpha^2$\,CVn with a spectral type of Ap\,SrEuCr \citep{1982mcts.book.....H}. The star is reported to have a mean magnetic field modulus of $7.9\pm0.5$\,kG \citep{2019ApJ...873L...5C}.

\begin{figure}
    \centering
     \includegraphics[width=\columnwidth]{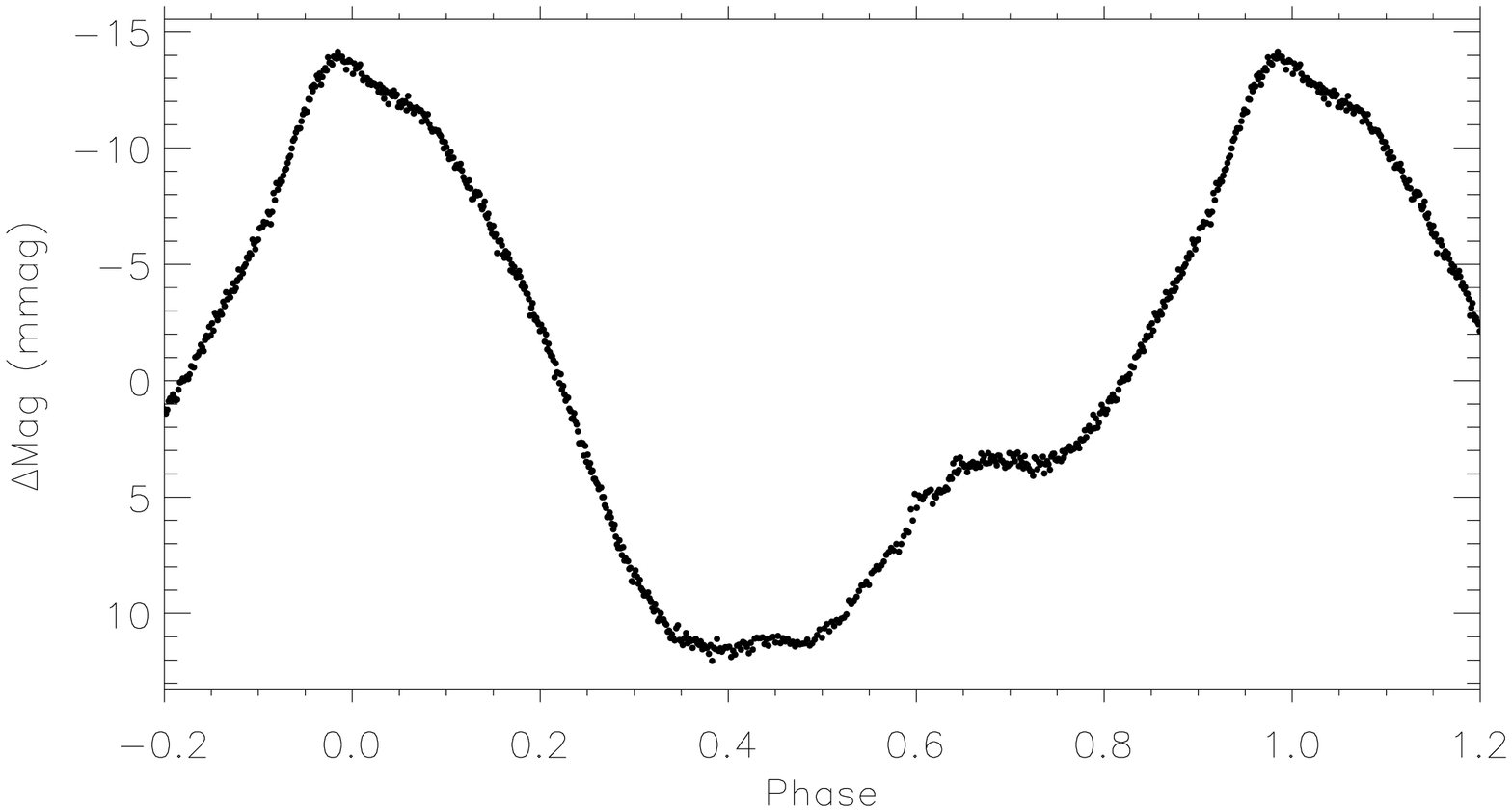}
    \includegraphics[width=\columnwidth]{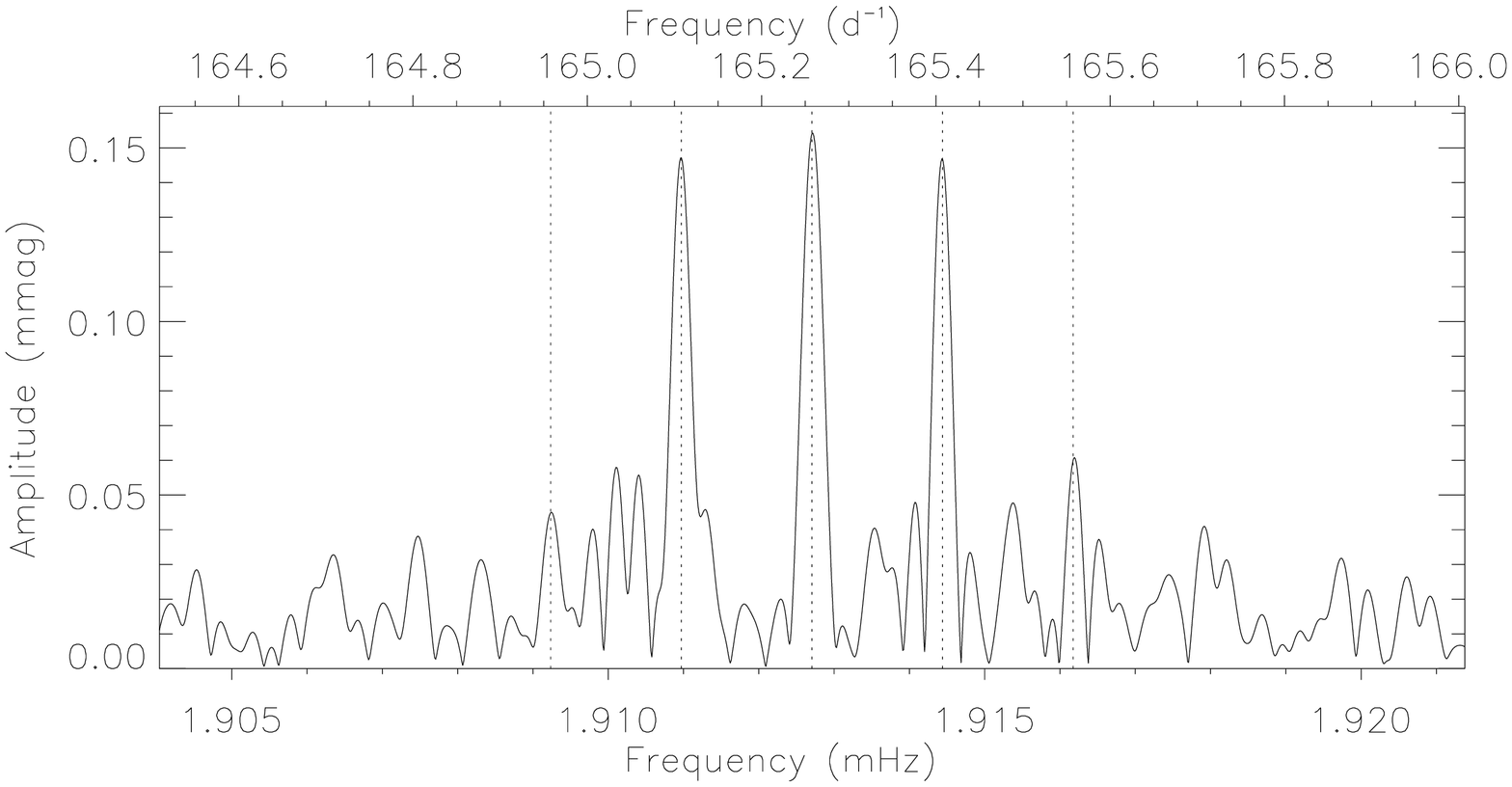}
    \caption{Top: light curve of TIC\,96315731 phase folded on the derived rotation period of 6.67134\,d. Bottom: the pulsation quintuple arising from a distorted quadrupole mode. The vertical dotted lines indicate the quintuplet components.}
    \label{fig:tic96315731}
\end{figure}

TIC\,96315731 was observed by TESS during sectors 6 and 7, and shows clear signatures of rotation and pulsation, making this the first report that TIC\,96315731 is a new roAp star, although \citep{Kobzar2020} listed it as a candidate roAp star. From the combined TESS light curve, we measure a rotation period of $6.6713\pm0.0007$\,d which is similar to the value of $6.675$\,d presented by \citet{2017MNRAS.468.2745N}. We show a light curve phased on the rotation period in the top panel of Fig.\,\ref{fig:tic96315731}.

At high frequency, the star shows an obvious pulsation signal at $1.91271\pm0.00001$\,mHz ($165.258\pm0.001$\,\cd) which is flanked by two sidelobes with significant amplitude (S/N$\sim14$) separated from the pulsation by the rotation frequency, and two further sidelobes which have lower significance (S/N$\sim6$) separated by twice the rotation frequency. This is indicative of either a distorted dipole mode, or a quadrupole mode, although an analysis of the pulsation amplitude and phase as a function of rotation phase favours a distorted quadrupole mode. In the case of a distorted quadrupole mode, the pulsation phase does not change by the expected $\pi-$rad when a pulsation node crosses the line of sight, but rather results in a phase blip \citep[see e.g.,][Shi et al. submitted]{2016MNRAS.462..876H,2018MNRAS.476..601H}. These peaks are shown in the bottom panel of Fig.\,\ref{fig:tic96315731}.


\subsubsection{TIC\,119327278}

TIC\,119327278 (HD\,45698) was initially classified as an Ap star by \cite{1973AJ.....78..687B} due to the presence of Sr absorption, with a revised spectral type of A2\,SrEu provided by \cite{1991A&AS...89..429R}. There is little information about this star in the literature other than a rotation period of 1.085\,d \citep{2017MNRAS.468.2745N}, and a lack of pulsation signal reported by \citet{2016A&A...590A.116J}.

\begin{figure}
    \centering
     \includegraphics[width=\columnwidth]{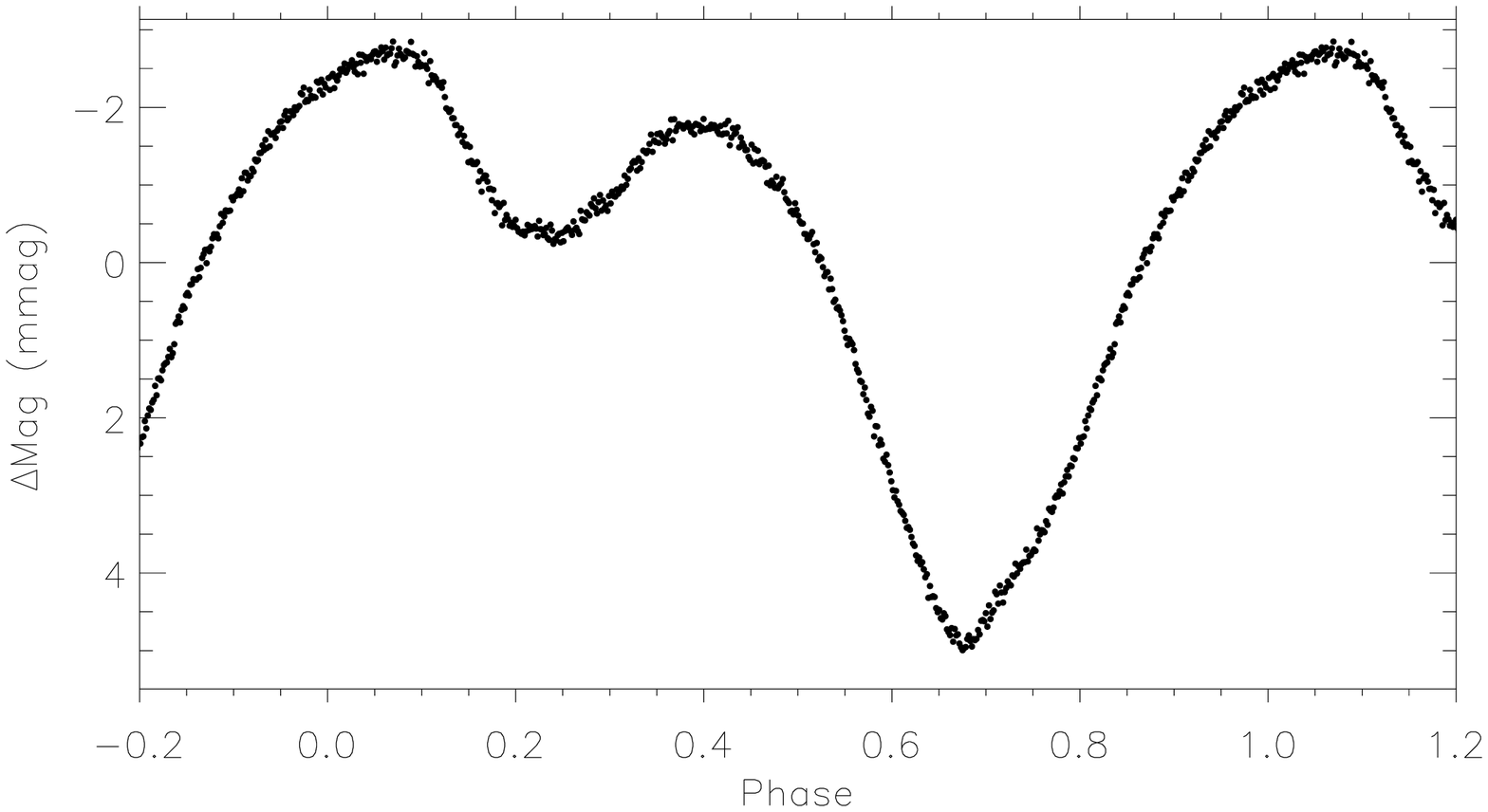}
      \includegraphics[width=\columnwidth]{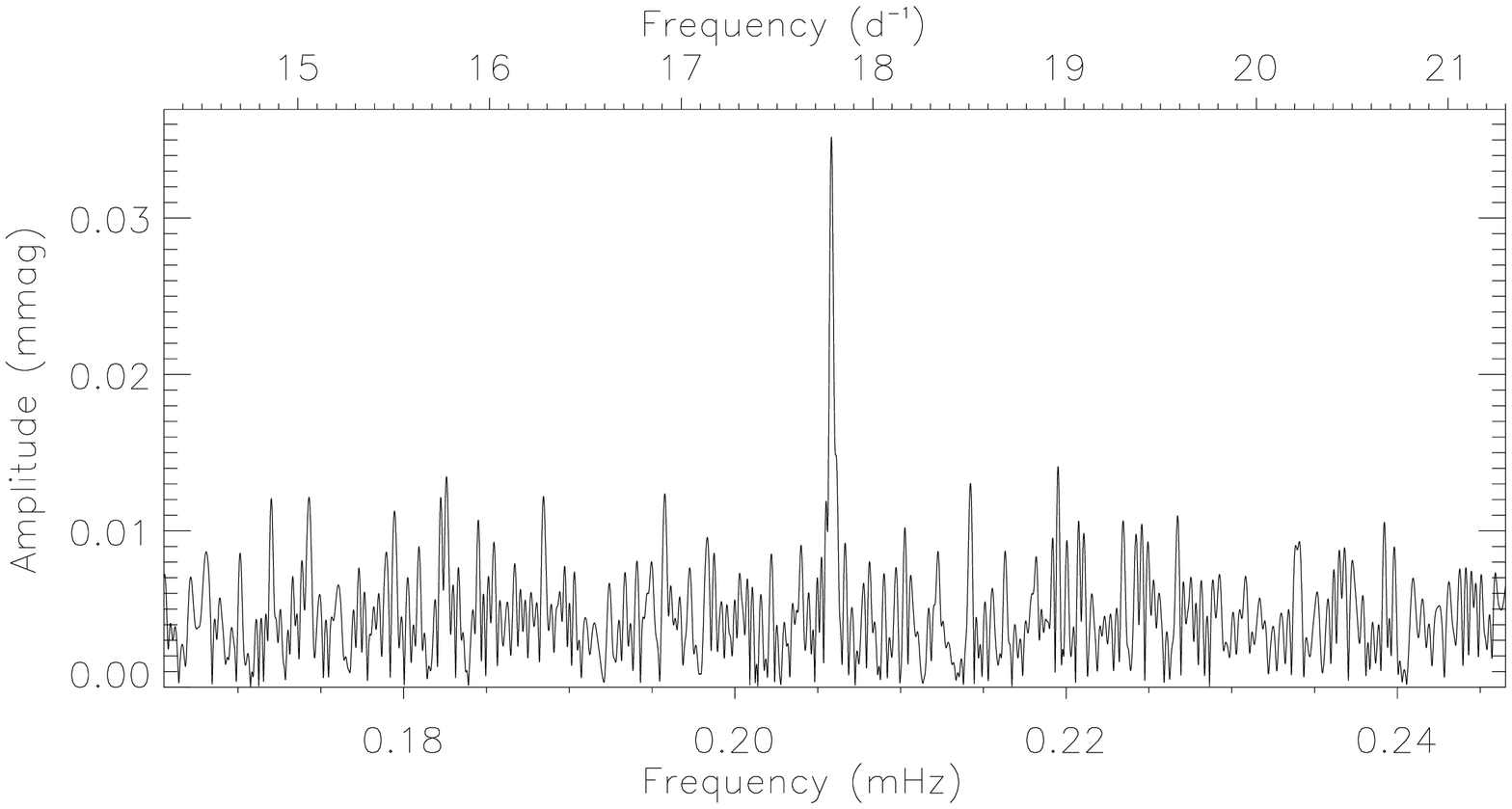}
    \includegraphics[width=\columnwidth]{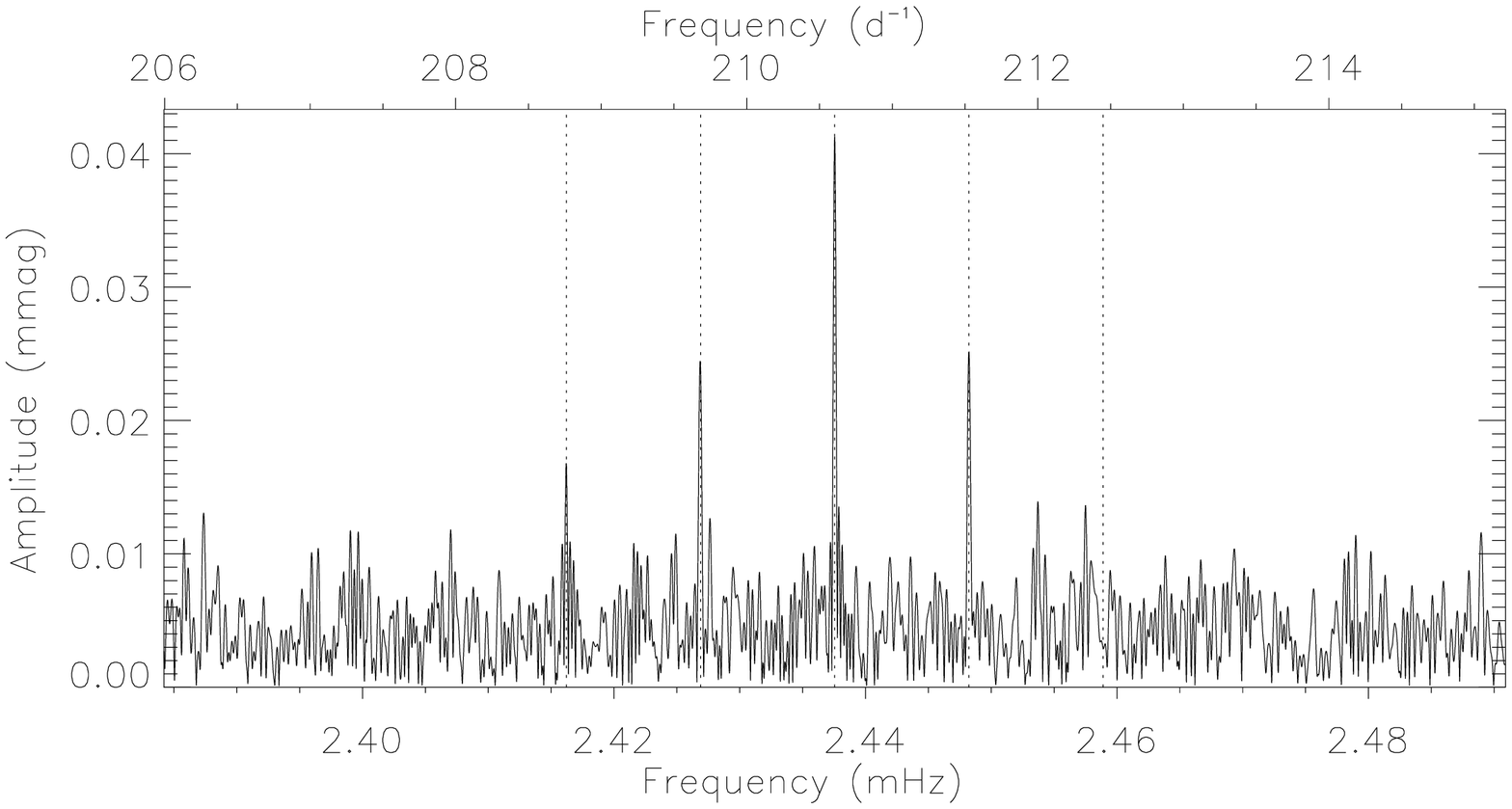}
    \caption{Top: light curve of TIC\,119327278 phase folded on the derived rotation period of 1.08457\,d. Middle: view of the $\delta$\,Sct mode in this star. Bottom: the pulsation multiplet signature. The vertical dotted lines indicate the quintuplet components.}
    \label{fig:tic119327278}
\end{figure}

TIC\,119327278 was observed during sectors 6 and 7. The data show the star to have a rotation period of $1.08457\pm0.00003$\,d (Fig.\,\ref{fig:tic119327278}) which is in agreement with the literature. Further to the harmonic series, we identify another mode that is typical of the $\delta$\,Sct stars, at a frequency of  $0.20583\pm0.00001$\,mHz ($17.783\pm0.001$\,\cd). Although it was not expected that Ap stars show these low-overtone modes as a result of magnetic suppression \citep{2005MNRAS.360.1022S}, \citet{2020MNRAS.498.4272M} have recently shown the existence of $\delta$\,Sct and roAp pulsations in the same star. 

The amplitude spectrum at high frequency shows a multiplet of 4 significant components ($\nu+2\nu_{\rm rot}$ is in the noise) of a presumed quintuplet (Fig.\,\ref{fig:tic119327278}), with the highest peak representing the pulsation mode at a frequency of $2.43755\pm0.00001$\,mHz ($210.605\pm0.001$\,\cd), which was also identified by \citet{Kobzar2020}. The components are split by the rotation frequency of the star as a result of oblique pulsation, thus confirming this star as a new roAp star. Given the presence of the low-overtone mode in TIC\,119327278, it may be the second example of a $\delta$\,Sct-roAp hybrid star. Given the low amplitude of the pulsation mode and the relatively short rotation period, analysing the pulsation amplitude and phase over the rotation period results in inconclusive results with regards to the identification of the mode degree. A detailed study, beyond the scope of this work, is needed to resolve this problem.


\subsubsection{TIC\,170586794}

TIC\,170586794 (HD\,107619) was classified as a possible metallic-line late-A type star by \citet{1982mcts.book.....H}. The star was later defined to be an F0 type by \citet{1993yCat.3135....0C} and an F0 type $\delta$\,Del star by \citet{1991PASP..103..494P}. It is not a frequently studied object with the literature being mainly concerned with $uvby$H$\upbeta$\footnote{Throughout we use the H$\upbeta$ notation to refer to the Str\"omgren-Crawford index to avoid confusion between angle of magnetic obliquity which is given the symbol $\beta$.} photometry. Among these studies, \citet{1996PASP..108..772P} gave dereddened Str\"omgren colour indexes and metallicity of the star, while \citet{1991PASP..103..494P} provided $(b-y)$, $m_1$, $c_1$ colour indexes and H$\upbeta$ index of 0.303, 0.226, 0.432 and 2.682, respectively. The temperature of the star was derived as $T_{\rm eff} = 6504$\,K by \citet{2017AJ....154..259S} and $T_{\rm eff}=6564$\,K by \citet{2006ApJ...638.1004A} from photometric investigations. 

\begin{figure}
    \centering
    \includegraphics[width=\columnwidth]{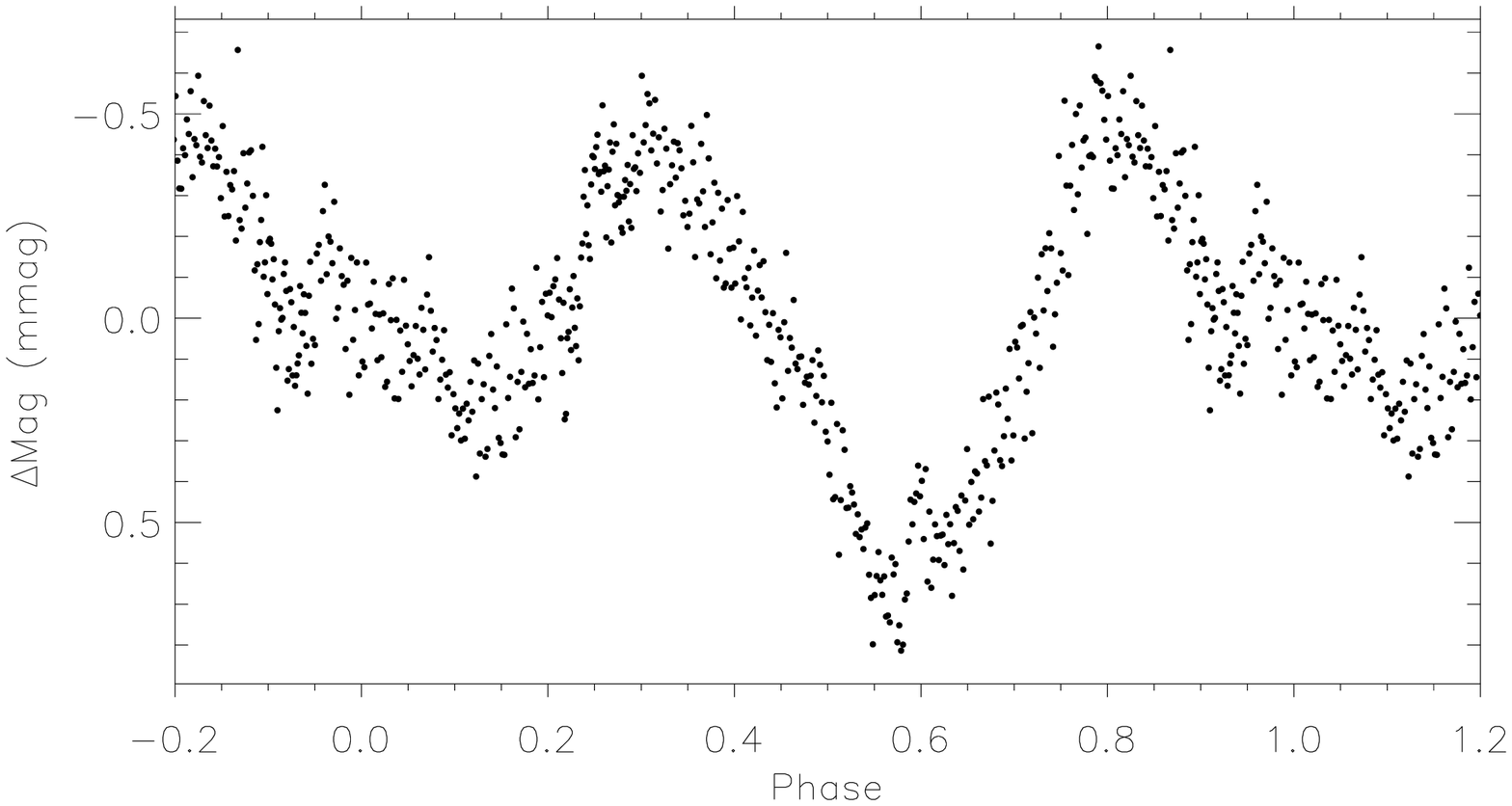}
    \includegraphics[width=\columnwidth]{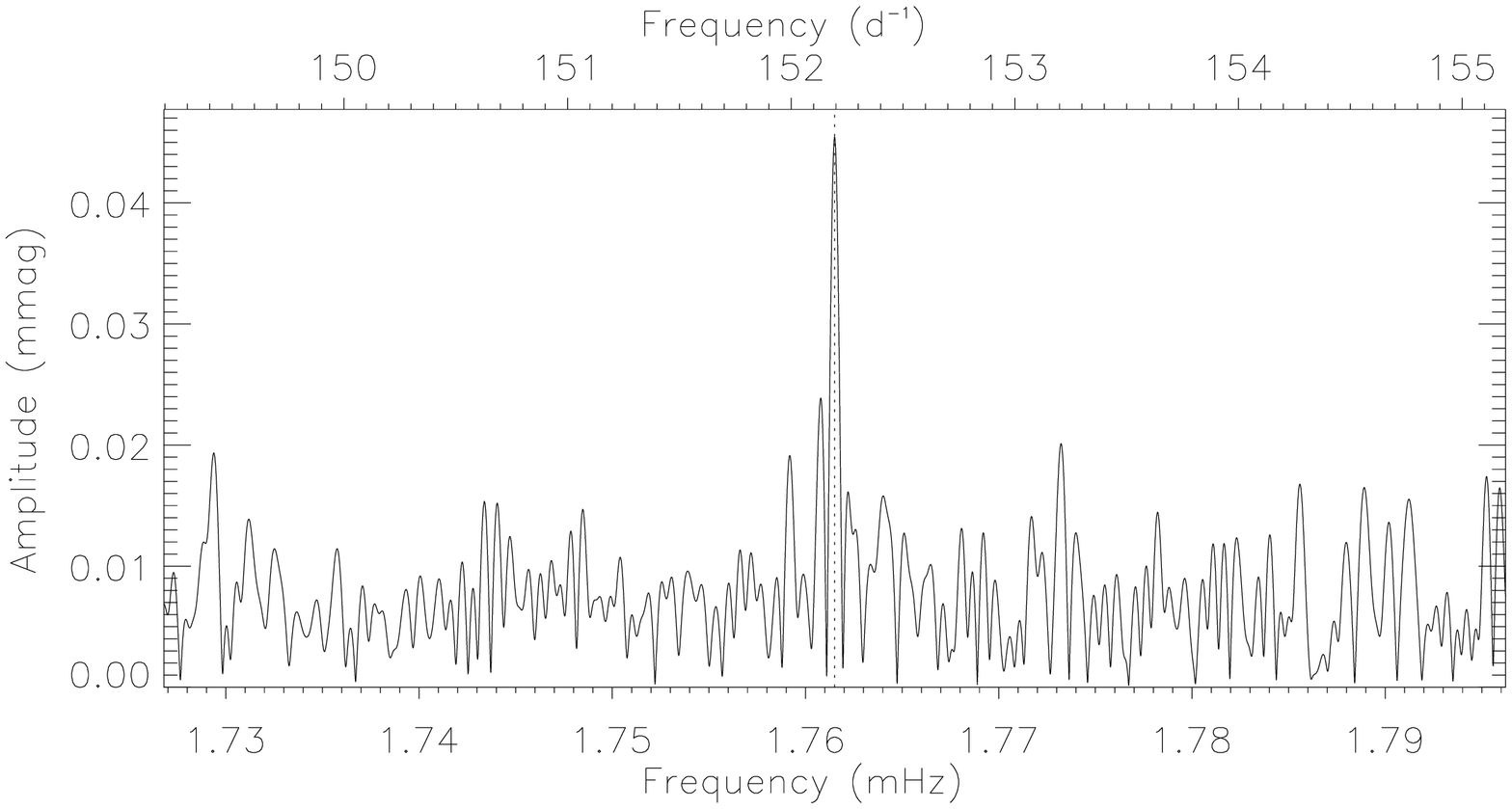}
    \caption{Top: the light curve of TIC\,170586794 phased on a period of 10.305\,d. Bottom: amplitude spectrum of the light curve showing a single pulsation mode.}
    \label{fig:tic170586794}
\end{figure}

TESS observed TIC\,170586794 during sector 10. The light curve shows the star to have a rotation period of $10.305\pm0.041$\,d, which has a double wave nature (Fig.\,\ref{fig:tic170586794}). At high frequency, this star shows a single pulsation mode at $1.76151\pm0.00003$\,mHz ($152.194\pm0.003$\,\cd; Fig.\,\ref{fig:tic170586794}). With both the presence of a rotationally modulated light curve, and a high frequency pulsation, we revisited this star spectroscopically. Our SAAO classification spectrum (Fig.\,\ref{fig:tic170586794spec}) shows this star to have overabundances of Eu\,{\sc{ii}} and Cr\,{\sc{ii}}, thus making it chemically peculiar. The Balmer lines in the spectrum are well matched to an F5\,V star, thus making this a cool F5p\,EuCr star, and confirming it as a new roAp star and making it one of the coolest roAp stars to date.

\subsubsection{TIC\,176516923}

TIC\,176516923 (HD\,38823) has several classifications in the literature: A5\,SrEu \citep{1979AAS...36..477V,1981A&AS...46..151H}; a strongly magnetic A5p\,Sr star \citep{2017AJ....153..218C}; and A5\,SrEuCr \citep{2009A&A...498..961R}. \citet{2019MNRAS.483.2300S} confirmed the star as a magnetic CP star and provided several abundance measures and estimated the age to be $\log t=8.58^{+0.56}_{-0.76}$\,(Gyrs). Estimates of the effective temperature range from 6600\,K to 7700\,K, the latter of which agrees with the TIC temperature provided in Table\,\ref{tab:stars}. $\log g$ values range from about 3.7\,\cms\ to 4.6\,\cms, with most rotational velocity measurements suggesting a $v\sin i\approx22$\,\kms.

There are several studies related to the  magnetic structure of TIC\,176516923. The strong magnetic field of the star was first discovered by \citet{2016AstBu..71..302R} after three observations made in 2009, with a $\langle B_z\rangle$ value of 1.74\,kG. They found a variable radial velocity within a small interval around $-10$\,\kms\ which is significantly different from the result of 1.40\,\kms\ by \citet{1999A&AS..137..451G}. Based on this result, they concluded that such differences in radial velocity could be an indication of a variability on a scale of up to several years or a decade. \citet{2007AstBu..62..147K} performed further spectroscopic observations and calculated radial velocities ranging from $-5$\,\kms\ to $-11$\,\kms. Therefore, they suspected that the star is in a binary system. 

The rotation period of the star has been under debate for several years, with fits to both photometric and magnetic data being used to determine it \citep{1981A&AS...46..151H,2006MNRAS.372.1804K,2012AN....333...41K}. Most recently, \citet{2020MNRAS.493.3293B} calculated the period of the star as $8.677\pm0.002$\,d which is representative of the other periods reported.

\begin{figure}
    \centering
     \includegraphics[width=\columnwidth]{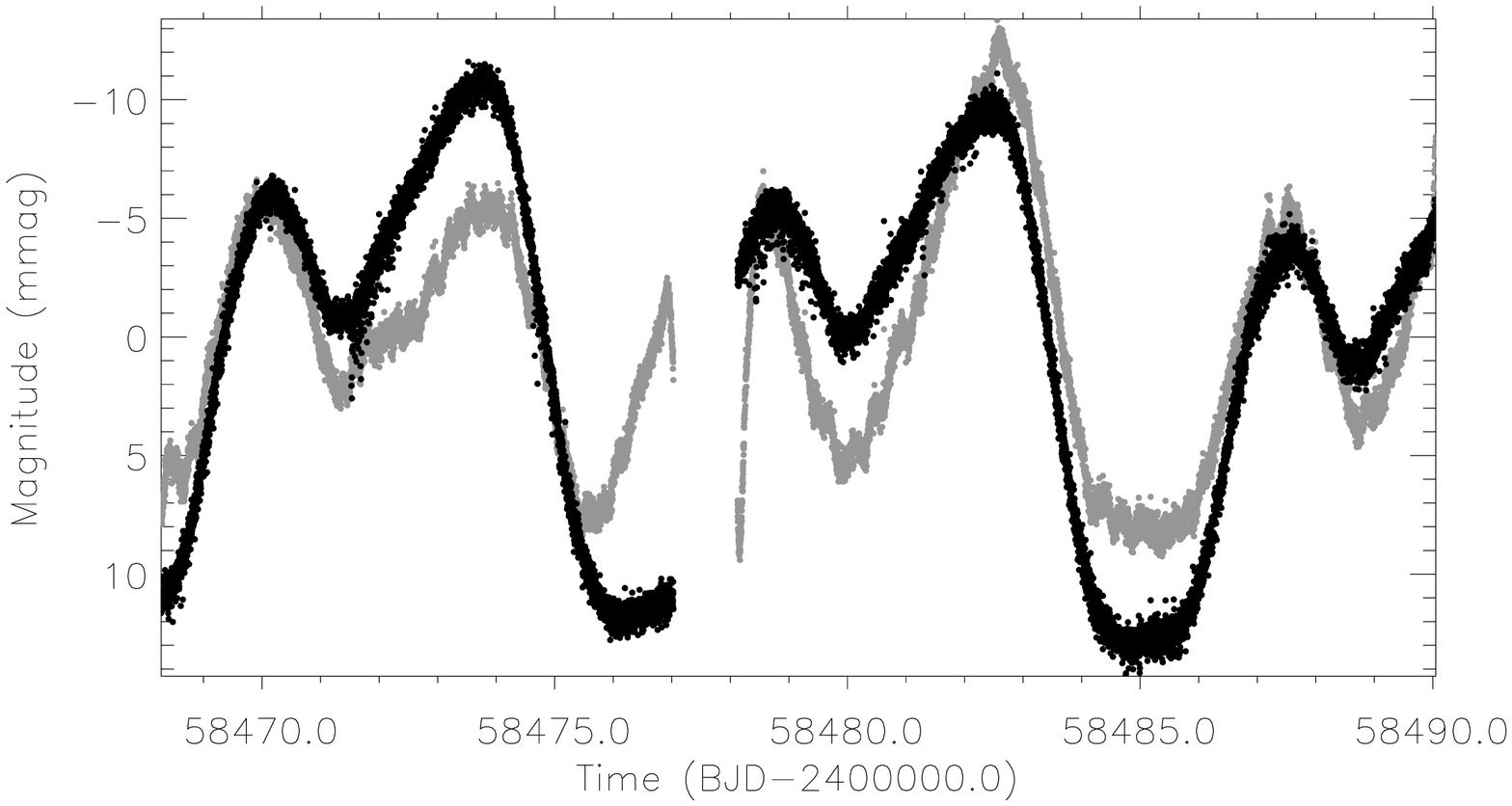}
      \includegraphics[width=\columnwidth]{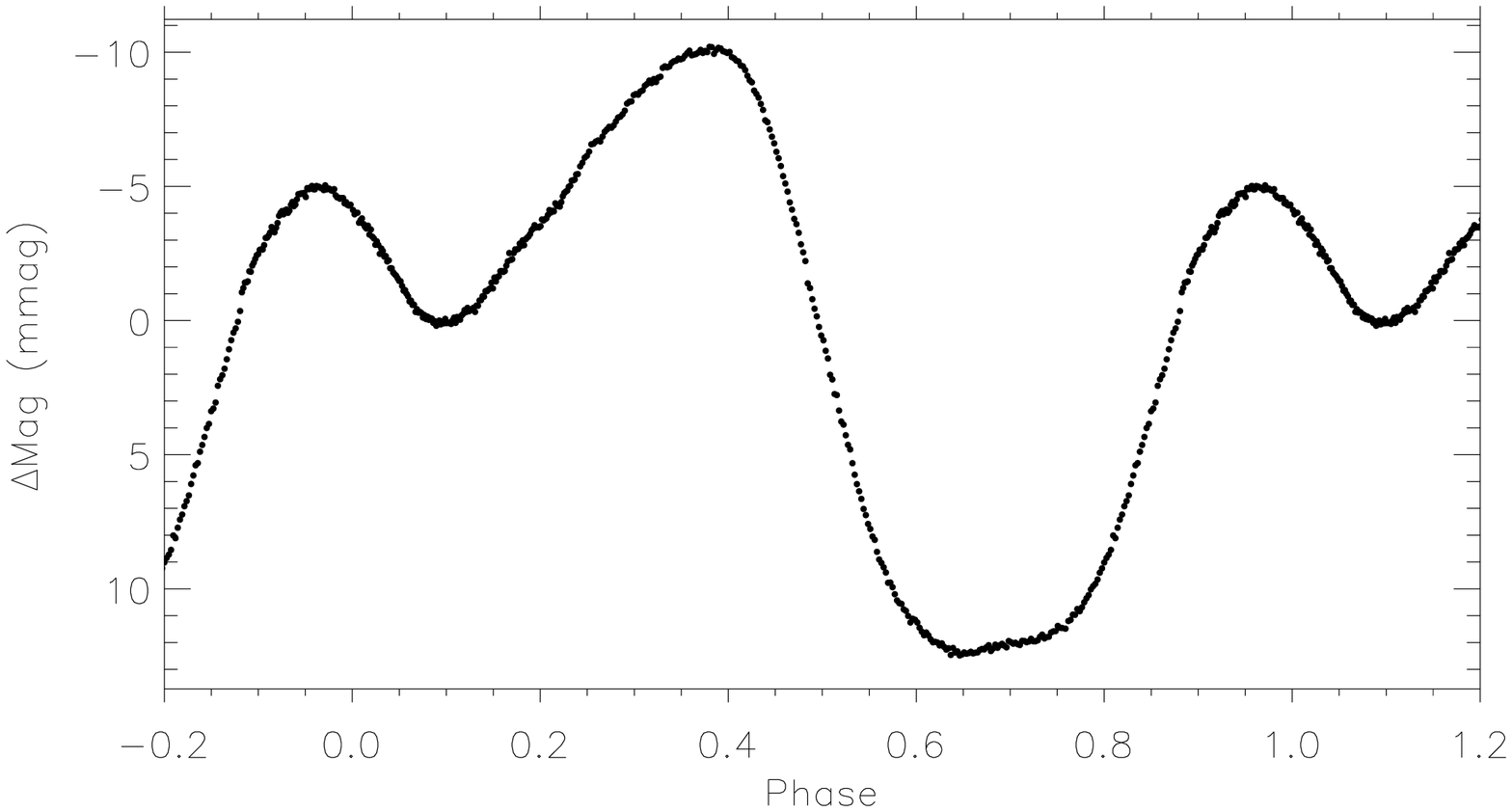}
    \includegraphics[width=\columnwidth]{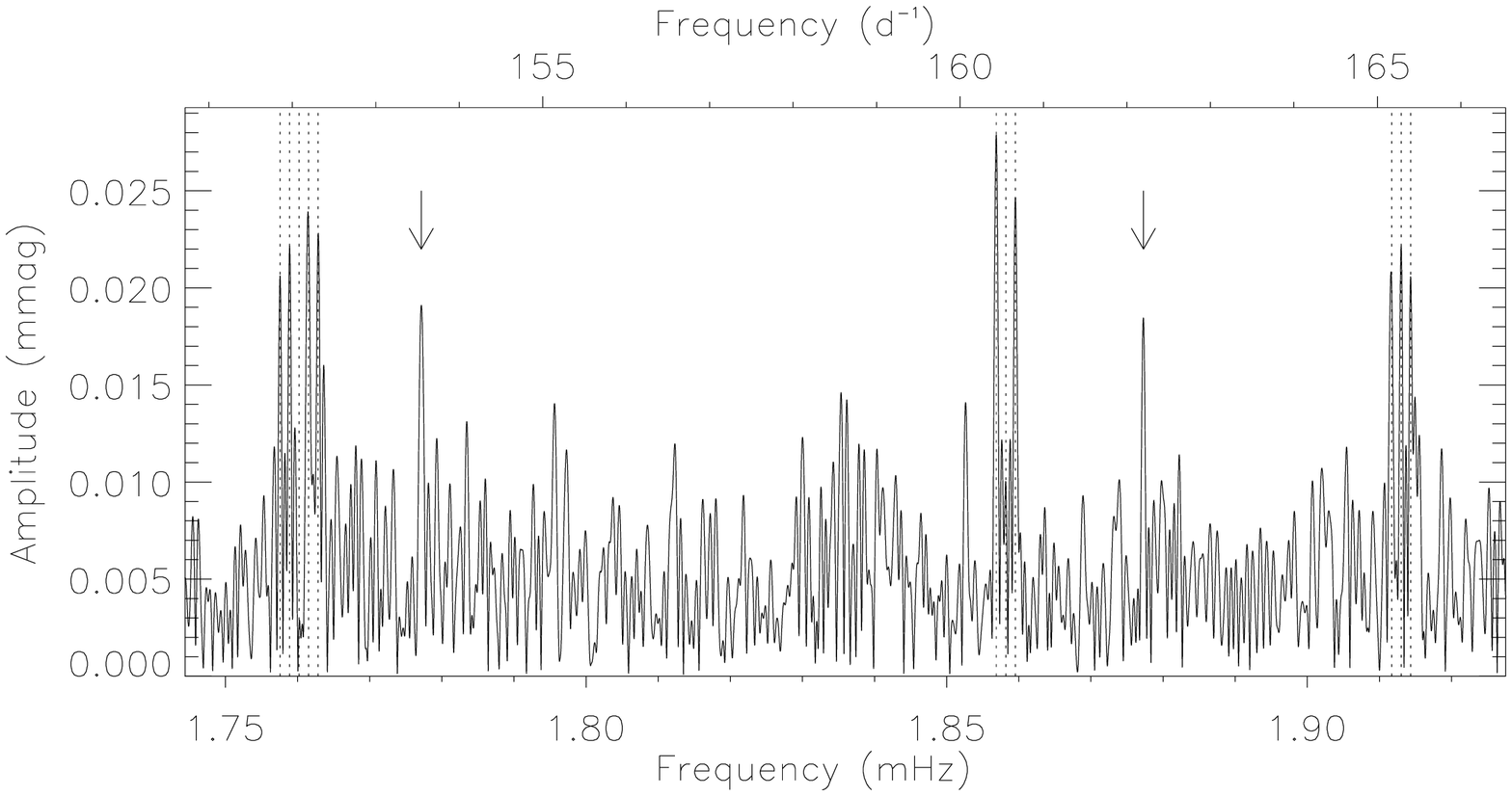}
    \caption{Top: comparison of the PDC\_SAP (grey) and AP (black) light curves for TIC\,176516923. Middle: SAP light curve phase folded on the derived rotation period of 8.7821\,d. Bottom: amplitude spectrum of the PDC\_SAP light curve showing the pulsational variability. The vertical dotted lines indicate the multiplet components, with the arrows indicating possible further modes.}
    \label{fig:tic176516923}
\end{figure}

TIC\,176516923 was observed by TESS in sector 6 only. The data for this star provides an example where the SPOC pipeline has interfered with the astrophysical signal of the rotation signature, and injected noise into the light curve. We demonstrate this in Fig.\,\ref{fig:tic176516923} where the top panel compares the PDC\_SAP (grey) data to the SAP data (black). Using the SAP data, we extract a rotation period of $8.782\pm0.003$\,d.

The pulsation spectrum of this star consists of a quadrupole quintuplet with a missing central component, a dipole triplet also with a missing central component, and a further dipole triplet with a significant central component. Furthermore, there are two peaks with significant amplitude (labelled by the arrows in Fig.\,\ref{fig:tic176516923}) that do not show any multiplet components. The pulsation modes are found at $1.76019\pm0.00005$\,mHz ($152.080\pm0.004$\,\cd), $1.85820\pm0.00004$\,mHz ($160.548\pm0.003$\,\cd), and $1.91302\pm0.00005$\,mHz ($165.285\pm0.004$\,\cd).

The difference in the multiplet structures is puzzling in this star. Although different modes have different depth pulsation cavities, one would expect both dipole modes to show very similar relative amplitudes since they are governed by the geometry of the star through $i$, the inclination angle, and $\beta$, the angle of magnetic obliquity. This difference may suggest different pulsation axes for the two dipole modes (as suggested for KIC\,10195926; \citealt{2011MNRAS.414.2550K}).  However, it is likely that we are seeing a depth dependence on the geometry which is being affected by different limb darkening weighting due to the stratified atmosphere of the star, as is proposed to be the case for HD\,6532 \citep{2020ASSP...57..313K}, and previously observed with time-resolved spectroscopic studies that resolved the vertical structure of pulsations by studying lines with different formation heights \citep{2006A&A...446.1051K,2009MNRAS.396..325F}. Another option is that the difference may result from the modes suffering different magnetic distortions in the atmosphere \citep{2011MNRAS.414.2576S,2018MNRAS.480.1676Q}, due to their slightly different frequencies. Either way, the nature of these different multiplet structures is not clear, which clearly warrants a further, in-depth, study with additional multi-colour observations and detailed modelling.


\subsubsection{TIC\,178575480}

TIC\,178575480 (HD\,55852) has a spectral classification of Ap\,SrEuCr \citep{1999MSS...C05....0H} and according to the TIC has an effective temperature of 7720\,K. \citet{2017MNRAS.468.2745N} provide a rotation period of 4.775\,d, with \citet{1998A&AS..128..265M} providing Str\"omgren-Crawford photometric indices of $b-y=0.164$, $m_1=0.237$, $c_1=0.727$ and H$\upbeta=2.859$. 

\begin{figure}
    \centering
      \includegraphics[width=\columnwidth]{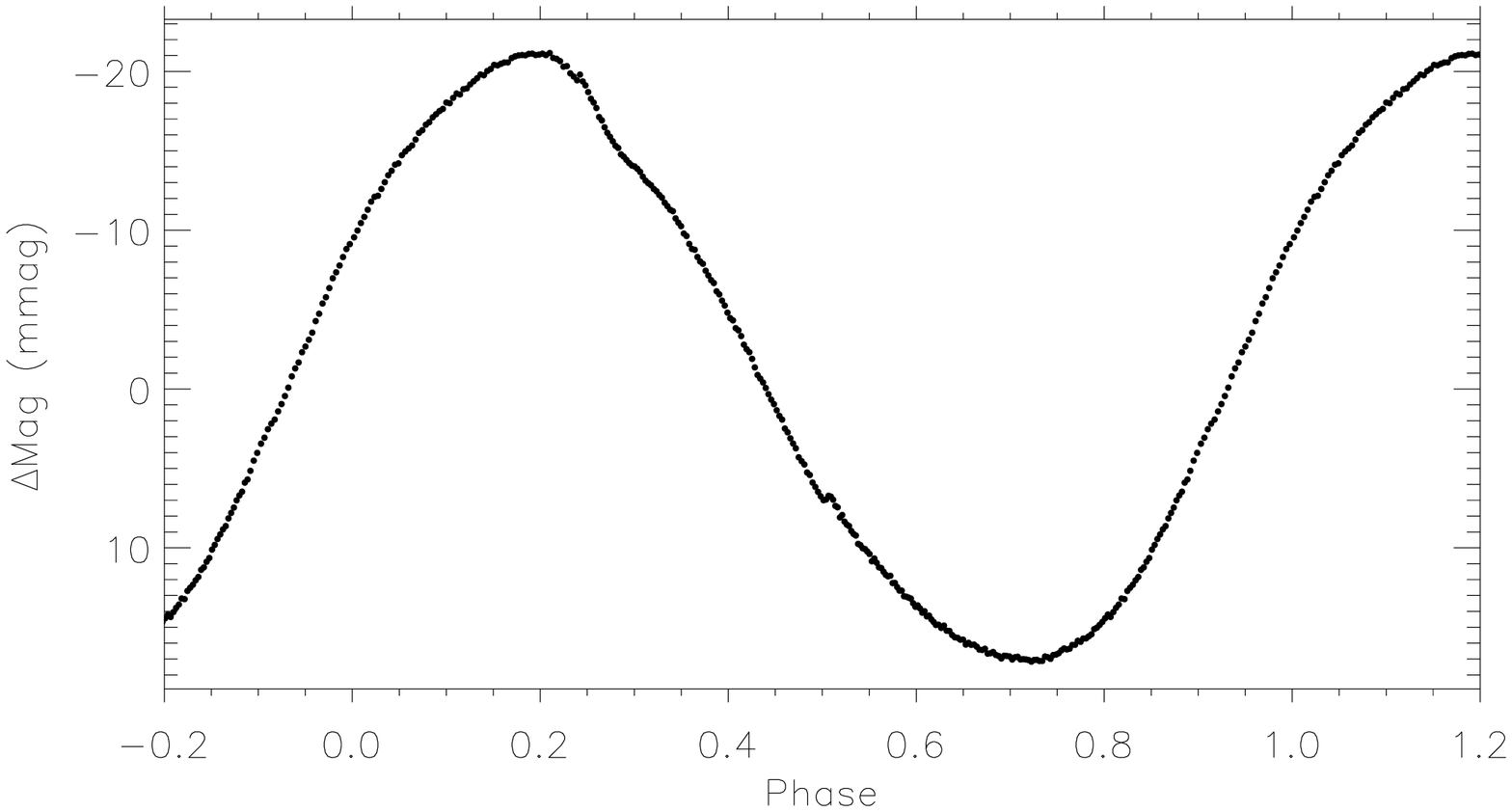}
    \includegraphics[width=\columnwidth]{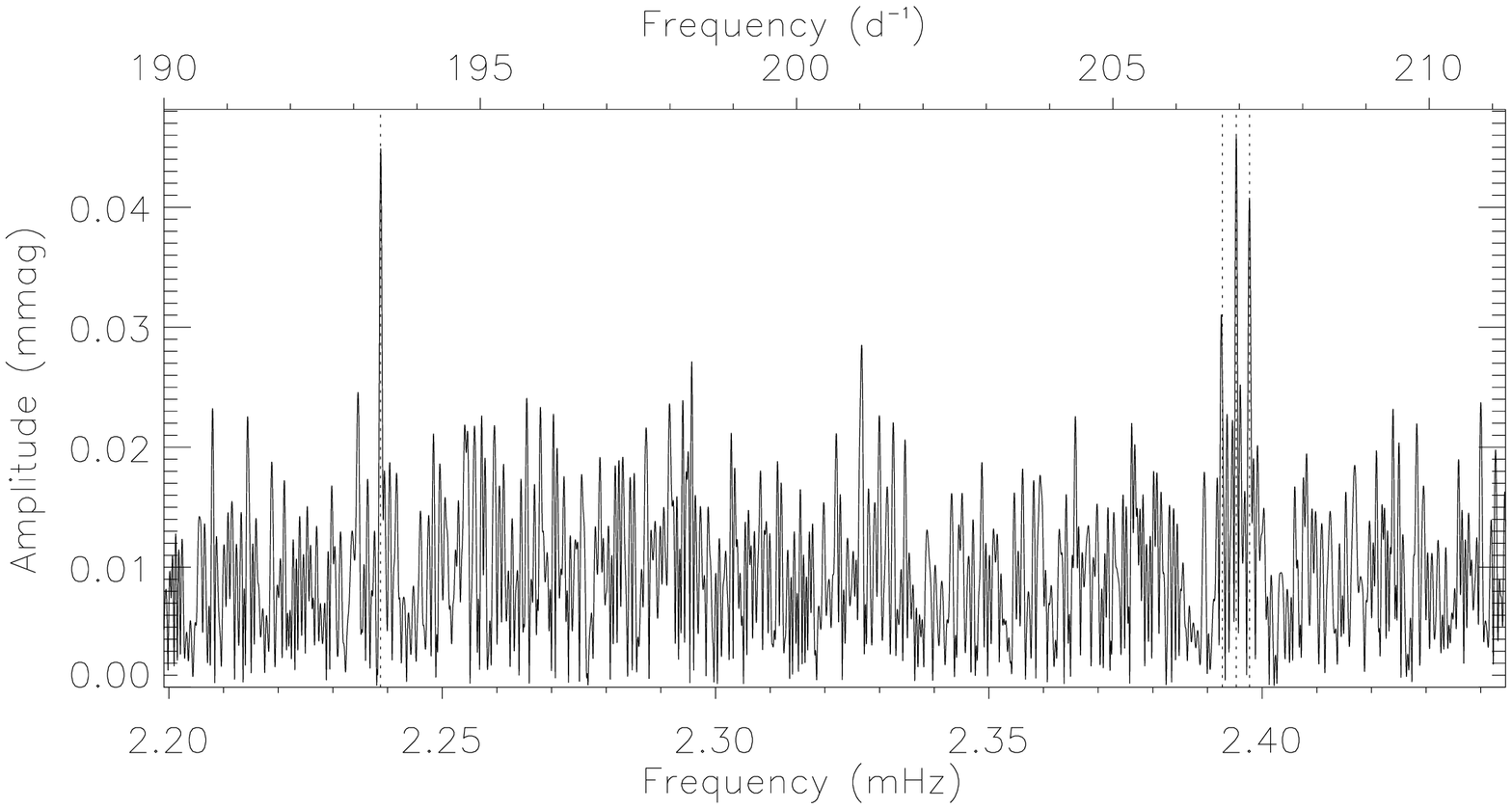}
    \caption{Top: the light curve of TIC\,178575480 phase folded on the derived rotation period of 4.7788\,d. Bottom: amplitude spectrum of the light curve showing the pulsational variability. The vertical dotted lines indicate the pulsation modes and the multiplet components of the high-frequency triplet.}
    \label{fig:tic178575480}
\end{figure}

This star was observed in sector 7, allowing us to measure a rotation period of $4.7788\pm0.0005$\,d. A phase folded light curve is shown in the top panel of Fig.\,\ref{fig:tic178575480}.

There are two pulsation modes seen in this star: a singlet at a frequency of $2.23870\pm0.00004$\,mHz ($193.424\pm0.004$\,\cd), and a triplet with a central frequency of $2.39524\pm0.00004$\,mHz ($206.949\pm0.003$\,\cd). Since the two modes show different multiplet structures, it is logical to assume that the modes are of different degree, however their identification is unclear. Given that the large frequency separation in roAp stars ranges between $\sim30-100\,\umu$Hz, the separation of the two modes, $\approx156.5\,\umu$Hz, is plausibly 1.5 times the large frequency separation. If this is the case, the two observed modes would be of different degree, as the different structures suggest. If the separation is twice the large frequency separation, the two modes seen would be of the same degree, with the same structure. We therefore suggest the large frequency separation in this star is about $100\,\umu$Hz, but note modelling of this star are required to confirm this.


\subsubsection{TIC\,294266638}
TIC\,294266638 (TYC\,6021-415-1) is a star with no mention in the literature, other than a {\it Gaia} parallax measurement of $3.018\pm0.017$\,mas, implying a distance of $331\pm2$\,pc \citep{2016A&A...595A...1G,2021A&A...649A...1G}.

We obtained a spectrum for this star with the SpUpNIC instrument to confirm its nature as an Ap star (Fig.\,\ref{fig:tic294266638_spec}). The spectrum shows enhancements of Sr and Eu leading to a classification of A7p\,SrEu. There are signs of Cr in the spectrum at 4111\,\AA, but no other lines of this element are clearly enhanced. 

\begin{figure}
    \centering
     \includegraphics[width=\columnwidth]{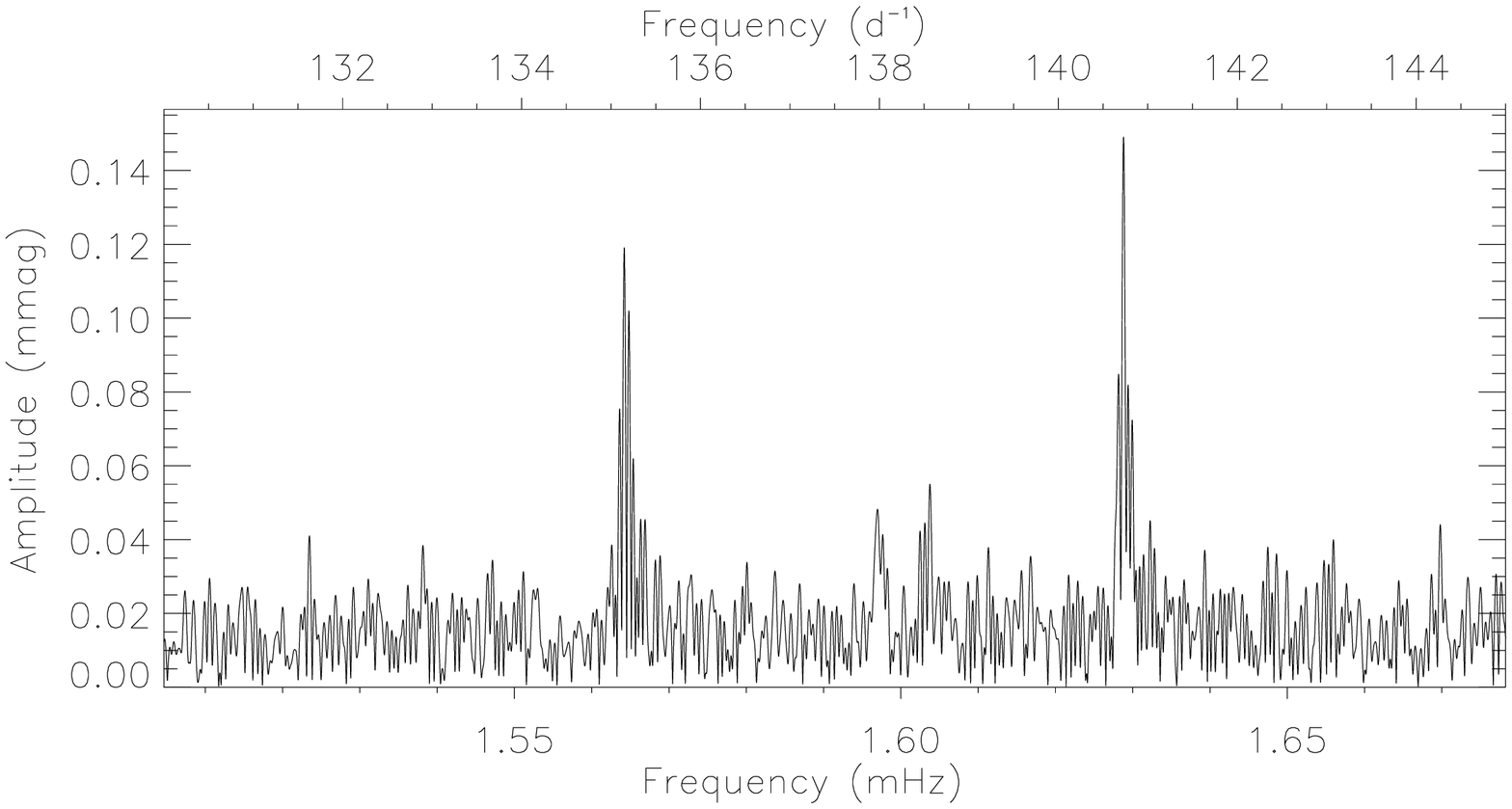}
    \includegraphics[width=\columnwidth]{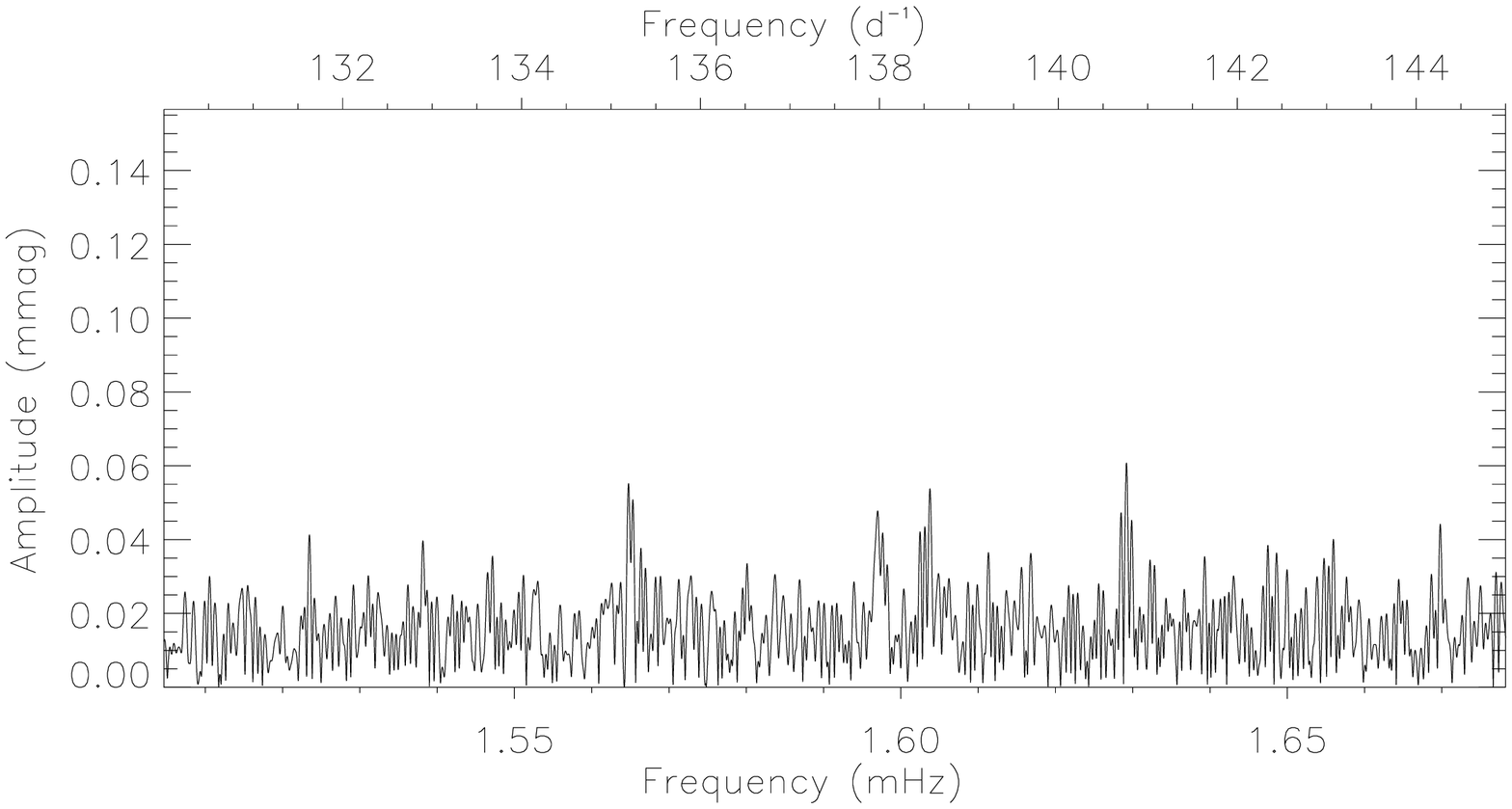}
    \caption{Top: pulsation modes in TIC\,294266638. Bottom: after removing the obvious pulsations, excess power is remaining which is suggestive of frequency variability.}
    \label{fig:tic294266638}
\end{figure}

This star was observed in sector 8. There is no evidence of rotational variability in the TESS light curve, however there are clear pulsation signals, as shown in Fig.\,\ref{fig:tic294266638}. We extract two pulsation modes from the amplitude spectrum, at $1.56421\pm0.00002$\,mHz ($135.148\pm0.002$\,\cd) and $1.62882\pm0.00002$\,mHz ($140.730\pm0.002$\,\cd). The separation of these two modes, $65\,\umu$Hz, is plausibly the large frequency separation for this star. After prewhitening these peaks, there is still evidence of excess power in the residual spectrum. This is indicative of phase/amplitude variability \citep[e.g.,][]{10.3389/fspas.2021.626398}. Furthermore, there is evidence of two additional modes at low amplitude around 1.6\,mHz which are significant. However, given the broad nature of the peaks, a clear frequency value cannot be determined.

\subsubsection{TIC\,294769049}

TIC\,294769049 (HD\,161423) was classified as Ap\,SrEu(Cr) by \citet{1975mcts.book.....H}, and listed in the catalogue of \citet{2009A&A...498..961R} with a spectral type of A2\,SrEu. TIC\,294769049 has been the target in two searches for rapid variability; both \citet{1994MNRAS.271..129M} and \citet{2016A&A...590A.116J} returned null results.

\begin{figure}
    \centering
     \includegraphics[width=\columnwidth]{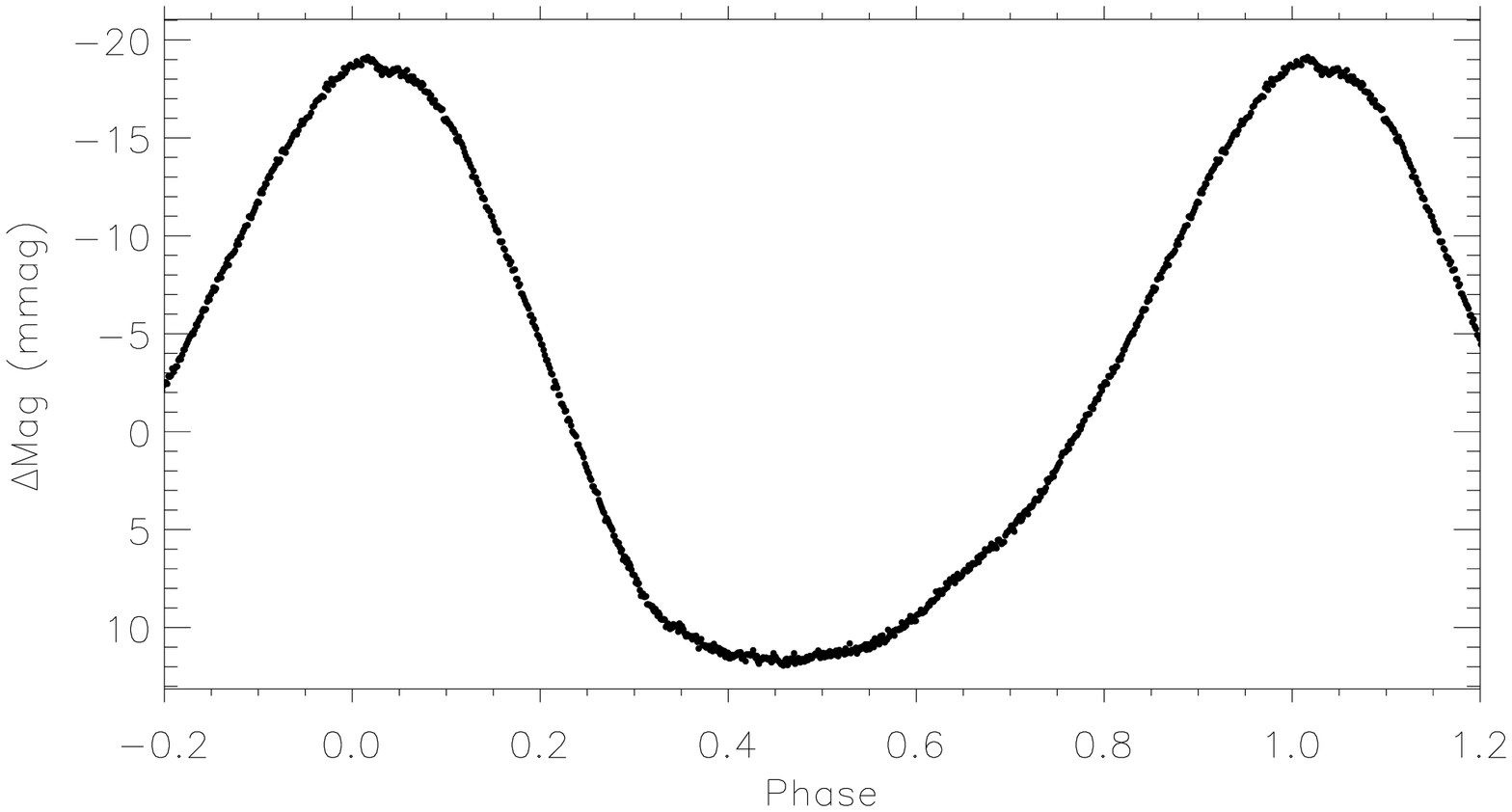}
    \includegraphics[width=\columnwidth]{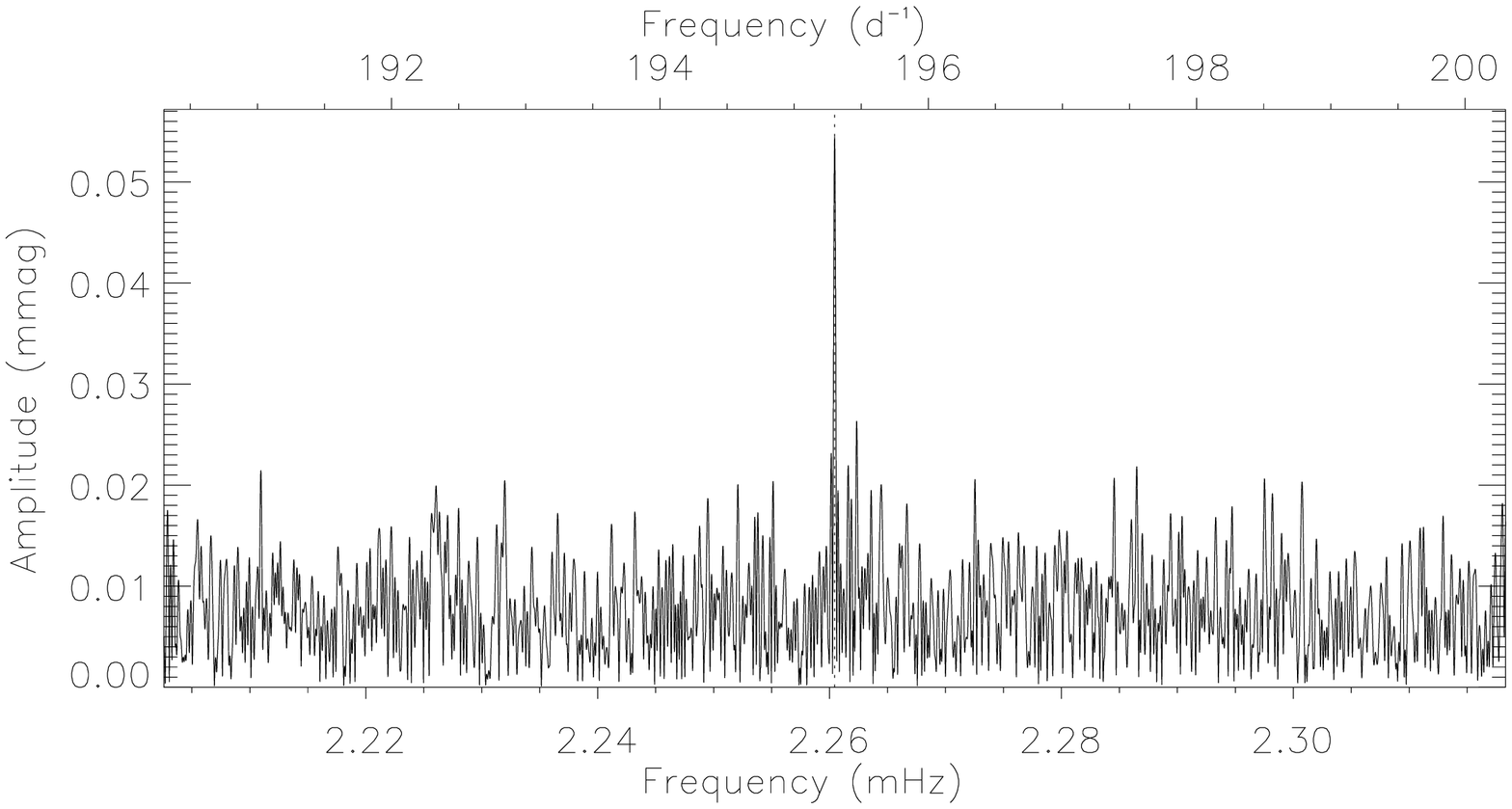}
    \caption{Top: phase folded light curve of TIC\,294769049, folded on a period of 10.4641\,d. Bottom: amplitude spectrum showing the only pulsation mode in this star, and the lack of a rotationally split multiplet.}
    \label{fig:tic294769049}
\end{figure}

TESS observed this star during sectors 12 and 13, providing a light curve with a time base of $57.4$\,d. From this we measure a rotation period of $10.4641\pm0.0007$\,d, and show a phase folded light curve in Fig.\,\ref{fig:tic294769049}. Analysis of the light curve at high frequency shows a single pulsation mode at $2.26044\pm0.00001$\,mHz ($195.302\pm0.001$\,\cd; Fig.\,\ref{fig:tic294769049}). The lack of an obliquely split multiplet suggests that either the pulsation axis is aligned with the rotation axis (thus $\beta=0$, assuming the original oblique pulsator model), despite the significant rotational variation, or the mode is an undistorted radial mode. The former would imply that the spots causing the mean light variation are not close to the pulsation pole \citep{2004A&A...424..935K,2010A&A...516A..53A}, or that the pulsation axis is aligned with the rotation axis, rather than the magnetic axis \citep{2002A&A...391..235B}.

\subsubsection{TIC\,310817678}

TIC\,310817678 (HD\,88507) was classified as Ap\,SrEu(Cr) by \citet{1978mcts.book.....H} and in the catalogue of \citet{2009A&A...498..961R} it is quoted as an Ap\,SrEu star. There is only one evaluation of the atmospheric parameters given by the TIC: $T_{\rm eff} = 8230$\,K and $\log g = 4.08$\,\cms.

\begin{figure}
    \centering
     \includegraphics[width=\columnwidth]{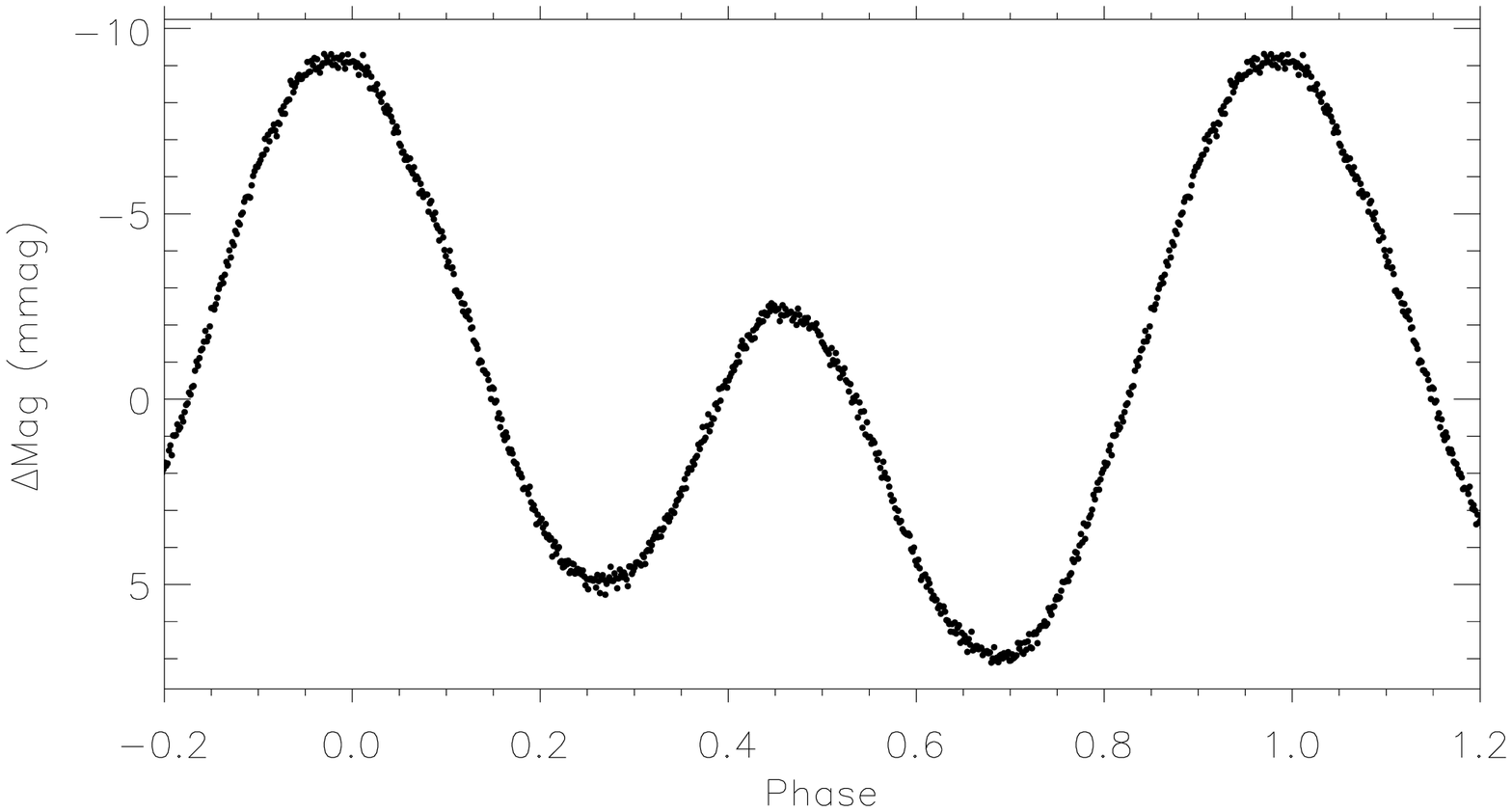}
    \includegraphics[width=\columnwidth]{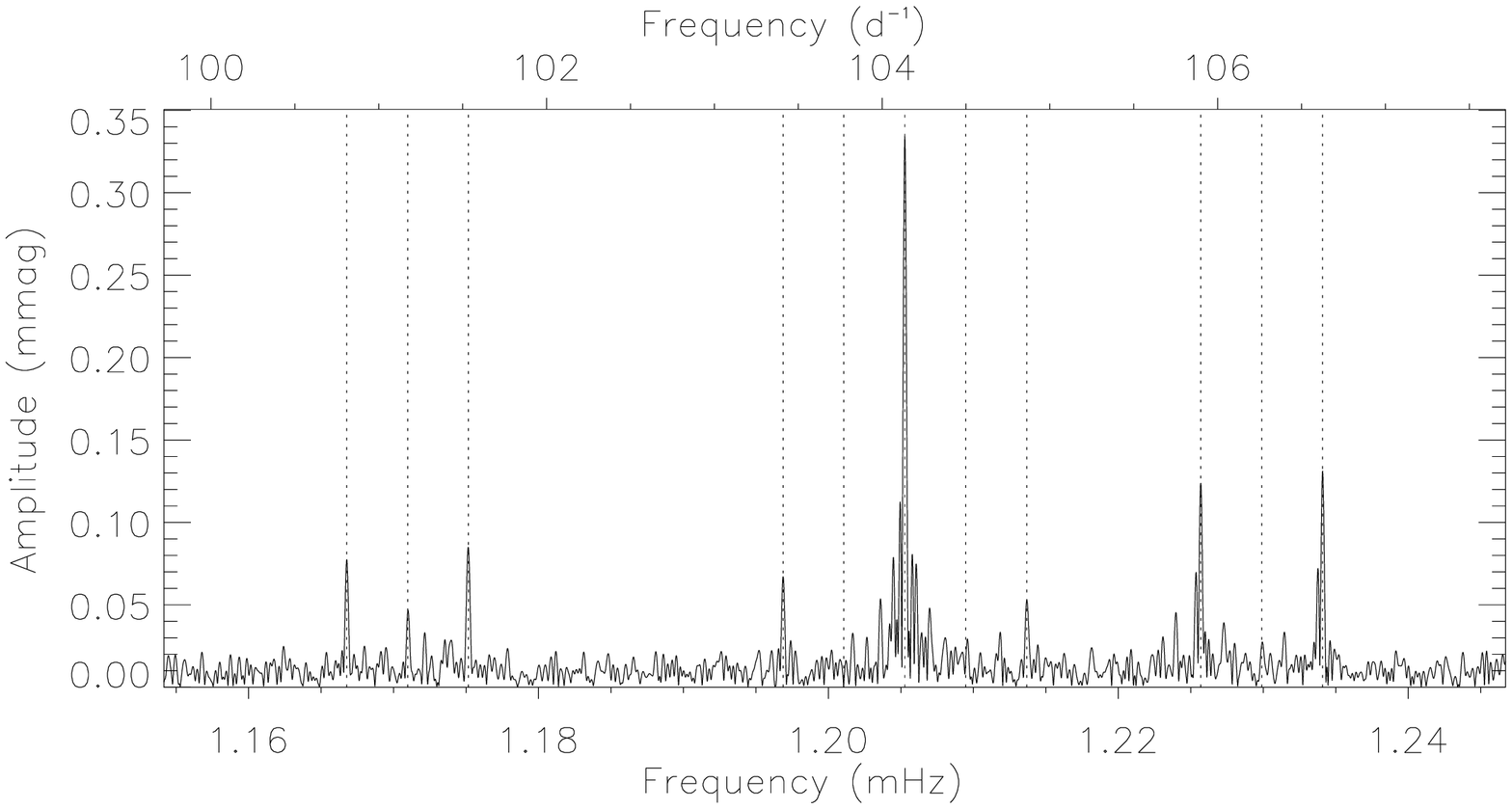}
    \caption{Top: phase folded light curve of TIC\,310817678, folded on a rotation period of $2.75003$\,d. Bottom: amplitude spectrum showing the rich pulsation pattern in this star. The vertical dotted lines denote the frequencies of rotationally split multiplet components.}
    \label{fig:tic310817678}
\end{figure}

TESS observed this star in sectors 9 and 10, providing a 51.8\,d time base. We measure a rotation period of $2.75003\pm0.00008$\,d (Fig.\,\ref{fig:tic310817678}). The pulsation spectrum in this star is rich, consisting of a dipole triplet centred on $1.17098\pm0.00002$\,mHz ($101.173\pm0.002$\,\cd), a quadrupole quintuplet with missing $\nu_{\rm rot}$ sidelobes centred on $1.205272\pm0.000003$\,mHz ($104.1355\pm0.0002$\,\cd), and another dipole with a central frequency at $1.22989\pm0.00003$\,mHz ($106.263\pm0.003$\,\cd). The two dipole modes are separated by $58.9\,\umu$Hz which is plausibly the large frequency separation for this star. \citet{Kobzar2020} identified this star as a candidate roAp star.

The high-frequency dipole sidelobes are unlike those seen in the other multiplets: they form doublets. We suspect that this is a signature of magnetic distortion since despite the sidelobes having greater amplitude than the low-frequency dipole mode, the central component is entirely absent. This star is another example of how the multiplet structures vary for the same degree mode.

\subsubsection{TIC\,356088697}

TIC\,356088697 (HD\,76460) was classified as an Ap\,Sr by \citet{1975mcts.book.....H} with \citet{2012MNRAS.420.2727E} measuring a magnetic field strength of $3.6$\,kG through partial Zeeman splitting of Fe lines. The authors also determined the effective temperature to be $7200$\,K and $v\sin i= 3$\,\kms, although this value is an upper limit due to the resolution limit of their data.

Observed in three consecutive TESS sectors (9-11), this star shows no signature of rotation. To confirm its Ap nature, we obtained a spectrum with SpUpNIC and derived a spectral class of A8p\,SrEuCr (Fig.\,\ref{fig:tic356088697spec}).

\begin{figure}
    \centering
     \includegraphics[width=\columnwidth]{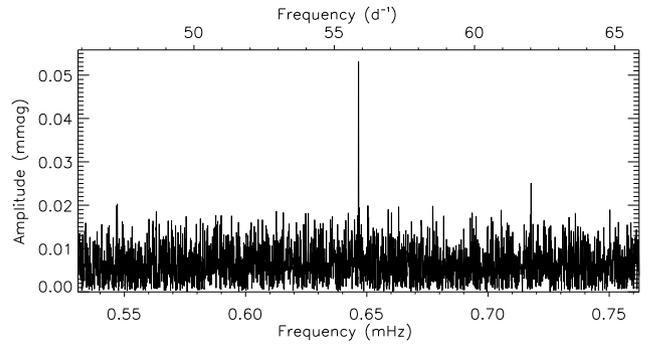}
    \caption{Amplitude spectrum of TIC\,356088697 showing the single pulsation mode in this star. }
    \label{fig:tic356088697}
\end{figure}

A single pulsation signal is found in this star at a frequency of $0.646584\pm0.000008$\,mHz ($55.8649\pm0.0007$\,\cd) with an amplitude of $0.053\pm0.006$\,mmag (Fig.\,\ref{fig:tic356088697}), which was also noted by \citet{2020A&A...639A..31M}. This mode does not show any amplitude or phase variability over the observing window, which would be consistent with a rotation period much longer than the observing window \citep{2020A&A...639A..31M}. The single mode in this star is at a frequency commonly associated with the $\delta$\,Sct stars \citep{2018MNRAS.476.3169B}, but it is uncommon to find just a single frequency. Therefore, with just a single mode and an Ap stellar classification, this star can be classed as an roAp star, making it the longest period roAp star, after HD\,177765, which was discovered by \citet{2012MNRAS.421L..82A} with the mode frequency later refined to be $0.702580\pm0.000006$\,mHz \citep[$60.7029\pm0.0005$\,\cd;][]{2016IBVS.6185....1H}. Given that the mode frequency overlaps with the $\delta$\,Sct frequency range, modelling of the pulsation is needed to confirm the overtone is representative of an roAp pulsation.

\subsubsection{TIC\,380651050}

TIC\,380651050 (HD\,176384) has no in-depth studies in the literature. \citet{2020ApJS..247...11R} provide a mass of $1.25$\,M$_\odot$ and $T_{\rm eff}=6531$\,K, with \citet{2015A&A...580A..23P} giving Str\"omgren-Crawford photometric indices of $b-y=0.285$, $m_1=0.162$, $c_1=0.446$ and H$\upbeta=2.679$, which are consistent with a cool Ap star. 

\begin{figure}
    \centering
     \includegraphics[width=\columnwidth]{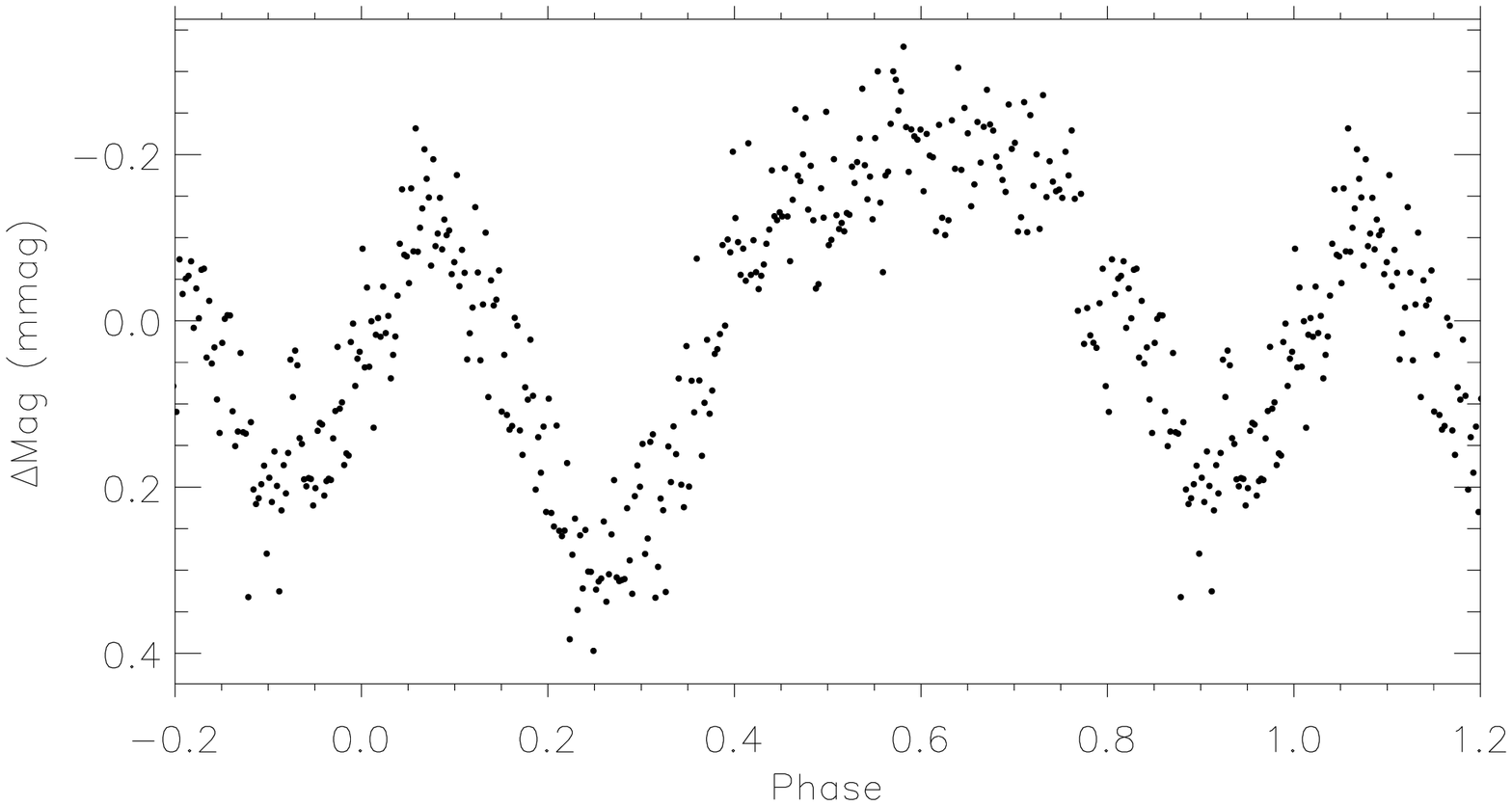}
      \includegraphics[width=\columnwidth]{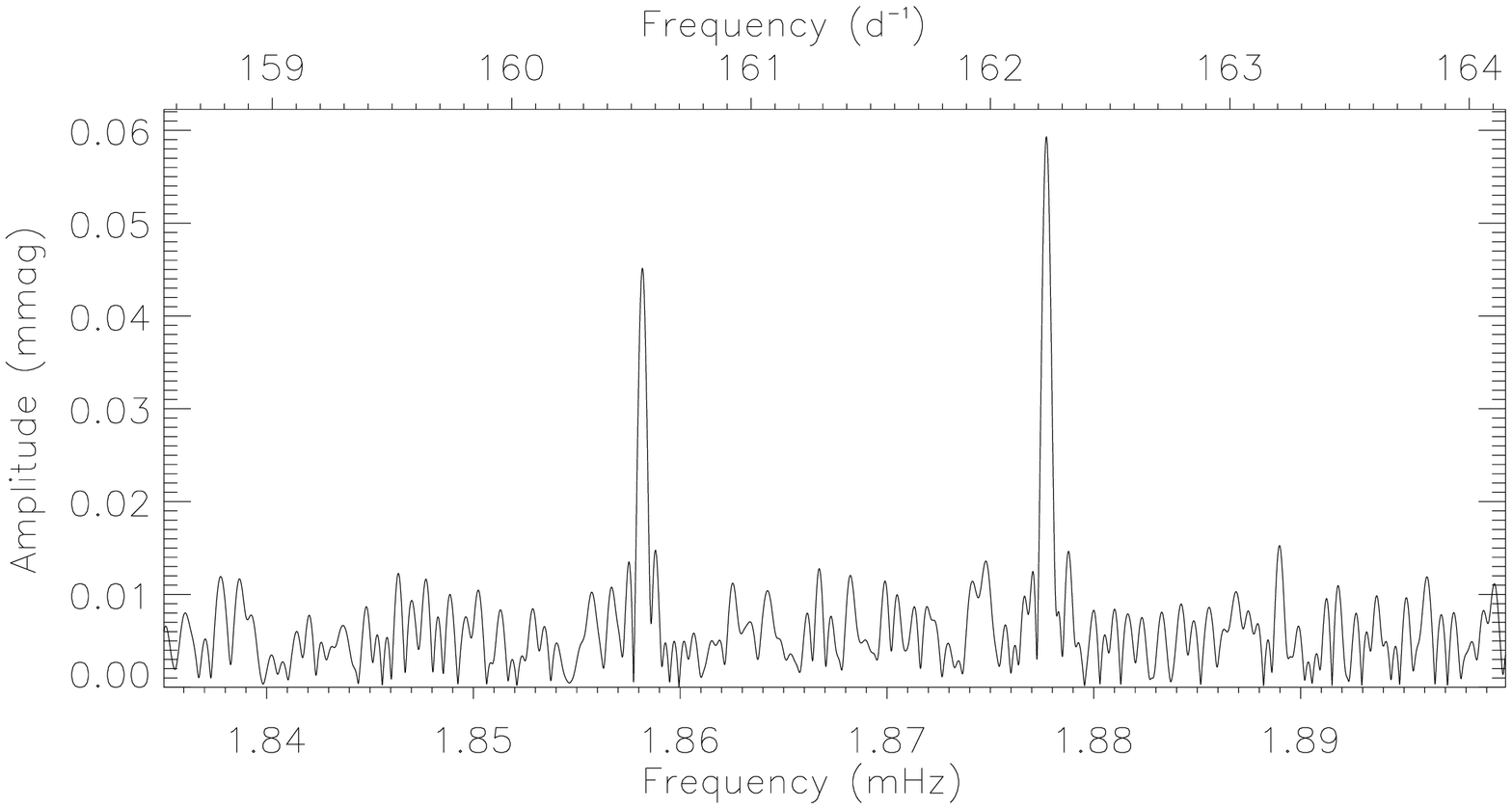}
    \caption{Top: phase folded light curve on the assumed rotation period of 4.189\,d. The significant scatter is a result of a second significant low frequency. Bottom: amplitude spectrum of TIC\,380651050 showing the two pulsation modes in this star.}
    \label{fig:tic380651050}
\end{figure}

TIC\,380651050 was observed in sector 13. There are two harmonic series at low frequency which both give plausible rotation periods; they are: $4.19\pm0.01$\,d and $3.51\pm0.02$\,d. We take the highest amplitude signal to be the rotation frequency (Fig.\,\ref{fig:tic380651050}), but suggest further observations are obtained to corroborate this. One of the two periods may be the result of binarity or contamination by a background star; spectroscopic observations will provide insight into this.

Two high frequency pulsation modes are found in this star: $1.85817\pm0.00002$\,mHz ($160.546\pm0.002$\,\cd) and $1.87771\pm0.00002$\,mHz ($162.234\pm0.002$\,\cd) which are separated by $19.53\pm0.03$\,$\umu$Hz (Fig.\,\ref{fig:tic380651050}). This separation is plausibly half of the large frequency separation, and would imply the that the star is quite evolved, possibly beyond the main sequence. This star was also identified in \citet{Kobzar2020} as a candidate roAp star.


\subsubsection{TIC\,387115314}

TIC\,387115314 (TYC\,9462-347-1) has a spectral classification of A5 \citep{1954AnCap..17.....J}, and has atmospheric parameters $T_{\rm eff} = 7765$\,K, surface gravity $\log g = 4.08$\,\cms, and metallicity [M/H]$\ = 0.18$ determined by \citet{2013AJ....146..134K} on the basis of RAdial Velocity Experiment (RAVE) data. 

\begin{figure}
    \centering
     \includegraphics[width=\columnwidth]{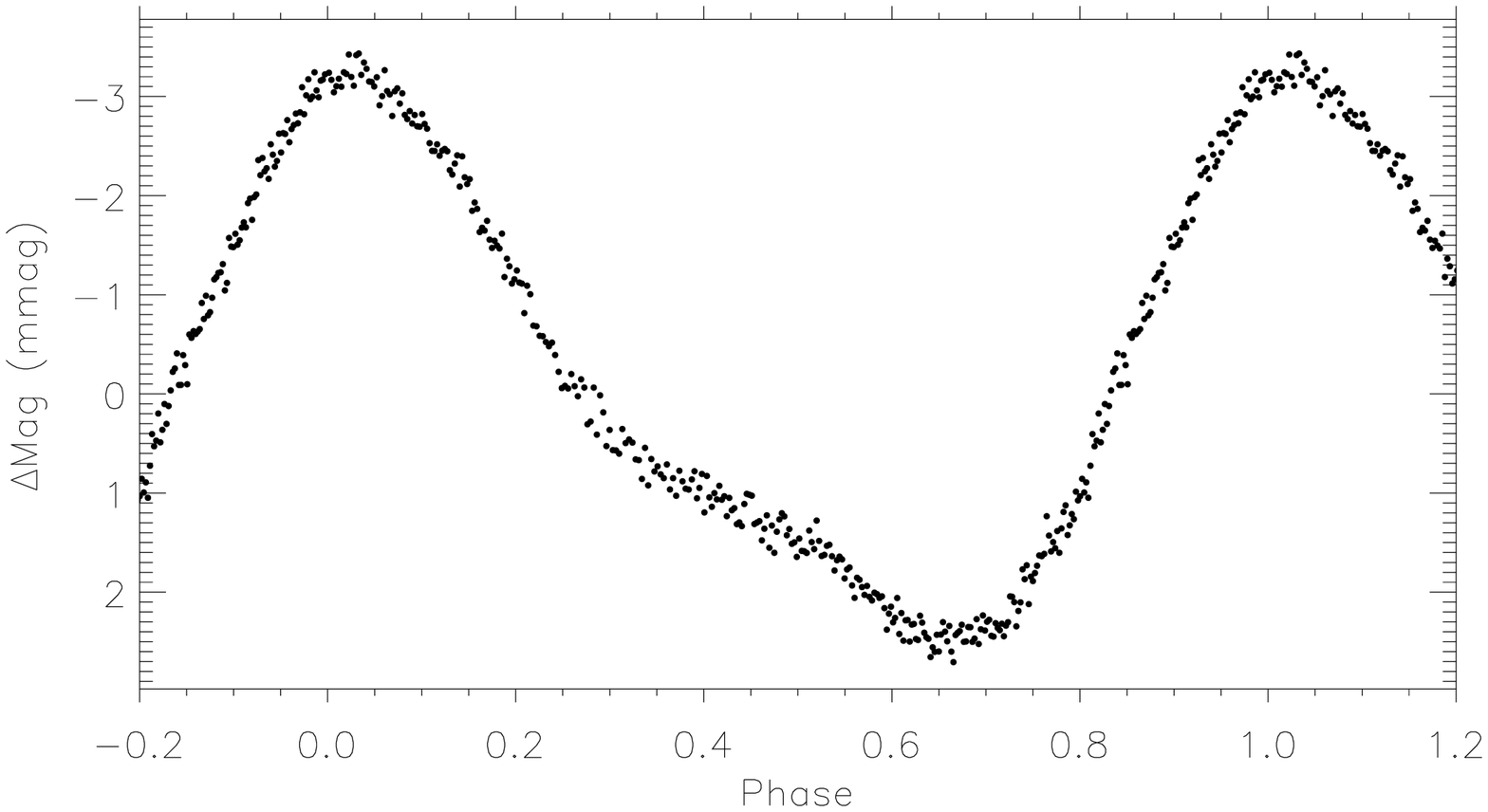}
      \includegraphics[width=\columnwidth]{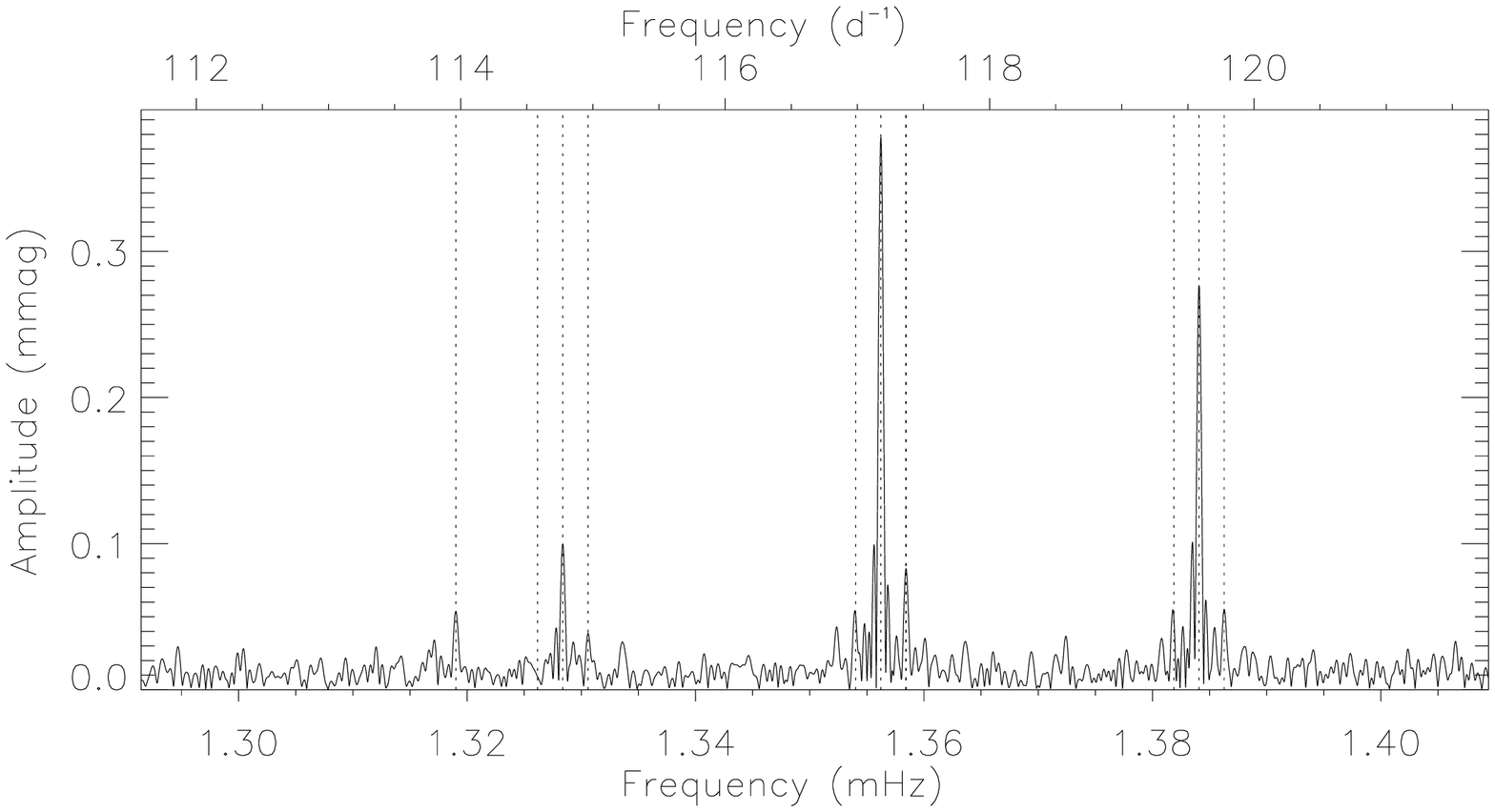}
    \caption{Top: phase folded light curve of TIC\,387115314, folded on the rotation period of 5.264\,d. Bottom: amplitude spectrum showing the pulsation modes in this star. }
    \label{fig:tic387115314}
\end{figure}

The star was observed in sector 13, from which we derive a rotation period of $5.264\pm0.002$\,d (Fig.\,\ref{fig:tic387115314}). We find a rich pulsation spectrum for this star at high frequencies, also seen by \citet{Kobzar2020}. Although we have a well determined rotation period, and significant mean light variations, the modes do not show significant rotationally split sidelobes which implies we see mostly one pulsation pole over the entire rotation period. The pulsation frequencies that we identify are: a singlet at $1.31902\pm0.00004$\,mHz ($113.963\pm0.003$\,\cd), two components of a triplet with the mode at $1.32838\pm0.00002$\,mHz ($114.772\pm0.002$\,\cd), a triplet centred on $1.356225\pm0.000006$\,mHz ($117.1778\pm0.0005$\,\cd) and another triplet centred on $1.384083\pm0.000007$\,mHz ($119.5848\pm0.0006$\,\cd). These modes, with their multiplet components, are shown in Fig.\,\ref{fig:tic387115314}.

The separation of the three highest modes is $\sim27.85$\,$\umu$Hz, with the lowest frequency two modes separated by $9.36$\,$\umu$Hz. Since we cannot determine the degree of the modes showing multiplets, due to the low amplitude of the obliquely split sidelobes, we are unable to determine if the separation of $\sim27.85$\,$\umu$Hz is the large frequency separation, or half of it. The separation of $9.36$\,$\umu$Hz is plausibly the small frequency separation for this star.


\subsubsection{TIC\,466260580}

TIC\,466260580 (TYC\,9087-1516-1) is a relatively obscure star with no previous reports of any variability. The RAVE survey \citep{2017AJ....153...75K} obtained $T_{\rm eff}=6780\pm170$\,K, $\log g=4.03\pm0.27$\,\cms\ and [M/H] $=-0.7\pm0.2$ which are consistent with the TIC values, although the metallicity is strikingly low for an Ap star. These suggest the star is spectral type F3, which is consistent with the Tycho $B-V=0.40\pm0.07$. However, the $T_{\rm eff}$ is uncertain; for example, \cite{2017MNRAS.471..770M} obtained $T_{\rm eff} = 6420\pm170$\,K from SED fitting, while \citet{2018A&A...616A...1G} give a value around 7000\,K.

We have also obtained a spectrum of this star with the HRS on SALT. From this, we obtain $T_{\rm eff}=6800\pm200$\,K, $\log g=4.0\pm0.2$\,\cms\ and [Fe/H] $=-0.6\pm0.2$, which are consistent with the RAVE results. We estimate a spectral type of Ap\,EuCr, and note that rare earth elements are enhanced in the region around 6150\,\AA, confirming the star to be an Ap star. A full abundance analysis is beyond the scope of this work, but is required to address the low metallicity in this star.

\begin{figure}
    \centering
     \includegraphics[width=\columnwidth]{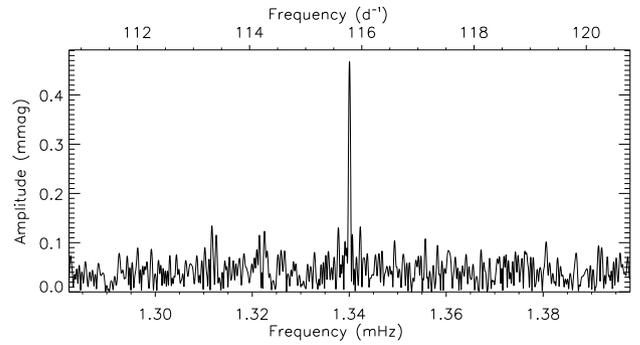}
    \caption{Amplitude spectrum showing the pulsation mode in TIC\,466260580.}
    \label{fig:tic466260580}
\end{figure}

The TESS data, obtained in sector 13, do not show clear signs of rotational variability, but exhibit many low-frequency variations. The pulsation signal is clear in this star, at a frequency of $1.34006\pm0.00002$\.mHz ($115.781\pm0.002$\,\cd; Fig.\,\ref{fig:tic466260580}).

\subsection{roAp stars previously discovered by TESS} 
\subsubsection{TIC\,12968953}
TIC\,12968953 (HD\,217704)  star was observed in TESS sector 2 and was analysed by \citet{2019MNRAS.487.3523C}. We find no significant differences between our analysis here and that previously presented. We refer the reader to the aforementioned publication for the details on TIC\,12968953. However, we obtained a new spectrum of this target. TIC\,12968953 was classified as Ap\,Sr by \citet{ 1982mcts.book.....H}; with our SAAO spectrum (Fig.\,\ref{fig:tic12968953spec}), we classify this star as A7p\,SrEuCr, and determine a temperature of $T_{\rm eff} =7800\pm200$\,K through comparison with synthetic spectra.

\subsubsection{TIC\,17676722}

TIC\,17676722 (HD\,63773) was observed in sector 7, and reported as an roAp star by \citet{2019MNRAS.487.2117B}.  We present a revised rotation period of $1.5995\pm0.0003$\,d, which was derived by fitting a 10-component harmonic series to the data. As in \citet{2019MNRAS.487.2117B} we find four significant peaks which we interpret to be a single quadrupole pulsation mode ($1.941545\pm0.000002$\,mHz; $167.7495\pm0.0002$\,\cd) and 3 sidelobes split by the rotation frequency of the star. 

This star has been classified as A2 \citep{1993yCat.3135....0C}, however our SAAO classification spectrum shows the star to be A3p\,SrEuCr (Fig.\,\ref{fig:tic17676722spec}).

\subsubsection{TIC\,41259805}

\begin{figure}
    \centering
     \includegraphics[width=\columnwidth]{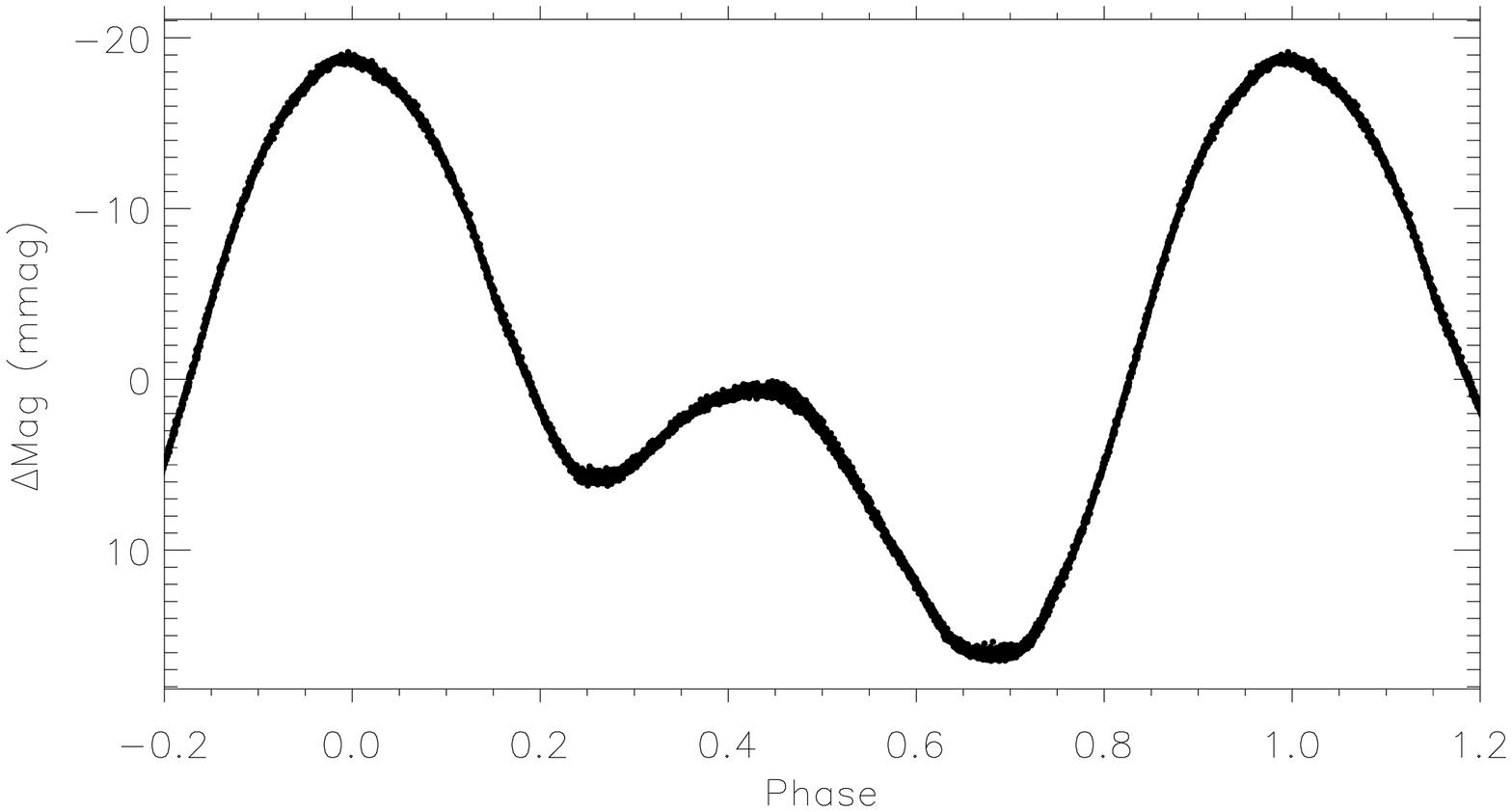}
     \includegraphics[width=\columnwidth]{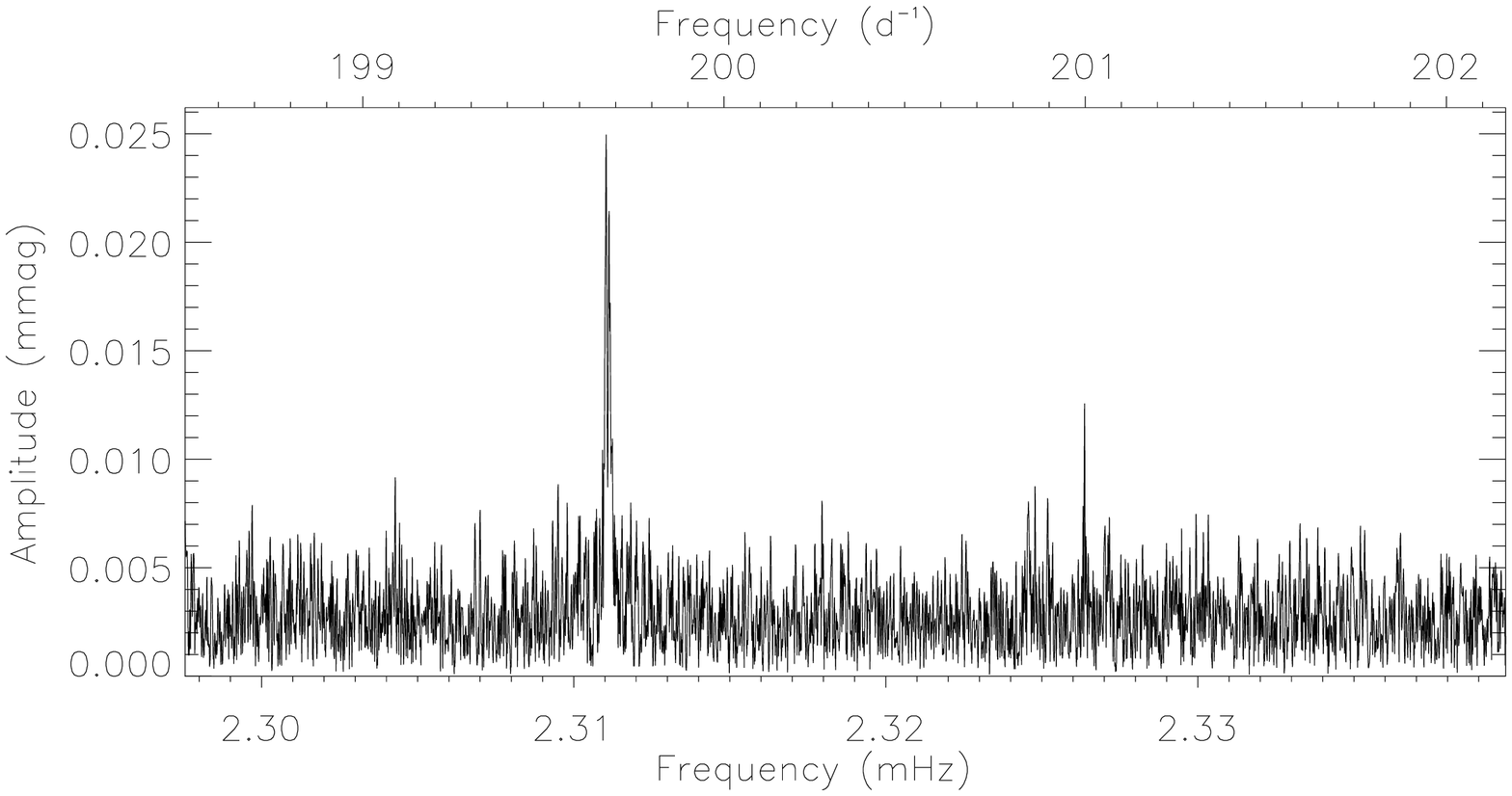}
    \caption{Top: Phase folded light curve of TIC\,41259805, folded on a period of 1.714489\,d. Bottom: the two pulsation modes found in this star, with the lower frequency mode showing evidence of frequency variability.}
    \label{fig:tic41259805}
\end{figure}

TIC\,41259805 (HD\,43226) was reported by both \citet{2019MNRAS.487.3523C} and \citet{2019MNRAS.487.2117B} to be an roAp star, based on 7 sectors of data. There are now 12 sectors of data available for this star from Cycle\,1, with no sector 9 data. With this extended time base, we are able to report a more precise rotation period of $1.714489\pm0.000002$\,d; Fig.\,\ref{fig:tic41259805}).  

There is clear frequency variability in this star, as seen by the ragged nature of the pulsation mode at $2.311039\pm0.000001$\,mHz ($199.6738\pm0.0001$\,\cd) in Fig.\,\ref{fig:tic41259805}, but not in the mode at $2.326368\pm0.000003$\,mHz ($200.9982\pm0.0003$\,\cd). The transient nature of the modes explains the presence of the $\nu-\nu_{\rm rot}$ peak found by \citet{2019MNRAS.487.3523C}, and the lack of its detection by \citet{2019MNRAS.487.2117B}. We note that here the amplitude is about half that reported in \citet{2019MNRAS.487.3523C} as a result of the power being spread over a broad frequency range. Since the frequency variability is only seen in one mode, it is likely to be intrinsic to the star, rather than caused by binary motion, for example. This type of frequency variability seems common in the roAp stars studied with high-precision data \citep{10.3389/fspas.2021.626398}, with a possible explanation of stochastic perturbations of classical pulsators \citep{2020MNRAS.492.4477A,2020MNRAS.499.4687C}, although this is currently untested.

We have obtained a classification spectrum of this star, since the literature class of A0\,SrEu implies a hot star which is uncommon amongst the roAp stars. With our spectrum (Fig.\,\ref{fig:tic41259805spec}) we derive a spectral class of A6p\,SrEu(Cr), with Balmer and metal lines indicating an A6p star.


\subsubsection{TIC\,49818005}

TIC\,49818005 (HD\,19687) was given the spectral type A9\,IV/V in the Michigan Catalog \citep{1999mctd.book.....H}. Conversely, \citet{2019MNRAS.487.2117B} list a spectral type of A3, which is inconsistent with the 6829\,K temperature they cite \citep[from][]{2017AJ....154..259S}. There is no information on its peculiarity, if any, but \citet{2019MNRAS.487.2117B} found a significant high frequency at 1.6424\,mHz (141.9\,\cd), suggesting this could be an roAp star. We obtained a new low-resolution classification spectrum to clarify this point (Fig.\,\ref{fig:tic49818005spec}). We find the star to be an F2p\,SrEu(Cr), with Balmer and metal lines indicating an F1p star, and confirm the presence of rare earth element lines in a high-resolution SALT spectrum. With the high-resolution spectrum, we determine $T_{\rm eff}=7100\pm100$\,K, which is consistent with the TIC value and previous determinations.

\begin{figure}
    \centering
     \includegraphics[width=\columnwidth]{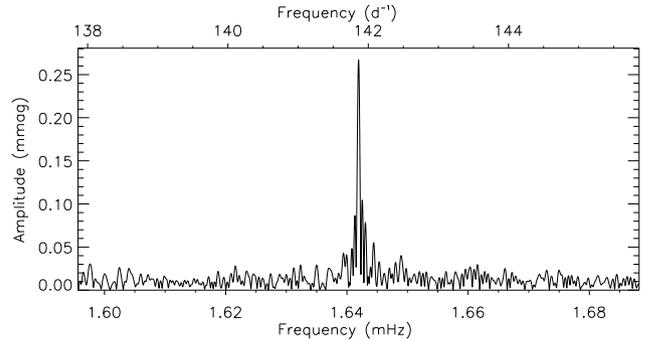}
    \caption{The pulsation mode in TIC\,49818005.}
    \label{fig:tic49818005}
\end{figure}

The TESS observations of this star were collected in sector 4, which were analysed by \citet{2019MNRAS.487.2117B}. We present a slightly different value for the pulsation frequency in this star, $1.641910\pm0.000008$\,mHz ($141.8619\pm0.0007$\,\cd), as shown in Fig.\,\ref{fig:tic49818005}. We suspect this difference is a result of the data treatment at low-frequency, which is not discussed in \citet{2019MNRAS.487.2117B}.

\subsubsection{TIC\,152808505}

TIC\,152808505 (HD\,216641) was observed in TESS sector 1 and was analysed by \citet{2019MNRAS.487.3523C}. We find no significant differences between our analysis here and that previously presented, so we refer the reader to the aforementioned publication for the details on this star, but provide the derived rotation period and pulsation frequencies in Table\,\ref{tab:stars}.

Since the only classification for this star is F3\,IV/V \citep{1978mcts.book.....H}, we obtained a new spectrum of this target with the SpUpNIC instrument at SAAO (Fig.\,\ref{fig:tic152808505spec}). We classify this star as F3p\,EuCr, and determine a temperature of $T_{\rm eff} =6900\pm300$\,K through comparison with synthetic spectra. This confirms this star to be an roAp star.


\subsubsection{TIC\,156886111}

TIC\,156886111 (HD\,47284) was observed in sectors 6 and 7, and originally reported as an roAp star by \citet{2019MNRAS.487.2117B}. The spectral type of this star is commonly given in the literature as A5\,SiEuCr \citep{1978mcts.book.....H}. However, we have obtained a new spectrum of this star (Fig.\,\ref{fig:tic156886111spec}) which shows this star to be an A8p\,SrEuCr star (showing Sr rather than Si).

We precisely determine the rotation period of this star to be $6.8580\pm0.0003$\,d through analysis of the mean light variations, which is in line with the literature value \citep{2006SASS...25...47W}. After the rotation signal, the dominant mode in this star is at $0.063251\pm0.000002$\,mHz ($5.4649\pm0.0002$\,\cd). This may be another example of a $\delta$\,Sct-roAp hybrid, as discussed for TIC\,158637987.

We find no more pulsation modes in the data than were presented by \citet{2019MNRAS.487.2117B}; we find a singlet at $1.27457\pm0.00002$\,mHz ($110.123\pm0.002$\,\cd), a doublet which is presumed to be a dipole triplet with a missing central component at $1.30274\pm0.00001$\,mHz ($112.557\pm0.001$\,\cd), and a quadrupole quintuplet centred on a frequency of $1.317816\pm0.000008$\,mHz ($113.8593\pm0.0007$\,\cd).  


\subsubsection{TIC\,259587315}

TIC\,259587315 (HD\,30849) was classified as an Ap\,SrEuCr by \citet{1978mcts.book.....H}, and has $T_{\rm eff}$ measures of 7720\,K in the TIC, 7250\,K \citep{1994BSAO...38..152G}, 8187\,K \citep{1997ESASP.402..239N} and 8000\,K \citep{2017MNRAS.468.2745N}.

The TESS data for this star, collected in sectors 4 and 5, have been analysed independently by \citet{2019MNRAS.487.2117B} and \citet{2019MNRAS.487.4695S} who both announced this star to be an roAp star, but provided conflicting rotational periods. Previous searches for high-frequency variability by \citet{1994MNRAS.271..129M} and \citet{2000A&A...355.1031H} were unsuccessful. 

Here, we use the SAP data from sector 4 (since the pipeline has distorted the rotation signal) and the PDC\_SAP data from sector 5 to determine a rotation period of $15.776\pm0.005$\,d. This is similar to the first reported period in the literature ($15.865\pm0.005$\,d; \citealt{1981A&AS...46..151H}) and that reported by \citet{2019MNRAS.487.2117B}, but is twice the period reported by \citet{2019MNRAS.487.4695S}, who have detected the first harmonic of the true period. It is not ideal to mix the SAP and PDC\_SAP data, but in this scenario such a procedure provides better results.

\begin{figure}
    \centering
     \includegraphics[width=\columnwidth]{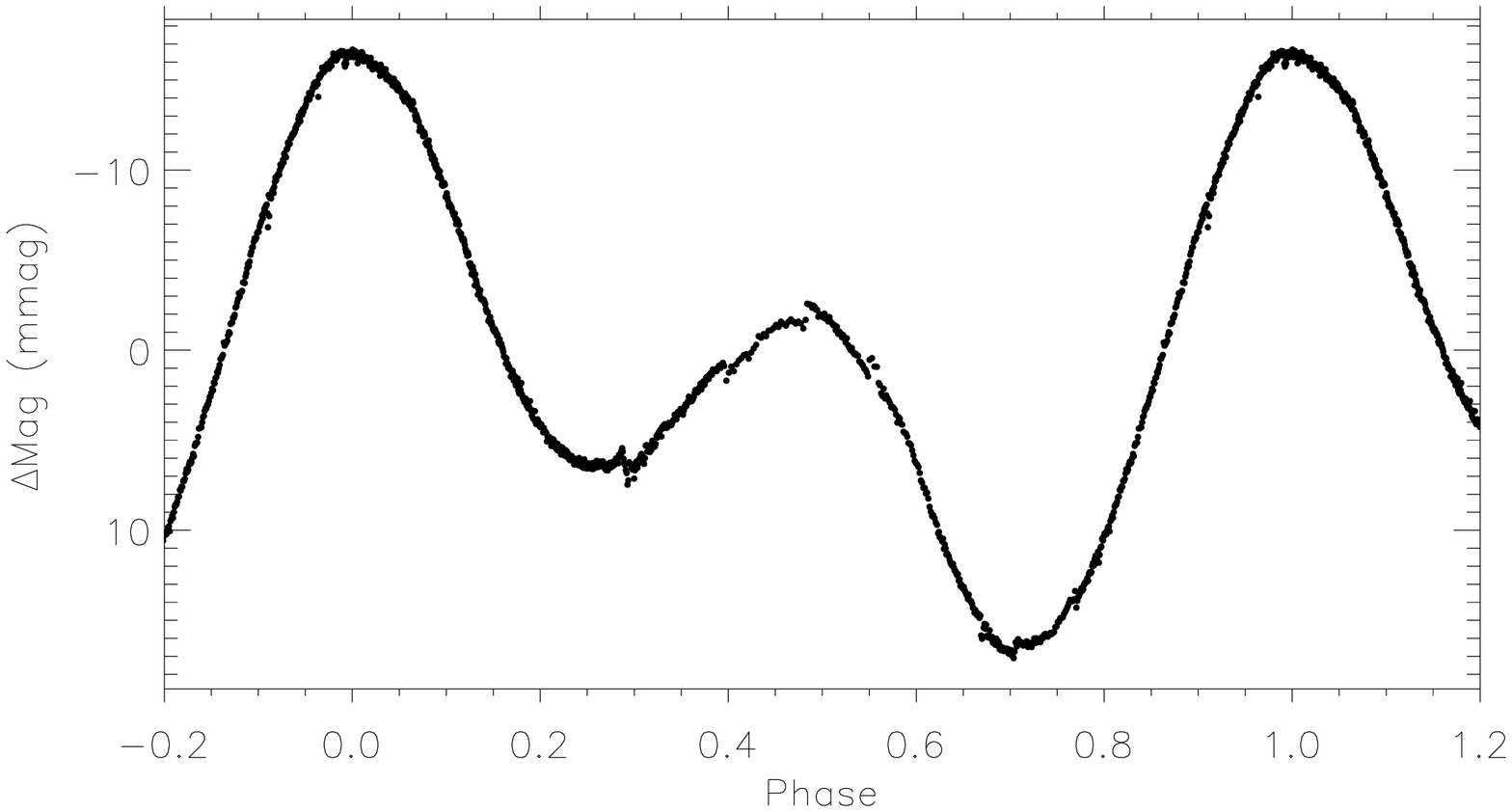}
     \includegraphics[width=\columnwidth]{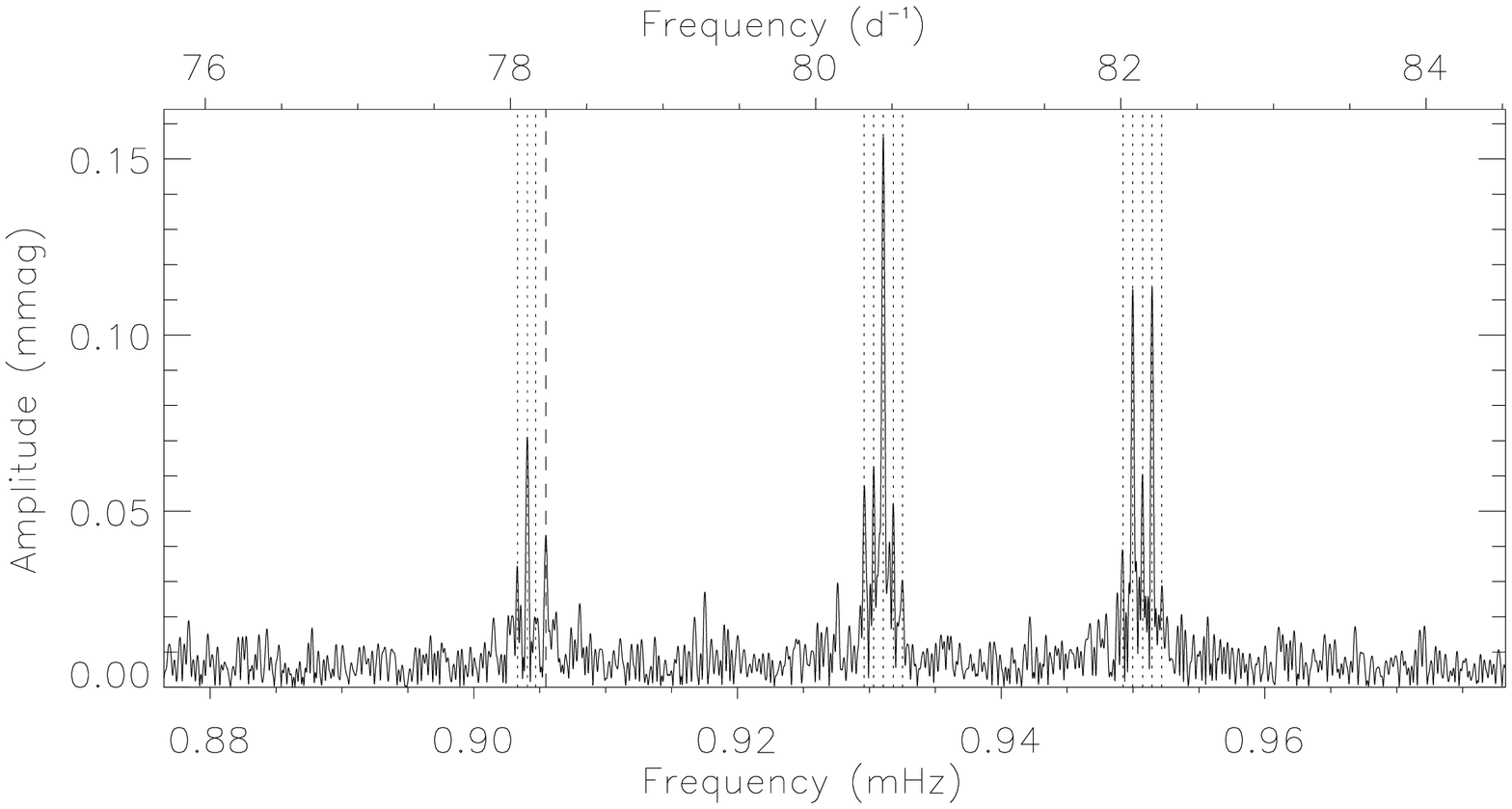}
    \caption{Top: phase folded light curve of TIC\,259587315, folded on a period of 15.776\,d. Bottom: the pulsation modes in TIC\,259587315. The vertical dotted lines show the components of obliquely split multiplets, with the dashed vertical line indicating an independent mode.}
    \label{fig:tic259587315}
\end{figure}

The star shows several pulsation modes: a triplet with a missing $+\nu_{\rm rot}$ sidelobe, a closely spaced singlet (denoted by the dashed vertical line in Fig.\,\ref{fig:tic259587315}), followed by two quintuplets. There appears to be further low amplitude modes between the multiplet groups, but at low S/N. The separation between the triplet and the singlet is $1.41\,\umu$Hz. The spacing between the two quintuplets is $\approx19.7\,\umu$Hz which is plausibly the large frequency separation for this star, indicating it is evolved.

\subsubsection{TIC\,349945078}

TIC\,349945078 (HD\,57040) was classified as A2\,EuCr by \citet{2009A&A...498..961R}. The TIC reports $T_{\rm eff} = 7203$\,K, though literature values for this parameter range from as low as $6518$\,K \citep{2019ApJS..241...12S} to as high as $8412$\,K \citep{2006ApJ...638.1004A}. \citet{2012MNRAS.420.2727E} reported the presence of a $7.5$\,kG magnetic field on the star. \citet{1994MNRAS.271..129M} were unable to detect oscillations from ground-based observations, so it was not identified as an roAp star until \citet{2019MNRAS.487.2117B} used TESS data to detect two oscillation frequencies.

\begin{figure}
    \centering
     \includegraphics[width=\columnwidth]{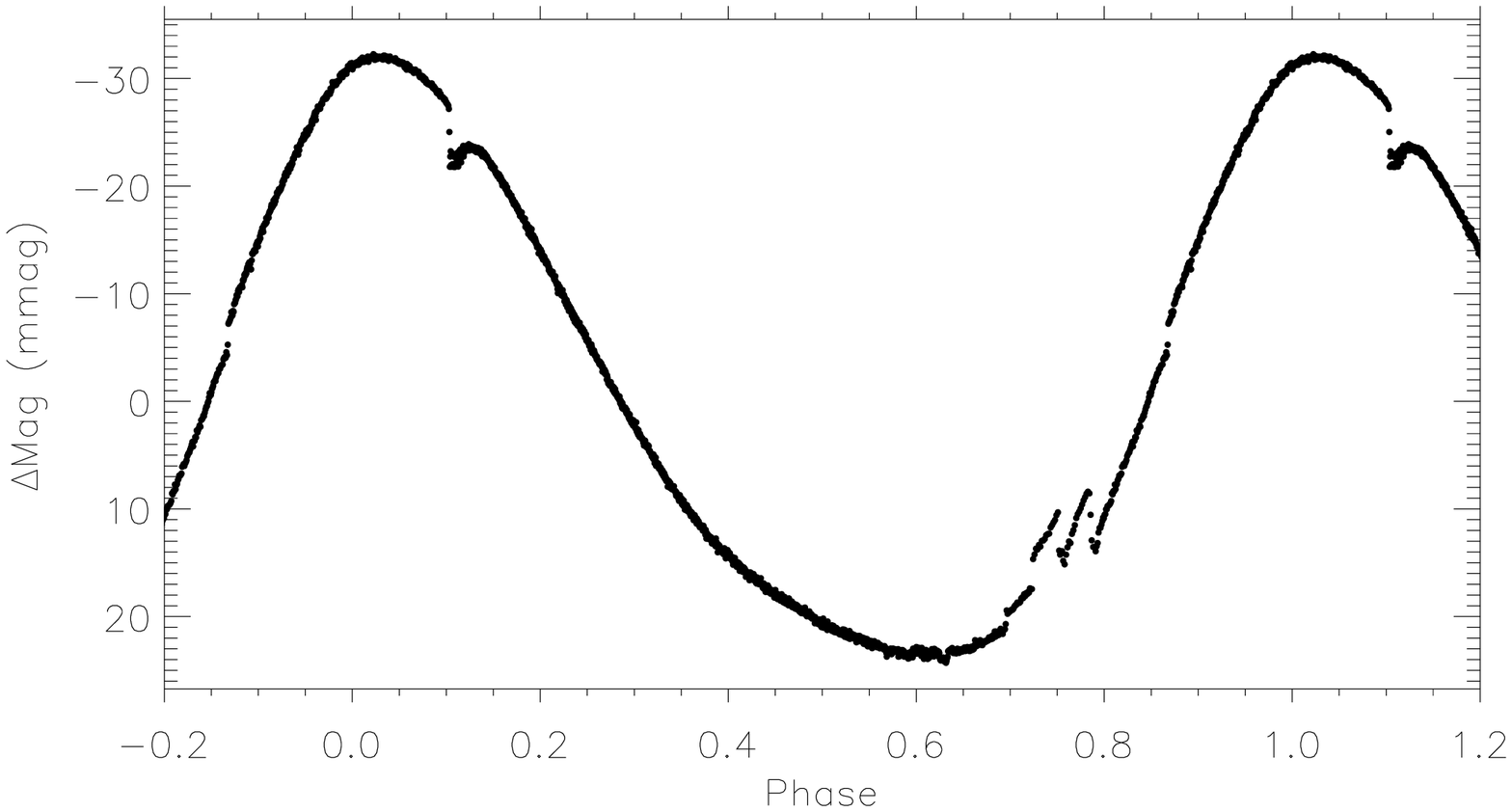}
     \includegraphics[width=\columnwidth]{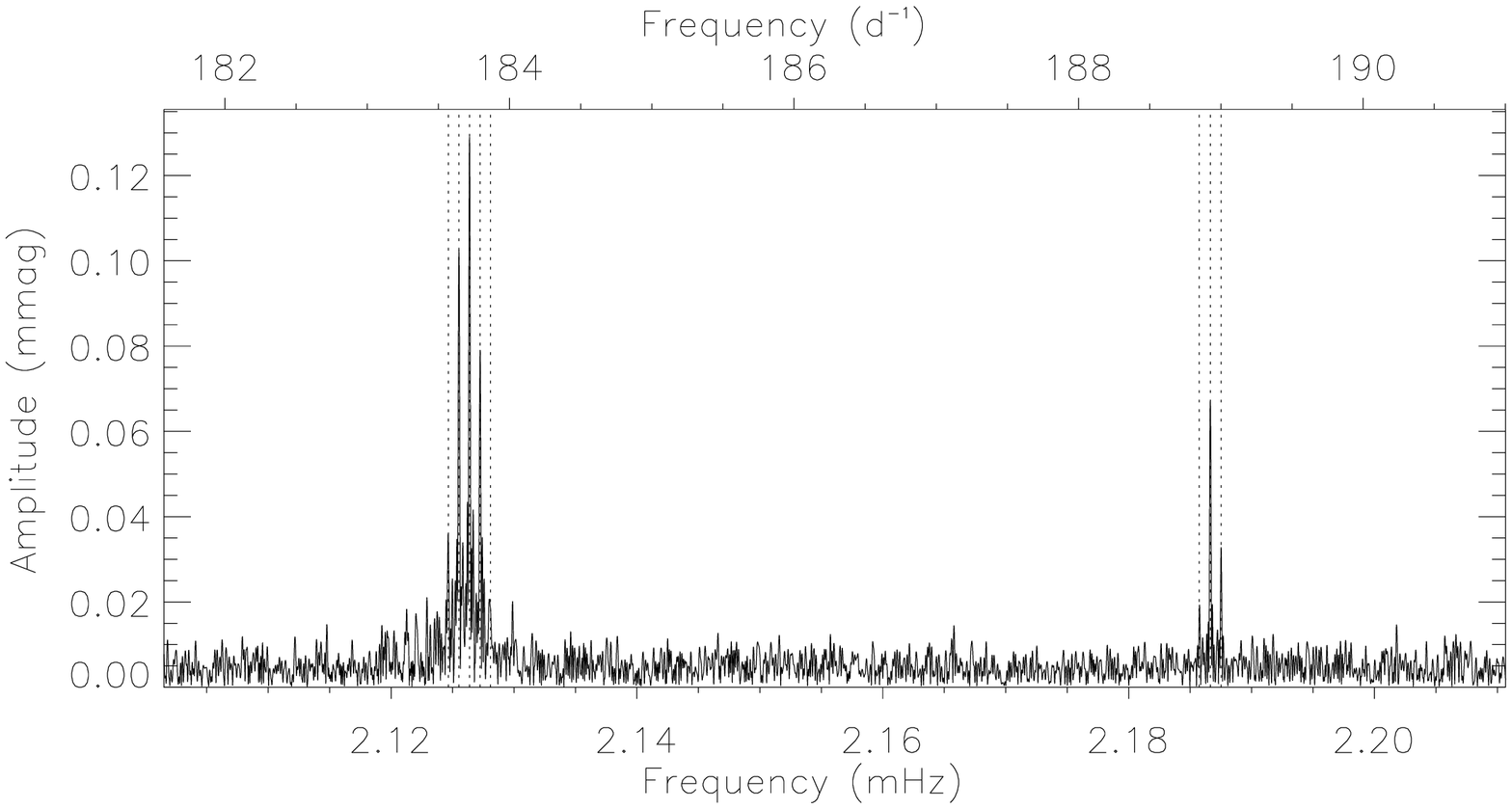}
    \caption{Top: phase folded SAP light curve of TIC\,349945078, folded on a period of 13.4256\,d. The discontinuities, and the dip, are instrumental artefacts. Bottom: the pulsation modes in TIC\,349945078. The vertical dotted lines show the components of obliquely split multiplets.}
    \label{fig:tic349945078}
\end{figure}

This star was observed in sectors 6-9, thus we revisit this star to provide a more detailed analysis of the variability with the availability of more data. To extract the rotation signal, we used the SAP data since the rotation signature in the PDC\_SAP data has been significantly altered. This was evident upon visual inspection. We fitted a four element harmonic series to derive a rotation period of $13.4256\pm0.0008$\,d (Fig.\,\ref{fig:tic349945078}). This is different from that presented by \citet{2019MNRAS.487.2117B} by a factor of two. We confirm our rotation period through the analysis of the oblique pulsations below.

For the pulsation analysis, we used the PDC\_SAP data and prewhitened the light curve to remove the rotation signal and instrumental artefacts. As a final correction, we removed outlying points. The resulting amplitude spectrum shows two groups of peaks composed of a quintuplet and a triplet (Fig.\,\ref{fig:tic349945078}). The two modes in this star are separated by $\approx60$\,$\umu$Hz which is plausibly the large frequency separation. 


\subsubsection{TIC\,350146296}

TIC\,350146296 (HD\,63087) was classified as A7\,IV by \citet{1975mcts.book.....H}. However, due to the clearly spotted nature of this star, and the pulsation variability, we obtained several spectra with the SpUpNIC instrument, and find this star to be an F0p\,EuCr star (Fig.\,\ref{fig:tic350146296spec}).

\begin{figure}
    \centering
     \includegraphics[width=\columnwidth]{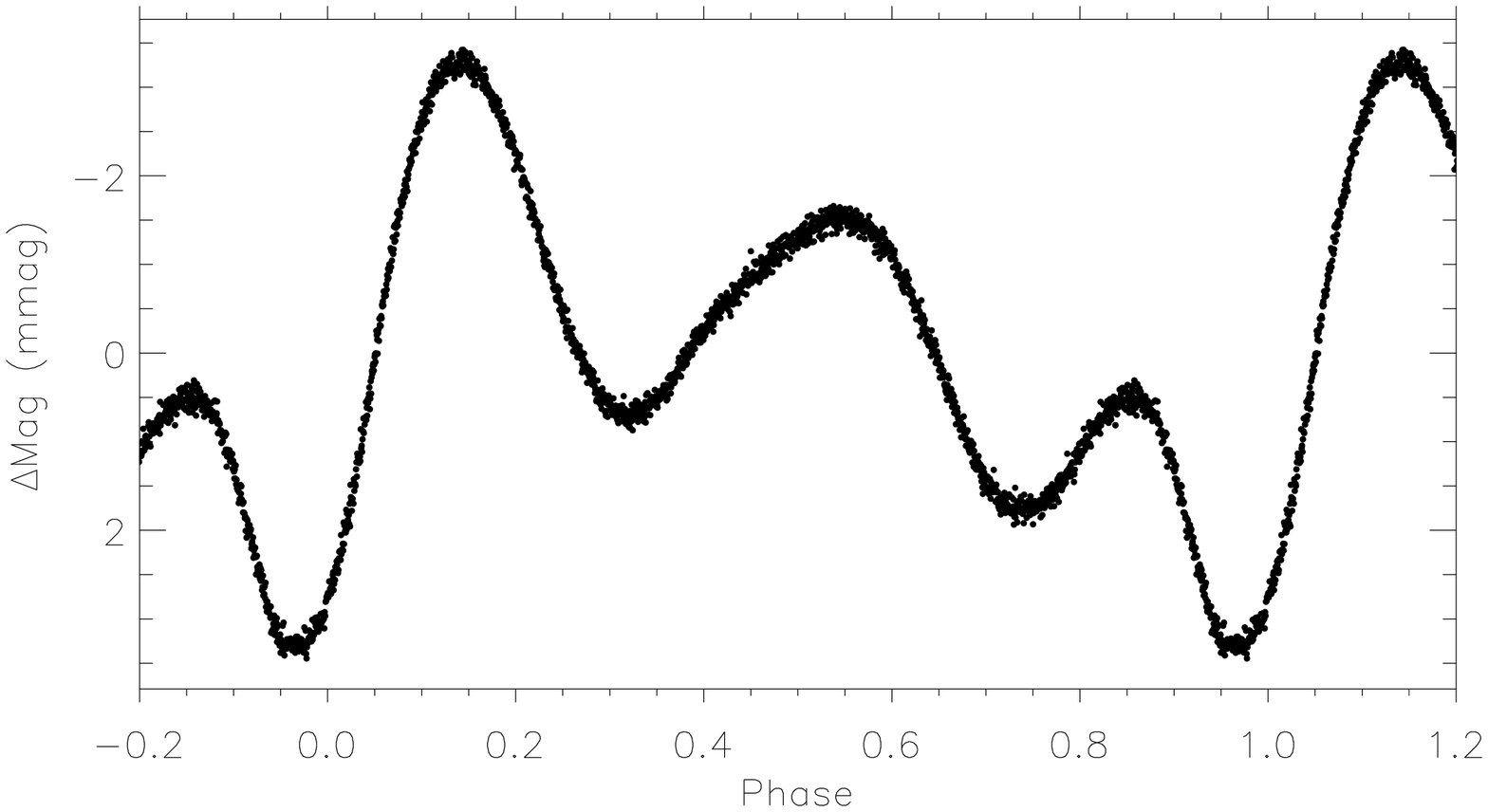}
     \includegraphics[width=\columnwidth]{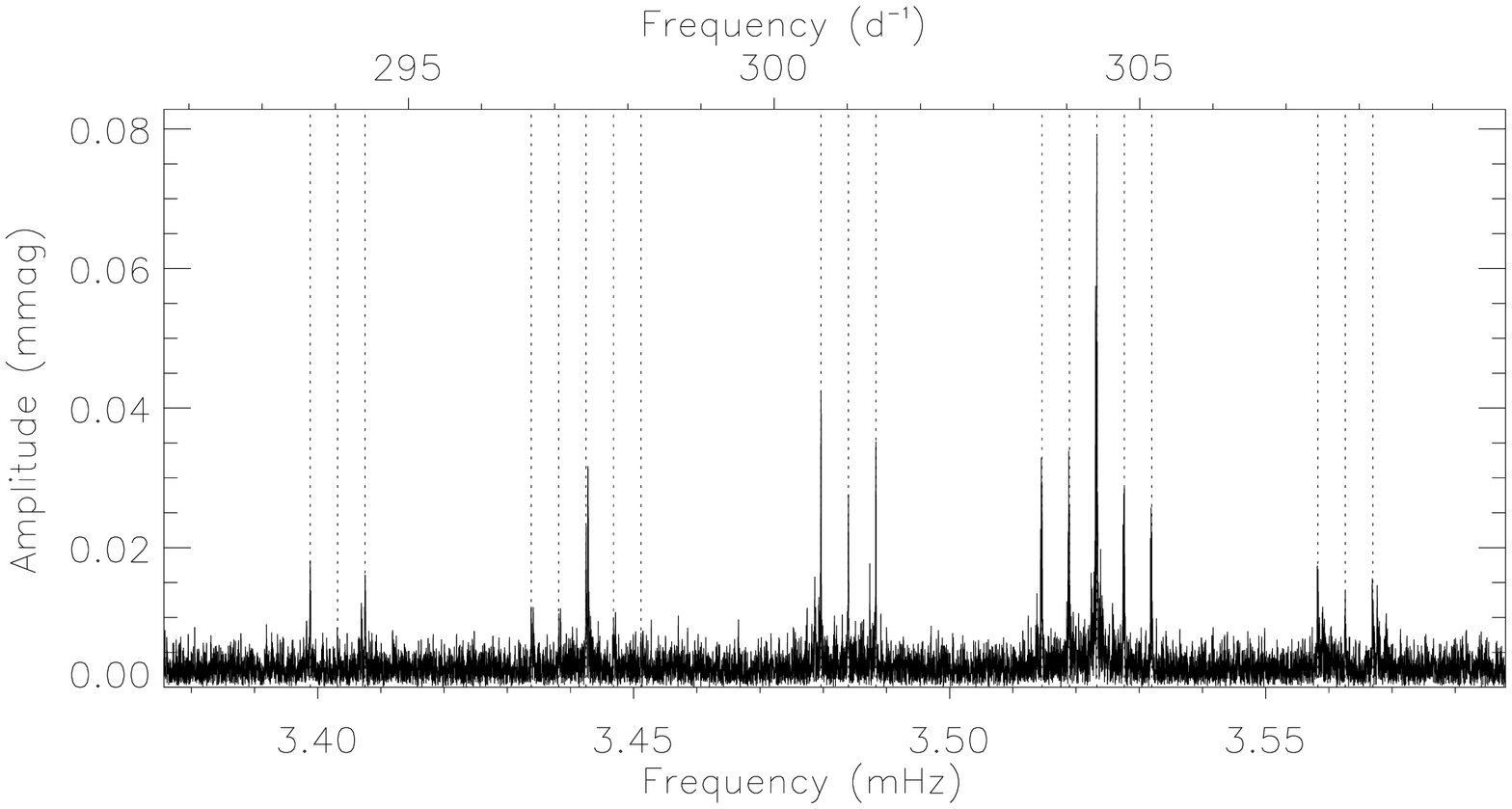}
    \caption{Top: phase folded light curve of TIC\,350146296, folded on a period of 2.66387\,d. Bottom: the rich pattern of pulsation modes in TIC\,350146296. The vertical dotted lines show the components of obliquely split multiplets.}
    \label{fig:tic350146296}
\end{figure}

TIC\,350146296 was first reported as an roAp star by \citet{2019MNRAS.487.3523C}, with \citet{2019MNRAS.487.2117B} providing an analysis of extra sectors of data. There are now 13 sectors for this star from Cycle\,1 which we analyse here. With the extended time base (357\,d) we derive a rotation period of $2.66387\pm0.00002$\,d. The phased light curve, shown in Fig.\,\ref{fig:tic350146296}, reveals a complex spot structure on this star.

There are five pulsation modes in TIC\,350146296, all of which have been split into a multiplet by oblique pulsation. The multiplet structures allow us to conclude that the modes are alternating dipole and quadrupole modes, with three dipole modes and two quadrupole modes. All frequencies in this star show signs of variability, making precise frequency extraction difficult. All the modes are split by $\approx40\,\umu$Hz which, given the multiplet structures, we understand to be half the large frequency separation. Shi et al. (submitted) have conducted an extensive study on this star, to which we refer the reader for further details, although we provide the pulsation frequencies and rotation period in Table\,1 here.


\subsubsection{TIC\,431380369}

TIC\,431380369 (HD\,20880) was identified as a chemically peculiar star by \citet{1975mcts.book.....H} who gave the classification of Ap\,Sr(EuCr). It was later reported by \citet{2019MNRAS.487.3523C} to be a new roAp star from the analysis of the TESS sector 2 data, after \citet{2016A&A...590A.116J} presented a null result in the search for pulsations. The star was subsequently observed in sectors 6 and 13, where the significant gaps serve to produce a complicated window function. With the extended data, we are able to derive a more precise rotation period for this star of $5.19716\pm0.00005$\,d. 

\begin{figure}
    \centering
     \includegraphics[width=\columnwidth]{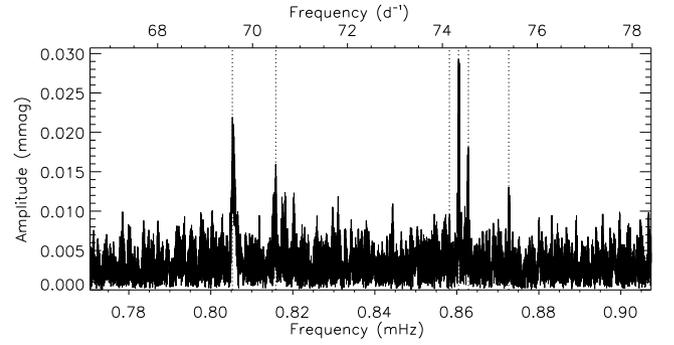}
    \caption{Pulsation modes in TIC\,431380369. The window function gives fine structure to the peaks. However, there is also significant amplitude and frequency variability in this star, as can be seen in the mode at 0.860412\,mHz. The vertical dotted lines denote the extracted modes, apart from the $\nu-\nu_{\rm rot}$ sidelobe of the highest amplitude peak which is lost in the noise.}
    \label{fig:tic431380369}
\end{figure}

This star shows a complex pulsation spectrum, as can be seen in Fig.\,\ref{fig:tic431380369}, which is not helped by the significant gaps in the data. The mode at $0.805252\pm0.000002$\,mHz ($69.5737\pm0.0002$\,\cd) is significantly broad, suggesting frequency and/or amplitude variability. The mode at $0.860412\pm0.000001$\,mHz ($74.3396\pm0.0001$\,\cd) also shows amplitude variability, but includes a rotationally split sidelobe at $\nu+\nu_{\rm rot}$. The four significant modes are listed in Table\,\ref{tab:stars}. We extract two extra modes, and the sidelobe, in addition to those reported by \citet{2019MNRAS.487.3523C}. This is likely a combination of the longer data set where the noise is slightly lower, and the mode lifetime of the pulsations seen in this star. A detailed study, on a rotation period-by-rotation period basis, of this star is required to extract the full information from the TESS data.

\subsection{Known roAp stars prior to TESS launch}

In this section, we present an analysis of the roAp stars which were known prior to the launch of the TESS mission. As with the new roAp star discoveries, we present only new information on stars which have already been discussed in the literature. 


\subsubsection{TIC\,6118924}

TIC\,6118924 (HD\,116114) was observed in sector 10. The star was classified as F0Vp\,SrCrEu by \citet{1979PASP...91..176A} and was identified as an roAp star by \citet{2005MNRAS.358..665E} who discovered variability in the Eu\,{\sc{ii}} lines with a period around 21\,min (0.790\,mHz; 68.26\,\cd). A search for photometric variability after the spectroscopic detection by \citet{2005IBVS.5651....1L} returned a null result at an amplitude level of about 0.5\,mmag in Johnson $B$.

\begin{figure}
    \centering
    \includegraphics[width=\columnwidth]{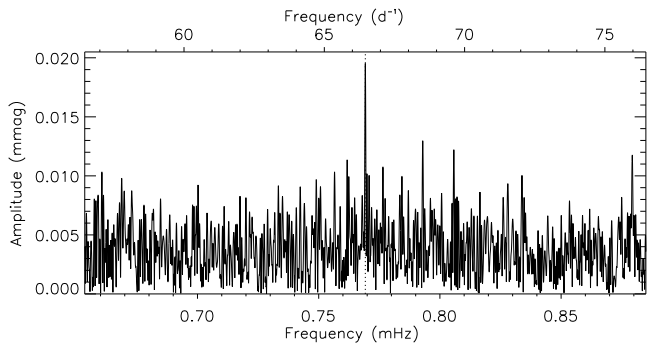}
    \caption{Amplitude spectrum of TIC\,6118924 showing the single known pulsation mode.}
    \label{fig:tic6118924}
\end{figure}

The rotation period of TIC\,6118924 was determined to be 27.6\,d by \citet{2000A&A...359..213L} which is longer than the TESS data span. We see no evidence of this period in the light curve. We detect one significant frequency at $0.76923\pm0.00004$\,mHz ($66.462\pm0.003$\,\cd) which is consistent with that found in the spectroscopic study (Fig.\,\ref{fig:tic6118924}). This is the first photometric detection of the pulsation in this star.

\subsubsection{TIC\,33601621}

TIC\,33601621 (HD\,42659) was classified as Ap\,SrCrEu by \citet{1988mcts.book.....H}. There are several occurrences of the stellar parameters in the literature with $T_{\rm eff}$ ranging between $7450-7940$\,K \citep{2006A&A...450..763K,2019MNRAS.487.2117B}. \citet{2006A&A...450..763K} also provides $\log L/{\rm L_{\odot}}=1.48\pm0.09$ and $M=2.1\pm0.1$\,M$_{\odot}$. There are also several measures of $\langle B_z\rangle$ which provide an average of $0.40\pm0.67$\,kG \citep{2006A&A...450..763K,2006AN....327..289H,2015A&A...583A.115B}.

This star was discovered to be an roAp star by \citet{1993IBVS.3844....1M} with later observations providing more detailed results by \citet{1994MNRAS.271..118M}. The TESS sector 6 observations have been previously analysed by \citet{2019MNRAS.487.2117B}, with a more detailed analysis presented by \citet{2019MNRAS.489.4063H}, analysing the dipole triplet. We do not present a re-analysis of the same data here, but refer the reader to the previous literature. It is worthy of note, however, that TIC\,33601621 is the only roAp star known to be a member of a spectroscopic binary (SB1) system \citep{2015A&A...582A..84H,2019MNRAS.489.4063H}. 


\subsubsection{TIC\,35905913}

TIC\,35905913 (HD\,132205) was classified as an Ap\,EuSrCr star by \citet{1975mcts.book.....H} and later discovered by \citet{2013MNRAS.431.2808K} to be an roAp star through time-resolved spectroscopic analysis. Their data revealed a pulsation at $201.68\pm0.59$\,\cd\ with an amplitude less than 100\,\ms. The star had previously been monitored photometrically for pulsations many times \citep{1994MNRAS.271..129M}, but no signal was detected which is unsurprising given the small amplitude in the spectroscopic data.

TESS observed TIC\,35905913 during sectors 11 and 12. The PDC\_SAP light curves for these two sectors differ significantly. A clear rotation signature can be seen the sector 12 data, whereas the sector 11 data have a greatly reduced amplitude attributed to this variability. On inspection, the contamination factor for this star is high, at 0.28, with the contaminating star about 18\,arcsec away. According to {\it Gaia} DR2 \citet{2018A&A...616A...1G}, the \textit{G}-band magnitude difference is 3.25\,mag which suggests, under the assumption of negligible extinction differences, that the contaminant star is a mid G-type star.

\begin{figure}
    \centering
    \includegraphics[width=\columnwidth]{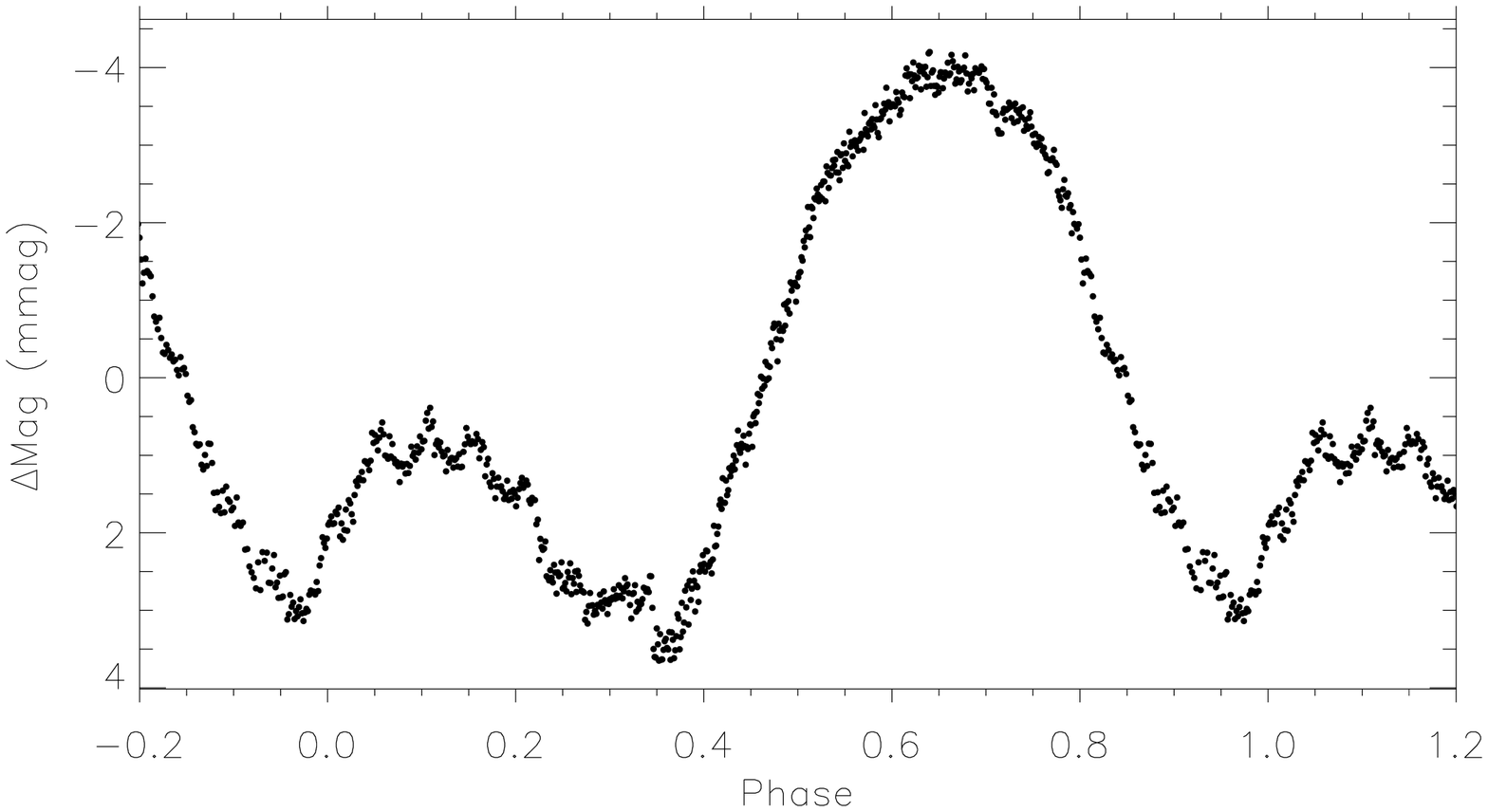}
    \includegraphics[width=\columnwidth]{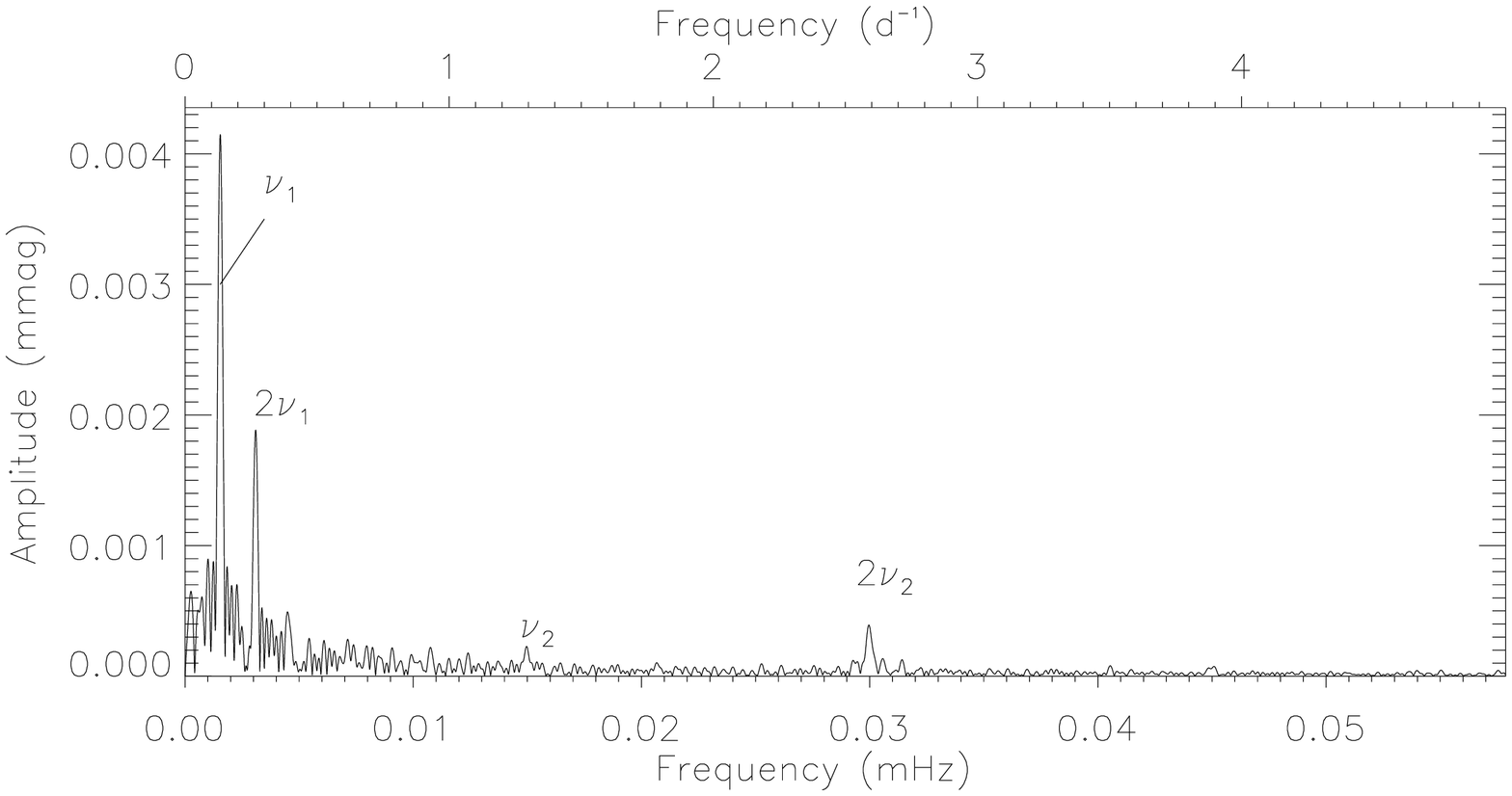}
    \includegraphics[width=\columnwidth]{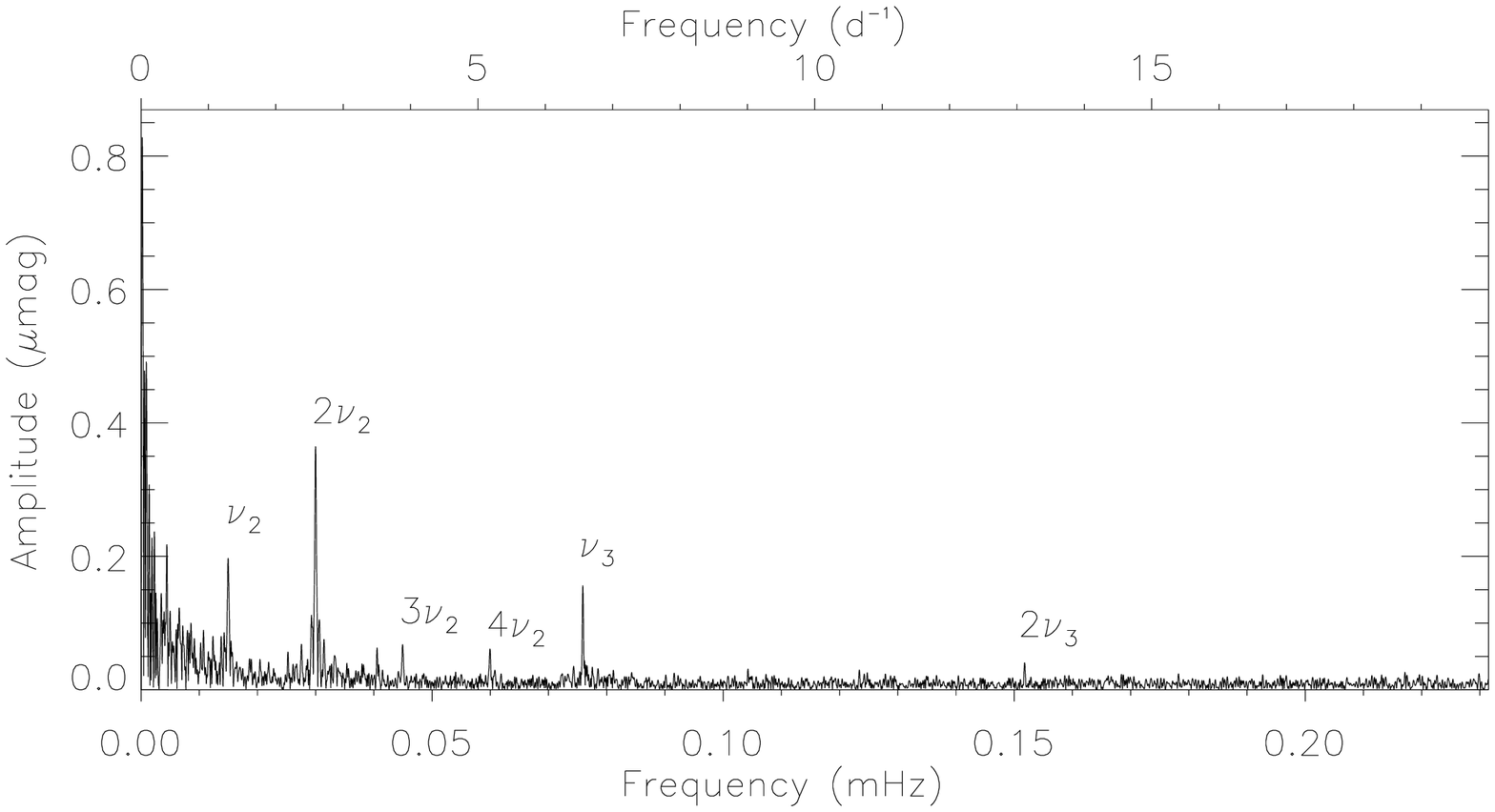}
    \caption{Top: rotation curve for TIC\,35905913. There is clear variability on a shorter time scale. Middle: low-frequency amplitude spectrum showing the rotation frequency of the Ap star ($\nu_1$) and its harmonic. $\nu_2$ and its harmonic are also indicated. Bottom: low-frequency amplitude spectrum after removing the Ap star rotation signature. Further harmonics of $\nu_2$ are evident, with an additional mode, $\nu_3$, and its harmonic labelled.}
    \label{fig:TIC_35905913_rot}
\end{figure}

By fitting a harmonic series of 5 components to the SAP sector 11 and PDC\_SAP sector 12 data simultaneously, we derive a rotation period of $7.513\pm0.001$\,d (derived from $\nu_1$ in Fig.\,\ref{fig:TIC_35905913_rot}). Again, we reiterate that mixing the raw and pipeline corrected data is not ideal, but provides the best results in this case. After the removal of this harmonic series, there are clear signs of variability remaining in the data at low frequency ($\nu_2$ and $\nu_3$ in the figure). The source of variability is uncertain given the proximity of the contaminating source. In fitting and removing a harmonic series to $\nu_2$, power remains around these frequencies in the amplitude spectrum, indicating an unstable frequency/amplitude which is common with star spots that change in size. This leads us to conclude that the contaminant is the source of the $\nu_2$ harmonic series. The origin of $\nu_3$ required further investigation, but could be a g\,mode in the Ap or contaminant star.

We detect no pulsation in the roAp range to a limit of about 25\,$\muup$mag in the combined PDC\_SAP flux data. This is unsurprising given the broad red bandpass of the TESS data, coupled with the low pulsation amplitude in spectroscopy.


\subsubsection{TIC\,44827786}
TIC\,44827786 (HD\,150562) was classified as F5Vp\,SrCrEu by \cite{1995msct.book.....B} and discovered to be an roAp star by \cite{1992IBVS.3750....1M} with a period of 10.75\,min.

The TIC gives $T_{\rm eff}$ = 7350\,K, which is consistent with that obtained using $uvby$H$\upbeta$ photometry from \citet{2016A&A...590A.116J} and the \citet{1985MNRAS.217..305M} calibration. However, several literature sources give a significantly lower value: 6390\,K \citep{2018A&A...616A...1G}, 6620\,K \citep{2017MNRAS.471..770M} and 6820\,K \citep{2006ApJ...638.1004A}. The star appears to be slowly rotating, with \citet{2008CoSka..38..317E} obtaining a $v \sin i = 1.5\pm0.5$\kms\ and \citet{2020A&A...639A..31M} finding no obvious signature of rotational variation from TESS data. This star shows variation of the mean magnetic field modulus from 4.7 to 5.0\,kG \citep{2017A&A...601A..14M}, with the mean longitudinal field changing between 1.2 and 1.7\,kG \citep{2015A&A...583A.115B,2017A&A...601A..14M}.

The mean magnetic field modulus was found to be 4.8\,kG \citep{2003Ap.....46..234K}, while more recently \citet{2015A&A...583A.115B} obtained a value of $\sim$2\,kG.

\begin{figure}
    \centering
    \includegraphics[width=\columnwidth]{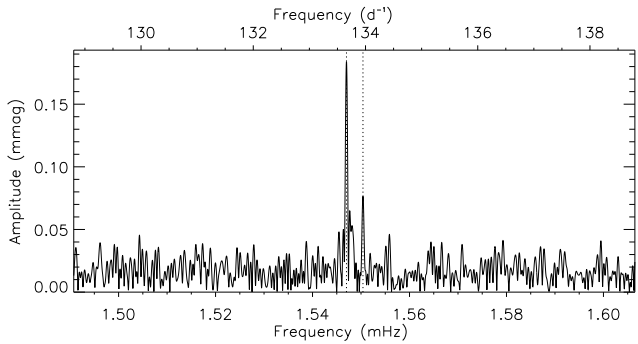}
    \caption{The pulsation signatures seen in TIC\,44827786. This is the first detection of the highest amplitude mode, whilst the low amplitude mode is reported in the literature.}
    \label{fig:tic44827786}
\end{figure}

We detect two significant modes of pulsation in the TESS sector 12 data (Fig.\,\ref{fig:tic44827786}). The highest amplitude mode at a frequency of $1.54700\pm0.00002$\,mHz ($133.661\pm0.002$\,\cd) is separated from the previously known pulsation mode ($1.55040\pm0.00005$\,mHz; $133.955\pm0.004$\,\cd) by $3.40\pm0.05\,\umu$Hz which is plausibly the small frequency separation.

\subsubsection{TIC\,49332521}   

TIC\,49332521 (HD\,119027) is classified on SIMBAD as an $\alpha^2$\,CVn star, but \citet{1998Obs...118..153M} found no rotational signal, and suggested a period of greater than 6\,months if there is any rotational variation. \citet{2020A&A...639A..31M} also found no rotational variation. Its spectral type is Ap\,SrEu(Cr) \citep{1982mcts.book.....H}. It has been known to be an roAp star for thirty years \citep{1991IBVS.3611....1M}, and has a rich p-mode oscillation spectrum \citep{1998MNRAS.300..188M}. Magnetic field properties have been measured for HD\,119027 \citep{2017A&A...601A..14M}, with a measured value for $B_z$ of 0.5\,kG, and multiple $\langle B\rangle$ measurements around 3.17\,kG.

TESS observed HD\,119027 in sector\,11, with multiple oscillation frequencies discussed by \citet{2020A&A...639A..31M}. Here we detect 7 independent modes in TIC\,49332521. Six of these modes are almost the same as presented by \citet{1998MNRAS.300..188M}: their five roughly equally split modes (separated by $\approx26\,\umu$Hz) plus their $\nu_{2p}$ mode. Our additional mode is found at a lower frequency ($1.57070\pm0.00005$\,mHz; $135.709\pm0.004$\,\cd) which is separated from the next mode by $264\,\umu$Hz. This means that modes spanning about five radial orders are not excited to a detectable amplitude in this star, which although is unusual, has been seen in other roAp stars, an excellent example of this being TIC\,139191168 (HD\,217522; Section\,\ref{sec:139191168}).

\begin{figure}
    \centering
    \includegraphics[width=\columnwidth]{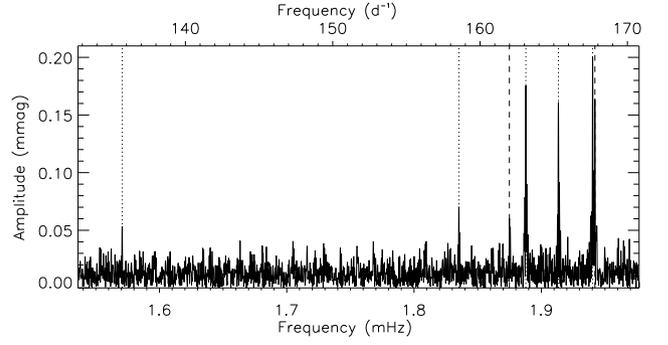}
    \caption{The pulsation signatures seen in TIC\,49332521. Modes marked with dotted lines are separated by multiples of $\approx26\,\umu$Hz while those with dashed lines do not fit this pattern.}
    \label{fig:tic49332521}
\end{figure}

We show in Fig.\,\ref{fig:tic49332521} the amplitude spectrum and the identified modes. Those marked by vertical dotted lines are separated by multiples of $\approx26\,\umu$Hz while those with dashed lines do not fit this pattern. We suspect, given the stellar parameters, that this separation is half of the large frequency separation, such that $\Delta\nu\approx52\,\umu$Hz for TIC\,49332521. The separation between the two highest amplitude modes ($1.78\,\umu$Hz) could represent the small frequency separation in this star, although this value is different from the value of 1.95\,$\umu$Hz proposed by \citet{1998MNRAS.300..188M}.

\subsubsection{TIC\,69855370}

TIC\,69855370 (HD\,213637) was observed in sector 2. The star is classified as A\,(pEuSrCr) by \citet{1988mcts.book.....H} and was identified as an roAp star by \citet{1998A&A...334..606M} who reported the discovery of two pulsation modes with periods around 11-12\,min,  while acknowledging that the frequency values identified needed to be confirmed due to the presence of 1\,\cd\ aliases in the data. Based on the TESS data, \citet{ 2019MNRAS.487.3523C} confirmed the presence of two pulsation modes in the same period range, the first corresponding to one of the modes identified by \citet{1998A&A...334..606M} and the other differing from their second mode by 1\,\cd. 

No sign of rotational modulation was found by \citet{ 2019MNRAS.487.3523C} in their analysis of the TESS data for this star. Nevertheless, based on the measurement of a projected rotational velocity of $v\sin i = 3.5\pm0.5$\,\kms, \citet{2003A&A...404..669K} suggested a rotation period of 25\,d. The lack of a rotational modulation in the TESS data could result from an unfavourable alignment ($i$ or $\beta$ close to zero), as discussed by \citet{2019MNRAS.487.3523C}, or be a result of a long rotation period \citep{2020A&A...639A..31M}. 

Since the TESS sector 2 data have been analysed in detail, we refer the reader to \citet{2019MNRAS.487.3523C} for further discussion, but include the pulsation frequencies in Table\,\ref{tab:stars} for completeness.


\subsubsection{TIC\,93522454}

TIC\,93522454 (HD\,143487) was observed in sector 12. The star is classified as A3\,SrEuCr in \citet{2009A&A...498..961R}. It was identified as an roAp star by \citet{2010MNRAS.404L.104E}, following the analysis of high time resolution spectra obtained with UVES from which a low amplitude pulsation with a period around 10\,min was inferred. The roAp nature of this star was later confirmed by \citet{2013MNRAS.431.2808K} based on the analysis of a larger set of UVES spectra which allowed the detection of the pulsations in the lines of rare-earth elements as well as in the core of H$_\alpha$ and the determination of a pulsation period of 9.63\,min (1.730\,mHz; 149.47\,\cd) with amplitudes up to $110$\,\ms. No rotation period is reported in the literature for this star.

\begin{figure}
    \centering
    \includegraphics[width=\columnwidth]{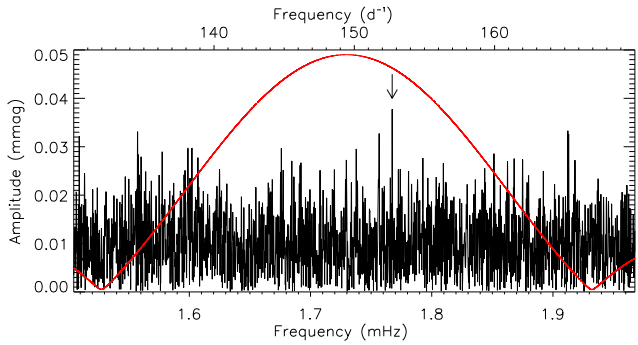}
    \caption{The amplitude spectrum of TIC\,93522454 in the frequency region of interest. We see no clear indication of pulsation, but mark with an arrow a possible signature. The smooth solid red curve represents the shape of the peak used by \citet{2013MNRAS.431.2808K} to provide a pulsation frequency of 1.73\,mHz (149.47\,\cd), demonstrating the broad frequency range where the mode may be found.}
    \label{fig:tic93522454}
\end{figure}

The TESS data do not show a clear pulsation signal. We show, in Fig.\,\ref{fig:tic93522454}, the amplitude spectrum in the frequency region where pulsations have been previously found. We indicate the possible location of the pulsation in this star with an arrow, but caution its presence. The lack of an obvious detection is not surprising given the low amplitude radial velocity variations reported. 

\subsubsection{TIC\,125297016}

TIC\,125297016 (HD\, 69013) was observed in sector 7. The star is classified as Ap\,EuSr by \citet{1988mcts.book.....H}. It was identified as an roAp star with a pulsation period around 11\,min by \citet{2011MNRAS.411..978E}, following the analysis of high time resolution spectra obtained with UVES. Pulsational variability with a similar period was also detected in photometric data published in the same work. The roAp nature of this star was confirmed by \citet{2013MNRAS.431.2808K} based on the analysis of a larger set of UVES spectra which allowed the detection of the pulsations in the lines of rare-earth elements and the determination of a pulsation period of 11.22\,min (1.4854\,mHz; 128.34\,\cd). No rotation period is reported in the literature and no sign of rotation variability was found by \citet{2020A&A...639A..31M} in their analysis of the TESS data for this star.

\begin{figure}
    \centering
    \includegraphics[width=\columnwidth]{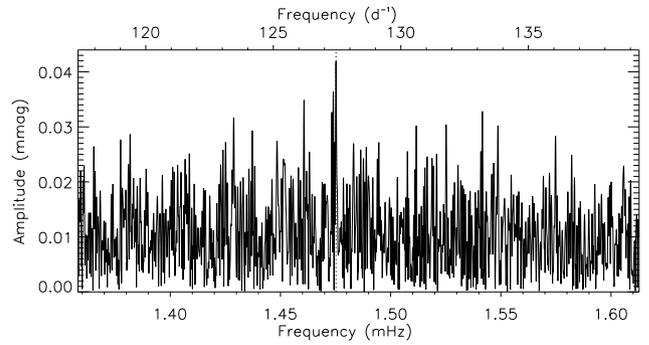}
    \caption{The amplitude spectrum of TIC\,125297016 in the frequency region of interest. The highest amplitude peak is at a frequency of $1.47519$\,mHz which is similar to that reported in the literature, however we are not confident in the detection.}
    \label{fig:tic125297016}
\end{figure}

The TESS data of this star do not clearly show the known pulsation frequency. A significant peak is detected at a frequency of $1.47519\pm0.00005$\,mHz ($127.456\pm0.004$\,\cd; Fig.\,\ref{fig:tic125297016}) which is similar to the spectroscopic literature value, but further TESS observations are required to reduce the noise in the data to allow a confident photometric detection of the pulsation.

\subsubsection{TIC\,136842396}

TIC\,136842396 (HD\,9289) is classified as A(p)\,SrEuCr by \citet{1999mctd.book.....H} and its mean magnetic field modulus was determined to be $\langle B\rangle=2.0$\,kG \citep{2007A&A...473..907R}. The star was discovered to be an roAp star by \citet{1994MNRAS.271..421K} who identified 3 pulsation modes in the amplitude spectrum of photometric data. Later, \citet{2011A&A...530A.135G} observed TIC\,136842396 with the MOST satellite and found a greater number of frequencies, but with few in agreement with the first study. TESS observed TIC\,136842396 during sector 3, with \citet{2019MNRAS.487.2117B} presenting the first analysis of those data.

\begin{figure}
    \centering
    \includegraphics[width=\columnwidth]{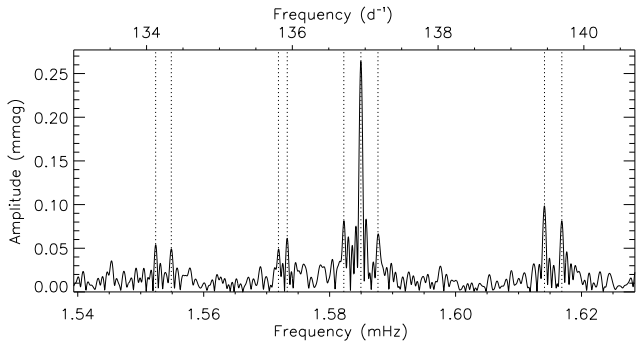}
    \caption{The pulsation signatures seen in TIC\,136842396. The vertical dotted lines represent the extracted peaks, all of which are, in their groups, separated by integer multiples of the rotation frequency.}
    \label{fig:tic136842396}
\end{figure}

Here, we derive a rotation period of the star from the SAP data since the pipeline has clearly altered the astrophysical signal. We measure a rotation period of $8.660\pm0.006$\,d, which is different from that provided by \citet{2019MNRAS.487.2117B}. The pulsation spectrum for this star poses questions due to the presence, and lack thereof, of many rotationally split sidelobes. We extract 9 significant peaks from the prewhitened light curve of this star, as shown in Fig.\,\ref{fig:tic136842396}. Given the lack of complete multiplets, we investigated the frequency separation between the {\it assumed} mode frequencies and found a reoccurrence of a separation of $\approx31\,\umu$Hz. Under the assumption that the highest amplitude mode is a quadrupole mode with missing $\pm\nu_{\rm rot}$ sidelobes, the pulsation spectrum consists of: a quadrupole mode showing only the $\nu_{\rm rot}$ sidelobes, a dipole doublet split by the rotation frequency (with the pulsation mode being the highest amplitude peak), the quadrupole mode that is missing the $\pm\nu_{\rm rot}$ sidelobes, and a final quadrupole mode showing only the $\nu_{\rm rot}$ sidelobes. These tentative mode identifications imply a large frequency separation of $\approx62\,\umu$Hz.

\subsubsection{TIC\,139191168}
\label{sec:139191168}

TIC\,139191168 (HD\,217522) was observed in sector 1. The star is classified as A5\,SrEuCr in \citet{2009A&A...498..961R} and has an upper limit on $\langle B\rangle$ of 1.5\,kG \citep{2008A&A...480..811R}. It was first identified as an roAp star with pulsation periods around 13.72\,min (1.2151\,mHz; 104.98\cd) by \citet{1983MNRAS.205....3K}, based on photometric data. The authors noted a significant night-to-night pulsation amplitude variability which was confirmed by a multi-site photometric campaign conducted by \citet{1991MNRAS.250..477K} in 1989. In addition to the pulsational variability around the period detected earlier, the latter data revealed a new pulsation with a period around 8.26\,min (2.0174\,mHz; 174.23\cd). In 2008, further photometric data, as well as high time resolution UVES/ESO spectra were collected. The analysis of these data, published by \citet{2015MNRAS.446.1347M}, showed that pulsational variability around both periods was still present.  

The Cycle\,1 TESS data for this star were analysed by \citet{ 2019MNRAS.487.3523C} who detected four modes with frequencies in the interval $1.2008- 1.2152$\,mHz, thus in the frequency region where all earlier data sets exhibited pulsations. However, the higher frequency mode, detected in the 1989 and 2008 data sets, was not detected in the TESS data. Similar to the earlier ground-based data, the TESS data showed evidence for amplitude and/or frequency modulation. No rotation period is reported in the literature for this star with \citet{2020A&A...639A..31M} classifying the star as a super slowly rotating Ap (ssrAp) star, although the longitudinal field measurements reported in the catalogue by  \citet{2015A&A...583A.115B} do show some indication of variability on the time scale of a few years. No sign of rotation variability was found in any of the published analyses of the TESS data \citep{2019MNRAS.487.3523C,2019MNRAS.487..304D,2020A&A...639A..31M}.

Since the TESS data for this star have already been analysed in detail, we refer the reader to those works, but include the pulsation frequencies in Table\,\ref{tab:stars} for completeness.

\subsubsection{TIC\,146715928}

TIC\,146715928 (HD\,92499) was classified as A2p\,SrEuCr \citep{1978mcts.book.....H} with $T_{\rm eff}$ values ranging from 7200\,K to 7810\,K \citep{2007MNRAS.378L..16H,2008MNRAS.389..441F,2010MNRAS.404L.104E}, and $\log g= 4.1\pm0.2$\,\cms. \citet{2010MNRAS.404L.104E} also provided a full chemical abundance analysis of this star. A mean magnetic field modulus of $8.5$\,kG was derived from FEROS spectra by \citet{2007MNRAS.378L..16H} which was corroborated by \citet{2017A&A...601A..14M} who also suggested the rotation period of TIC\,146715928 must be greater than 5\,yr. TIC\,146715928 has a $v\sin i\approx3.5$\,\kms\ \citep{2007MNRAS.378L..16H,2010MNRAS.404L.104E}.

\begin{figure}
    \centering
    \includegraphics[width=\columnwidth]{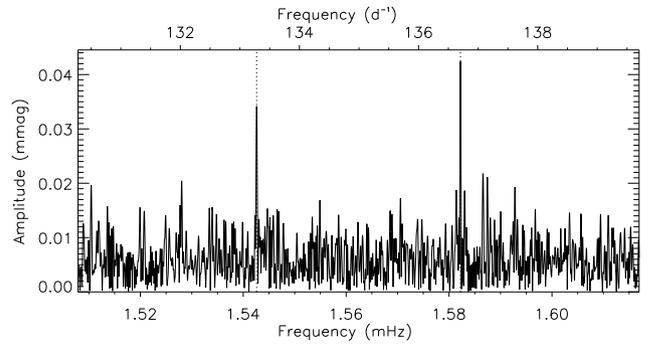}
    \caption{The pulsation signatures seen in TIC\,146715928. The vertical dotted lines represent the extracted peaks, which are separated by $40\,\muup$Hz.}
    \label{fig:tic146715928}
\end{figure}

Rapid oscillations were found in the spectra of TIC\,146715928 by \citet{2010MNRAS.404L.104E} with a period of $10.4\pm0.3$\,min. TESS observed the star during sectors 9 and 10. \citet{2020A&A...639A..31M} used the data to classify the star as an ssrAp star from the lack of rotation signal. They also provided a brief analysis of the two pulsation modes in this star that are separated by $40\,\muup$Hz, which could be the large frequency separation, or half of it. We present the extracted mode frequencies in Table\,\ref{tab:stars}, and show a plot of the amplitude spectrum in Fig.\,\ref{fig:tic146715928}.

\subsubsection{TIC\,167695608}
TIC\,167695608 (TYC\,8912-1407-1) was observed in 12 of the 13 sectors of Cycle\,1 (all but sector 5). The star was classified as F0p\,SrEu(Cr) by \citet{2014MNRAS.439.2078H}, who collected spectra to confirm its peculiar nature following the detection of rapid variability, with a period around 11\,min (1.532\,mHz; 132.38\,\cd) in a survey conducted using the SuperWASP archive \cite{2006PASP..118.1407P}.  

The first two sectors of Cycle\,1 TESS data were analysed by \citet{2019MNRAS.487.3523C} who confirmed the pulsation detected previously and identified two more modes at slightly higher frequencies. A follow-up analysis of TESS data collected up to sector 7 revealed yet another pulsation mode at slightly lower frequency than the modes detected previously \citep{2019MNRAS.487.2117B}. Altogether, the TESS data up to sector 7 had revealed four pulsation frequencies. No sign of rotational variability was found for this star in the SuperWASP \citep{2014MNRAS.439.2078H} or in the year-long TESS Cycle\,1 data \citep{2020A&A...639A..31M}. 

\begin{figure}
    \centering
    \includegraphics[width=\columnwidth]{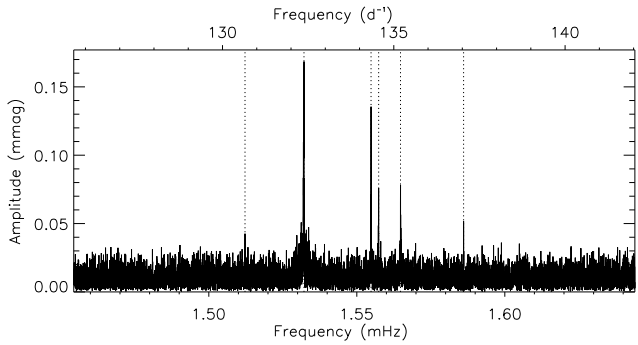}
    \caption{The pulsation signatures seen in TIC\,167695608. There is significant frequency variability in most modes.}
    \label{fig:tic167695608}
\end{figure}

With the longer data set, we are now able to detect 6 pulsation frequencies in this star (Fig.\,\ref{fig:tic167695608}), most of which show significant frequency variability, evidenced by broad ragged peaks in the amplitude spectrum that are not cleanly extracted when prewhitened. There is no clear pattern to the splitting of the pulsation modes, thus inhibiting the determination of the large or small frequency separation in this star. 

\subsubsection{TIC\,168383678}

TIC\,168383678 (HD\,96237) is an roAp star discovered by \citet{2011MNRAS.411..978E} and confirmed by \citet{2013MNRAS.431.2808K}  who determined the following stellar parameters: $T_{\rm eff}=7800$\,K, $\log g= 4.3$\cms, $v \sin i=6$\,\kms\ and a mean magnetic field modulus of $\langle B\rangle=2.9$\,kG. TIC\,168383678 has a spectral classification of A4\,SrEuCr \citep{2009A&A...498..961R} and a rotation period of 20.91\,d \citep{2008MNRAS.389..441F}. The dominant pulsation mode was reported at 1.2\,mHz \citep{2011MNRAS.411..978E, 2013MNRAS.431.2808K}.

\begin{figure}
    \centering
    \includegraphics[width=\columnwidth]{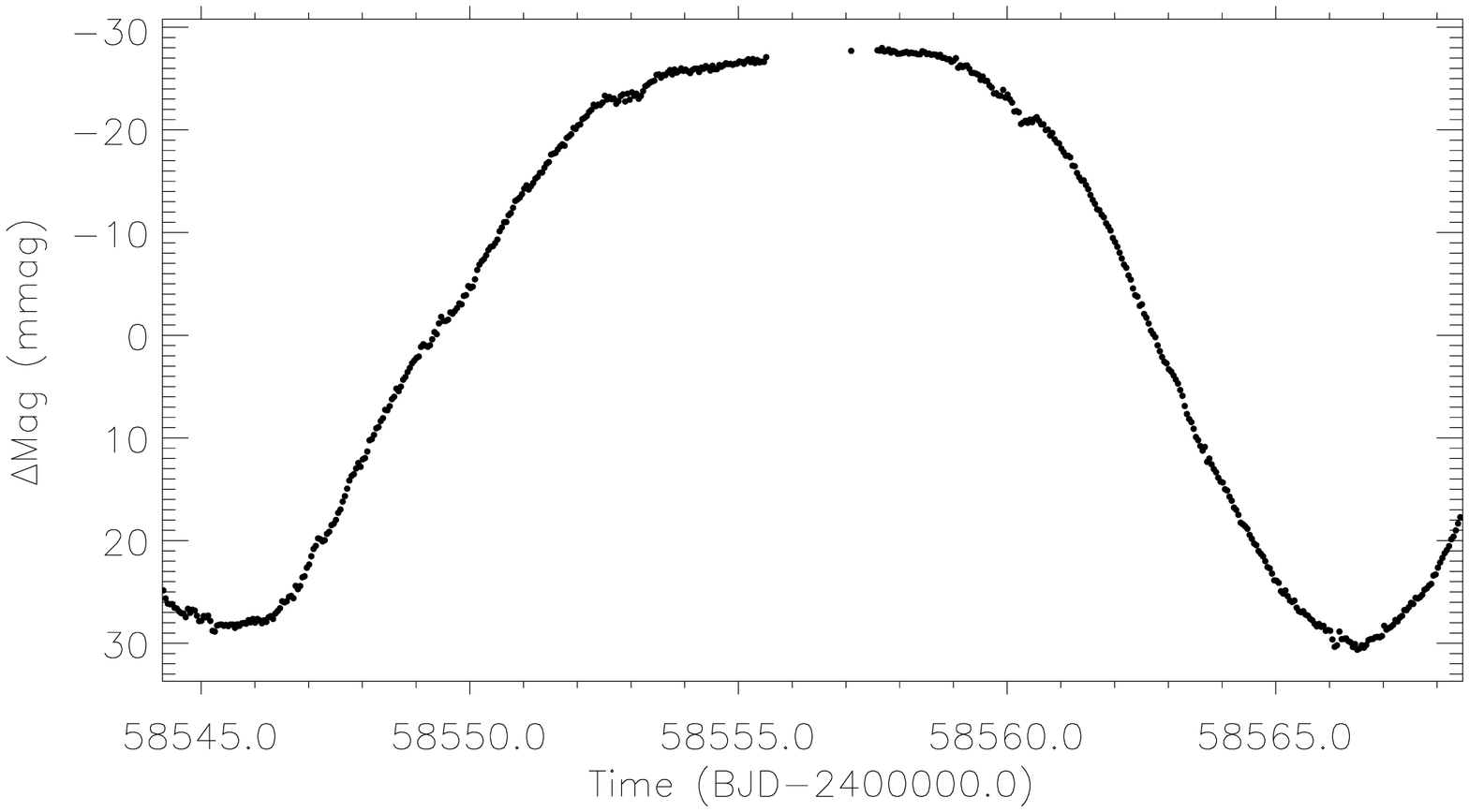}
    \includegraphics[width=\columnwidth]{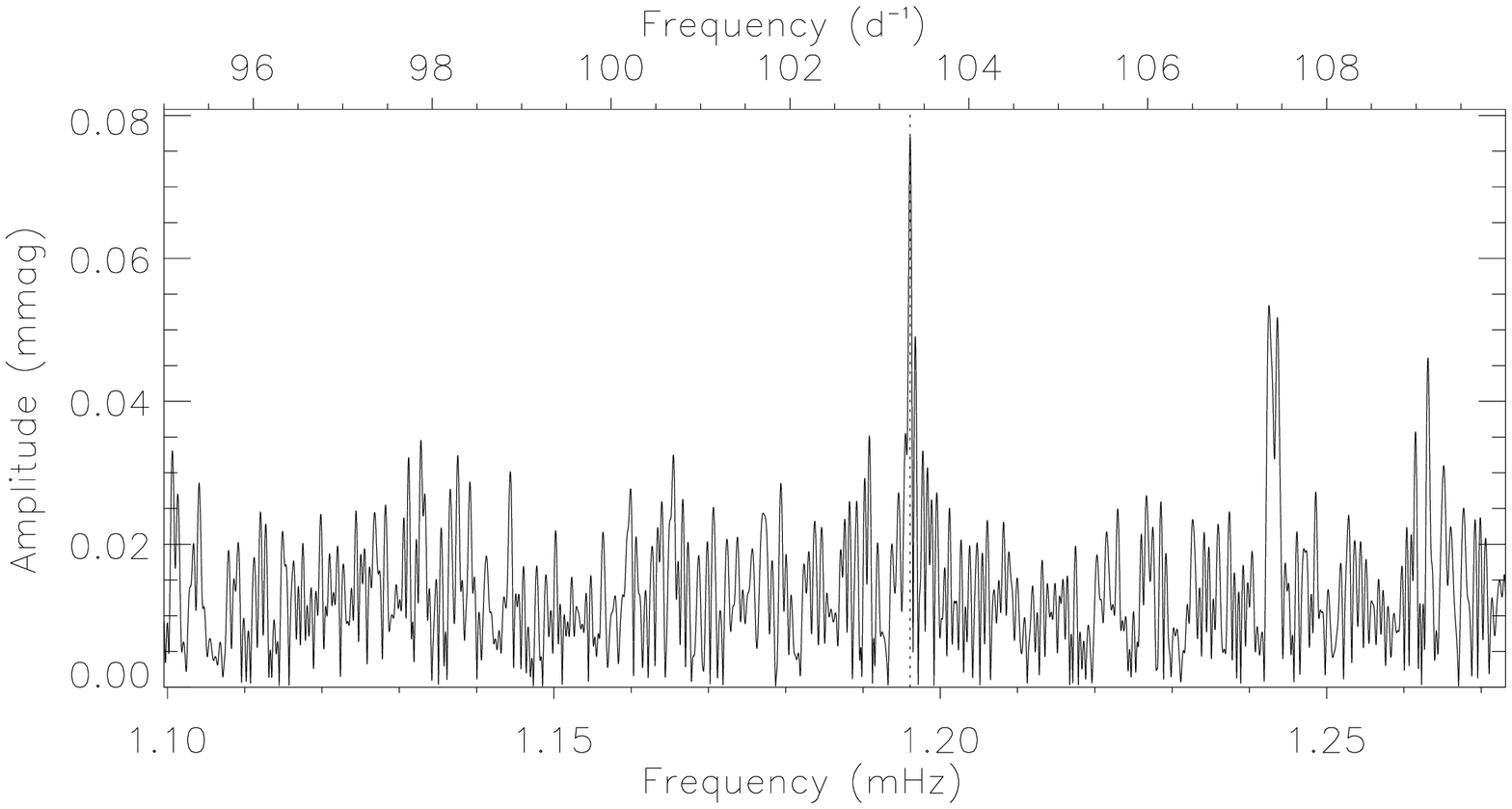}
    \caption{Top: the SAP light curve of TIC\,168383678. It appears that one cycle is completed, but inspection of the unequal light minima suggests we may be seeing only half of a rotation cycle. Bottom: the amplitude spectrum of TIC\,168383678 with the known pulsation mode indicated. There may be further modes at higher frequency, but with low significance in this data set.}
    \label{fig:tic168383678}
\end{figure}

TESS observed the star during sector 9. The SPOC pipeline has removed the astrophysical signal associated with the rotation period in this star, so we use the SAP data. Upon inspection of the light curve, it is not clear if we have observed one rotation cycle or half of it. Inspection of the two observed light minima show subtly different shapes (Fig.\,\ref{fig:tic168383678}). The data for this star span just 24.2\,d which is just a little longer than the rotation period reported in the literature. From the light curve, we estimate a rotation period of either $\sim21$\,d or $\sim42$\,d.

We detect a significant peak in the amplitude of the SAP light curve at a frequency of $1.19610\pm0.00003$\,mHz ($103.343\pm0.003$\,\cd) which has a S/N of 7.6. There are, perhaps, further modes in this star at slightly higher frequencies than that detected, but more data are required to confirm their presence.


\subsubsection{TIC\,170419024}

TIC\,170419024 (HD\,151860) was classified as Ap\,SrEu(Cr) \citep{1975mcts.book.....H}. \citet{2013MNRAS.431.2808K} measured a mean magnetic field modulus of 2.5\,kG, and detected rapid pulsations through the analysis of high-resolution spectra collected with UVES, after the null detection in photometry \citep{1994MNRAS.271..129M}. An average pulsation period of $12.30\pm0.09$\,min was established from the radial velocity of lines of rare earth elements. Some lower amplitude peaks at other frequencies were detected indicating this star to be a multiperiodic roAp star. 

\begin{figure}
    \centering
    \includegraphics[width=\columnwidth]{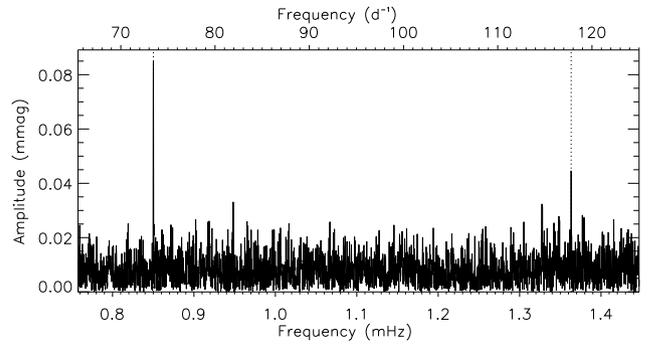}
    \caption{The pulsation signatures seen in TIC\,170419024. The vertical dotted lines represent the extracted peaks.}
    \label{fig:tic170419024}
\end{figure}

The TESS sector 12 data show this star to be an ssrAp star \citep{2020A&A...639A..31M} since the data show no rotation signature. \citet{2020A&A...639A..31M} also commented on the pulsation modes in this star, as detected in the TESS data. There is a significant peak at $0.85014\pm0.00002$\,mHz ($73.452\pm0.002$\,\cd) which was not reported by \citet{2013MNRAS.431.2808K}, and a second peak at $1.36330\pm0.00004$\,mHz ($117.789\pm0.003$\,\cd; Fig.\,\ref{fig:tic170419024}) which is consistent with the spectroscopic analysis.

\subsubsection{TIC\,173372645}

TIC\,173372645 (HD\,154708), with a rotation period of 5.37\,d, an effective temperature of 7200\,K \citep{2016A&A...590A.116J}, $\log g= 4.11$\,\cms, and $v \sin i=4.0$\,\kms, is classified as an Ap\,EuSrCr star \citep{2009A&A...498..961R}. The dominant pulsation mode has a frequency of 2.088\,mHz (180.4\,\cd), which was discovered by \citet{2006MNRAS.372..286K} through the analysis of high time resolution UVES spectra at an amplitude of just $\approx60$\,\ms. TIC\,173372645 has a magnetic field of 24.5\,kG, the strongest ever measured for an roAp star \citep{2005A&A...440L..37H}, and one of the strongest for an Ap star. 

\begin{figure}
    \centering
    \includegraphics[width=\columnwidth]{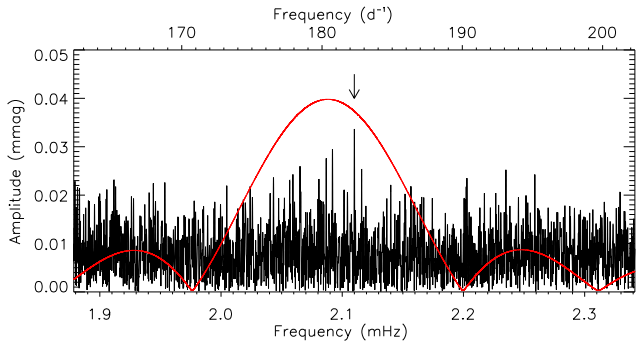}
    \caption{The amplitude spectrum of TIC\,173372645. The arrow indicates a potential peak, whereas the broad smooth red curve represents the shape of the peak used by \citet{2006MNRAS.372..286K} to identify the pulsation mode in this star.}
    \label{fig:tic173372645}
\end{figure}

TESS observed TIC\,173372645 in sector 12. The light curve is rotationally modulated with a period of $5.363\pm0.001$\,d. There is no clear indication of pulsation in this star in the TESS data, only a few peaks in the amplitude spectrum in the frequency region where the pulsation was previously found \citep[Fig.\,\ref{fig:tic173372645};][]{2006MNRAS.372..286K}. The low amplitude in the spectroscopic data seems to have translated to an undetectable peak in the TESS photometry. 

\subsubsection{TIC\,189996908}

TIC\,189996908 (HD\,75445) was classified as an Ap\,SrEu(Cr) star by \citet{1982mcts.book.....H} and has a magnetic field strength of $\langle B\rangle = 2.99\pm0.04$\,kG  \citep{1997A&AS..123..353M} . The stellar parameters were measured to be  $T_{\rm eff}=7700$\,K ,  $\log g= 4.3$\,\cms\ \citep{2004A&A...423..705R},  $v \sin i \leq 2$\,\kms  \citep{2009A&A...493L..45K} and the Hipparcos parallax resulted in $\log L = 1.17\pm0.06$\,L$_\odot$ \citep{2006A&A...450..763K}.  Based on radial velocity measurements,  \citet{2009A&A...493L..45K} discovered rapid oscillations in this star, with a main pulsation frequency at 1.85\,mHz (159.8\,\cd). 

\begin{figure}
    \centering
    \includegraphics[width=\columnwidth]{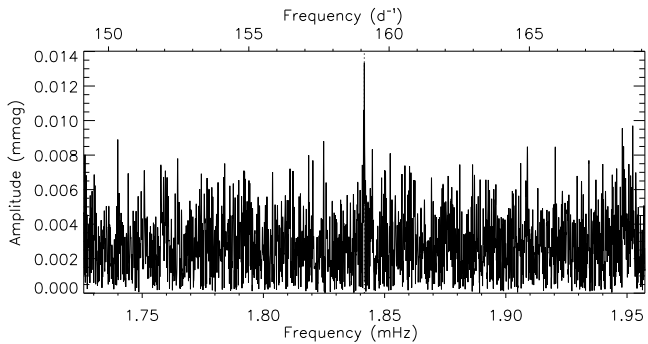}
    \caption{The amplitude spectrum of TIC\,189996908. The low S/N peak detection agrees with that found in spectroscopic observations.}
    \label{fig:tic189996908}
\end{figure}

TESS observed TIC\,189996908 in sectors 8 and 9. The data show no rotational modulation which implies a long rotation period, or an unfavourable view. The magnetic field measurements presented by \citet{2017A&A...601A..14M} show changes over short time periods, implying a low inclination angle, rather than slow rotation. The pulsation signal is weak in this star, and would perhaps be classed as noise if it was not previously known. We find a peak in the amplitude spectrum at $1.84163\pm0.00002$\,mHz ($159.117\pm0.002$\,\cd) with a S/N of 5.3 (Fig.\,\ref{fig:tic189996908}). This is another example of how low amplitude modes in spectroscopic observations are difficult to observe photometrically. 

\subsubsection{TIC\,211404370}
TIC\,211404370 (HD\,203932) is classified as A5\,SrEu in the catalogue of \citet{2009A&A...498..961R}. \citet{1997A&A...319..630G} derived $T_{\rm eff}=7540\pm100$\,K and  $\log g= 4.1$\cms\ for this star, while the average magnetic field, $\langle B\rangle$, was estimated to be below 1\,kG \citep{2008A&A...480..811R}, while the root mean square longitudinal field was reported to be $0.25\pm0.15$\,kG \citep{2009MNRAS.394.1338B} and the $v \sin i$ was measured to be 4.7\,\kms\ \citep{2015MNRAS.452.3334S}. Pulsations were first detected in TIC\,211404370 by \citet{1984MNRAS.209..841K} and it was later shown to have 4 significant peaks in the amplitude spectrum between 1.280 and 2.737\,mHz \citep{1990MNRAS.246..699M}. 

This star was observed in TESS's sector 1, with the data analysed in \citet{2019MNRAS.487.3523C}. We find no significant difference in our analysis and that previously presented, so we refer the reader to that work. We include the rotation frequency, and the frequency of the two pulsation modes, in Table\,\ref{tab:stars}.


\subsubsection{TIC\,237336864}

TIC\,237336864 (HD\,218495), with a spectral type of Ap\,EuSr \citep{2009A&A...498..961R}, was shown to have a $v \sin i$ of 16\,\kms\ \citep{2015MNRAS.452.3334S} with a mean longitudinal magnetic field, $B_z$, of about $-1$\,kG \citep{2015A&A...583A.115B}. \citet{2019MNRAS.487.3523C} derived $T_{\rm eff}=7950\pm 160$\,K and $\log g= 4.1$\cms. Rapid oscillations were first detected by \citet{1990IBVS.3509....1M} in this star. 

This star was observed in sector 1, with the data analysed in \citet{2019MNRAS.487.3523C} and by \citet{Kobzar2020}. The data show a complex pulsation pattern with, in increasing frequency, a singlet, a doublet which is split by twice the rotation frequency, a quintuplet with components split by the rotation frequency, and a triplet of frequencies with each component split by twice the rotation frequency. Given the variety of different multiplet structures, a simple understanding of this star is not possible, but to develop a complete understanding is beyond the scope of this work. We provide the inferred pulsation frequencies and amplitudes in Table\,\ref{tab:stars}.


\subsubsection{TIC\,268751602}
\citet{1988mcts.book.....H} classified TIC\,268751602 (HD\,12932) as Ap\,SrEuCr, with \citet{2007MNRAS.376..651K} providing parameters of $T_{\rm eff}=7620$\,K, $\log g=4.15$\,\cms, and $v\sin i=3.5$\,\kms, and \citet{2007A&A...473..907R} deriving $\langle B\rangle=1.7$\,kG.

\citet{1990IBVS.3520....1S} and \citet{1990IBVS.3539....1K} independently collected high time resolution photometric data of TIC\,268751602 and discovered variability with a period of about 11.6\,min (1.4368\,mHz; 124.14\,\cd), thus announcing TIC\,268751602 as an roAp star. This period was later refined by \citet{1992A&A...257..130S} to be 11.7\,min, with the authors suggesting a rotation period of 3.5295\,d. Analysis of photometric data by \citet{1994MNRAS.271..305M} showed that TIC\,268751602 was a single-mode pulsator with a period of 11.6\,min. The earlier identification of the rotation period was concluded to be likely incorrect. \citet{2007MNRAS.376..651K} identified variability in spectral lines of some elements in the atmosphere of TIC\,268751602.

\begin{figure}
    \centering
    \includegraphics[width=\columnwidth]{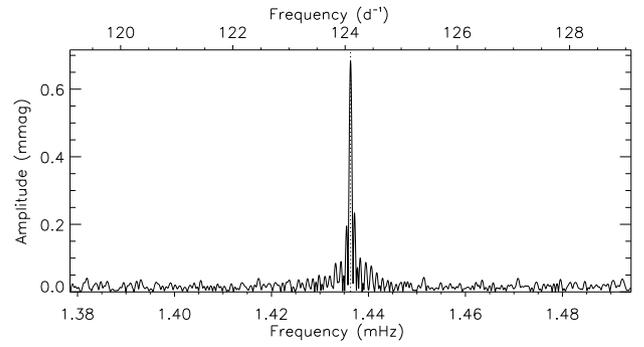}
    \caption{The single pulsation mode seen in TIC\,268751602.}
    \label{fig:tic268751602}
\end{figure}

TESS observed this star during sector 3, with an analysis of those data presented by \citet{2019MNRAS.487.2117B}. We confirm the lack of detection of rotation in this star, and agree with the pulsation frequency presented in that work, namely $1.436302\pm0.000006$\,mHz ($124.0965\pm0.0005$\,\cd). There is only one pulsation mode seen in these data of this star (Fig.\,\ref{fig:tic268751602}).

\subsubsection{TIC\,279485093}

TIC\,279485093 (HD\,24712, HR\,1217) was classified as A9Vp\,SrEuCr by \citet{1995ApJS...99..135A}. \citet{2016A&A...590A.117P} derived a temperature from bolometric flux and optical interferometry of $7235\pm280$\,K, while \citet{2019MNRAS.483.2300S} derived $7150\pm170$\,K from spectroscopy. The mean magnetic field modulus for TIC\,279485093 was determined to be 2.3\,kG \citep{2008A&A...480..811R}. Detailed topology of the magnetic field of this star was reconstructed by \citet{2010A&A...509A..71L} and \citet{2015A&A...573A.123R} - the only roAp for which such constraint is available. \citet{1987MNRAS.229..285K} obtained a rotation period from the mean light variation of 12.45733\,d, while \citet{2013A&A...558A...8R} derived 12.45812\,d from the longitudinal field measurements spanning 40\,yr and \citet{2020pase.conf..214K} found a signal in the Fourier spectrum of the TESS data that may correspond to a rotation period of $12.44\pm0.02$\,d. \citet{1981IBVS.1915....1K} first discovered 6.15\,min (2.71\,mHz; 234.1\,\cd) oscillations in this star, and \citet{2019MNRAS.487.2117B} list 10 independent frequencies.  Many of these modes show the expected rotation-modulated amplitude variations, but unexplained variations in this modulation also occur on shorter timescales of days \citep{2011MNRAS.415.1638W}.  This star is one of the first discovered and best studied roAp stars, appearing extensively in the literature.

\begin{figure}
    \centering
    \includegraphics[width=\columnwidth]{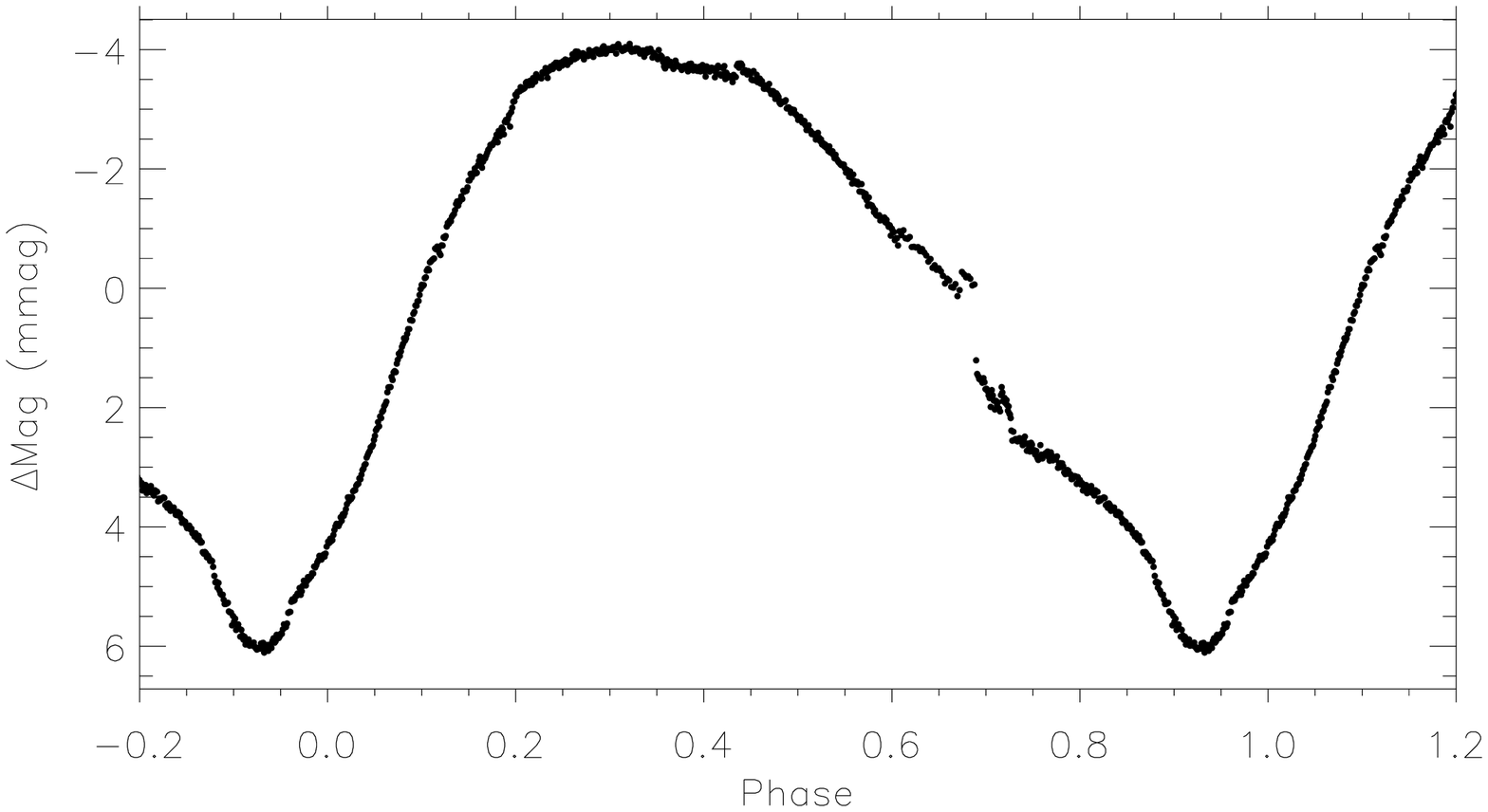}
    \includegraphics[width=\columnwidth]{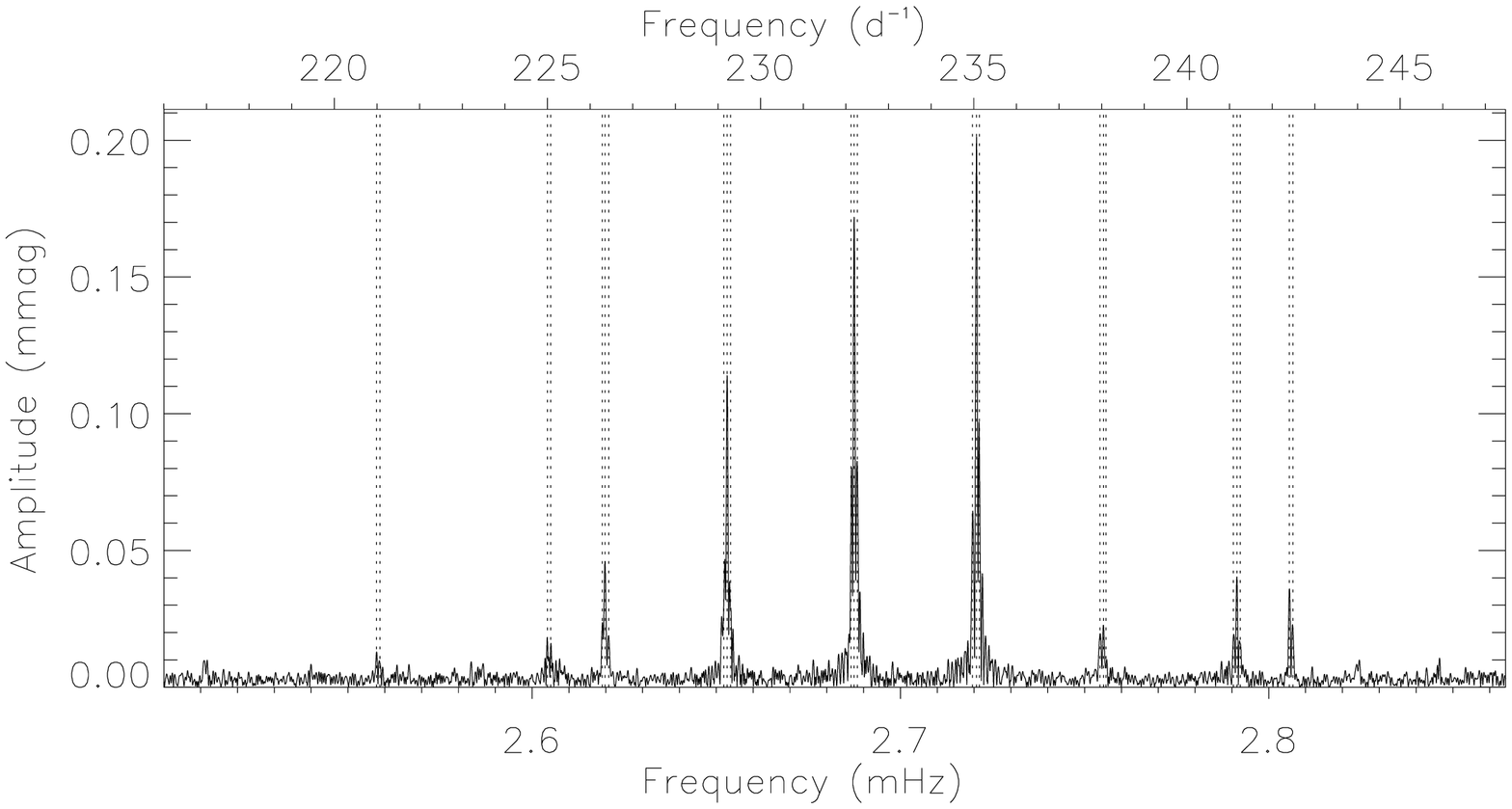}
    \includegraphics[width=\columnwidth]{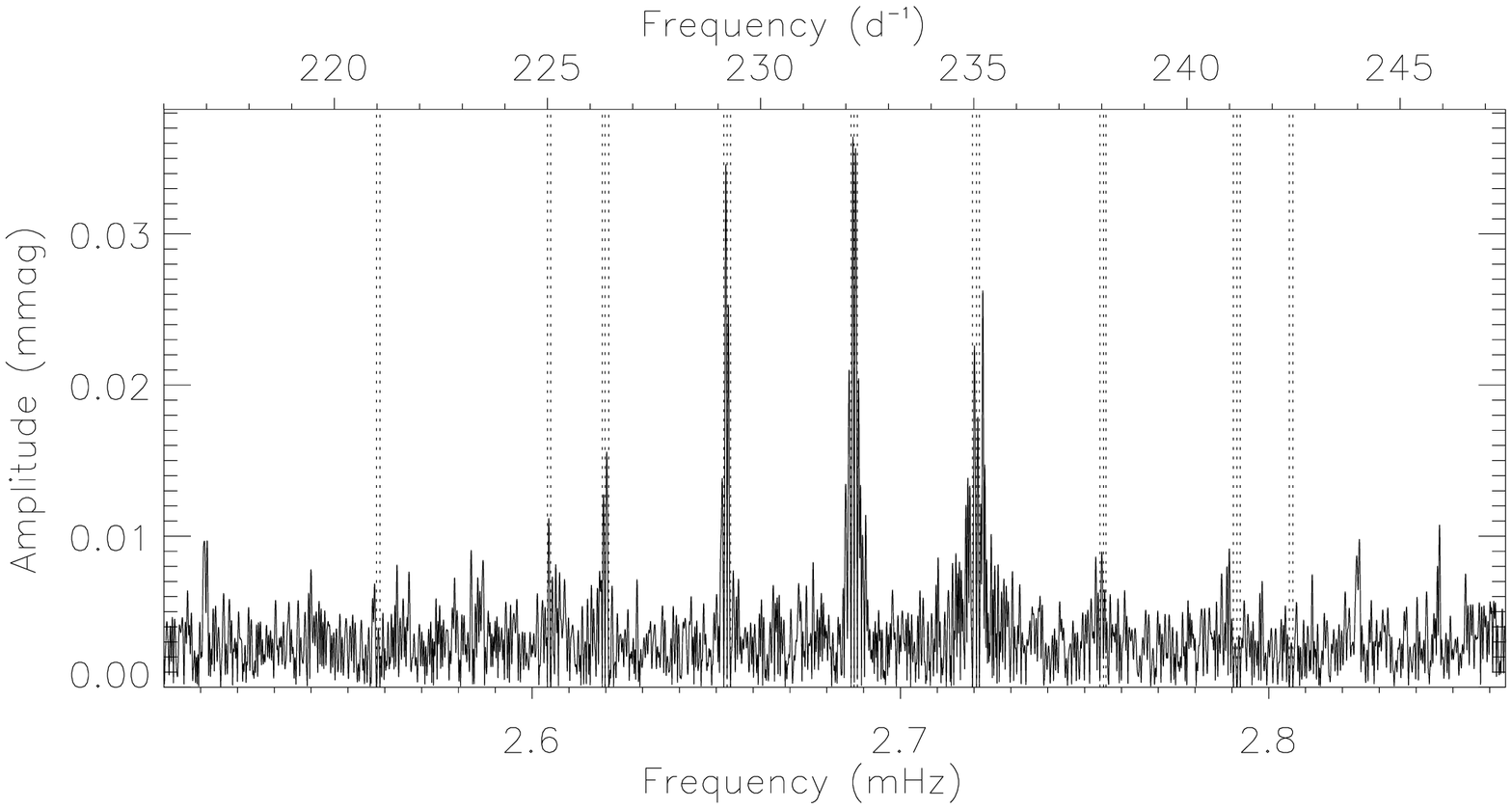}
    \caption{Top: the phase folded SAP light curve of TIC\,279485093, folded on a period of 12.578\,d. Middle: the amplitude spectrum of the PDC\_SAP data. The vertical dotted lines represent the frequencies extracted. Bottom: amplitude spectrum of the residuals after removing 24 frequencies (9 pulsation modes and significant sidelobes). There is power remaining around the dominant modes which suggests significant frequency variability in this star. It seems likely there are further modes close to the noise level, too.}
    \label{fig:tic279485093}
\end{figure}

As mentioned, TESS observed this star in sector 5. From the SAP data, we derive a rotation period of $12.578\pm0.008$\,d which is different from previous literature values (Fig.\,\ref{fig:tic279485093}). The sector 5 SAP data, after removing points obviously affected by instrumental artefacts, span $25.63$\,d -- a little over twice the derived rotation period, thus this value should be treated with caution.

To inspect the pulsations in this star, we used the PDC\_SAP light curve. The pulsation spectrum is known to be rich in this star, as evidenced by the middle panel in Fig.\,\ref{fig:tic279485093}. We are able to extract 9 pulsation modes (as listed in Table\,\ref{tab:stars}) and 15 sidelobes that are split from the pulsation frequency by oblique pulsation. In comparison to the best ground based photometric data set for this star \citep{2005MNRAS.358..651K}, we detect all but their $\nu_6$ frequency (2.789\,mHz; 240.97\,\cd) although we see a power excess in the bottom panel of Fig.\,\ref{fig:tic279485093} around this frequency. We detect two further modes (our lowest frequencies) which likely correspond to two modes detected by \citet{2005A&A...430..263M} via high-resolution spectroscopy.


\subsubsection{TIC\,280198016}

TIC\,280198016 (HD\,83368, HR\,3831) is a close visual binary, with the roAp star having spectral type A8V\,SrCrEu and the companion having type F9V \citep{1984ApJS...55..657C}. \citet{2019MNRAS.483.2300S} derive $T_{\rm eff}=7660\pm170$\,K which is consistent with the TIC value. TIC\,280198016 shows a polarity reversing magnetic field strength that varies about zero mean with amplitude $0.74\pm0.07$\,kG \citep{1991A&AS...89..121M}, and has a mean field modulus of 2.1\,kG \citep{2004A&A...424..935K}. The rotation period of TIC\,280198016 was precisely determined to be $2.851976\pm0.000003$\,d \citep{1997MNRAS.287...69K}, with the star showing a single dipole pulsation mode at 123.38\,\cd\ that was exactly split into a septuplet by pulsation amplitude modulation with the rotation frequency. \citet{1998MNRAS.300L..39B} found that the radial velocity variations were amplitude and phase modulated in the same manner as the photometric variations, first confirming the oblique pulsator model and ruling out the spotted pulsator model for roAp stars \citep{1985A&A...151..315M}. A detailed study of pulsational line profile variations based on high-resolution spectroscopic observations was carried out by \citet{2006A&A...446.1051K}. The surface distribution of the pulsation amplitude and phase was mapped by \citet{2004ApJ...615L.149K} with the help of the Doppler imaging method - the only such analysis available for an roAp star.

\begin{figure}
    \centering
    \includegraphics[width=\columnwidth]{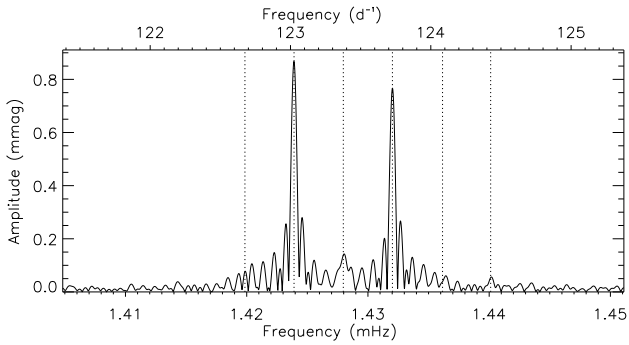}
    \caption{Amplitude spectrum of  PDC\_SAP light curve of TIC\,280198016. We detect 6 of the 7 septuplet components found by \citet{1997MNRAS.287...69K}.}
    \label{fig:tic280198016}
\end{figure}

The TESS sector 9 data of TIC\,280198016 allow us to derive a rotation frequency (Table\,\ref{tab:stars}) which agrees with that of \citet{1997MNRAS.287...69K}. We detect 6 significant peaks (Fig.\,\ref{fig:tic280198016}) of the reported septuplet in TIC\,280198016 and find that the mode shows frequency variability, as reported by \citep{1997MNRAS.287...69K}. We also detect the second harmonic of the pulsation mode which forms a triplet centred on twice the pulsation frequency, although \citet{1997MNRAS.287...69K} show four harmonics.

\subsubsection{TIC\,315098995}

TIC\,315098995 (HD\,84041) was classified as Ap\,SrEuCr by \citet{1982mcts.book.....H}, and has effective temperature measurements of 8100\,K \citep{1996A&A...311..901M}, 7800\,K and 7500\,K from \citet{2008CoSka..38..317E} obtained through the photometric calibrations of \citet{1985MNRAS.217..305M} and fitting synthetic spectra to the H$_\alpha$ profile. TIC\,315098995 is a magnetic star with \citet{2006A&A...450..763K} reporting the first detection of a longitudinal magnetic field of $0.50\pm0.09$\,kG. The latest evaluation given in the FORS1 catalogue of stellar magnetic field measurements was $0.64\pm0.05$\,kG measured from Balmer and metal lines \citep{2015A&A...583A.115B}.

\citet{1991IBVS.3621....1M} first reported the star as an roAp star with a pulsation period of about 14.6\,min (1.14\,mHz; 98.6\,\cd). Later \citet{1993MNRAS.263..273M} clarified that the star pulsated in at least three modes with periods near 15\,min and with frequency separations of about $30\,\umu$Hz. However, it was noted that these frequencies were not a complete description of the pulsation behaviour of TIC\,315098995. It exhibited mean light variations of a double-wave nature, indicating that both magnetic poles were visible. The rotation period determined from these observations was $3.69\pm0.01$\,d. \citet{2008CoSka..38..317E} reported maximum pulsation amplitude of $0.5$\,\kms and a pulsation amplitude from the H$_\alpha$ core of $0.20$\,\kms.

TESS observed TIC\,315098995 during sectors 8 and 9. We cautiously combined the PDC\_SAP data from sector 8 and the SAP sector 9 data to determine the rotation period since the SPOC pipeline has distorted the astrophysical signal in the latter. We measure a rotation period of $3.6884\pm0.0004$\,d, which is consistent with the literature value.

\begin{figure}
    \centering
    \includegraphics[width=\columnwidth]{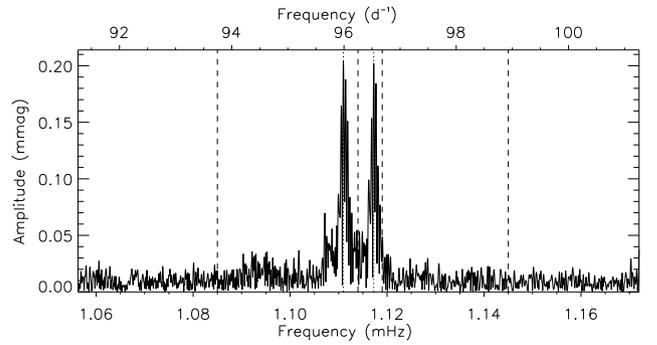}
    \caption{Amplitude spectrum of the two-sector light curve of TIC\,315098995. There are at least two very variable peaks in this star. The vertical dashed lines represent the positions of the frequencies found by \citet{1993MNRAS.263..273M}.}
    \label{fig:tic315098995}
\end{figure}

The TESS data show this star to have a very variable pulsation signal, as reported by previous studies. We find two peaks (Fig.\,\ref{fig:tic315098995}) that are split by twice the rotation frequency. We assume then, that we are seeing two sidelobes of a dipole triplet where the peak corresponding to the actual pulsation mode frequency is missing, implying that either $i$ or $\beta$ are close to $90^\circ$. This is corroborated when considering the results of \citet{1993MNRAS.263..273M}. We do not detect, in the combined light curve, the other modes reported to be present in this star. A careful inspection of the data, in shorter segments, is required to extract the full information on this star.

\subsubsection{TIC\,322732889}


TIC\,322732889 (HD\,99563) was classified as A9\,III star by \citet{1978mcts.book.....H} and with a Sr peculiarity added later \citep{1981AJ.....86..553B}, although the star appears in the catalogue of \citet{2009A&A...498..961R} as an F0\,Sr. From high-resolution spectra, \citet{2008CoSka..38..317E} found $T_{\rm eff} = 7700$\,K and $\log g =  4.2$\,\cms. The first measure of the longitudinal magnetic field gave $0.69\pm0.15$\,kG \citep{2004A&A...415..685H}, with two further measurements at different phases of $-0.24\pm0.07$\,kG and $0.67\pm0.04$\,kG \citep{2006AN....327..289H}. These detections were confirmed by \citet{2006A&A...450..763K} who measured a field strength of $-0.392$\,kG.

\citet{1998CoSka..27..338D} first announced the detection of pulsation in TIC\,322732889 with a period of about 11.17\,min (1.4921\,mHz; 128.9\,\cd). Later, \citet{1999A&AS..135...57H} confirmed the roAp nature of the star with a period of 10.7\,min. \citet{2005MNRAS.364..864E} discovered remarkably large amplitude pulsations in the star for some spectral lines with amplitudes up to 5\,\kms\ making this star the largest radial velocity amplitude roAp star known. A more detailed study by \citet{2006MNRAS.366..257H} using multi-site photometry found a single rotationally modulated mode with a frequency 1.558\,mHz (134.6\,\cd) connected with the rotation determined to be 2.912\,d. Further spectroscopic observations found little variation in the pulsation amplitude in TIC\,322732889 \citep{2006MNRAS.370.1274K}, while \citet{2009MNRAS.396..325F} claimed the presence of H$\alpha$ spots caused by the settling of He in the presence of the magnetic field. Based on MOST observations, \citet{2011A&A...530A.135G} found a new candidate frequency independent of the primary multiplet.

\begin{figure}
    \centering
    \includegraphics[width=\columnwidth]{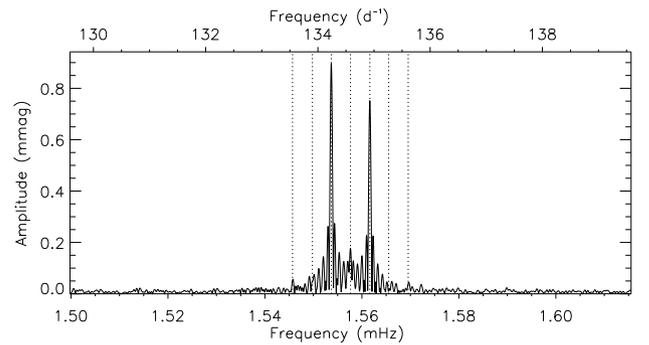}
    \caption{Amplitude spectrum of the light curve of TIC\,322732889. The vertical dotted lines represent the septuplet components split by oblique pulsation.}
    \label{fig:tic322732889}
\end{figure}

TESS observed TIC\,322732889 during sector 9. With those data, we derive a rotation period of $2.9114\pm0.0002$\,d which is consistent with that in the literature. As reported in previous works, we find a septuplet of peaks in the amplitude spectrum separated by the rotation frequency. The central peak, representing the pulsation mode, is at $1.557657\pm0.000002$\,mHz ($134.5816\pm0.0002$\,\cd; Fig.\,\ref{fig:tic322732889}), and has a smaller amplitude than the first rotational sidelobes. We detect none of the candidate modes of \citet{2011A&A...530A.135G}.

\subsubsection{TIC\,326185137}

TIC\,326185137 (HD\,6532) was classified as Ap\,SrCrEu by \citet{1982mcts.book.....H}, although there are a variety of different classifications \citep{2014yCat....1.2023S}. Effective temperature measurements for this star range from 7995\,K by \citet{2019AJ....158...93B} to 8382\,K by \citet{2017MNRAS.471..770M}. The $\log g$ value for this star is also poorly constrained in the literature, with an average value of $4.25$\,\cms. Several measurements of the magnetic field have been made, with the $\langle B\rangle_{\rm rms}$ being 0.411\,kG \citep{1997A&AS..124..475M,2009MNRAS.394.1338B,2015A&A...583A.115B}.

TIC\,326185137 was discovered to be an roAp star by \citet{1985MNRAS.216..987K} with a period of 6.9\,min (2.4155\,mHz; 208.7\,\cd). They found a frequency triplet split by the rotation frequency of the star. Subsequently, the star was the subject of many ground-based photometric campaigns, with \citet{1996MNRAS.281..883K} concluding the star pulsates with a distorted dipole mode. The rotation period of this star has been refined many times in the literature, with an average period of 1.94\,d \citep{1987MNRAS.228..141K,1996MNRAS.280....1K,2009A&A...498..961R,2019MNRAS.487.2117B}.

TESS observed this star in sector 3, with an initial analysis of the data being presented by \citet{2019MNRAS.487.2117B} where it was noted that the multiplet structure was vastly different from previous $B$ photometric data. \citet{2020ASSP...57..313K} explored this further, postulating two scenarios that could explain the striking difference between the two data sets: either the star has changed its pulsation axis or, more likely, we are seeing a depth dependence on the mode geometry revealed by the two wavelengths of observation. 

\begin{figure}
    \centering
    \includegraphics[width=\columnwidth]{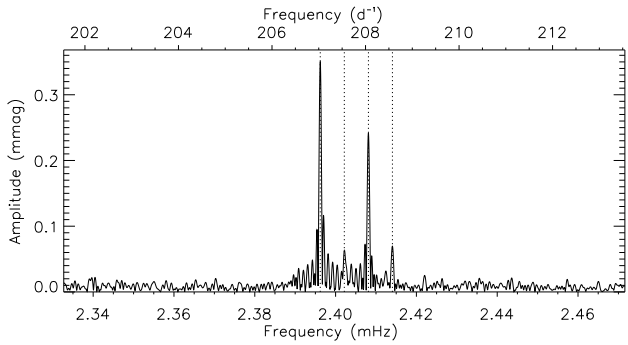}
    \caption{Amplitude spectrum of the light curve of TIC\,326185137, showing the single pulsation mode split into a multiplet. Note the different structure seen here compared with the photometric $B$ data of \citet{1996MNRAS.281..883K}.}
    \label{fig:tic326185137}
\end{figure}

From the data, we determine a rotation period of $1.9447\pm0.0001$\,d, and a pulsation frequency of $2.402163\pm0.000007$\,mHz ($207.5469\pm0.0006$\,\cd). We identify 4 of the 5 peaks of the distorted dipole multiplet (Fig.\,\ref{fig:tic326185137}) in the TESS Cycle\,1 data.

\subsubsection{TIC\,340006157}

TIC\,340006157 (HD\,60435) was classified as A3\,SrEu in \citet{1991A&AS...89..429R}. \citet{2019MNRAS.487.5922G} report abundances for the star and classify it as neither He-poor or rich. \citet{2006A&A...450..763K} detected a surface magnetic field with $B_z = -0.32\pm0.09$\,kG. The TIC gives $T_{\rm eff} = 8427\rm~K$, though values in the literature range as low as $7694\rm~K$ \citep{2012MNRAS.427..343M}. The rotation period was reported by \citet{1990MNRAS.243..289K} to be $7.68$\,d, with the light curve displaying a double wave of unequal depth.

The star was first identified as an roAp star by \citet{1984MNRAS.209..841K}. \citet{1987PhDT........68M} conducted a detailed study and identified 17 frequencies, which were identified as consecutive radial overtones of $\ell = 1$ and $\ell = 2$ modes, with amplitudes which were highly variable on timescales of days. \citet{2004A&A...425..179V} attempted modelling based on these frequencies, with results that suggested the presence of a He gradient inside the star. \citet{2019MNRAS.487.2117B} performed a preliminary analysis using TESS and detected 6 independent frequencies between 1.3079 and 1.4352 \,mHz ($113-124$\,\cd). All peaks were broad and poorly resolved, and many displayed variable amplitudes which both confirmed the results of \citet{1987PhDT........68M} and perhaps explained why not all the frequencies reported by the earlier author were detected in TESS data.

This star was observed in 7 sectors of Cycle\,1, with data spanning a total of 296\,d but with two 2-sector gaps. To derive the rotation period here, we used a combination of the raw and pipeline corrected data since the pipeline has altered the astrophysical signal in sectors 7 and 10 for this star. We derive a rotation period of $7.6797\pm0.0001$\,d.

\begin{figure}
    \centering
    \includegraphics[width=\columnwidth]{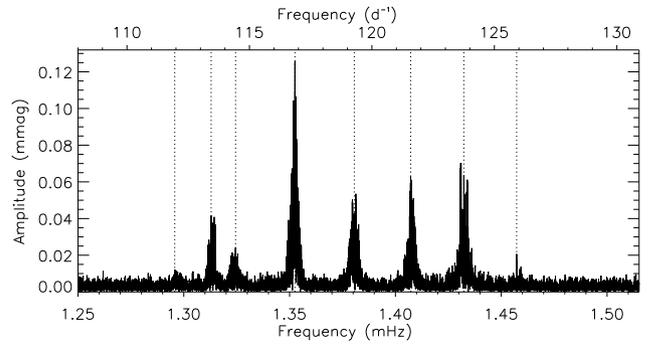}
    \caption{Amplitude spectrum of the light curve of TIC\,340006157, showing the many pulsation modes. The broad peaks indicate significant amplitude variability in this star, as reported in the literature.}
    \label{fig:tic340006157}
\end{figure}

The TESS amplitude spectrum is no different for this star as reported in the literature: it shows significant frequency and/or amplitude variability, making precise frequency extraction difficult. We are, however, able to identify 8 pulsation frequencies in the data, which we have listed in Table\,\ref{tab:stars}. Some of the extracted frequencies are the same as those reported by \citet{1987PhDT........68M}, but we do not detect the lowest frequency modes reported in that work. Only the second highest frequency mode shows rotationally split sidelobes in the amplitude spectrum. Due to the broad nature of the signal in the amplitude spectrum (Fig.\,\ref{fig:tic340006157}) we have inflated our uncertainties by a factor of 100 to reflect the uncertainty in the frequency determination. This star requires a detailed individual study in the light of significant amplitude variations and short mode lifetimes.

\subsubsection{TIC\,348717688}

TIC\,348717688 (HD\,19918) was observed in sectors 1,12 and 13, and has a spectral type of A5\,SrEuCr \citep{2009A&A...498..961R}. Its pulsations were first detected by \citet{1991IBVS.3553....1M} and revisited later by \citet{1995MNRAS.276.1435M} using ground-based photometry from SAAO. In addition to the dominant pulsation frequency at 1.510\,mHz (130.5\,\cd) and its harmonics, a second frequency at 1.4806\,mHz (127.9\,\cd) was detected. TIC\,348717688 has a known mean longitudinal magnetic field strength of $-0.63\pm0.09$\,kG \citep{2006AN....327..289H}, although a rotation period is not currently known. A mean field modulus of 1.6\,kG was derived by \citet{2007A&A...473..907R}. 

More recently, \citet{2019MNRAS.487.3523C} analysed TESS sector 1 data of TIC\,348717688 and extracted eight significant pulsation frequencies, including the dominant frequency originally reported by \citet{1991IBVS.3553....1M}. \citet{2020A&A...639A..31M}, using the complete TESS Cycle\,1 data set, classified this star as an ssrAp star through the lack of detection of a rotation signal, and very sharp spectral lines.

\begin{figure}
    \centering
    \includegraphics[width=\columnwidth]{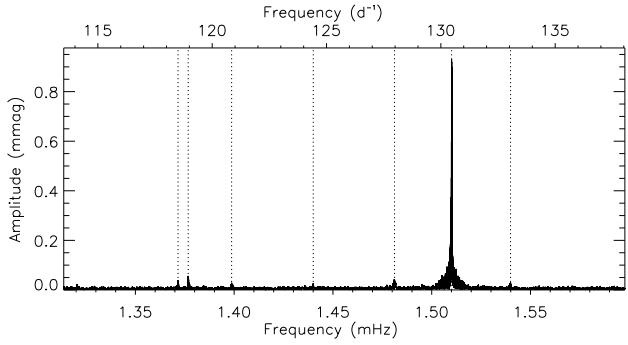}
    \caption{Amplitude spectrum of the light curve of TIC\,348717688, showing the multiple pulsation modes. All modes show signs of frequency variability.}
    \label{fig:tic348717688}
\end{figure}

Despite the large gap in the Cycle\,1 data set, the window function of the data does not significantly interfere with the frequency analysis of this star. However, we identify frequency variability in this star. We are able to extract seven pulsation frequencies from the Cycle\,1 light curve (Fig.\,\ref{fig:tic348717688}), and note the presence of the first harmonic of the highest amplitude mode. The frequency not detected here, but quoted in \citet{2019MNRAS.487.3523C}, is likely the result of the frequency variability of a nearby mode as this mode is now split into several peaks in the amplitude spectrum.

\subsubsection{TIC\,363716787}

TIC\,363716787 (HD\,161459) was observed in sector 13, and has a spectral type of Ap\,EuSrCr \citep{1978mcts.book.....H}. It has a dominant pulsation frequency of 1.39\,mHz \citep[120.1\,\cd;][]{1990IBVS.3507....1M,1991MNRAS.250..666M}. A single mean longitudinal magnetic field measurement indicates $B_z=-1.8$\,kG \citep{2015A&A...583A.115B}.

\begin{figure}
    \centering
    \includegraphics[width=\columnwidth]{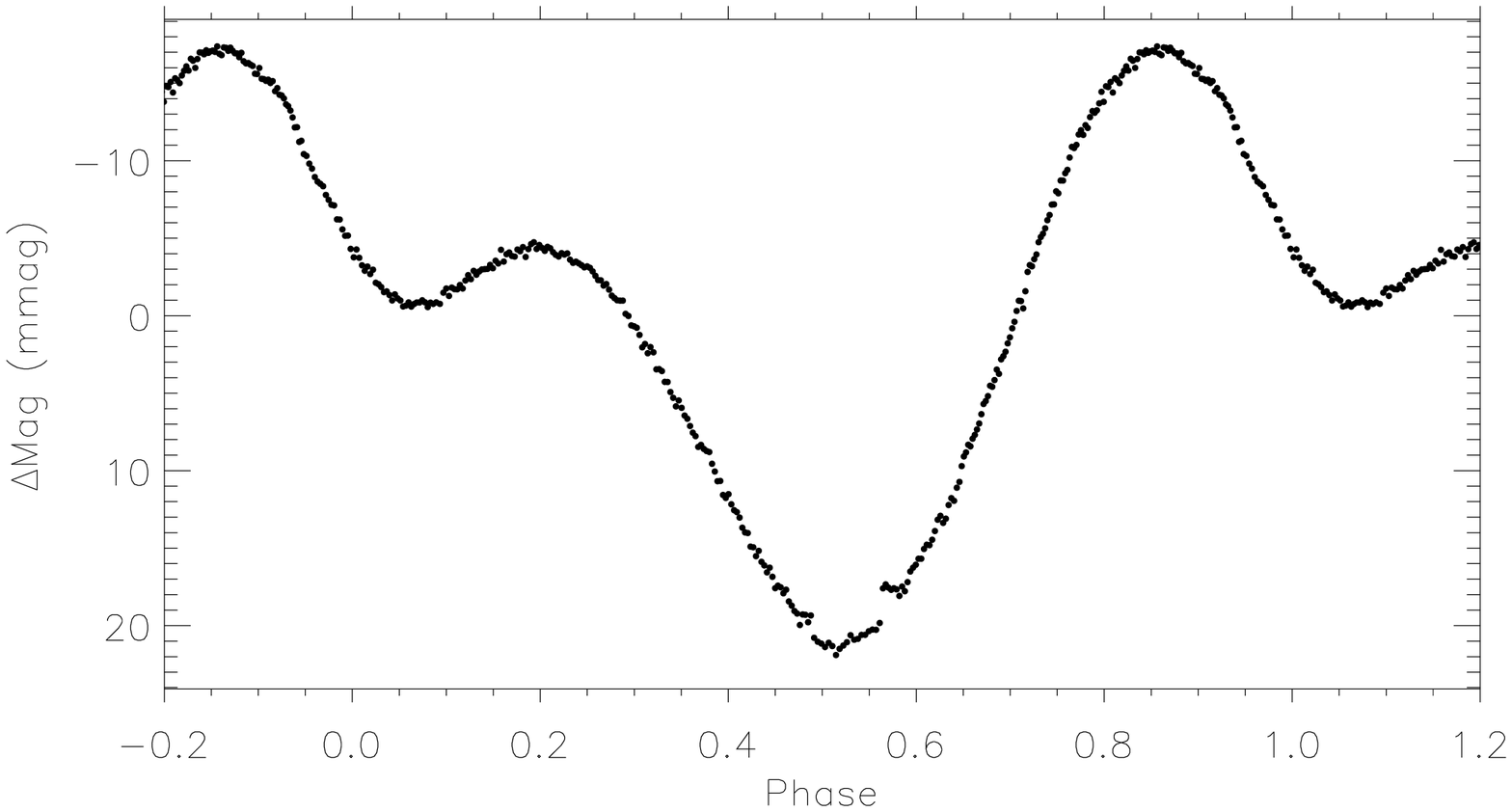}
    \includegraphics[width=\columnwidth]{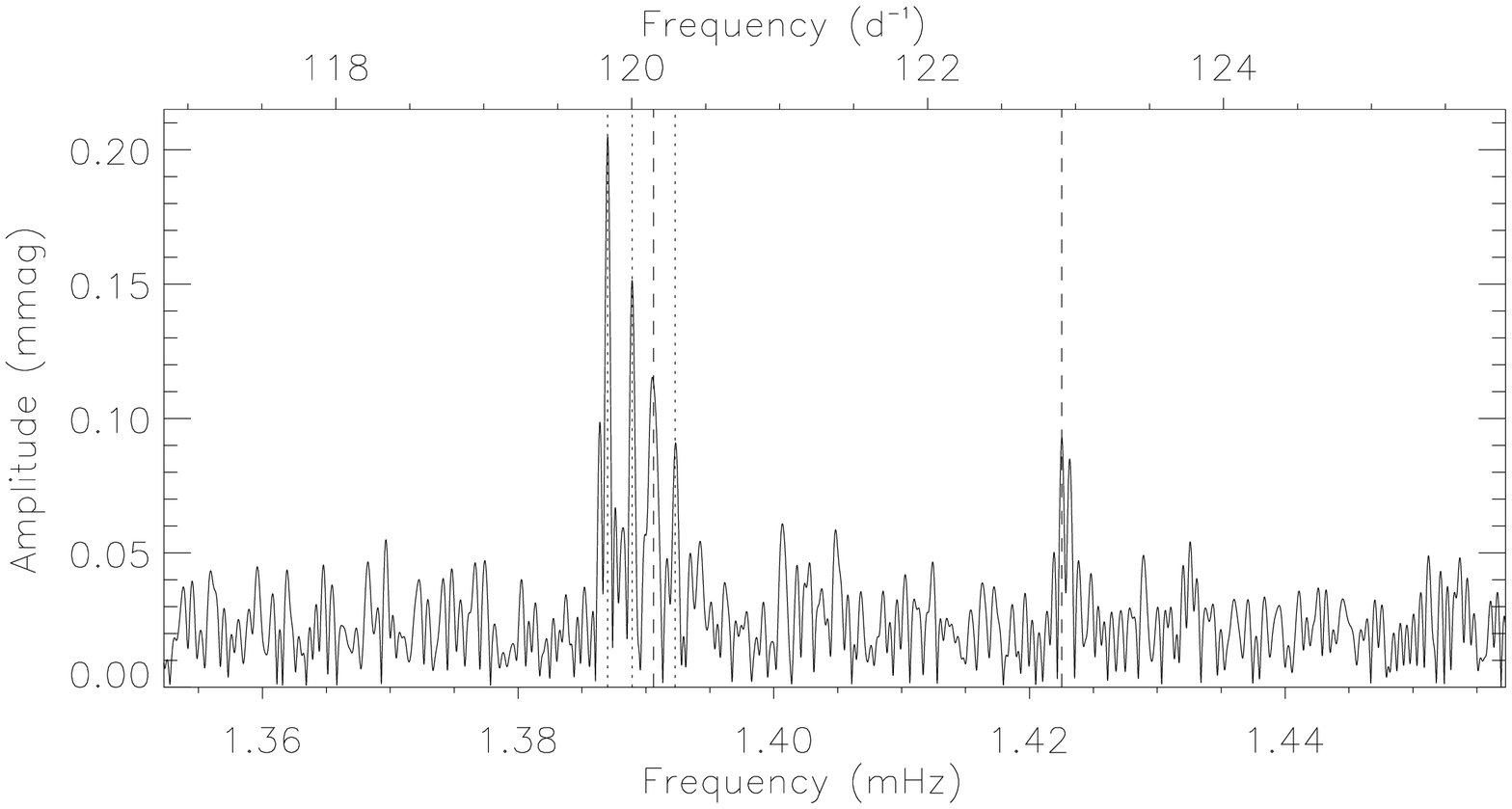}
    \caption{Top: light curve of TIC\,363716787 phased on its rotation period of 5.9660\,d. Bottom: amplitude spectrum of the light curve of TIC\,363716787. The dashed lines represents the two extracted modes, while the dotted lines show the multiplet components.}
    \label{fig:tic363716787}
\end{figure}

The TESS data allow us to provide a rotation period measurement of $5.966\pm0.001$\,d. The folded light curve, Fig.\,\ref{fig:tic363716787}, shows a double wave, implying that we see both magnetic poles, under the assumption that the chemical spots form at or near the magnetic poles. The pulsation spectrum of this star, as noted by \citet{1991MNRAS.250..666M}, shows amplitude variability of its modes. We detect two independent modes in this star, with one showing rotationally split sidelobes. It is unclear which of the multiplet peaks is the pulsation mode, but we follow \citet{1991MNRAS.250..666M}, taking the mode to be at 1.3909\,mHz ($120.1725$\,\cd). This mode is separated from the next by $\approx31\,\umu$Hz which is plausibly half the large frequency separation in this star.

\subsubsection{TIC\,368866492}

TIC\, 368866492 (HD\,166473) was classified as A5p\,SrCrEu in \citet{2009A&A...498..961R}. The TESS Asteroseismic Target List \citep[ATL;][]{2019ApJS..241...12S} reports $T_{\rm eff} = 6811$\,K, but this is the low end of the range of $T_{\rm eff}$ values from the literature, which range as high as $8500$\,K \citep{2006ApJ...638.1004A}. Abundance analysis of this star, conducted by \citet{2000A&A...356..200G}, indicated abundance anomalies typical of roAp stars. \citet{2020A&A...636A...6M} used magnetic field modulus measurements to determine the rotation period to be $3836\pm30$\,d, using high-resolution spectroscopy spanning the period $1992-2019$, and confirming an earlier report of $3514$\,d by \citet{2014IAUS..302..274S}. The magnetic field modulus varies from 5.6 to 8.6\,kG for this star \citep{2020A&A...636A...6M}.

\citet{1987MNRAS.226..187K} first detected photometric oscillations in TIC\,368866492, finding periods between 8.8 and 9.1\,min ($1.8315-1.8939$\,mHz; $158.2-163.6$\,\cd), while \citet{2003MNRAS.343L...5K} and \citet{2007MNRAS.380..181M} confirmed the presence of oscillations using radial velocity measurements. \citet{2007MNRAS.380..181M} reported three frequencies, 1.833, 1.886, and 1.928\,mHz, with amplitudes up to $110$\,\ms, but all with low signal-to-noise. They also suggested that their data hinted at periodic variability of the mean magnetic field modulus, with an amplitude of $0.021\pm0.005$\,kG. 

TESS observed TIC\,368866492 in sector 13. As discussed by \citet{2020A&A...639A..31M}, there is no sign of rotational variability in the light curve, only low-frequency noise suspected to arise from contamination. We detect three main pulsation frequencies in this star, but note there is power excess in the amplitude spectrum after they have been removed. This leads us to speculate there are further low-amplitude modes that are not fully resolved in the TESS data.

\begin{figure}
    \centering
    \includegraphics[width=\columnwidth]{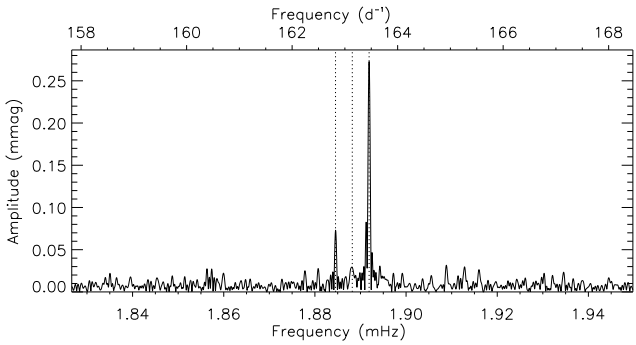}
    \caption{Amplitude spectrum of the light curve of TIC\,368866492 showing the three equally split modes.}
    \label{fig:tic368866492}
\end{figure}

The three peaks that we do see (Fig.\,\ref{fig:tic368866492}) have equal phases and are equally split by $3.69\,\umu$Hz which would correspond to a rotation frequency of $3.14$\,d if the multiplet is caused by oblique pulsation. However, from the literature, we know this star to have an extremely long rotation period. A different interpretation of this splitting is either the large or small frequency separation. With a value of $3.69\,\umu$Hz this is too small to be the large frequency separation, but is plausibly the small separation. However, in that situation, one would require modes of $\ell = 1,3,5$ or $\ell = 0,2,4$ to be visible and would need an explanation for why they would be equally split. This star needs careful consideration and modelling to solve these conundrums, which we leave to a future study.

\subsubsection{TIC\,369845536}

TIC\,369845536 (2MASS\,J19400781-4420093) is a well-studied, high-amplitude roAp star with a distorted quadrupole mode \citep{2018MNRAS.473...91H}. It was originally discovered to have high-frequency variability by \citet{2014MNRAS.439.2078H} using SuperWASP photometry, and a dominant pulsation frequency of 2.042\,mHz (176.4\,\cd). Spectroscopy obtained by \citet{2018MNRAS.473...91H} led them to classify its spectral type as A7Vp\,Eu(Cr). The ground-based photometry also allowed the rotation period to be determined as $9.5344\pm0.0012$\,d, and modelling of the amplitude and phase modulation of its quadrupole pulsation mode confirmed its distorted nature.

\begin{figure}
    \centering
    \includegraphics[width=\columnwidth]{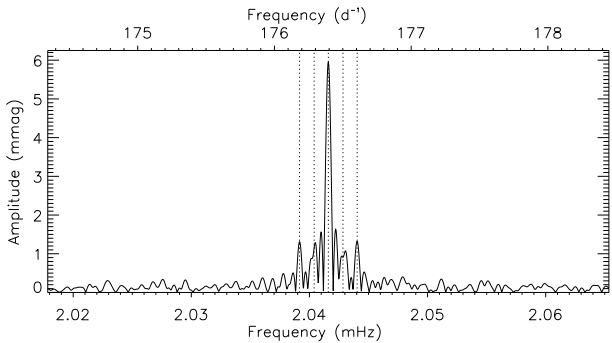}
    \caption{Amplitude spectrum of the light curve of TIC\,369845536 showing the quadrupole mode with its equally split quintuplet of peaks.}
    \label{fig:tic369845536}
\end{figure}

TESS observed TIC\,369845536 during sector 13. The data span 26\,d which makes the determination of the rotation period less certain in the TESS data, at $9.50\pm0.02$\,d over the previous measurement. The pulsation signature is strikingly large in this star; it is the roAp star with the highest amplitude mode seen in $B$ photometric observations, which is also the case here with the TESS observations (Fig.\,\ref{fig:tic369845536}). Beyond the known pulsation mode in this star, we detect no further modes. The multiplet structure of the $\ell=2$ mode is similar to that seen in $B$ data \citep{2018MNRAS.473...91H}.

\subsubsection{TIC\,394124612}

TIC\,394124612 (HD\,218994) was observed in sector 1, and previously analysed by \citet{2019MNRAS.487.3523C}. It has a spectral type of A3\,Sr \citep{2009A&A...498..961R}, but is also known to be a visual binary with a separation of 1.2\,arcsec \citep{1991A&AS...89..429R}. Pulsations consistent with a $\delta$\,Sct star and a longitudinal magnetic field strength of $0.44\pm0.02$\,kG were detected by \citet{2008MNRAS.386.1750K}. Despite attempts to detect roAp pulsations, none have so far been detected in photometry (see e.g. \citealt{1994MNRAS.271..129M}). However, analysis of spectroscopic time series data has discovered variability in its Nd\,{\sc{iii}} and Pr\,{\sc{iii}} lines \citep{2008MNRAS.384.1140G}, which showed a frequency of 1.17\,mHz (101.1\,\cd). 

Since an analysis of the TESS data has already been presented by \citet{2019MNRAS.487.3523C}, we refer the reader to that work while providing the rotation period and pulsation frequency in Table\,\ref{tab:stars}.


\subsubsection{TIC\,394272819}

TIC\,394272819 (HD\,115226) is a well-known $\alpha^2$\,CVn variable star. It was classified as Ap\,Sr(Eu) by \citet{1975mcts.book.....H}. The $T_{\rm eff}$ values in the literature for this star are mostly in agreement, with a value around $7640$\,K \citep{2006A&A...450..763K,2015MNRAS.452.3334S}. Using this value and the Hipparcos parallax, \citet{2006A&A...450..763K} determine $L = 7.2\pm1.9$\,L$_\odot$. Later, \citet{2008A&A...479L..29K} determined the projected rotational velocity $v\sin i = 25-30$\,\kms\ and conducted an abundance analysis, finding nearly solar abundances of Mg, Si, Ti, and Fe, an overabundance of Cr, Co, and rare earth elements, and an under-abundance of Ba. \citet{2015A&A...583A.115B} reported a $\langle B_z\rangle_{\rm rms}$ value of 0.828\,kG.

The roAp nature of TIC\,394272819 was discovered by \citet{2008A&A...479L..29K}. Using time-series high-resolution HARPS spectroscopy, they measured the high-amplitude 10.87-min (1.5333\,mHz; 132.5\,\cd) oscillation in this star from radial velocity variations. Pulsational variability was discovered in the cores of hydrogen lines and in doubly-ionized Pr, Nd, and Dy lines. Recently, the same 102 HARPS spectra were analysed by \citet{2020A&A...642A.146B} by using the unit-sphere representation periodogram (USuRPER), finding the same period.

Photometric ASAS-3 data of TIC\,394272819 were analysed by \citet{2016AJ....152..104H}. They detected a clear signal with a period of $2.9882$\,d. In addition they checked the Hipparcos data and found marginal variability with $P = 3.61$\,d, which is not present in the ASAS-3 data. \citet{2016A&A...590A.116J} confirmed the presence of 10.86-min oscillations analysing high-speed photometry from the Nainital-Cape Survey.

\begin{figure}
    \centering
    \includegraphics[width=\columnwidth]{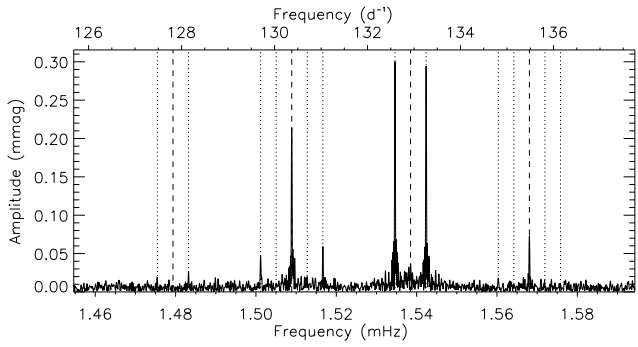}
    \caption{Amplitude spectrum of the light curve of TIC\,394272819 showing the multiple modes in this star. We mark the presumed modes with dashed lines, with rotationally split sidelobes marked by dotted lines.}
    \label{fig:tic394272819}
\end{figure}

TIC\,394272819 was observed during sectors 11 and 12. The combined PDC\_SAP data allow us to derive a rotation period of $2.98827\pm0.00008$\,d which is consistent with the ASAS-3 data. There is no evidence of the 3.61-d signal found in the Hipparcos data. We detect four pulsation modes in this star. In increasing frequency we see a singlet, three components of a quintuplet, a triplet and another singlet. Fig.\,\ref{fig:tic394272819} shows the amplitude spectrum of this star. We interpret this pattern as a series of alternating $\ell=1$ and $\ell=2$ modes with some components lost in the noise. This interpretation is reinforced when considering the spacing between the modes; under our assumption they are all separated by $29.58\,\umu$Hz which we understand to be the half of the large frequency separation for this star.

\subsubsection{TIC\,402546736}

TIC\,402546736 (HD\,128898; $\alpha$\,Cir) has many similar classifications in the literature. Here we quote that of \citet{2006AJ....132..161G}: A7Vp\,SrCrEu. This star is a well studied, naked eye object, and as such has a plethora of information in the literature. We provide here a very limited introduction to the star, and encourage the interested reader to explore the references provided.

Effective temperature measurements for TIC\,402546736 range from 7420\,K \citep[e.g.,][]{2008MNRAS.386.2039B,2015A&A...573A.126D} to around 8000\,K \citep[e.g.,][]{1993ASPC...44..561K,1999ApJ...511..422M} derived through a variety of different methods. $\log g$ determinations also vary, but are consistently around 4.1\,\cms\ within the quoted errors \citep[e.g.,][]{2009A&A...499..851K,2019MNRAS.483.2300S}, with $v\sin i$ values consistently around 13\,\kms\ \citep[e.g.,][]{1996A&A...308..886K,2012A&A...542A.116A}. The abundance pattern in TIC\,402546736 is typical for Ap stars \citep{1996A&A...308..886K,2008MNRAS.386.2039B}, with under-abundances of C, N, and O, and overabundances of rare-earth and some other heavy elements. \citet{2009A&A...499..851K} built a self-consistent empirical model atmosphere of TIC\,402546736 and conducted stratification modelling of some elements, showing that chemical stratification has an important effect on the model structure of TIC\,402546736. A $\langle B\rangle$ value of 0.9\,kG have been reported for this star \citep{2008MNRAS.386.2039B} \citet{2019MNRAS.483.3127S} derived a dipolar field strength of 1.4\,kG from the analysis of mean longitudinal field measurements.

TIC\,402546736 was classified as an roAp star showing a 6.8-min (2.4510\,mHz; 211.8\,\cd) oscillation by \citet{1981IBVS.1987....1K} and subsequently, 47 hours of high-speed, ground-based photometry was analysed by \citet{1981IBVS.2033....1K}. Notable results from studies of this star are those of \citet{1994MNRAS.270..674K} who found the star to pulsate in a dipole mode with a frequency of 2.442\,mHz (211.0\,\cd), and subsequent lower amplitude modes. They also provided a precise rotation period of 4.4790\,d. Later, \citet{2009MNRAS.396.1189B} used 84\,d of WIRE observations to refine the separation of the modes in this star to be $30.2\,\umu$Hz. \citet{2016A&A...588A..54W} and \citet{2020A&A...642A..64W} used 4 BRITE satellites to observe TIC\,402546736 for a total of 146\,d, providing them data to model the spot configuration of the star, and confirm the previously tentative detection of some low amplitude modes. 

Beside the photometric observations of TIC\,402546736, there have been substantial spectroscopic studies using high time resolution spectra. These studies \citep[e.g.,][]{1998MNRAS.295...33B,1999MNRAS.302..381B,2001A&A...377L..22K,2003MNRAS.344..242B,2007A&A...473..907R} have shown that the principal pulsation mode can have an amplitude up to 1\,\kms, with some spectral lines pulsating in anti-phase with others, suggesting a node, or false node, in the atmosphere of the star \citep{2011MNRAS.414.2576S,2018MNRAS.480.1676Q}.

\begin{figure}
    \centering
    \includegraphics[width=\columnwidth]{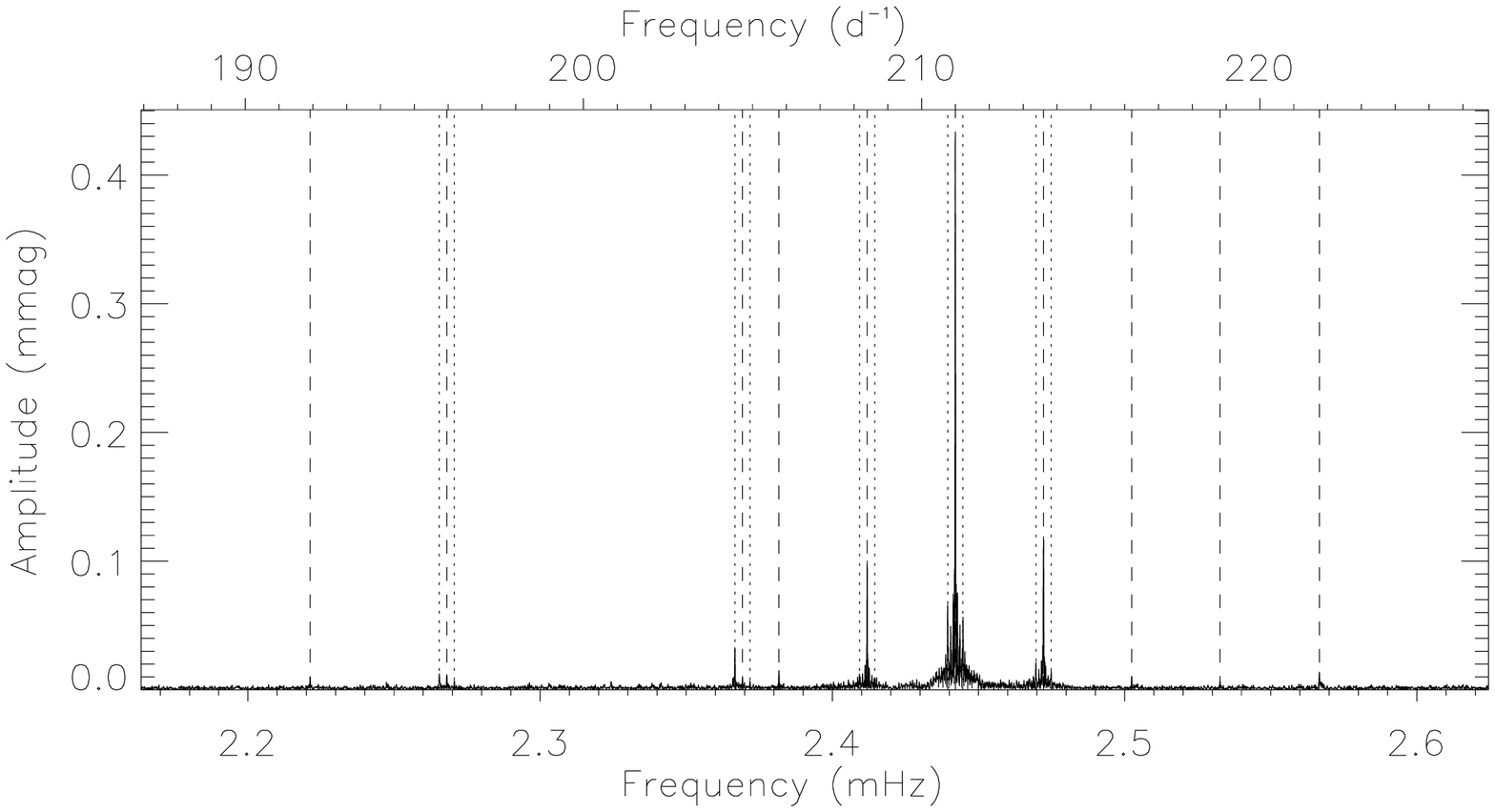}
    \includegraphics[width=\columnwidth]{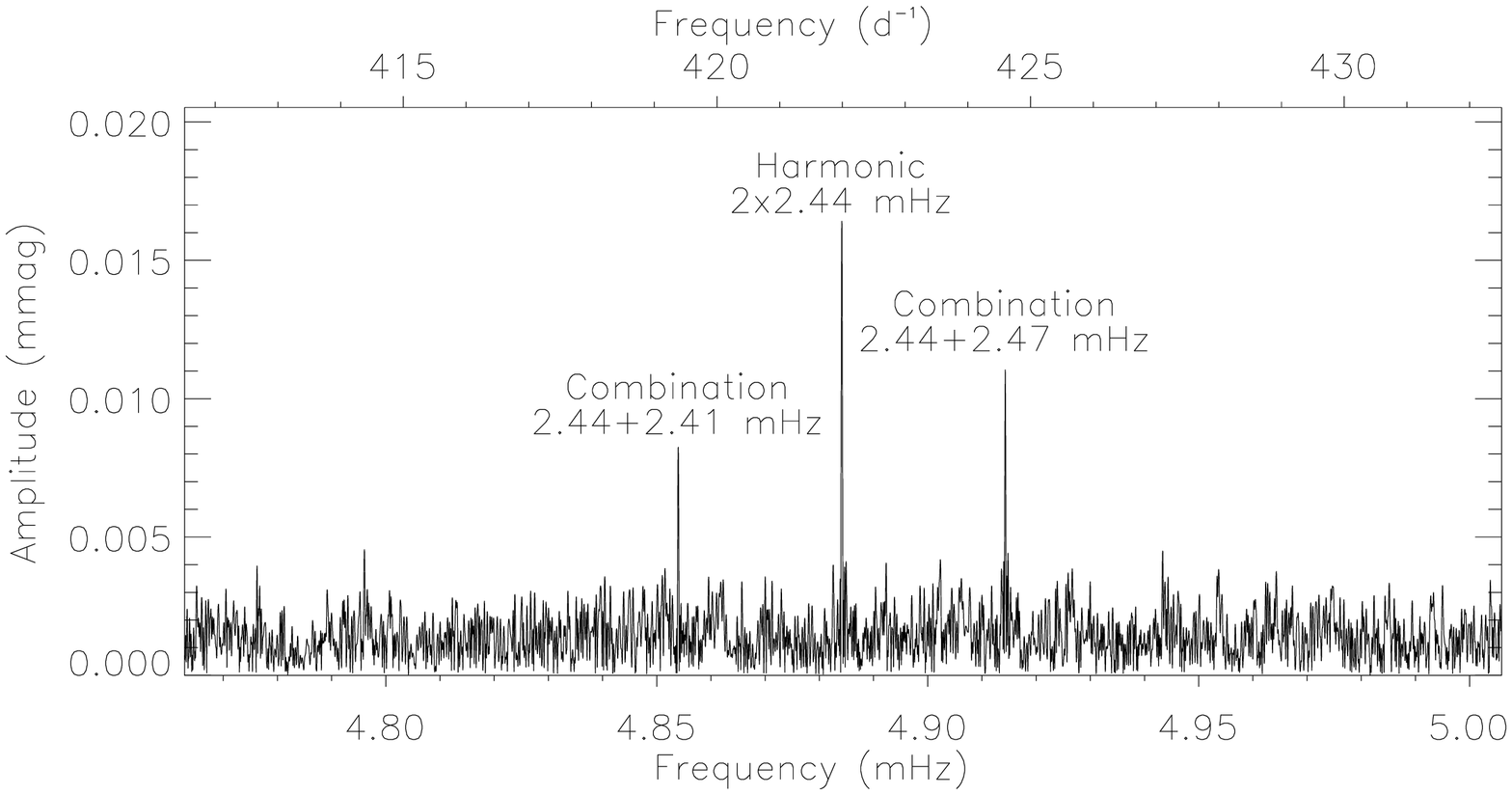}
    \caption{Amplitude spectrum of the SAP light curve of TIC\,402546736 ($\alpha$\,Cir) showing the multiple modes in this star. The top panel shows the region where the pulsation modes are detected. We mark the modes with dashed lines, with rotationally split sidelobes marked by dotted lines. The bottom panel shows the harmonic of the dominant mode, and two additional peaks that are the result of non-linear interactions in this star. Note the change in scale between the two panels.}
    \label{fig:tic402546736}
\end{figure}

TESS observed TIC\,402546736 during sectors 11 and 12. We use the SAP data to determine a rotation period of $4.4812\pm0.0005$\,d which is different from that of \citet{1994MNRAS.270..674K} by $4.5\sigma$. We also use the SAP data for the pulsation analysis that follows, since the SPOC pipeline has introduced significant noise to this bright star. As expected from the literature, the amplitude spectrum is rich in pulsation modes and rotationally split sidelobes (Fig.\,\ref{fig:tic402546736}). We extract 23 significant peaks from the amplitude spectrum, but note there is evidence of further modes seemingly present in the data. Of those 23 peaks, we identify 10 as pulsation modes (see Table\,\ref{tab:stars}), one as a harmonic and 2 as peaks due to non-linear interactions. The latter 3 peaks are all found above the Nyquist frequency and are reminiscent of those seen in the K2 data of 33\,Lib \citep[HD\,137949;][]{2018MNRAS.480.2976H}, making TIC\,402546736 only the second roAp star to show such non-linear interactions.

Although not all modes are separated by the same frequency, many are separated by $30.18\,\umu$Hz which we take to be half of the large frequency separation in this star, as did \citet{2009MNRAS.396.1189B}. For those modes where the frequency spacing is not a multiple of $30.18\,\umu$Hz, we postulate that these mode frequencies have been affected by the magnetic field, as was seen in TIC\,279485093 \citep[HR\,1217;][]{2005MNRAS.358..651K}.


\subsubsection{TIC\,434449811}

TIC\,434449811 (HD\,80316) was classified as Ap\,Sr(Eu?) with `extremely strong Sr\,{\sc ii}' by \citet{1974PASP...86..408B}. There is no detailed spectroscopic analysis of the fundamental properties in the literature, so  \Teff\ values are based on photometry with values ranging from 8110\,K \citep{2018A&A...616A...1G} to 9120\,K \citep{2006ApJ...638.1004A}. Using $uvby$H$\upbeta$ from \citet{2015A&A...580A..23P}, and the calibrations of \citet{1985MNRAS.217..305M}, gives $T_{\rm eff}$ = 8420\,K, which is consistent with 8280\,K obtained from the spectral energy distribution \citep{2012MNRAS.427..343M}. The mean longitudinal magnetic field strength has been determined to be $-0.18\pm0.04$\,kG by \citet{2006AN....327..289H} and $-0.25\pm0.09$\,kG by \citet{2015A&A...583A.115B}.

The rotation period was determined to be 2.09\,d \citep{1986A&AS...64....9M,1997MNRAS.289..645K}, although a recent value of 1.91\,d was given by \cite{2018AJ....155...39O}. With $v \sin i = 32\pm5$\,\kms\ \citet{1997A&A...325.1063W} estimated a rotational axis inclination of $i= 60\pm15\,^\circ$. The radial velocity is given as $9.5\pm1.1$\,\kms\ by \citet{2006AstL...32..759G}. However, five measurements with HARPS show a large variation over a 2-yr time-span, ranging from 10.6 to 46.7\,\kms\ \citep{2020A&A...636A..74T}, indicating this could be a spectroscopic binary star. The star is part of a very wide ($>70$\,arcsec) optical pair with HD\,80298 \citep[BD-19\,2673; WDS\,J09184-2022A; ][]{2001AJ....122.3466M}, but there is no physical association.

TIC\,434449811 is a known roAp star, with a 7.4-min (2.2523\,mHz; 194.6\,\cd) oscillation discovered by \citet{1990MNRAS.242..489K} in photometric observations. A later, more in depth photometric study was presented by \citet{1997MNRAS.289..645K} who found a frequency triplet with equal amplitudes in a $B$ filter. \citet{2008CoSka..38..317E} found a pulsation signal of the same period with an amplitude of 0.32\,\kms\ in spectral lines.

\begin{figure}
    \centering
    \includegraphics[width=\columnwidth]{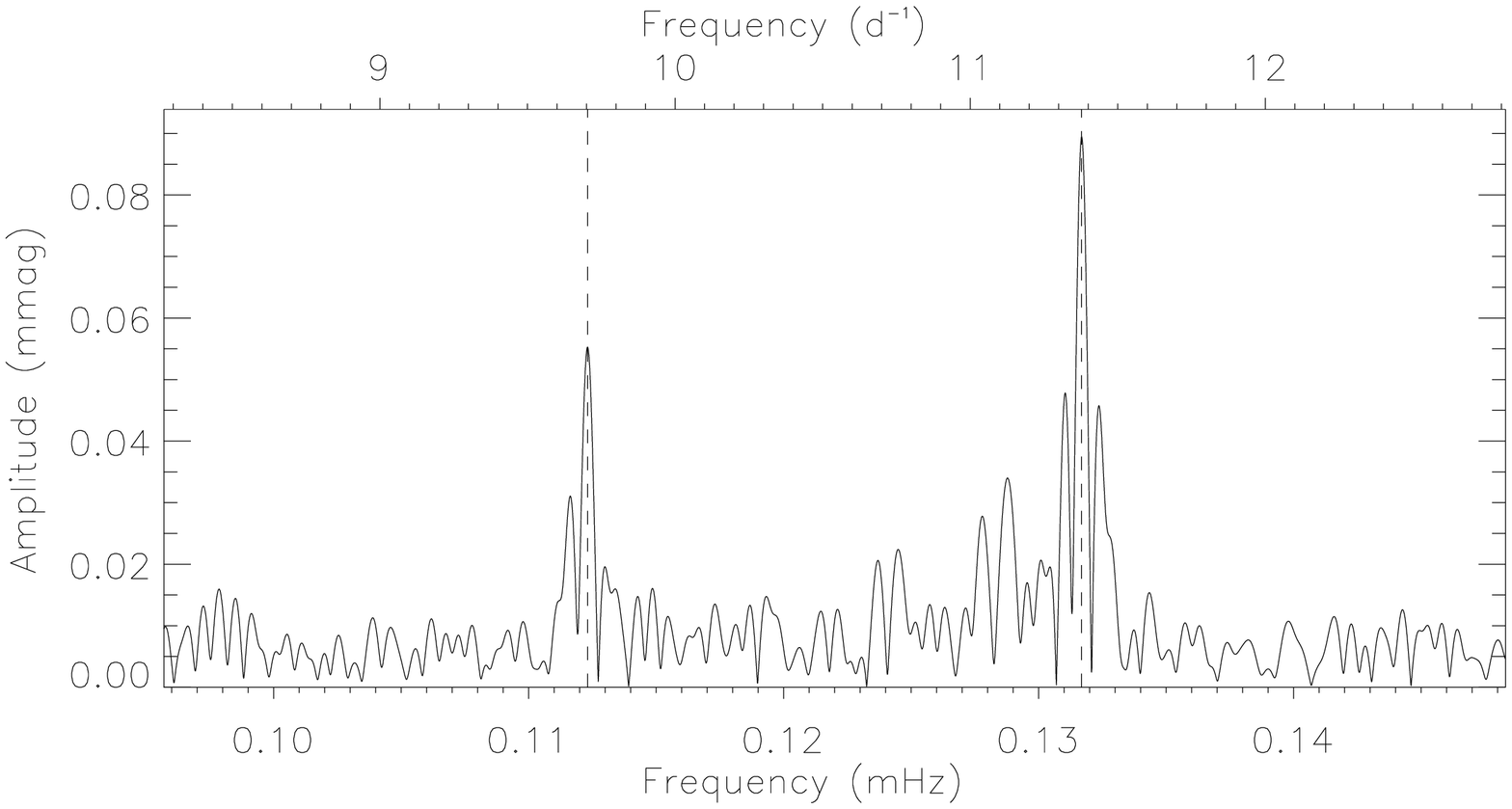}
    \includegraphics[width=\columnwidth]{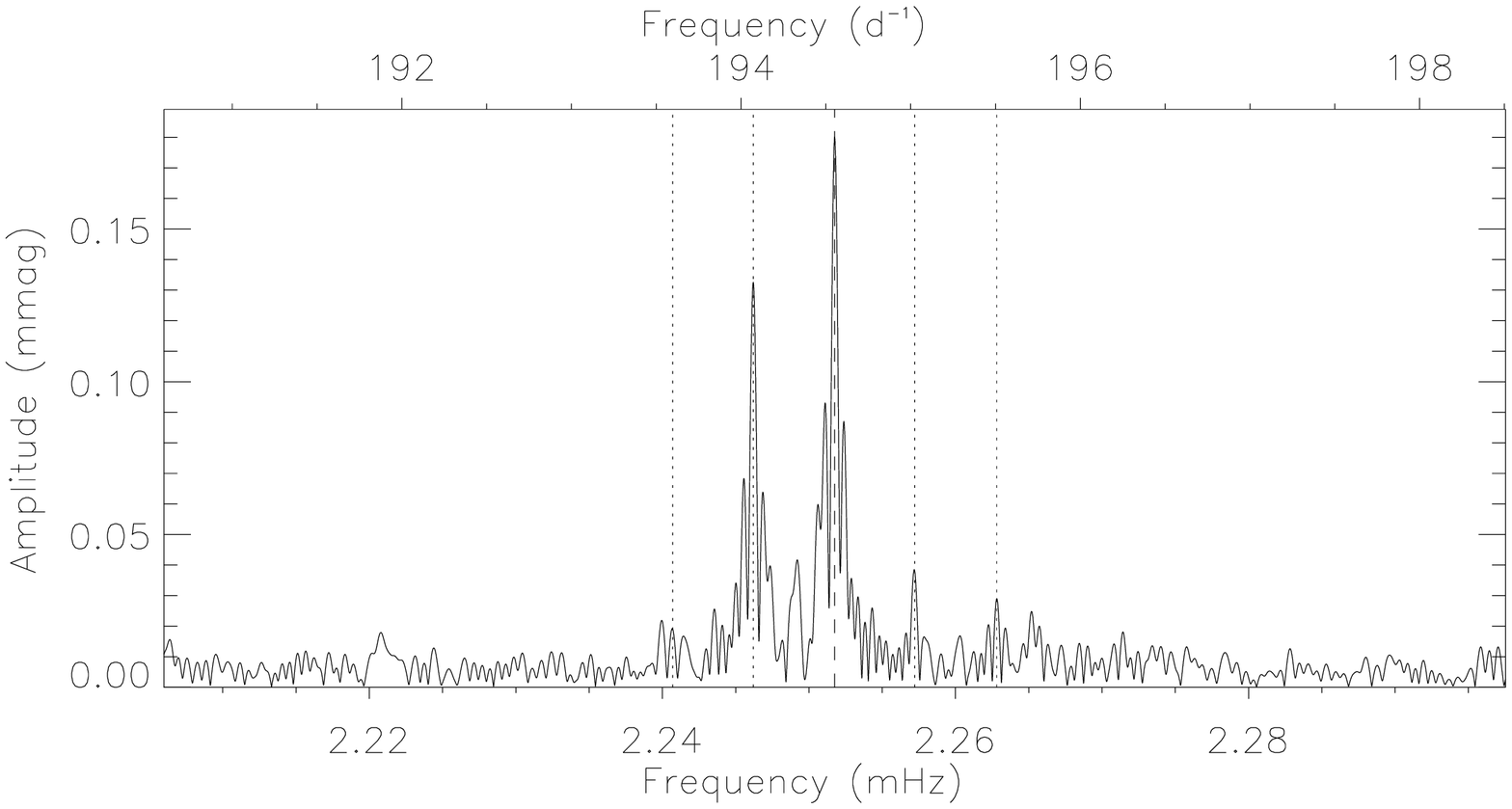}
    \caption{Amplitude spectrum of the light curve of TIC\,434449811. The top panel shows the low-frequency modes found in this star, with the bottom panel showing the multiplet around the known pulsation mode.}
    \label{fig:tic434449811}
\end{figure}

TESS observed the star during sector 8. The PDC\_SAP data, after the removal of some outlying points at the start of the second orbit, provide a rotation period for this star of $2.08862\pm0.00004$\,d, which is consistent with the literature value. After removal of the rotation signature and its harmonics, there are two peaks remaining in the $\delta$\,Sct frequency range (Fig.\,\ref{fig:tic434449811}). It is unclear whether these are mode frequencies in TIC\,434449811 or in a companion star. Given the relatively weak magnetic field in this star, it is possible that the $\delta$\,Sct frequencies are in the Ap star \citep{2020MNRAS.498.4272M}, but detailed modelling and the exclusion of other possibilities is required.

The roAp pulsation signature in this star has a different multiplet structure than previously seen \citep{1997MNRAS.289..645K}; there are, in the TESS data, two peaks with similar amplitudes, and a further 3 peaks with low signal-to-noise. This change in multiplet structure is similar to that seen in TIC\,326185137 \citep[HD\,6532; ][]{2020ASSP...57..313K} and may be a result of the different filters probing different depths in the stellar atmosphere. A detailed analysis of this star is required to fully understand its pulsation properties.


\subsubsection{TIC\,469246567}

TIC\,469246567 (HD\,86181) is listed as an $\alpha^2$\,CVn star on SIMBAD and has a spectral type Ap\,Sr \citep{1975mcts.book.....H}. \citet{1994IBVS.4013....1K} discovered 6.2-min (2.6882\,mHz; 232.3\,\cd) oscillations in TIC\,469246567, making it one of the most rapidly oscillating Ap stars. A longitudinal magnetic field strength of $536\pm58$\,kG was measured by \citet{2015A&A...583A.115B}. Despite the $\alpha^2$\,CVn classification, there is only one recent measure of the rotation period of this star (Shi et al. submitted).

TIC\,469246567 was observed by TESS during sectors 9 and 10. These data have been extensively analysed by Shi et al. (submitted) so we refer the reader there to a detailed discussion. In short, the pulsation spectrum consists of alternating dipole, quadrupole, and dipole modes, all rotationally split into multiplets. The TESS data also provide the first measure of the rotation period of this star to be $2.05115\pm0.00006$\,d. The pulsation frequencies, and rotation period, are included in Table\,\ref{tab:stars} for completeness.

\section{Candidate roAp stars}
\label{sec:cand}

In this section, we provide information on a number of stars which were flagged by the search teams as possible roAp stars. These stars are classed as candidate roAp stars since either their spectral classification does not indicate chemical peculiarity, their pulsation signature does not show signs of multiplets split by the rotation frequency, or their presumed pulsation signal is close to a signal-to-noise limit of 4.0.

\subsection{TIC\,1727745}
TIC\,1727745 (HD\,113414) was classified as an F7/8V type star by \citet{1988mcts.book.....H}. There have been numerous studies that attempt to calculate physical parameters of the star. Among these, \citet{2009ApJS..184..138H} identified the star as an x-ray source and \citet{2003ApJ...595.1206S} found its x-ray luminosity to be log$L_\mathrm{x}=29.45$ erg\,s$^{-1}$ in the investigation of x-ray properties of the ROSAT F stars. \citet{2003ApJ...595.1206S} also estimated the temperature and metallicity to be $T_{\rm eff}=6200$ K and Fe/H $= 0.04$, respectively.

Using precision radial velocities from HIRES on the Keck I telescope, \citet{2017AJ....153..208B} stated that the star showed chromospheric emission variations, which led to radial velocity variations, due to either stellar rotation or long-term activity cycles.  Additionally, $(b-y)$, $m_1$, $c_1$ colour indexes and H$\upbeta$ index of the star were provided to be 0.357, 0.182, 0.386 and 2.624, respectively \citep{2015A&A...580A..23P,1998A&AS..129..431H}, which do not imply this star is chemically peculiar.

\begin{figure}
    \centering
    \includegraphics[width=\columnwidth]{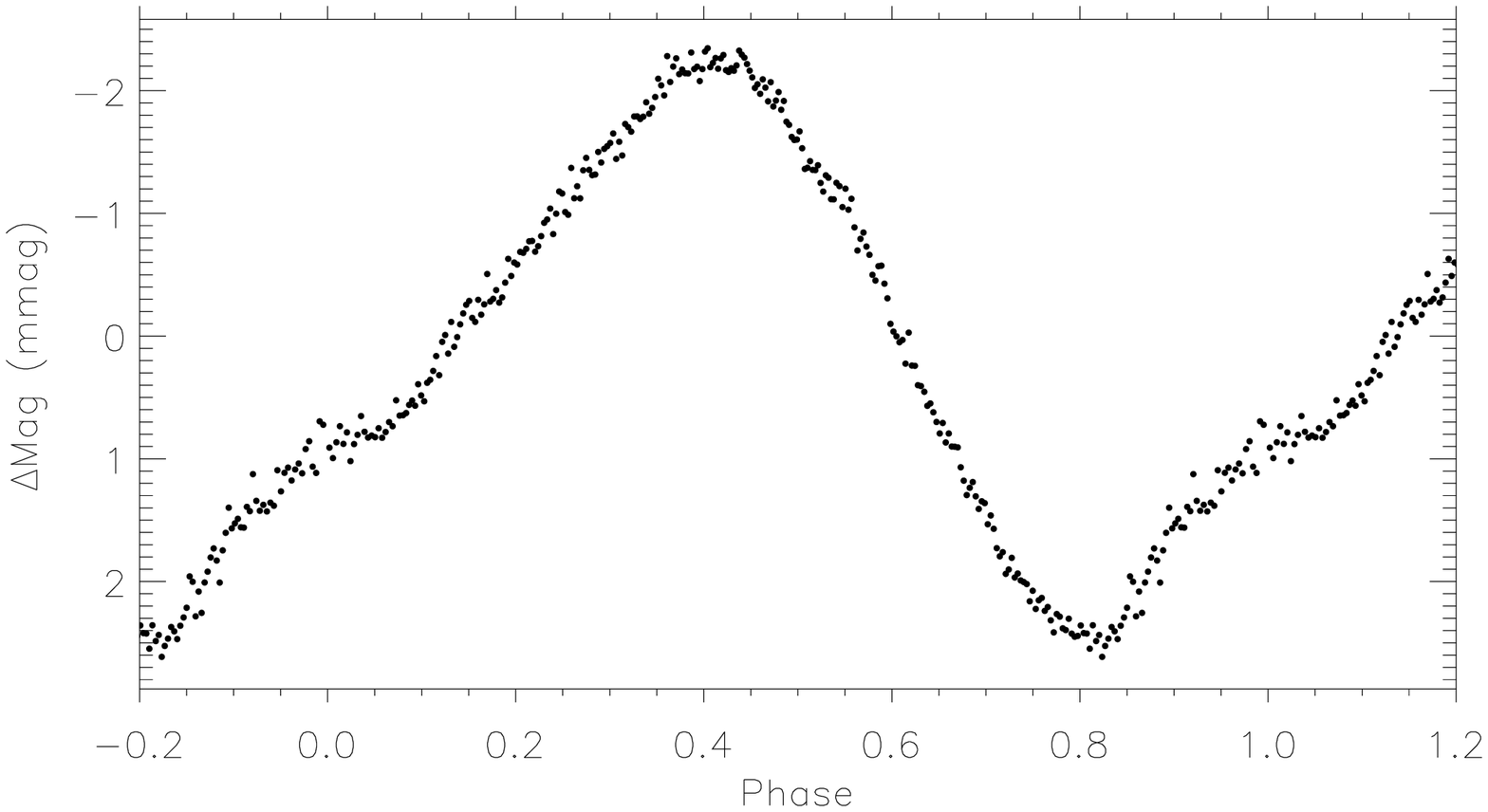}
    \includegraphics[width=\columnwidth]{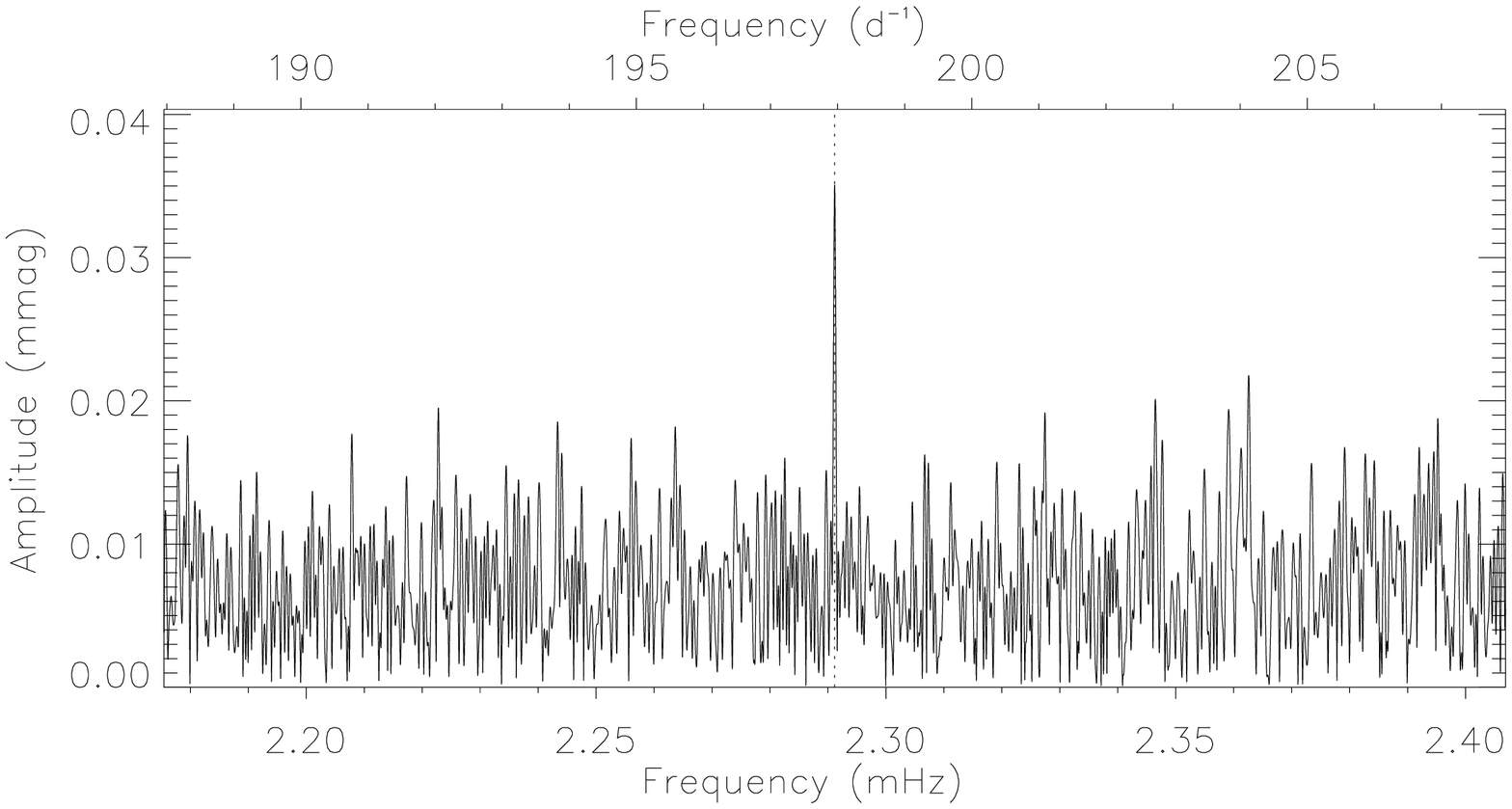}
    \caption{Top: phase folded light curve of TIC\,1727745, phased on a period of 3.172\,d. Bottom: the high frequency pulsation mode found in this star.}
    \label{fig:tic1727745}
\end{figure}

Observed in sector 10, the TESS data show a rotationally modulated light curve, which has a period of $3.172\pm0.001$\,d (Fig.\,\ref{fig:tic1727745}). The well defined phased light curve suggests that the surface inhomogeneities are stable over the observations period. The pulsation signal in the star occurs at $2.29116\pm0.00004$\,mHz ($197.956\pm0.004$\,\cd) which is in the frequency range of known roAp pulsations. 

The contamination factor, in the TIC, for this star is very small (0.0001) which implies all of the light in the photometric aperture is originating from the target star. 
  
\subsection{TIC\,3814749}

TIC\,3814749 (HD\,3748) was also listed in both \citet{2019MNRAS.487.2117B} and \citet{Kobzar2020} as an roAp candidate. The star is a known double system (with a contamination factor of 1.431) which have spectral types of A0/1\,IV/V and A9/F0V \citep{1999mctd.book.....H}, but with also A5 quoted in the literature. We obtained a classification spectrum from SAAO and confirm the lack of strong chemical peculiarities in this star, suggesting it is not an Ap star (Fig.\,\ref{fig:tic3814749spec}). However, the Ca\,{\sc{ii}}\,K line is weak in the star, as is the Mg\,{\sc{ii}} 4481\,\AA\ line. 

\begin{figure}
    \centering
    \includegraphics[width=\columnwidth]{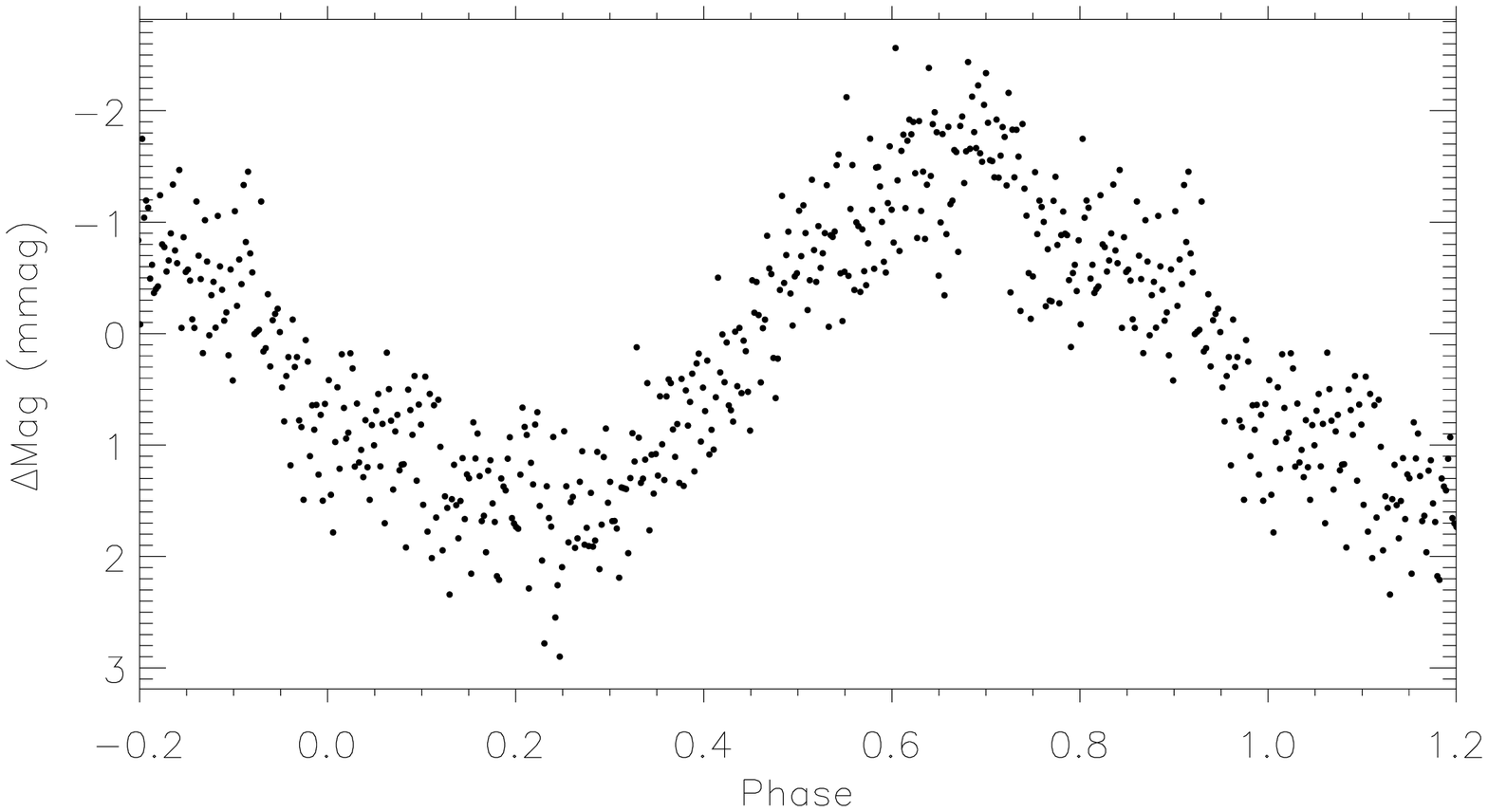}
    \includegraphics[width=\columnwidth]{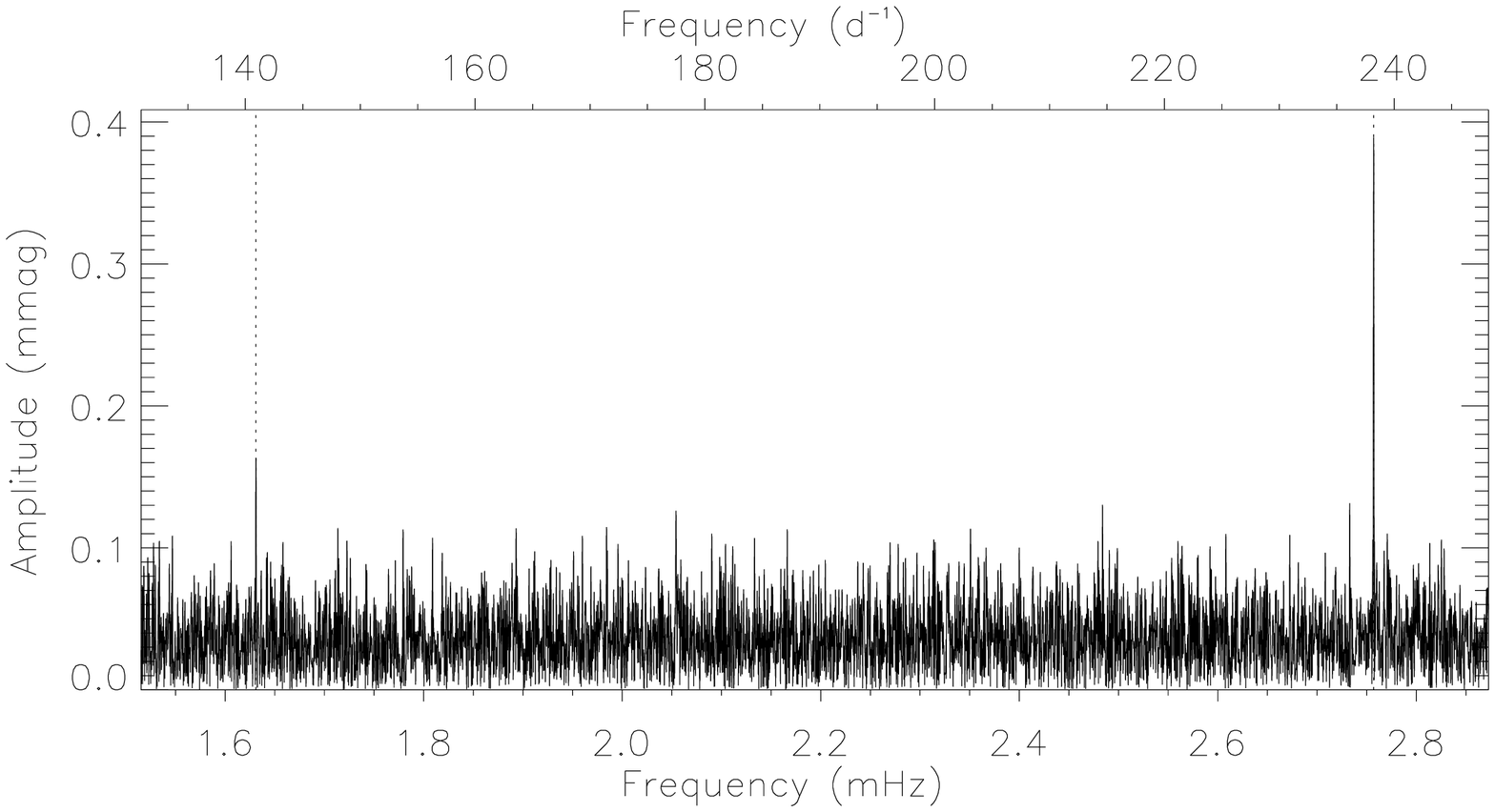}
    \caption{Top: phase folded light curve of TIC\,3814749, phased on a period of 1.689\,d. Bottom: the two high frequency peaks found in this star, as marked by dotted lines.}
    \label{fig:tic3814749}
\end{figure}

TESS observed TIC\,3814749 during sector 3. With those data, we detect a low frequency signal corresponding to a period of $1.689\pm0.002$\,d. We also find two peaks at higher frequency which we assume are a result of pulsational variability (Fig.\,\ref{fig:tic3814749}), although the lower frequency peak is at low significance. These two peaks are well separated in frequency, by 1.12\,mHz, which requires a significant range of unexcited overtones as has been seen in TIC\,139191168 (HD\,217522) and now also in TIC\,49332521 (HD\,119027).

However, the star is listed as an Algol type binary \citep{2003AstL...29..468S}. It is possible that the low frequency signal is the orbital frequency of the A star and a compact companion. In this case, a large gap between the observed frequencies is less uncommon \citep[e.g.,][]{2020MNRAS.499.5508S}. High-resolution spectroscopic monitoring of this system is needed to solve this problem.

\subsection{TIC\,158637987}

There is little information in the literature about TIC\,158637987 (HD\,10330). \cite{1975mcts.book.....H} gave the spectral type as A9V, indicating a chemically normal star. \citet{2015A&A...580A..23P} provided Str\"omgren-Crawford photometric indices of $b-y=0.207$, $m_1=0.143$, $c_1=0.702$, and H$\upbeta=2.732$. Using the calibrations of \citet{1979AJ.....84.1858C}, these indices suggest the star is significantly metal weak.

\begin{figure}
    \centering
    \includegraphics[width=\columnwidth]{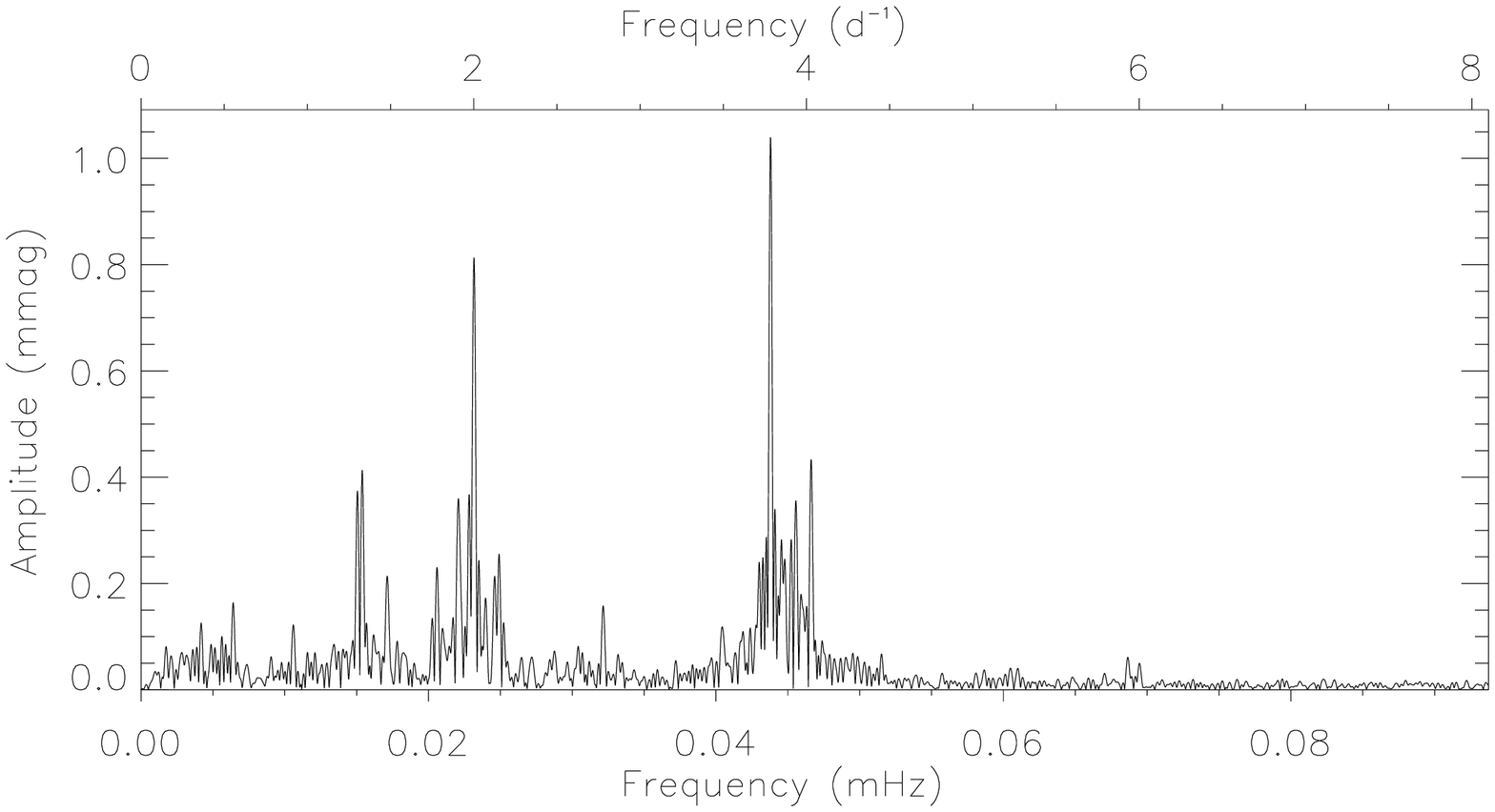}
    \includegraphics[width=\columnwidth]{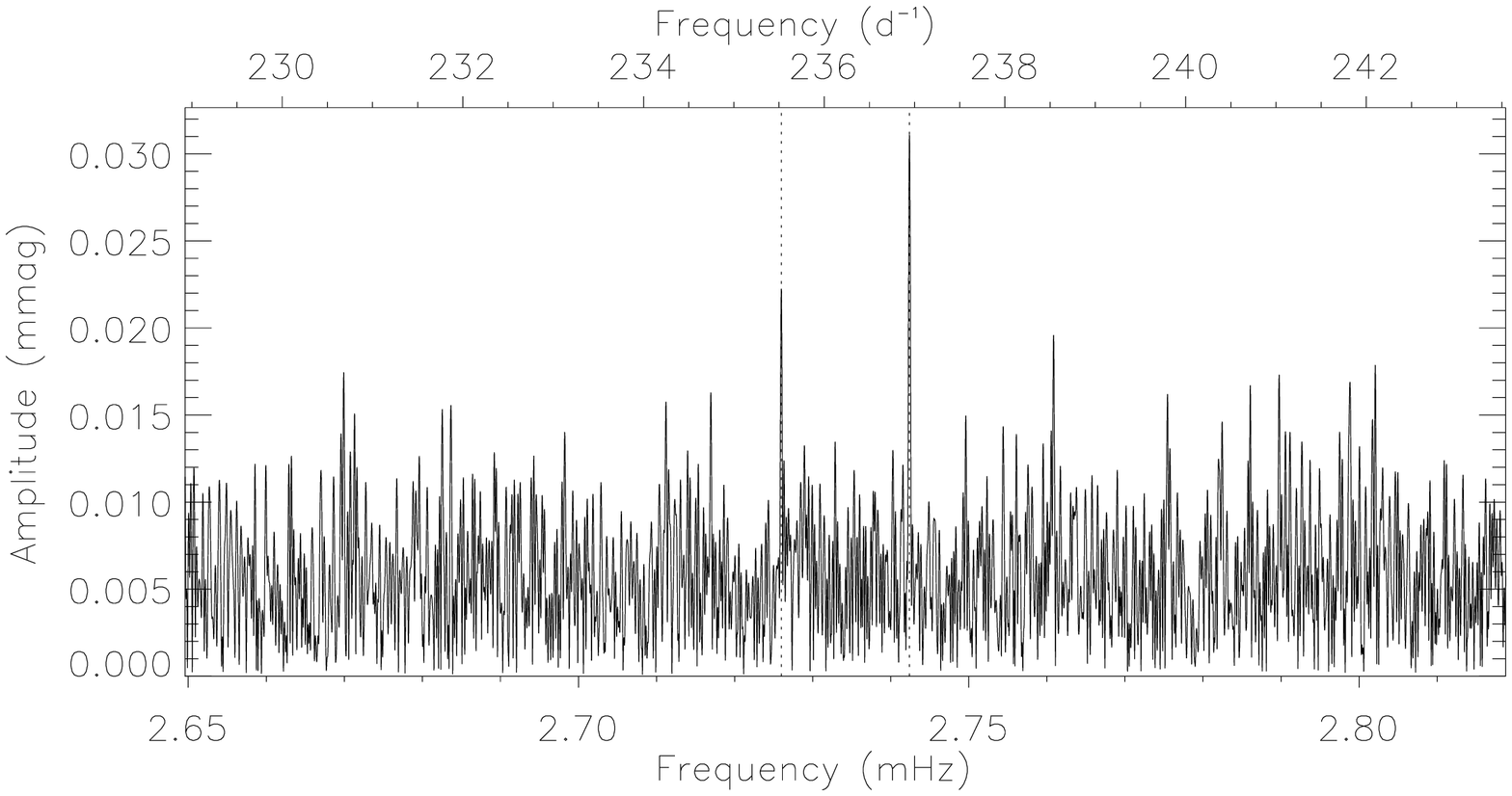}
    \caption{Top: g-mode frequencies seen in TIC\,158637987. Bottom: the two pulsation modes found in this star, as marked by dotted lines.}
    \label{fig:tic158637987}
\end{figure}

The TESS observations, collected during sectors 2 and 3, show significant variability in the low-frequency range which we take to be g mode pulsations (Fig.\,\ref{fig:tic158637987}). The high-frequency variability in this star occurs at about 2.71\,mHz (236\,\cd; Fig.\,\ref{fig:tic158637987}), where we identify two peaks of marginal significance. The separation of these two peaks (0.01642\,mHz; 1.4187\,\cd) is close to, but not the same as, a peak in the low frequency range (0.01713\,mHz; 1.4797\,\cd), thus we cannot deduce whether the peaks are split by the stellar rotation frequency. Spectroscopic observations are required of this star to test the existence of chemical peculiarities and/or a magnetic field. If either exist, this star provides another example of a $\gamma$\,Dor-roAp hybrid star. If this is an roAp star, it is unique in having apparently low metallicity.

The TIC provides a contamination factor of 0.048 for TIC\,158637987, suggesting there is very little contamination from a nearby (on the sky) star.


\subsection{TIC\,324048193}

TIC\,324048193 (HD\,85892) was classified as Ap\,Si by \citet{1975mcts.book.....H}. The literature provides \Teff\ values of between $7770$\,K \citep{2019AJ....158...93B} to $10110$\,K \citep{2019A&A...628A..94A}, but with most estimates around $9300$\,K \citep{2012MNRAS.427..343M,2016AJ....151...59C,2006ApJ...638.1004A}. The TIC estimate is by far the highest temperature estimate at $11950$\,K, and we treat that with caution. The presence of Si in the stellar spectrum suggests a late B early A star which is consistent with the higher temperature range. The mass and radius estimates for the star, 5.2\,M$_\odot$ and 4.5\,R$_\odot$ \citep{2019A&A...623A..72K,2017MNRAS.471..770M}, suggest it has evolved away from the zero-age main sequence. A $v\sin i$ value of $45\,$\kms\ was quoted by \citet{2009A&A...498..961R}, with \citet{2006SASS...25...47W} measuring a rotation period of 4.295\,d.

\begin{figure}
    \centering
    \includegraphics[width=\columnwidth]{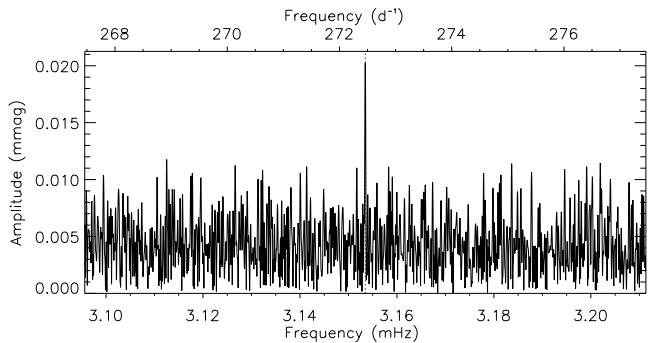}
    \caption{Suspected pulsation mode in TIC\,324048193.}
    \label{fig:tic324048193}
\end{figure}

TESS observed TIC\,324048193 during sectors 11 and 12. The light curve shows clear modulation due to spots, as expected for a magnetic CP star. From the variations, we derive a rotation period of $4.2953\pm0.0001$\,d which is consistent with that in the literature. The pulsation signal in this star is at a high frequency of $3.15347\pm0.00002$\,mHz ($272.460\pm0.002$\,\cd; Fig.\,\ref{fig:tic324048193}). This is among the highest frequencies known in the roAp stars. This star is also at the upper temperature range of the known roAp stars, after TIC\,123231021 \citep[KIC\,7582608;][]{2014MNRAS.443.2049H}, but it is well within the theoretical instability strip for this class of pulsator \citep{2002MNRAS.333...47C}.

We classify this as a candidate roAp star since the signal is detected at marginal significance (S/N$=5.4$), and lacks signs of rotationally split sidelobes (which would be lost in the noise). Further TESS observations of this star are due, which will hopefully allow this star to be confirmed as a new roAp star. Since the contamination factor is small (0.0032) we are confident that the signal is from this star. 

\subsection{TIC\,410163387}

TIC\,410163387 (HD\,76279) was classified as an Ap\,SrEuCr star by \citet{1988mcts.book.....H}, although they note it as a `weak case'. There are differing \Teff\ values in the literature, with most in the region of 7100\,K \citep[e.g.,][]{2006ApJ...638.1004A,2019AJ....158...93B}. There has been one previous published attempt to find pulsations in this star, however the authors returned a null result at a magnitude limit of 0.8\,mmag in 2-hrs of photometric $B$ observations \citep{1993AJ....105.1903N}.

\begin{figure}
    \centering
    \includegraphics[width=\columnwidth]{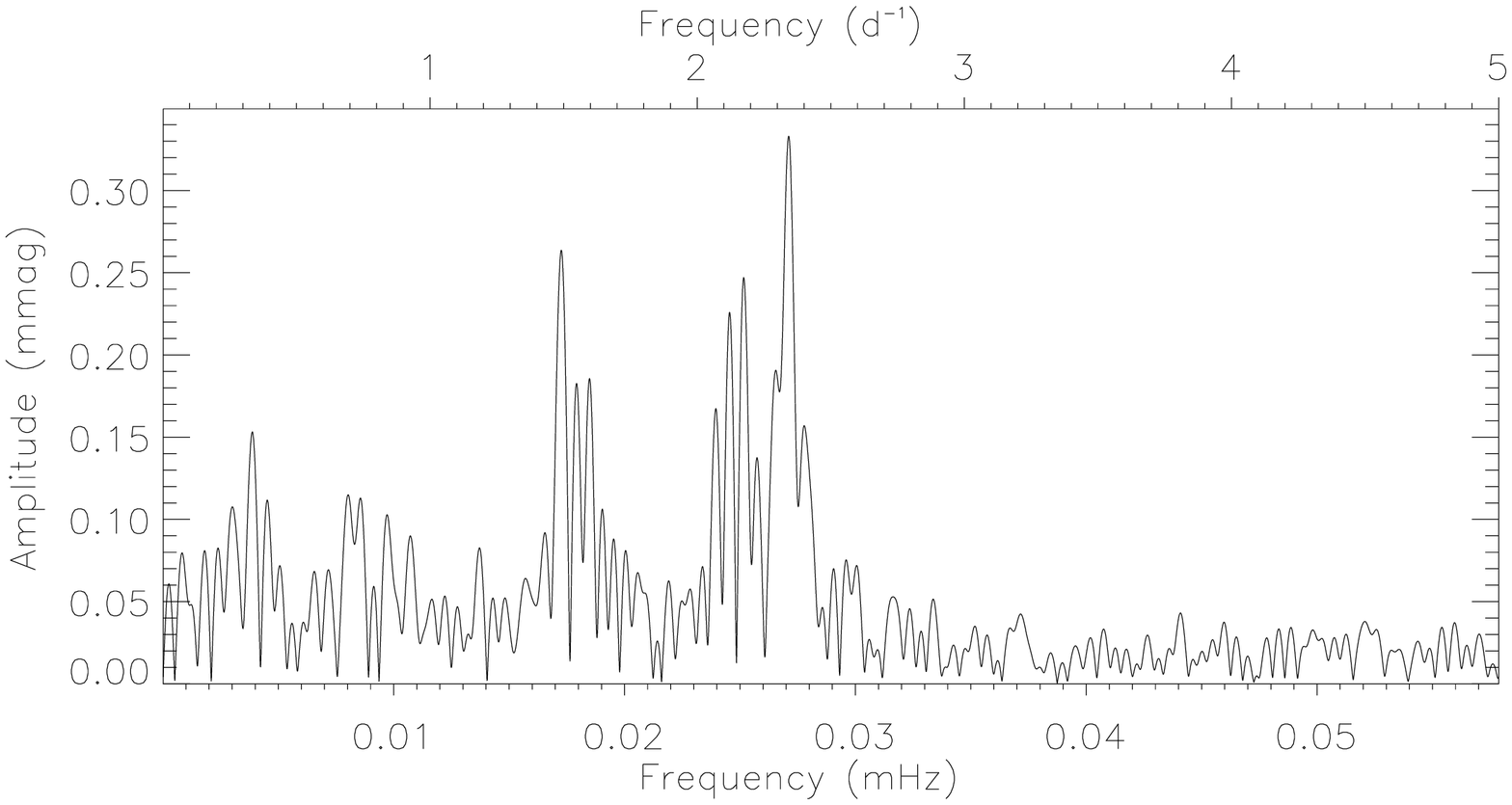}
    \includegraphics[width=\columnwidth]{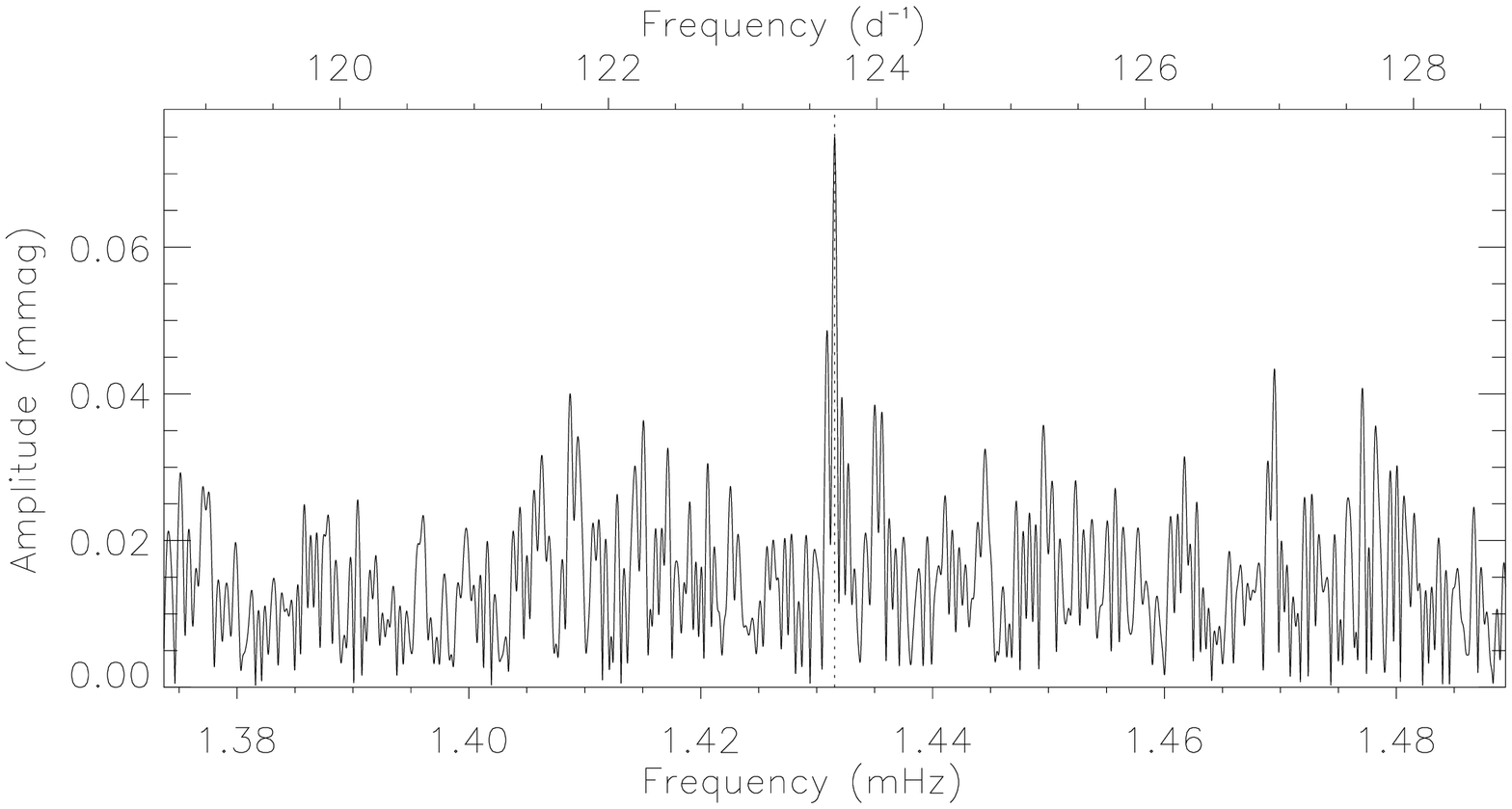}
    \caption{Top: suspected g-mode pulsations in TIC\,410163387. Bottom: suspected roAp pulsation mode in TIC\,410163387.}
    \label{fig:tic410163387}
\end{figure}

TESS observed TIC\,410163387 in sector 8. An amplitude spectrum of the light curve at low frequency (Fig.\,\ref{fig:tic410163387}) reveals the presence of probable g-mode pulsations in this star. At high frequency, we detect a single peak at $1.43155\pm0.00003$\,mHz ($123.686\pm0.003$\,\cd) which has a S/N of $7.0$. We cannot conclude, with confidence, that this peak is a true pulsation mode, and, coupled with the presence of the g-modes, we rate this star as a candidate roAp star until further observations are made to confirm the existence of the mode. We also note that the TIC contamination factor for this star is non-negligible at 0.01, implying that some of the signals may be from a contaminant source.

These factors led us to obtain a high resolution spectrum of TIC\,410163387 with SALT/HRS. The single spectrum reveals this star to be a spectroscopic binary of two components of similar temperature given the relative depths of the H$_\alpha$ absorption line (Fig.\,\ref{fig:410163387_halpha}). We suspect that the two pulsation signals are therefore originating on two separate stars, but confirmation of the high frequency variability is still required.

\section{Conclusions}
\label{sec:conc}

We have presented the results of a systematic survey of hot ($>6000$\,K) stars observed during the first Cycle of TESS observations, covering the Southern Ecliptic hemisphere, in 2-min cadence. We selected over 50\,700 stars in this temperature range, and searched for pulsational variability with frequencies greater than 0.52\,mHz, resulting in over 6713 detections. After the removal of obvious contaminant stars, the final sample consisted of 163 stars for detailed frequency analysis. The main discoveries from this work can be summarised as follows:
\begin{enumerate}
\item We have reported the discovery of 12 new roAp stars. Six of these stars show multiple pulsation frequencies in their light curves, making them of utmost importance for asteroseismic modelling. In addition to these stars, another (TIC\,119327278) shows the presence of a $\delta$\,Sct mode, suggesting this star belongs to the rare $\delta$\,Sct-roAp hybrid stars. This star also shows the shortest rotation period among the class of roAp stars, at 1.085\,d. We have identified the roAp star with the lowest frequency pulsation, TIC\,356088697, which shows a single pulsation mode at $0.646584$\,mHz ($P=25.8\,$min), allowing us to test the lowest end of the radial orders excited in the roAp stars. Of the 12 new roAp stars, we provided the first measures of the rotation period for five, and find three to show no rotational variability. 

\item We have presented the results of TESS observations of a further 44 roAp stars that were known in the literature. Of these 44, we find three (TIC\,35905913 -- HD\,132205, TIC\,93522454 -- HD\,143487, and TIC\,173372645 -- HD\,154708) that show no clear pulsation signal in the high-precision TESS data and note that these three stars were discovered to have low amplitude pulsations through the analysis of time-resolved high resolution spectroscopic observations. Furthermore, we present the tentative detection of modes in 2 other roAp stars which were discovered through spectroscopic observations.

\item Of the entire sample of 56 confirmed roAp stars observed, 68\,per\,cent (38) show rotationally modulated light curves, with rotation periods in the range $1.085-21 (\rm{or}\ 42)$\,d. It is not yet clear if there is a causal link between rotation and pulsation, but these observations will provide new results to test this relationship.

\item There are many instances where the rotationally split multiplets show different structures, in the same star, for the same degree mode (e.g., TIC\,176516923 -- HD\,38823, TIC\,310817678 -- HD\,88507, TIC\,259587315 -- HD\,30849). This provides a problem for the application of the oblique pulsator model where the stellar inclination angle, $i$, and the magnetic obliquity, $\beta$, can be derived or constrained. This, coupled with the difference in structure between observations made in different filters \citep[e.g., TIC\,326185137 -- HD\,6532;][]{2020ASSP...57..313K}, and spectral lines \citep[e.g., TIC\,280198016 -- HD\,83368][]{2006A&A...446.1051K}, provides a significant challenge in understanding the results of the oblique pulsator model.

\item We have reported the presence of non-linear interactions in the pulsation spectrum of TIC\,402546736 (HD\,128898) around the harmonic of the principal mode. This is only the second star \citep[after 33\,Lib;][]{2018MNRAS.480.2976H} where this phenomenon is seen.

\end{enumerate}
The results presented here have provided new insights into the roAp class of pulsationally variable stars. We have identified new pulsation frequencies in many stars, and shown frequency variability is a common feature of the roAp stars. These observational results will drive forward the development of state-of-the-art theoretical models of these stars, providing insight into the physical processes at play in these magnetic, chemically peculiar stars. Puzzling frequency separations, high-frequency modes, and a lack of roAp stars probing the blue edge of the instability strip, mean these results will have a significant impact on the theoretical understanding of the rapidly oscillating Ap stars. We will use these results to conduct a homogeneous study of the properties of roAp stars and explore any correlations between stellar and pulsation properties, and investigate any differences between the observed frequencies and those predicted by models.

\section*{Acknowledgements}

We thank the anonymous referee for a careful reading of the manuscript. DLH acknowledges financial support from the Science and Technology Facilities Council (STFC) via grant ST/M000877/1. MSC acknowledges the support by FCT/MCTES through the research grants UIDB/04434/2020, UIDP/04434/2020 and PTDC/FIS-AST/30389/2017, and by FEDER - Fundo Europeu de Desenvolvimento Regional through COMPETE2020 - Programa Operacional Competitividade e Internacionalização (grant: POCI-01-0145-FEDER-030389). MSC is supported by national funds through FCT in the form of a work contract. VA was supported by a research grant (00028173) from VILLUM FONDEN. Funding for the Stellar Astrophysics Centre is provided by The Danish National Research Foundation (Grant agreement no.: DNRF106). The research leading to these results has received funding from the Research Foundation Flanders (FWO) by means of a senior postdoctoral fellowship to DMB with grant agreement No. 1286521N. This project has been supported by the Lend\"ulet Program  of the Hungarian Academy of Sciences, project No. LP2018-7/2020 and by the EU's MW-Gaia COST Action (CA18104). ASG acknowledges financial support from the Max Planck Society under grant ``Preparations for PLATO science'' and from ALMA-CONICYT under grant \#31170029. ZsB acknowledges the support by the J\'anos Bolyai Research Scholarship of the Hungarian Academy of Sciences. LFM acknowledges the financial support from the UNAM under grant
PAPIIT IN100918. CCL acknowledges the support of the Natural Sciences and Engineering Research Council of Canada (NSERC). DM acknowledges the support of the National Astronomical Research Institute of Thailand (NARIT). JPG acknowledges funding support from Spanish public funds for research from project PID2019-107061GB-C63 from the `Programas Estatales de Generaci\'on de Conocimiento y Fortalecimiento Cient\'ifico y Tecnol\'ogico del Sistema de I+D+i y de I+D+i Orientada a los Retos de la Sociedad', and from the State Agency for Research through the ``Center of Excellence Severo Ochoa'' award to the Instituto de Astrof\'isica de Andaluc\'ia (SEV-2017-0709), all from the Spanish Ministry of Science, Innovation and Universities (MCIU). TRY acknowledges support from the NSF REU program, grant number PHY-1359195. The material is based upon work supported by NASA under award number 80GSFC21M0002. AD was supported by the \'UNKP-20-5 New National Excellence Program of the Ministry of Human Capacities and the J\'anos Bolyai Research Scholarship of the Hungarian Academy of Sciences. AD would like to thank the City of Szombathely for support under Agreement No. 67.177-21/2016. AGH acknowledges support from ``FEDER/Junta de Andaluc\'{\i}a-Consejer\'{\i}a de Econom\'{\i}a y Conocimiento'' under project E-FQM-041-UGR18 by Universidad de Granada and from Spanish public funds (including FEDER funds) for research under project ESP2017-87676-C5-2-R. ARB acknowledges funding support from Spanish public funds for research through the research grant PRE2018-084322, through projects ESP2017-87676-C5-5-R and PID2019-107061GB-C63, and from the State Agency for Research through the ``Center of Excellence Severo Ochoa'' award to the Instituto de Astrof\'isica de Andaluc\'ia (SEV-2017-0709), all from the Spanish Ministry of Science, Innovation and Universities (MCIU). We are grateful to Sowgata Chowdhury, supported by
Polish NCN grant 2015/18/A/ST9/00578, for conducting some spectroscopic observations that are reported in this work. This paper includes data collected by the TESS mission, which are publicly available from the Mikulski Archive for Space Telescopes (MAST). Funding for the TESS mission is provided by NASA’s Science Mission directorate. Funding for the TESS Asteroseismic Science Operations Centre is provided by the Danish National Research Foundation (Grant agreement no.: DNRF106), ESA PRODEX (PEA 4000119301) and Stellar Astrophysics Centre (SAC) at Aarhus University. We thank the TESS team and staff and TASC/TASOC for their support of the present work. This research has made use of the SIMBAD database, operated at CDS, Strasbourg, France. This work has made use of data from the European Space Agency (ESA) mission {\it Gaia} (\url{https://www.cosmos.esa.int/gaia}), processed by the {\it Gaia} Data Processing and Analysis Consortium (DPAC, \url{https://www.cosmos.esa.int/web/gaia/dpac/consortium}). Funding for the DPAC has been provided by national institutions, in particular the institutions participating in the {\it Gaia} Multilateral Agreement.

\section*{Data availability}
Data used in this work are available upon reasonable request from the corresponding author.



\bibliographystyle{mnras}
\bibliography{references} 



\appendix

\section{Author Affiliations}
\label{app:affiliations}
$^{1}$Jeremiah Horrocks Institute, University of Central Lancashire, Preston PR1 2HE, UK\\
$^{2}$Instituto de Astrof\'isica e Ci\^encias do Espa\c co, Universidade do Porto CAUP, Rua das Estrelas, PT4150-762 Porto, Portugal\\
$^{3}$Centre for Space Research, Physics Department, North West University, Mahikeng, 2745, South Africa\\
$^{4}$DTU Space, National Space Institute, Technical University of Denmark, Elektrovej 328, 2800 Kgs., Lyngby, Denmark\\
$^{5}$Stellar Astrophysics Centre, Aarhus University, DK-8000 Aarhus C, Denmark\\
$^{6}$School of Physics, Sydney Institute for Astronomy (SIfA), The University of Sydney, NSW 2006, Australia\\
$^{7}$Institute of Astronomy, KU Leuven, Celestijnenlaan 200D, 3001 Leuven, Belgium\\
$^{8}$D\'epartement de Physique et d'Astronomie, Universit\'e de Moncton, 18 avenue Antonine-Maillet, Moncton, NB E1A 3E9, Canada\\
$^{9}$Department of Chemistry and Physics, Florida Gulf Coast University, 10501 FGCU Blvd. S., Fort Myers, FL 33965 USA\\
$^{10}$Department of Physics and Astronomy, Uppsala University, Box 516, 75120 Uppsala, Sweden\\
$^{11}$Astronomical Institute of University of Wroc\l{}aw, ul. Kopernika 11, 51-622 Wroc\l{}aw, Poland\\
$^{12}$Ankara University, Faculty of Science, Dept. of Astronomy and Space Sciences, 06100, Tandogan - Ankara / Turkey\\
$^{13}$Department of Astronomy, School of Physics, Peking University, Beijing 100871, People's Republic of China\\
$^{14}$Kavli Institute for Astronomy and Astrophysics, Peking University, Beijing 100871, People's Republic of China\\
$^{15}$Konkoly Observatory, Research Centre for Astronomy and Earth Sciences, ELKH, H-1121 Konkoly Thege Mikl\'os \'ut 15-17, Budapest, Hungary\\
$^{16}$MTA CSFK Lend\"ulet Near-Field Cosmology Research Group\\
$^{17}$ELTE E\"otv\"os Lor\'and University, Institute of Physics, Budapest, Hungary\\
$^{18}$Max-Planck-Institut f\"ur Sonnensystemforschung, Justus-von-Liebig-Weg 3, 37077 G\"ottingen, Germany\\
$^{19}$N\'ucleo de Astronom\'ia, Facultad de Ingenier\'ia y Ciencias, Universidad Diego Portales, Av. Ej\'ercito Libertador 441, Santiago, Chile\\
$^{20}$Instituto de Astronom\'{\i}a--Universidad Nacional Aut\'onoma de M\'exico, Ap. P. 877, Ensenada, BC 22860, Mexico\\
$^{21}$Instituto de Astrof\'isica de Andaluc\'ia~(CSIC), Glorieta de la Astronom\'ia s/n, E-18008 Granada, Spain\\
$^{22}$Physics Department, Mount Allison University, Sackville, NB, Canada, E4L1E6\\
$^{23}$National Astronomical Research Institue of Thailand, 260 Moo 4, T. Donkaew, A. Maerim, Chiangmai, 50180 Thailand\\
$^{24}$Department of Theoretical Physics and Astrophysics, Masaryk University, Kotl\'a\v rsk\'a 2, 611 37 Brno, Czech Republic\\
$^{25}$School of Earth and Space Exploration, Arizona State University, Tempe, AZ 85281, USA\\
$^{26}$Physics and Astronomy Department, Bishop's University, 2600 College Street, Sherbrooke, QC J1M 1Z7, Canada\\
$^{27}$School of Physics, Xi'an Jiaotong University, 710049 Xi'an, People's Republic of China\\
$^{28}$Department of Physics, University of York, Heslington, York, YO10 5DD, UK\\
$^{29}$School of Physical and Chemical Sciences, Te Kura Mat\={u}, University of Canterbury, Christchurch 8020, New Zealand\\
$^{30}$Department of Physics and Astronomy, Howard University, Washington, DC 20059, USA\\
$^{31}$Center for Research and Exploration in Space Science and Technology, and X-ray Astrophysics Laboratory, NASA/GSFC, Greenbelt, MD 20771, USA\\
$^{32}$ELTE E\"otv\"os Lor\'and University, Gothard Astrophysical Observatory, Szombathely, Hungary\\
$^{33}$MTA-ELTE Exoplanet Research Group, 9700 Szombathely, Szent Imre h. u. 112, Hungary\\
$^{34}$Dept. F\'\i sica Te\'orica y del Cosmos, Universidad de Granada, Campus de Fuentenueva s/n, E-18071 Granada, Spain\\
$^{35}$Los Alamos National Laboratory, MS T-082, Los Alamos, NM 87545 USA\\
$^{36}$Department of Astronomy, University of Cape Town, Private Bag X3, Rondebosch 7701, South Africa\\
$^{37}$N\'ucleo de Astronom\'\i a de la Facultad de Ingenier\'\i a y Ciencias, Universidad Diego Portales, Av. Ej\'ercito Libertador 441, Santiago, Chile\\
$^{38}$Koninklijke Sterrenwacht van Belgi\"e, Ringlaan 3, 1180 Brussel, Belgium\\
$^{38}$LESIA, UMR 8109, Observatoire de Paris et Universit\'e Pierre et Marie Curie Sorbonne Universit\'es, place J. Janssen, Meudon, France\\
$^{40}$Departamento de F\'isica, Universidad T\'ecnica Federico Santa Mar\'ia, Avda. Espa\~{n}a  1680, Valpara\'iso, Chile\\
$^{41}$Astrophysics Group, Keele University, Staffordshire ST5 5BG, UK\\
$^{42}$Institute of Astronomy and NAO, Bulgarian Academy of Sciences, blvd. Tsarigradsko chaussee 72, Sofia 1784, Bulgaria\\
$^{43}$Department of Physics and Kavli Institute for Astrophysics and Space Research, Massachusetts Institute of Technology, Cambridge, MA 02139, USA

\section{Spectral classification plots}
Here we provide the spectral classification plots used to provide new or updated spectral classifications. We have used the SpUpNIC long slit spectrograph mounted on the SAAO 1.9-m telescope to record these spectra. We used grating 4 with a grating angle of 4.2$^\circ$, with observations typically made with a 1\,arcsec slit to reach a resolving power of $R\sim3000$, but was widened in poor seeing or low sky transparency situations. The data were flat field and bias corrected, cleaned of cosmic rays, and had a wavelength solution applied using a CuAr reference spectrum obtained directly after the science observation. The spectra were normalised using an automated spline fitting procedure in i{\sc{spec}} spectral analysis software \citep{2014A&A...569A.111B,2019MNRAS.486.2075B}.
We also provide, in Table\,\ref{tab:spec_obs}, a log of all spectroscopic observations related to this work.

\begin{table}
    \caption{Log of the spectroscopic observations collected for stars presented in this paper. The resolution refers to the canonical resolving power of the instrument in the configuration used. The final column shows the number of observations per instrument, with a `B' or `R' referring to the \#4 or \#6 grating of the SpUpNIC instrument. For full details see Section\,\ref{sec:spec}. The table is split into three sections to reflect Table\,\ref{tab:stars}, but with the known roAp stars prior to TESS launch section excluded.}
    \label{tab:spec_obs}
    \centering
    \begin{tabular}{lcrc}
    \hline
    \multicolumn{1}{c}{TIC}     & Spectrograph  & \multicolumn{1}{c}{Resolution} & Nobs\\
    \hline
119327278	&	HRS	    &	63000	&	1	\\
170586794	&	SpUpNIC &	7500    &	1B	\\
            &   HRS     &   63000   & 1     \\
176516923	&	SpUpNIC	&	7500	&	1B	\\
294266638	&	SpUpNIC	&	7500	&	1B	\\
310817678	&	HRS	    &	43000	&	4	\\
356088697	&	SpUpNIC	&	7500	&	1B	\\            
380651050	&	HRS	    &	63000	&	1	\\
466260580	&	HRS	    &	63000	&	1	\\
\hline
12968953	&   SpUpNIC	&	7500	&	1B	\\
            &   HRS	    &	63000	&	2	\\
17676722	&	SpUpNIC	&	7500/4700	&	2B,1R	\\
41259805	&	SpUpNIC &	7500	&	4B	\\
            &   HRS     &   63000   &   1   \\
49818005	&	SpUpNIC &	7500	&	3B	\\
            &   HRS     &   63000   & 1     \\
152808505	&	SpUpNIC	&	7500	&	1B	\\
            &   HRS     &   63000   & 1     \\
156886111	&	SpUpNIC	&	7500/4700	&	1B,1R	\\
            &   HRS     &   63000   & 1     \\
259587315	&	SpUpNIC	&	7500	&	3B	\\
            &   HRS     &   63000   & 1     \\            
349945078	&	SpUpNIC	&	7500/4700	&	2B,3R	\\
            &   HRS     &   63000   & 1     \\
350146296	&	SpUpNIC &	7500/4700	&	5B,1R	\\
            &   HRS     &   63000   & 2     \\
            &   HERCULES&   41000   & 6     \\
431380369 	&	SpUpNIC	&	7500	&	2B	\\
            &   HRS     &   63000   & 1     \\            
\hline
3814749	    &	SpUpNIC	&	7500	&	1B	\\ 
410163387	&	SpUpNIC	&	7500	&	1B	\\ 
            &   HRS     &   63000   & 1     \\
\hline
    \end{tabular}
\end{table}

\begin{figure}
    \centering
     \includegraphics[width=\columnwidth]{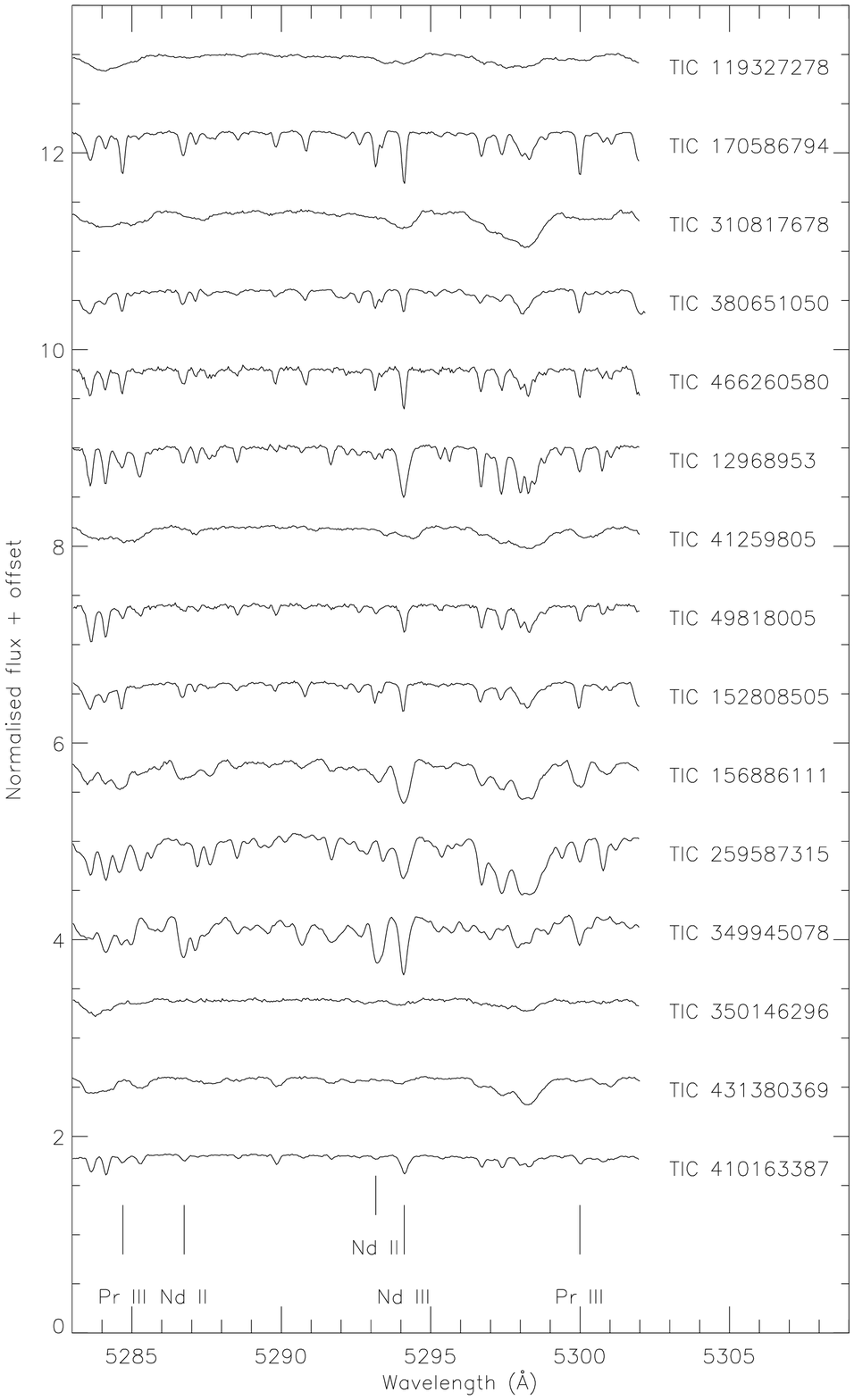}
    \caption{SALT/HRS spectra of new and candidate roAp stars, as listed in Table\,\ref{tab:spec_obs}, showing a region in the blue part of the spectrum of use in identifying chemical peculiarities.}
    \label{fig:all_HRS_B}
\end{figure}

\begin{figure}
    \centering
     \includegraphics[width=\columnwidth]{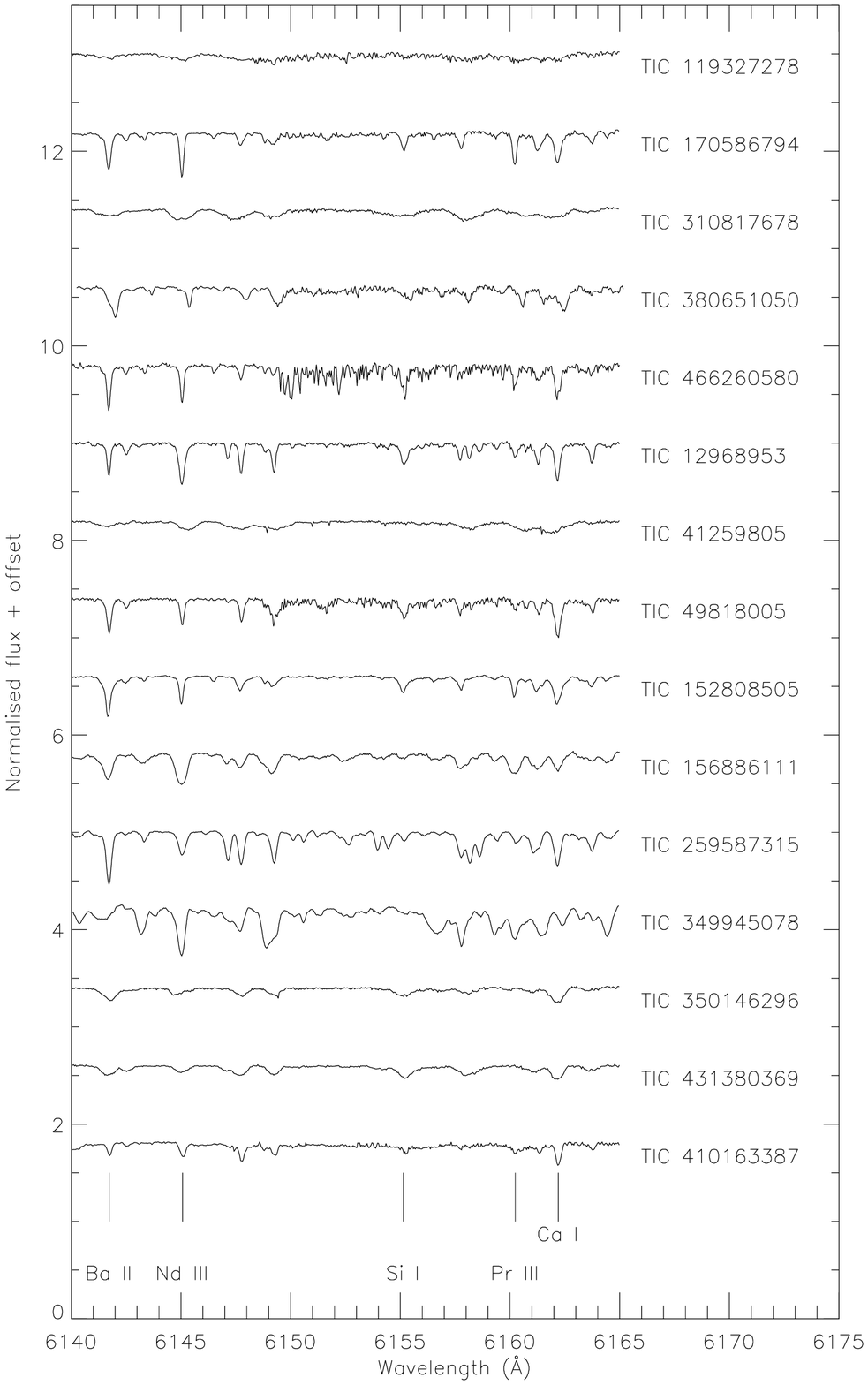}
    \caption{SALT/HRS spectra of new and candidate roAp stars, as listed in Table\,\ref{tab:spec_obs}, showing a region in the red part of the spectrum of use in identifying chemical peculiarities.}
    \label{fig:all_HRS_R}
\end{figure}

Where we have high resolution data available, we plot two regions of useful in identifying chemical peculiarities in Ap stars in Fig.\,\ref{fig:all_HRS_B} and Fig.\,\ref{fig:all_HRS_R}. 

\begin{figure}
    \centering
     \includegraphics[width=\columnwidth]{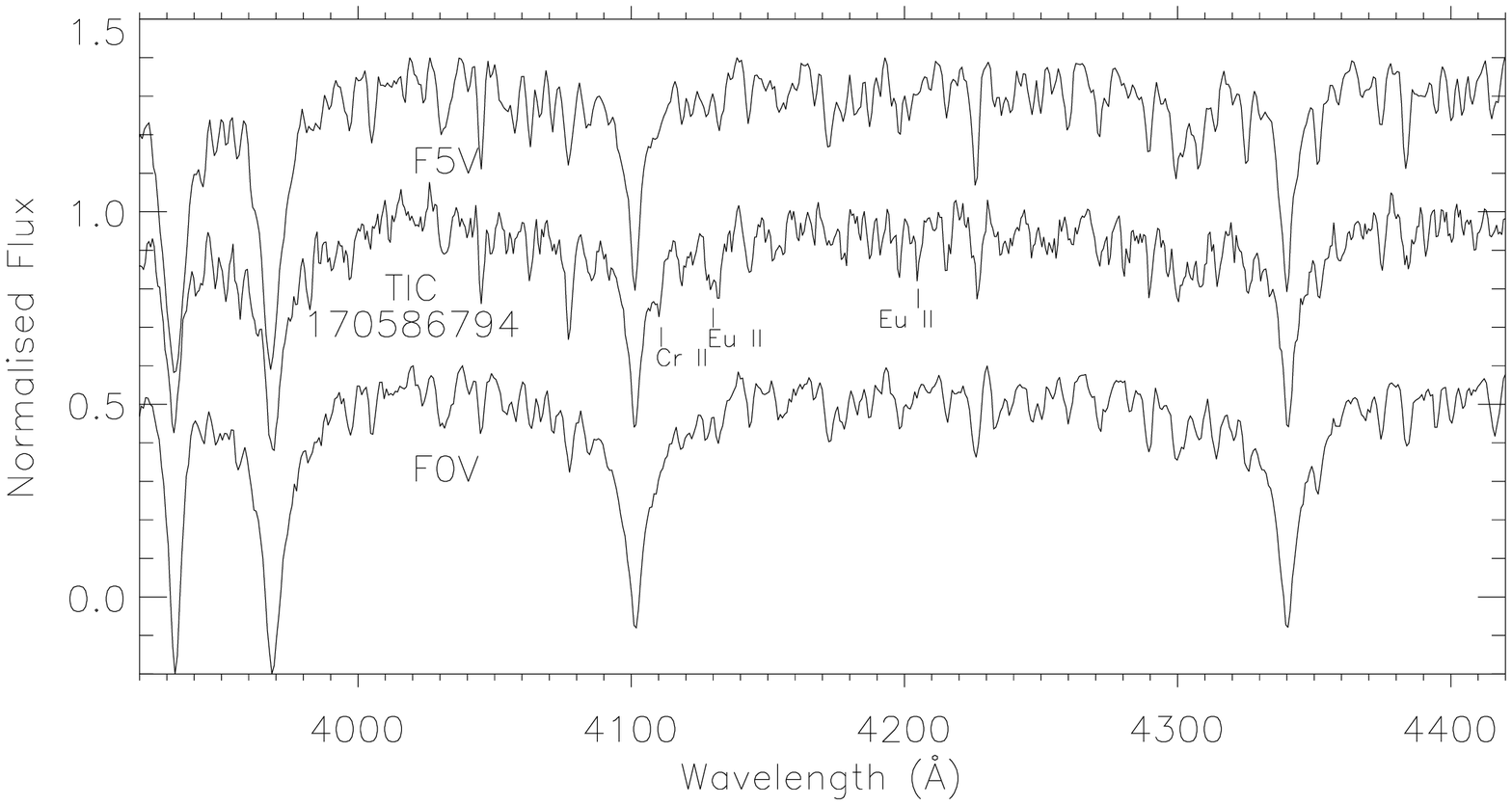}
    \caption{SAAO spectrum of TIC\,170586794 (centre), with stars of spectral class F5V (above) and F0V (below) for comparison. Lines of Eu and Cr are enhanced in this star, leading to the classification of F5p\,EuCr.}
    \label{fig:tic170586794spec}
\end{figure}

\begin{figure}
    \centering
     \includegraphics[width=\columnwidth]{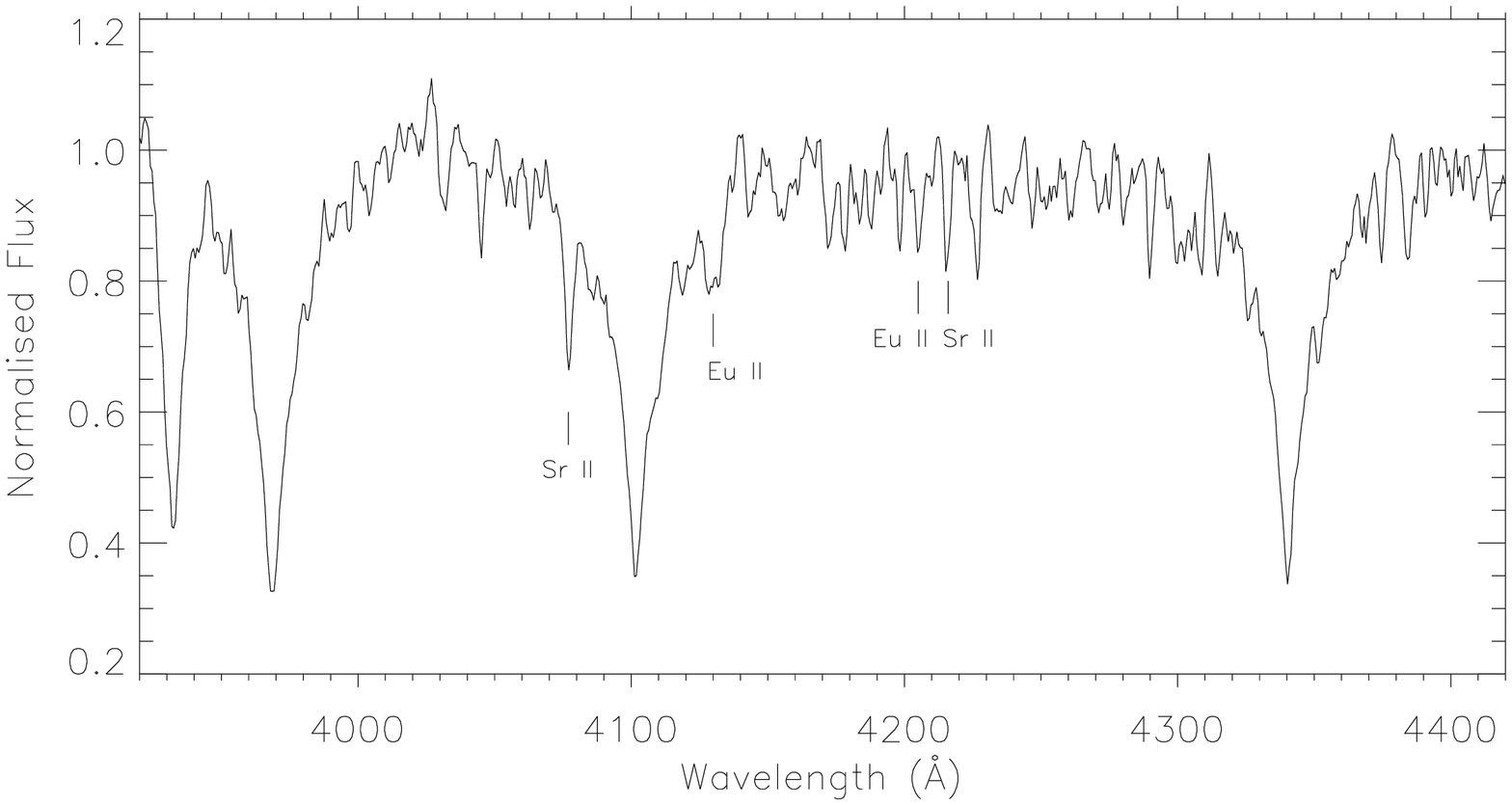}
    \caption{SAAO spectrum of TIC\,294266638. Lines of Sr and Eu are noted as being pronounced in this star, leading to the classification of A7p\,SrEu.}
    \label{fig:tic294266638_spec}
\end{figure}

\begin{figure}
    \centering
     \includegraphics[width=\columnwidth]{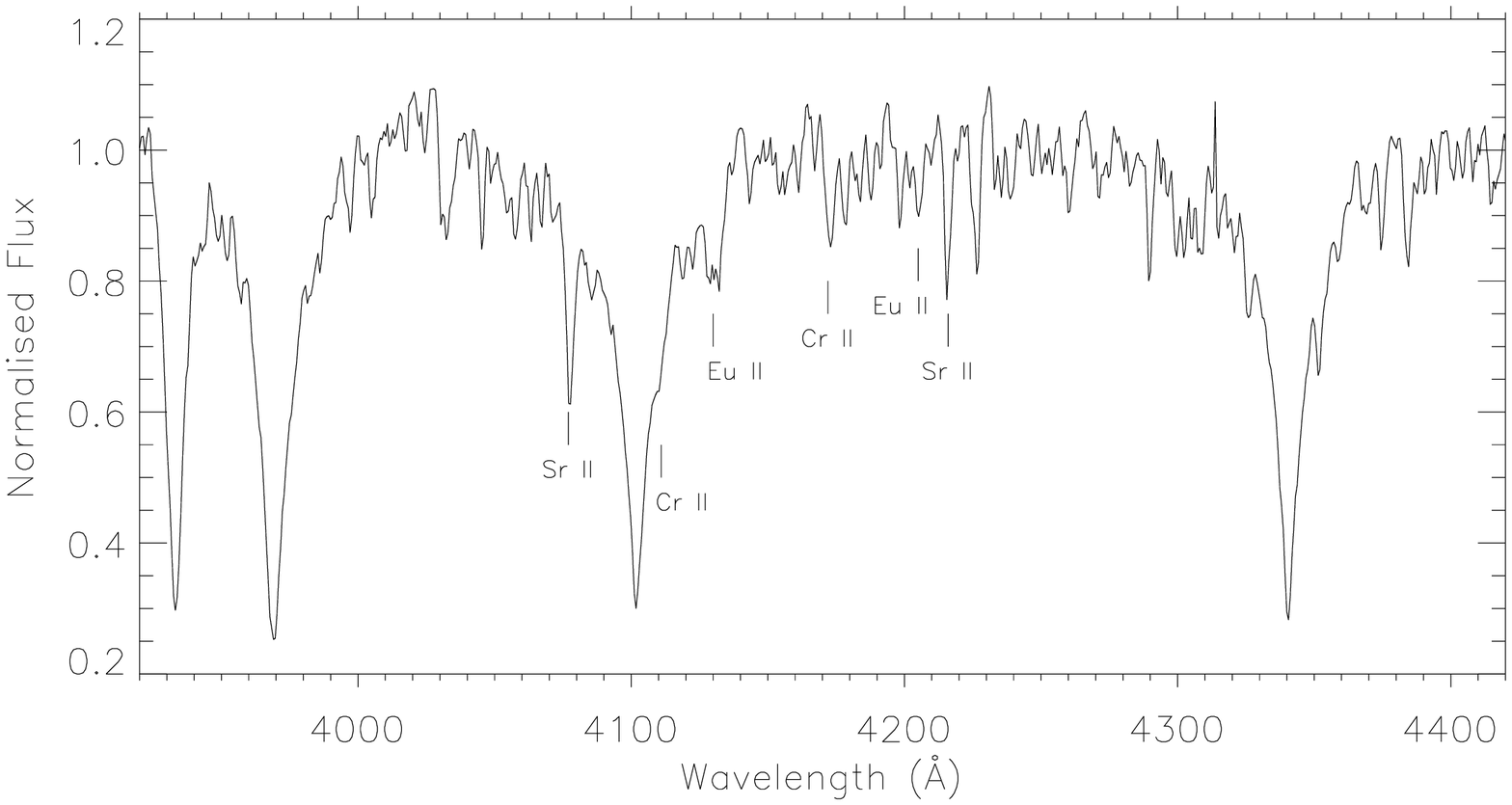}
    \caption{The SAAO spectrum of TIC\,356088697 with lines of Sr, Eu and Cr identified. We refine the spectral classification to A8p\,SrEuCr.}
    \label{fig:tic356088697spec}
\end{figure}

\begin{figure}
    \centering
     \includegraphics[width=\columnwidth]{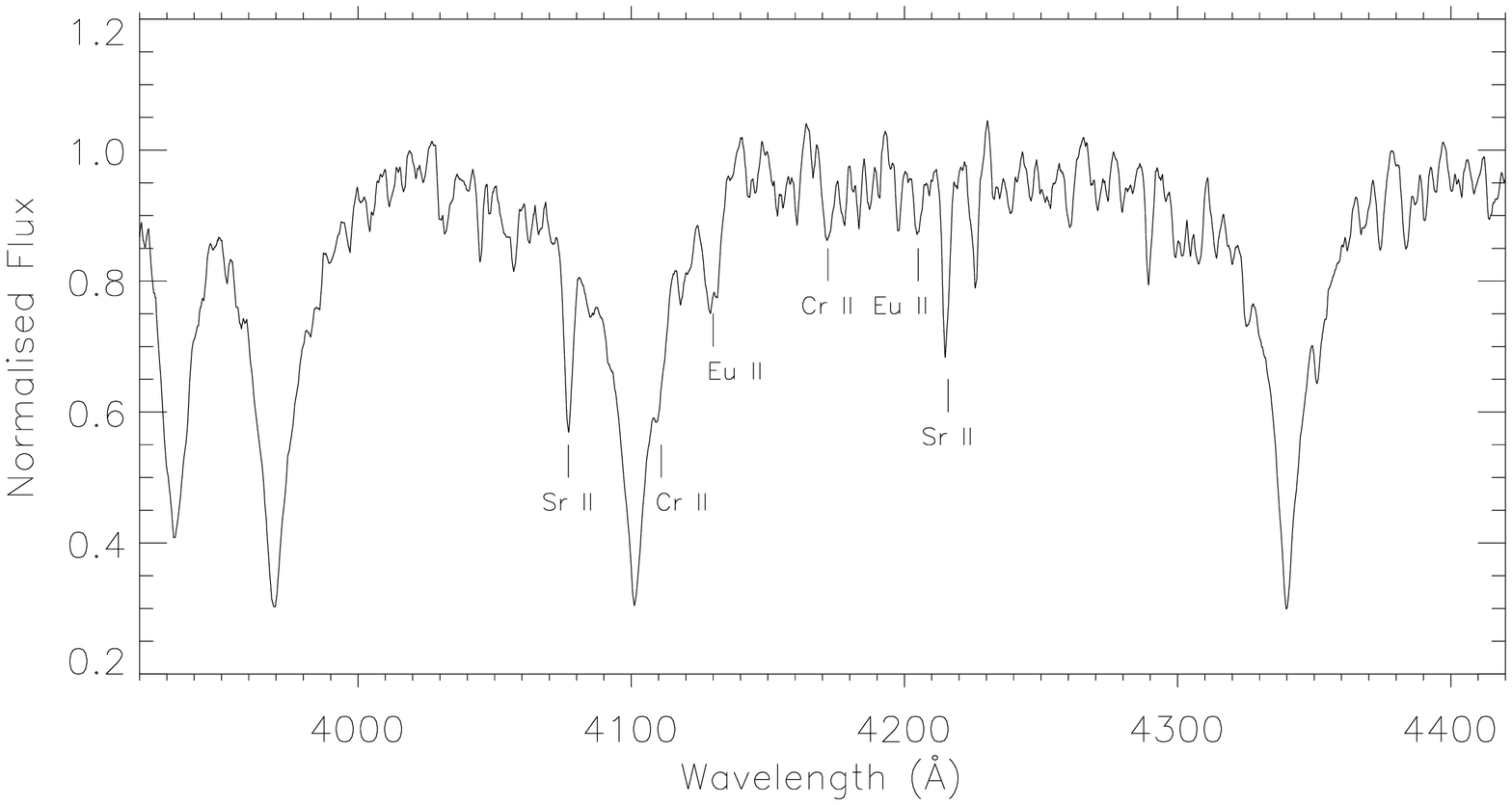}
    \caption{The SAAO spectrum of TIC\,12968953 with lines of Sr, Eu and Cr identified. We refine the spectral classification to A7p\,SrEuCr.}
    \label{fig:tic12968953spec}
\end{figure}

\begin{figure}
    \centering
     \includegraphics[width=\columnwidth]{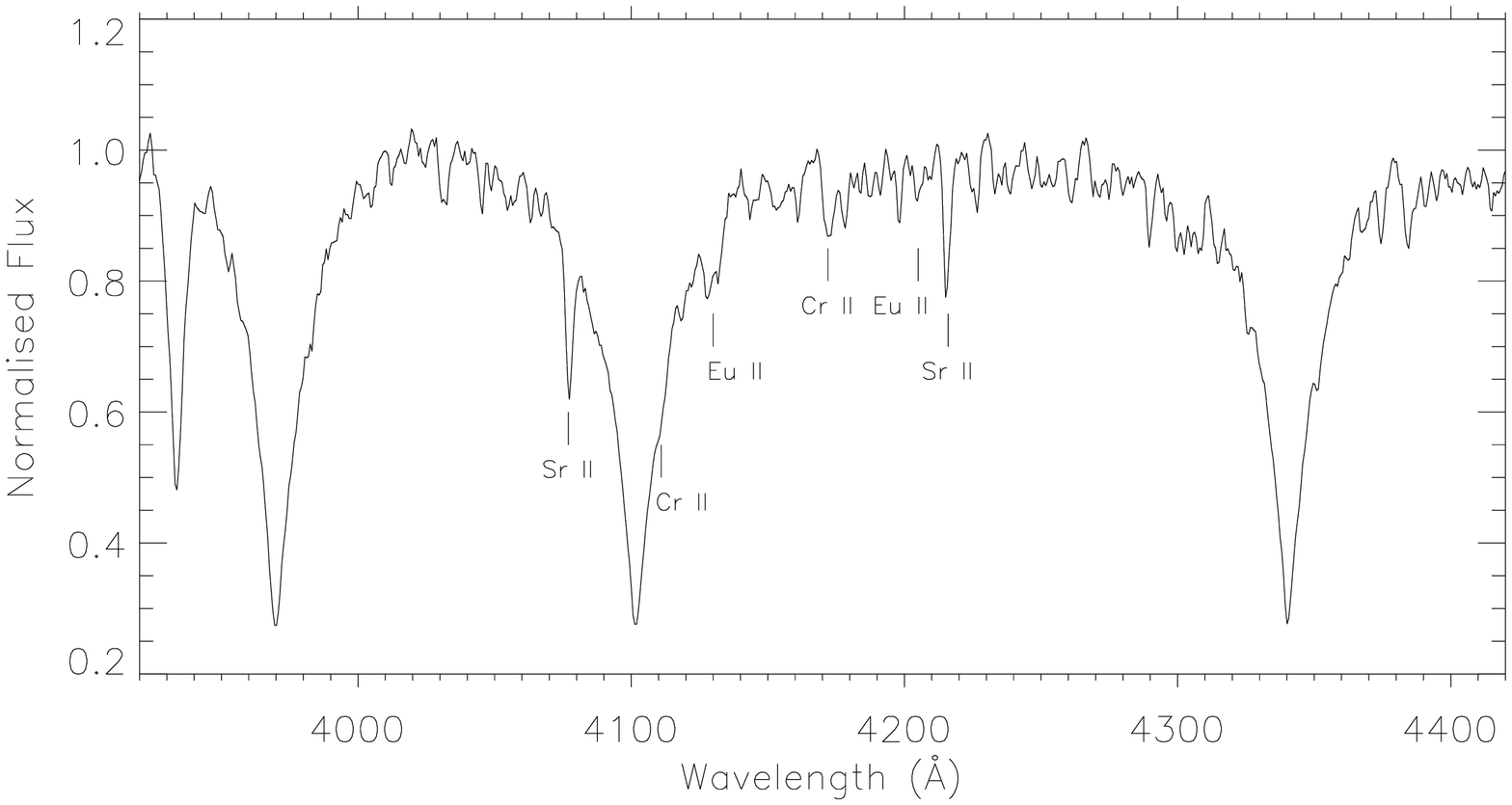}
    \caption{The SAAO spectrum of TIC\,17676722 with lines of Sr, Eu and Cr identified. We refine the spectral classification to A3p\,SrEuCr.}
    \label{fig:tic17676722spec}
\end{figure}

\begin{figure}
    \centering
     \includegraphics[width=\columnwidth]{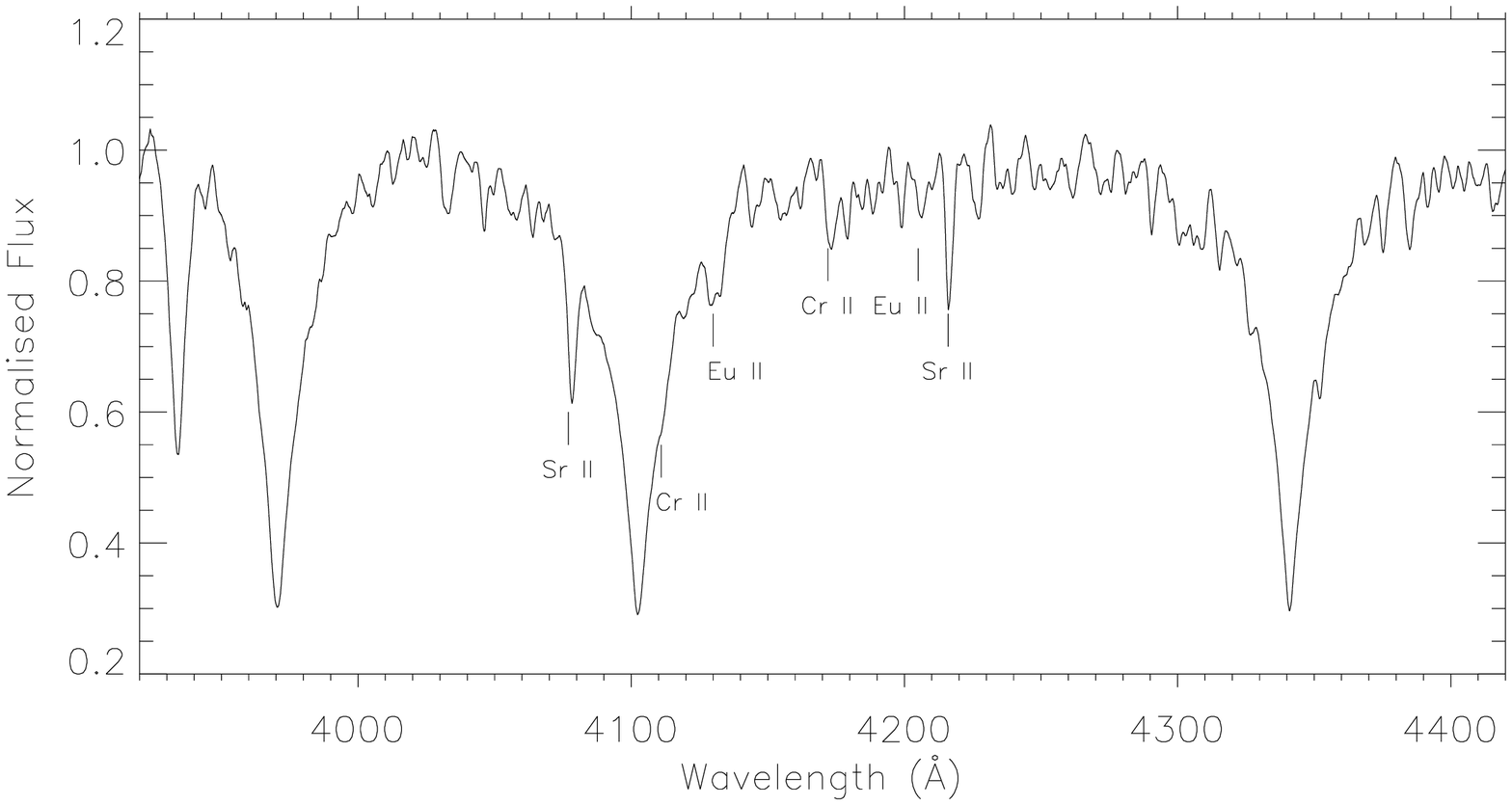}
    \caption{The SAAO spectrum of TIC\,41259805 with lines of Sr, Eu and Cr identified. We refine the spectral classification to A6p\,SrEu(Cr).}
    \label{fig:tic41259805spec}
\end{figure}

\begin{figure}
    \centering
     \includegraphics[width=\columnwidth]{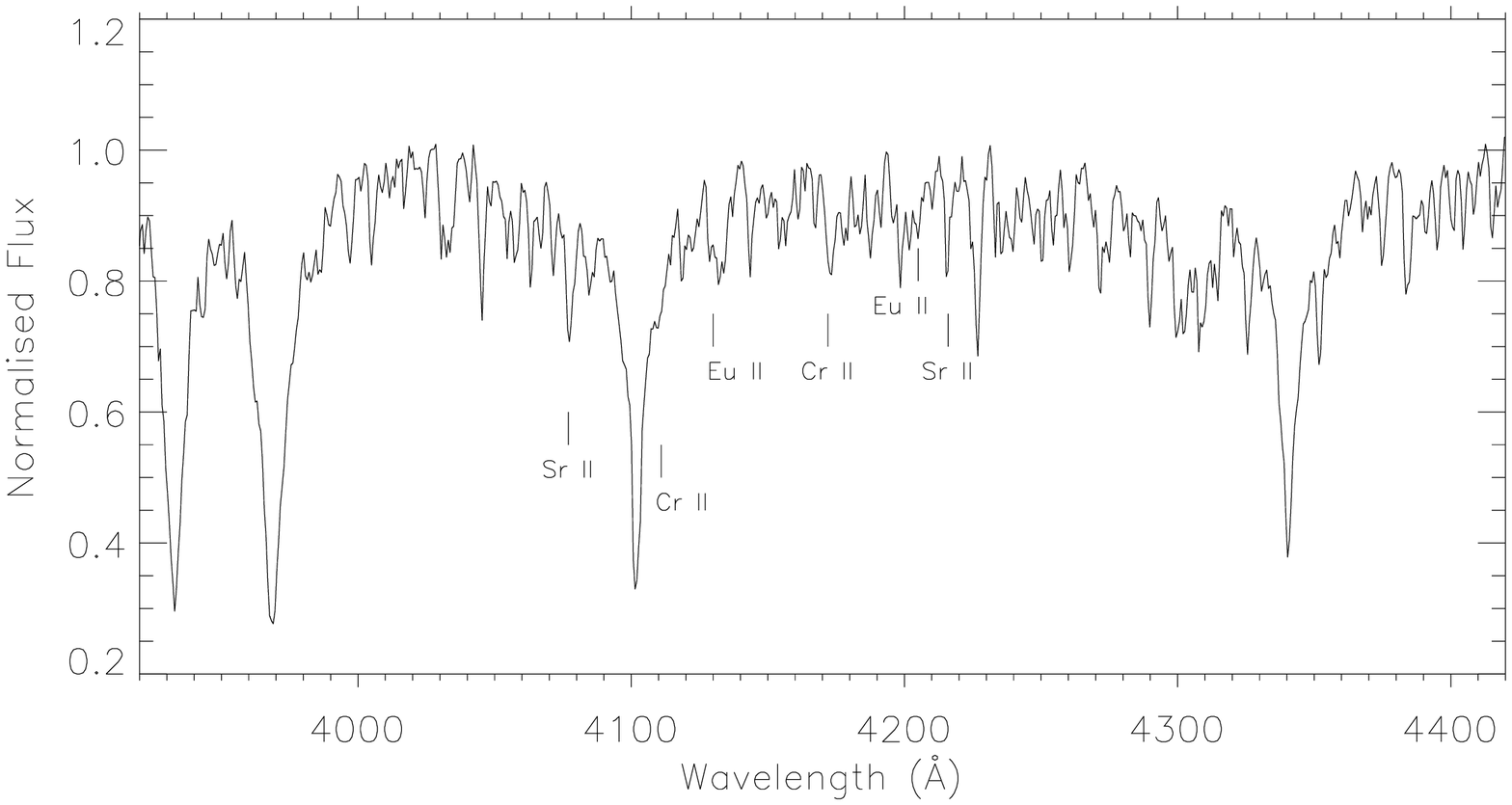}
    \caption{The SAAO spectrum of TIC\,49818005 with lines of Sr, Eu and Cr identified. We provide a spectral classification of F2p\,SrEu(Cr).}
    \label{fig:tic49818005spec}
\end{figure}

\begin{figure}
    \centering
     \includegraphics[width=\columnwidth]{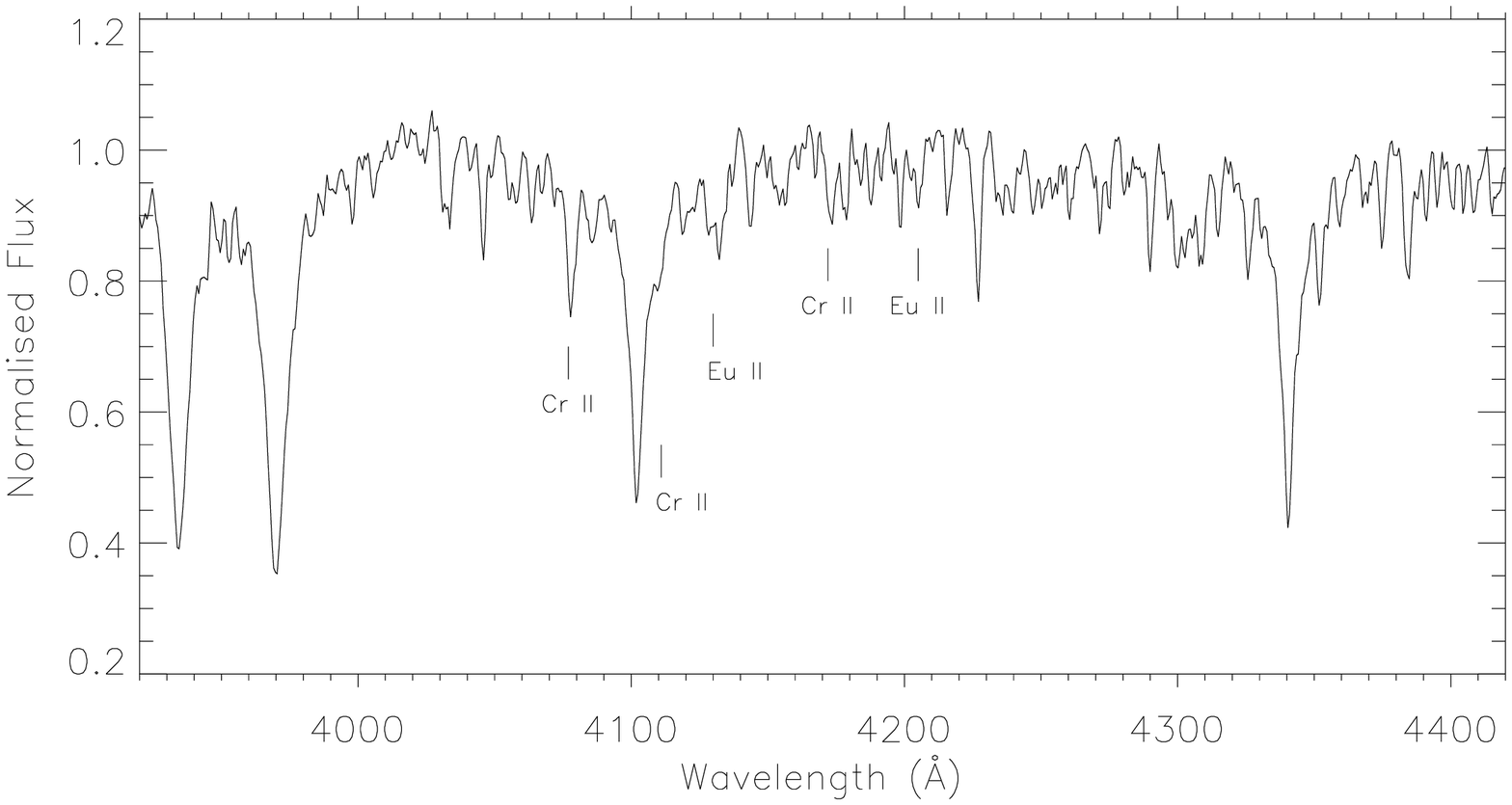}
    \caption{The SAAO spectrum of TIC\,152808505. We provide a spectral classification of F3p\,EuCr.}
    \label{fig:tic152808505spec}
\end{figure}

\begin{figure}
    \centering
     \includegraphics[width=\columnwidth]{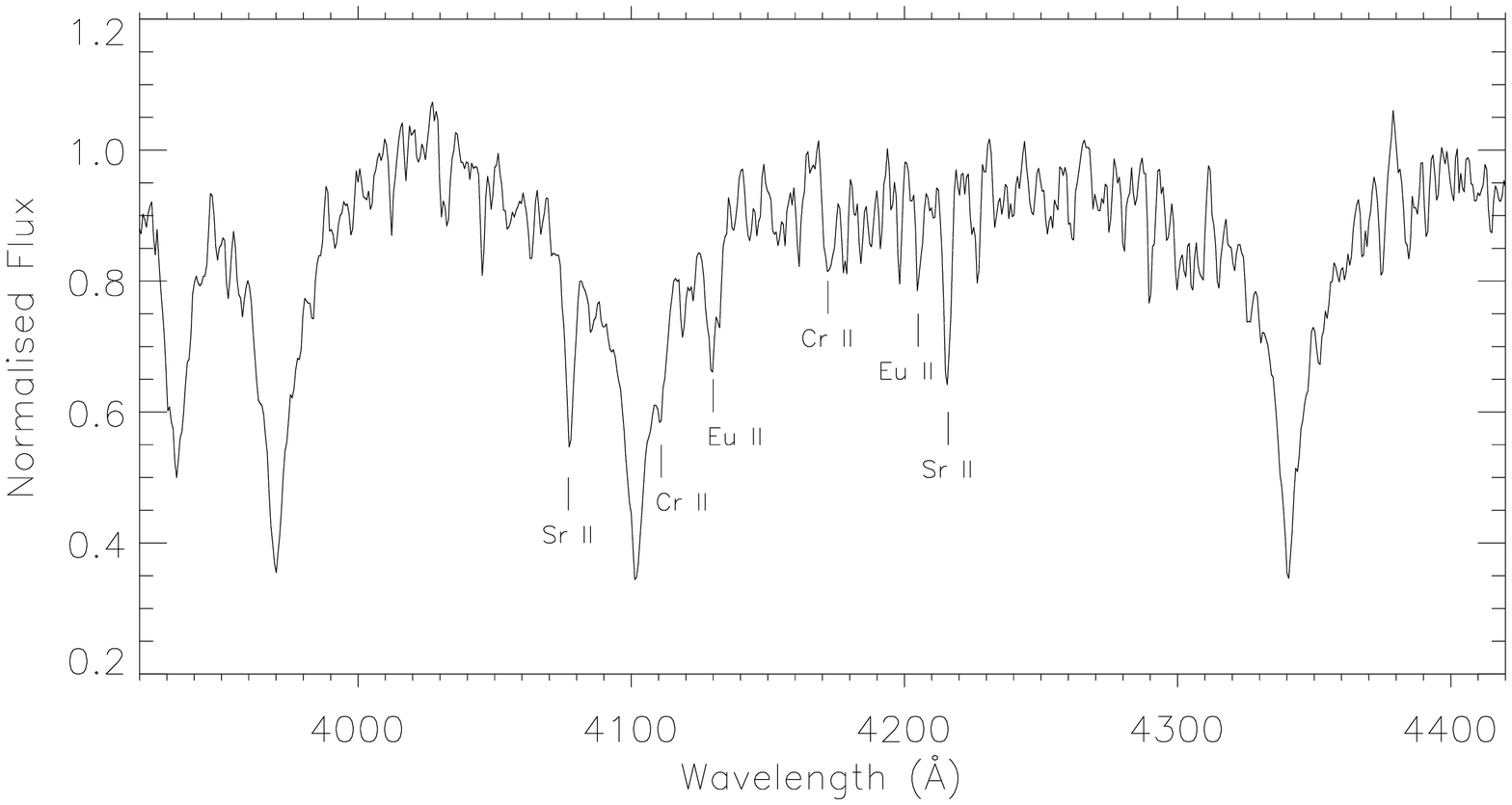}
    \caption{The SAAO spectrum of TIC\,156886111, from which we deduce a classification of A8p\,SrEuCr.}
    \label{fig:tic156886111spec}
\end{figure}

\begin{figure}
    \centering
     \includegraphics[width=\columnwidth]{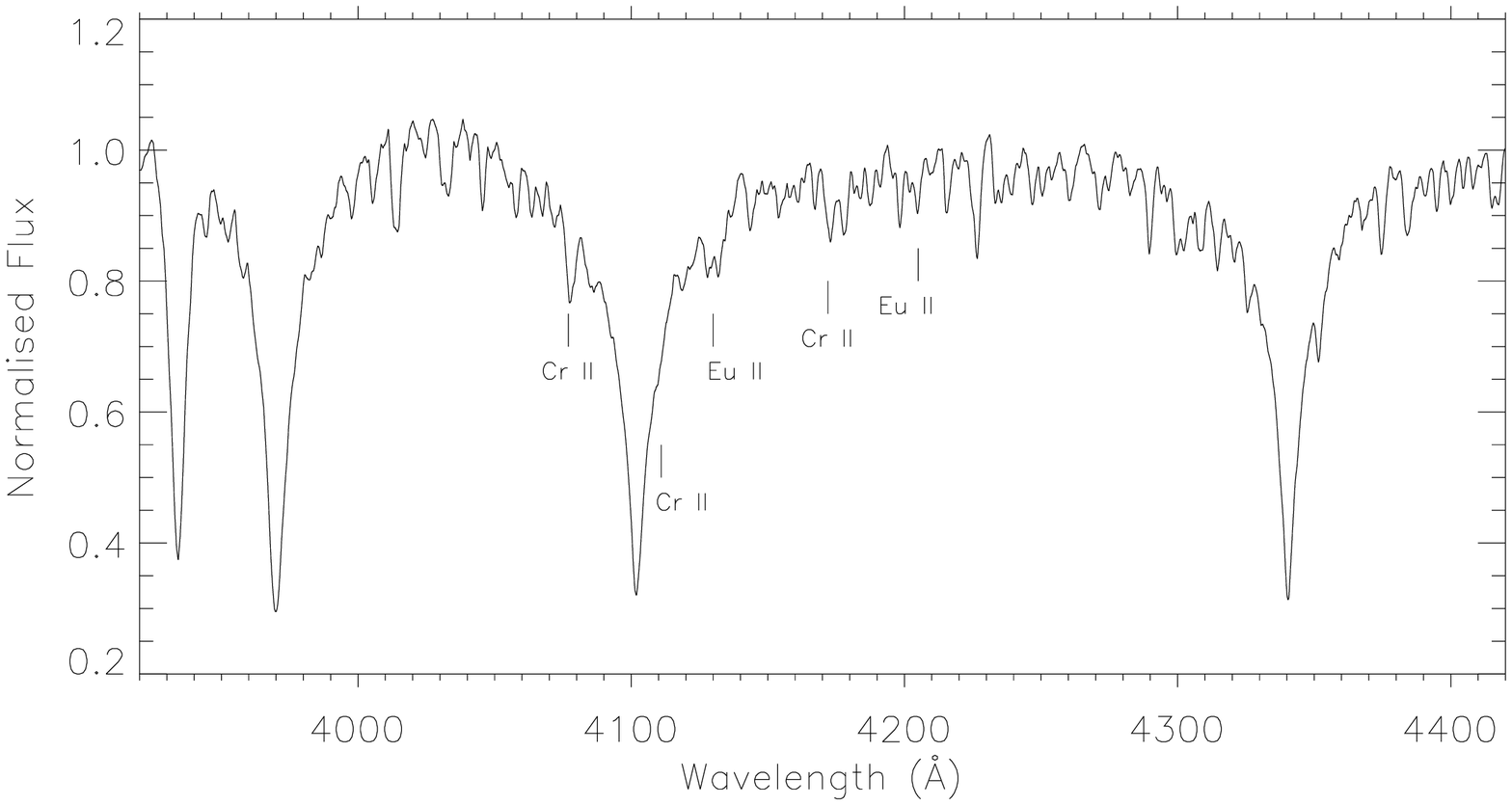}
    \caption{The SAAO spectrum of TIC\,350146296, from which we deduce a classification of F0p\,EuCr.}
    \label{fig:tic350146296spec}
\end{figure}

\begin{figure}
    \centering
     \includegraphics[width=\columnwidth]{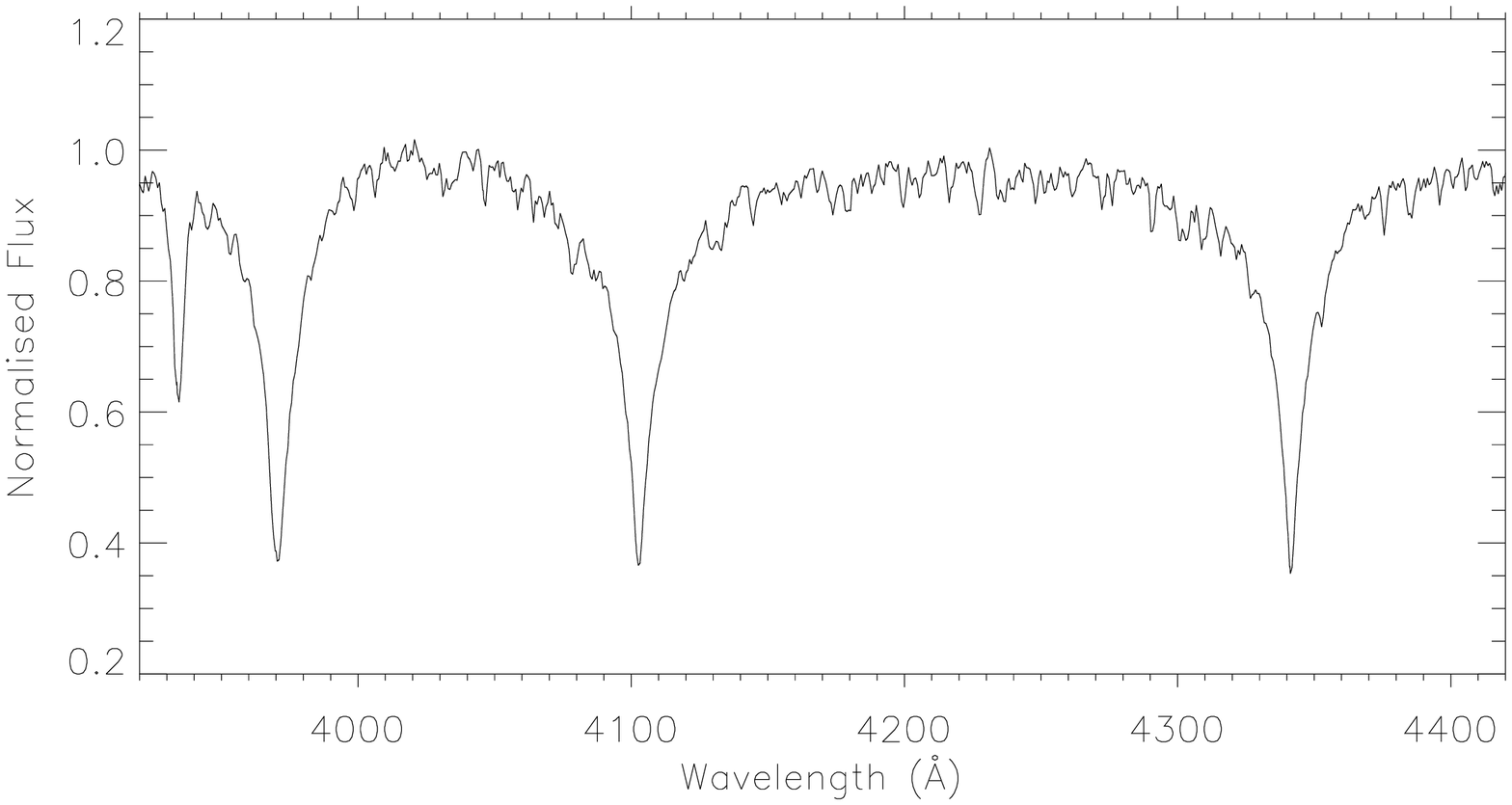}
    \caption{The SAAO spectrum of TIC\,3814749 showing the lack of strong chemical peculiarities in this star.}
    \label{fig:tic3814749spec}
\end{figure}

\begin{figure}
    \centering
     \includegraphics[width=\columnwidth]{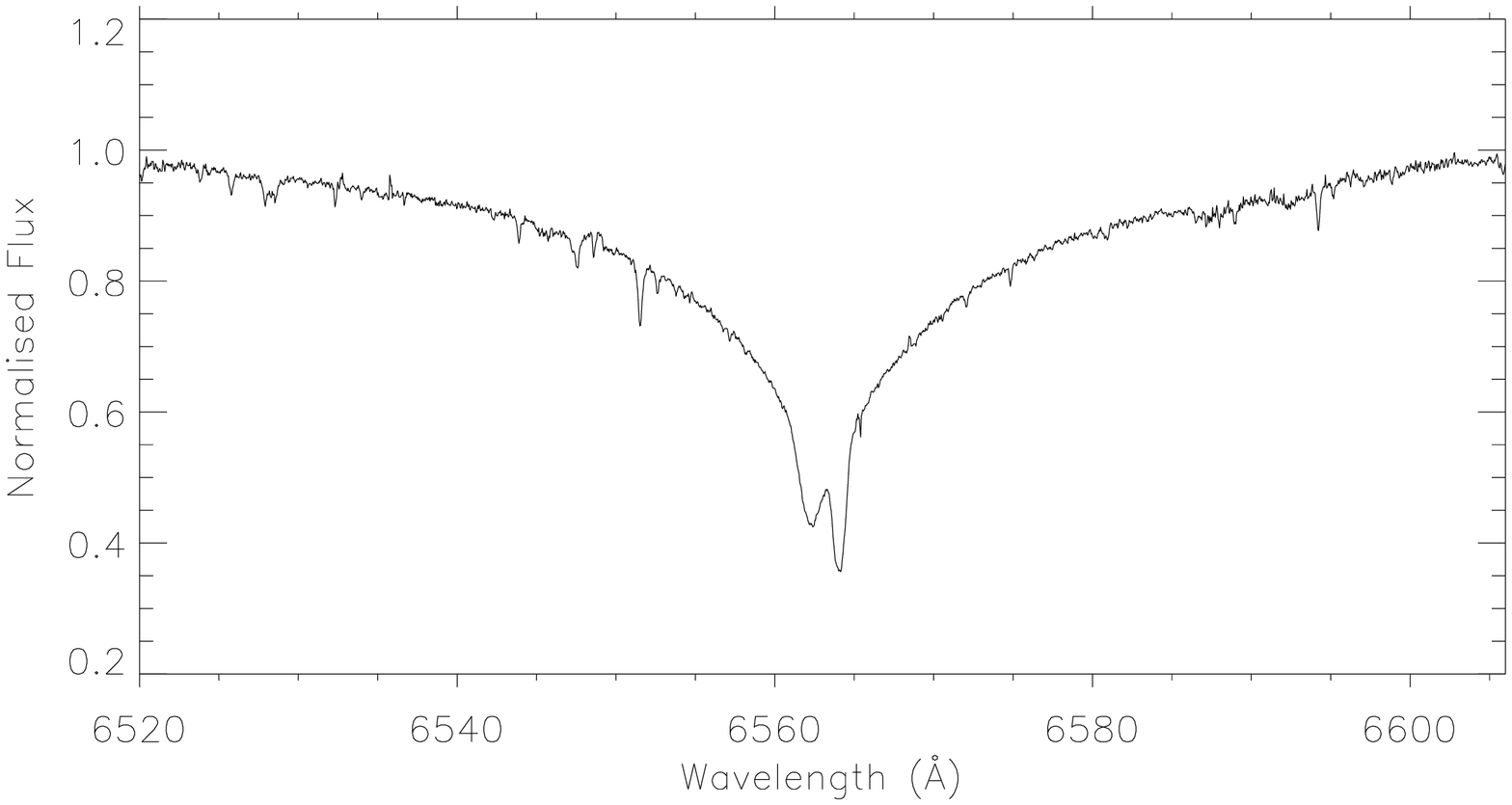}
    \caption{The SALT/HRS spectrum of the H$_\alpha$ line in TIC\,410163387 showing the star to be a spectroscopic binary system.}
    \label{fig:410163387_halpha}
\end{figure}

\bsp
\label{lastpage}
\end{document}